UNIVERSITY OF SOUTHAMPTON

# Modified Chalcogenide Glasses for Optical Device Applications

by

## Mark A. Hughes

A thesis submitted for the degree of Doctor of Philosophy

Faculty of Engineering, Science & Mathematics
Optoelectronics Research Centre

May 2007

MODIFIED CHALCOGENIDE GLASSES FOR OPTICAL DEVICE
APPLICATIONS

by Mark A. Hughes

This thesis focuses on two different, but complementary, aspects of the modification of gallium lanthanum sulphide (GLS) glasses. Firstly the addition of transition metal ions as dopants is examined and their potential for use as active optical materials is explored. It is also argued that the spectroscopic analysis of transition metal ions is a useful tool for evaluating the local environment of their host. Secondly femtosecond (fs) laser modification of GLS is investigated as a method for waveguide formation.

Vanadium doped GLS displays three absorption bands at 580, 730 and 1155 nm identified by photoluminescence excitation measurements. Broad photoluminescence, with a full width half maximum of ~500 nm, is observed peaking at 1500 nm when exciting at 514, 808 and 1064 nm. The fluorescence lifetime and quantum efficiency at 300 K were measured to be 33.4 µs and 4% respectively. Analysis of the emission decay, at various vanadium concentrations, indicated a preferentially filled, high efficiency, oxide site that gives rise to characteristic long lifetimes and a low efficiency sulphide site that gives rise to characteristic short lifetimes. X-ray photoelectron spectroscopy measurements indicated the presence of vanadium in a broad range of oxidation states from $V^+$ to $V^{5+}$. Tanabe-Sugano analysis indicates that the optically active ion is $V^{2+}$ in octahedral coordination and the crystal field strength (Dq/B) was 1.84. Titanium and nickel doped GLS display a single absorption band at 590 and 690 nm, and emission lifetimes of 97 and 70 µs respectively. Bismuth doped GLS displays two absorption bands at 665 and 850 nm and lifetime components of 7 and 47 µs. Based on comparisons to other work the optically active ions are proposed to be $Ti^{3+}$, $Ni^+$ and $Bi^+$, all of these displayed emission peaking at ~900 nm.

Through optical characterisation of fs laser written waveguides in GLS, a formation mechanism has been proposed. Tunnelling has been identified as the dominant nonlinear absorption mechanism in the formation of the waveguides. Single mode guidance at 633 nm has been demonstrated. The writing parameters for the minimum propagation loss of 1.47 dB/cm are 0.36 µJ pulse energy and 50 µm/s scanning speed. The observation of spectral broadening in these waveguides indicates that they may have applications for nonlinear optical devices. Fs laser written waveguides in transition metal doped GLS could lead to broadband active optical devices.

# Contents

# Nomenclature

## Symbols

| | |
|---|---|
| A | Absorbance |
| a | Absorption coefficient [cm$^{-1}$] |
| b | Confocal parameter [μm] |
| B | Racah B parameter [cm$^{-1}$] |
| C | Racah C parameter [cm$^{-1}$] |
| Dq | Crystal field strength [cm$^{-1}$] |
| E | Energy [cm$^{-1}$] |
| FOM | Nonlinear figure of merit |
| I | Intensity [Wcm$^{-2}$] |
| L$_w$ | Walk off length [m$^{-1}$] |
| n | Refractive index |
| n$_2$ | Nonlinear refractive index [m$^2$W$^{-1}$] |
| NA | Numerical aperture |
| Pcr | Power for critical self focusing [MW] |
| Q | Configurational coordinate |
| QE | Quantum efficiency |
| S | Huang-Rhys parameter |
| SS | Stokes shift [cm$^{-1}$, nm] |
| T$_0$ | Pulse duration [ps] |
| W, W$_r$, W$_{nr}$ | total, radiative, non-radiative decay rate [s$^{-1}$] |
| W$_{PI}$, W$_{tun}$ | Photoionisation, tunnelling rate [s$^{-1}$] |
| α | Loss coefficient [cm$^{-1}$] |
| β | Stretch factor |
| β$_{TPA}$ | Two photon absorption coefficient [cmW$^{-1}$] |
| γ | Keldysh parameter |
| Γ | Loss [dBcm$^{-1}$] |
| Δ | Bandgap energy [eV] |
| λ | Wavelength [nm] |
| ν | Frequency [s$^{-1}$] |
| σ$_{em}$ | Emission cross section [cm$^2$, m$^2$] |
| τ, τ$_r$, τ$_{nr}$ | Total, radiative, non-radiative lifetime [μs] |
| ω | Angular frequency [rads$^{-1}$] |

## Acronyms

| | |
|---|---|
| AOM | Acousto optic modulator |
| AOS | All optical switching |
| CW | Continuous wave |
| DIC | Differential interference contrast |
| EDFA | Erbium doped fibre amplifier |
| EDX | Energy dispersive X-ray |
| EPR | Electron paramagnetic resonance |
| EXAFS | Extended X-ray absorption fine structure |
| FRFL | Frequency resolved fluorescence lifetime |
| FWHM | Full width half maximum |
| GLS | Gallium lanthanum sulphide glass |
| GLSO | Gallium lanthanum sulphide oxide glass |
| GVD | Group velocity dispersion |
| HMO | Heavy metal oxide |
| IR | Infrared |
| MZI | Mach-Zehnder interferometer |
| NIF | National ignition facility |
| OSA | Optical spectrum analyser |
| OTDM | Optical time division multiplexing |
| PL | Photoluminescence |
| PLE | Photoluminescence excitation |
| QPM | Quantitative phase microscopy |
| SCCM | Single configurational coordinate model |
| SNR | Signal to noise ratio |
| SOA | Semiconductor optical amplifier |
| SOP | State of polarisation |
| SPM | Self phase modulation |
| SRS | Stimulated Raman scattering |
| TM | Transition metals |
| TRFL | Temporally resolved fluorescence lifetime |
| UV | Ultraviolet |
| XPS | X-ray photoelectron spectroscopy |
| YAG | Yttrium aluminium garnet |
| ZBLAN | Fluorozirconate (Zr-Ba-La-Al-Na fluoride) glass |
| ZPL | Zero phonon line |

# List of figures

# List of tables

# Acknowledgements

I would like to thank everyone who helped me with this project, especially after the Mountbatten fire, their name are too numerous to mention here. In particular I would like to thank my supervisor Prof. Dan Hewak for his technical support and guidance and my co-supervisor Dr. Richard Curry for his continued support after moving to another institution one year into my PhD. I would also like to thank Prof. Harvey Rutt for assisting my work his seemingly bottomless knowledge of spectroscopy, Weijia Yang for his assistance with the waveguide fabrication and characterisation. Sincere thanks to Dr. Eleanor Tarbox for going through my thesis with a fine tooth comb. Thanks to Kenton Knight for helping me to melt glass, Dr. N. Blanchard for his assistance with the XPS measurements and Dr. Giampaolo D'Alessandro for his assistance with the continuous lifetime distribution model.

# Chapter 1

# Introduction

## 1.1 Motivation

Over the past few decades silica fibre has revolutionised the way in which we communicate by allowing low cost, high bandwidth transmission of data over long distances. This has enabled millions of people around the world access to data resources like the World Wide Web as well as voice and video phone. Further improvements to the data bandwidth available to home users could revolutionise the way media such as news, films and music are accessed. An example of this is the growing implementation of media-on-demand.[21] The invention of the erbium doped fibre amplifier (EDFA) in 1985[22] was instrumental in allowing long distance data transmission through silica fibre and was a significant improvement on electronic repeaters which required the conversion between optical and electronic signals. By a quirk of nature an emission band of erbium, which dictates the gain bandwidth of the EDFA, sits neatly in the low loss window of silica. The technology behind silica fibre is now mature; its structure and properties are well understood and the loss achievable in silica fibre comes close to its theoretical minimum.

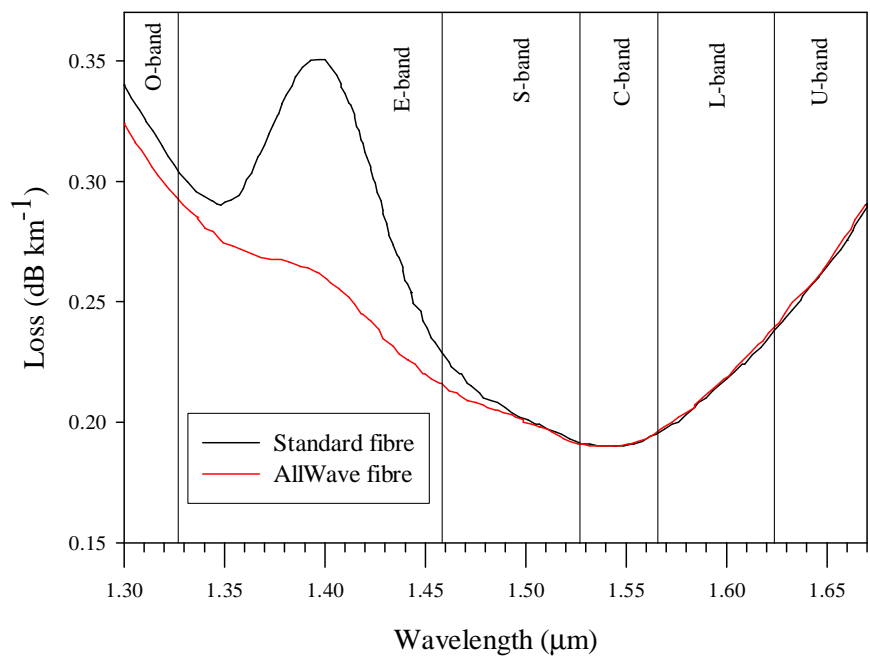

FIGURE 1.1 Loss of standard and AllWave silica fibres showing the region of minimum attenuation and the six conventional bands of optical telecommunications. After[1].

Until recently the presence of the hydroxyl impurity in silica introduced an overtone absorption around 1400 nm which effectively divided the region of lowest attenuation into two separate windows. In 1998, Lucent technologies introduced the AllWave fibre which contains less than one part per billion hydroxyl ions.[23] This ultra dry fibre effectively gives one continuous low loss window spanning ~1260 to 1670 nm, as shown in figure 1.1. The gain bandwidth of the EDFA spans only a small fraction of this continuous low loss window. Doping with other rare earths could allow wavelengths not covered by the EDFA to be used, but problems associated with the vibrational structure of silica prevent its use as an amplifier medium when doped with ions such as praseodymium, thulium and dysprosium.[24]

Silica glass is by a long way the most favourable material to use for long distance optical fibre telecommunications. However, there are aspects of the properties of silica which make it unsuitable for certain applications. For instance the low rare earth solubility of silica means that the interaction length of active devices based on rare-earth doped silica is relatively long. The high phonon energy of silica means that transitions of many rare earth dopants decay non-radiatively. The relatively low non-linear refractive index of silica means that non-linear devices based on silica require relatively high intensities in order to function. The transmission wavelength of silica is also limited to 2 μm. These shortcomings of silica have merited the investigation of a variety of novel glasses for optical device applications. These include phosphate, heavy metal oxide (HMO), fluoride and chalcogenide glasses.

The high energy storage and extraction characteristics of phosphate glasses make them suitable for high power laser applications such as fusion research.[25] The national ignition facility (NIF) at the Lawrence Livermore National Laboratory, (California, USA) uses Nd doped phosphate glass as a gain medium. When completed the NIF is expected to produce ~ns pulses with energies of ~2 MJ and peak powers of ~500 TW.[26]

HMO glasses are arbitrarily defined as those glasses containing over 50 cation % of bismuth and/or lead. They are believed to have the highest refractive indexes of any oxide glass and are also characterised by high density, high thermal expansion, low transformation temperature and excellent infrared transmission up to ~6 μm.[27] A 2R regenerator (see section 6.1.3.3 for a description) that exploits the nonlinearity of bismuth oxide fibre has been demonstrated.[28]

The fluoride glass ZBLAN is based on the fluorides of Zr, Ba, La, Al and Na. A loss of 0.45 dB km$^{-1}$ at 2.35 μm has been achieved for ZBLAN fibre.[29] Such fibres area now used for various passive applications requiring the handling of IR signal. In this respect, fluoride fibres are complementary to silica fibres when the wavelength exceeds 2 μm. Laser power delivery is another field of application for these fibres, for example Er:YAG laser at 2.9 μm attracts a growing interest for dental applications.[30]

Chalcogenide glasses contain a chalcogen element (sulphur, selenium or tellurium) as a substantial constituent. Oxygen is also a chalcogen but it is not usually included in the definition of chalcogenide glasses because oxide glasses form a large group with distinctly different properties to glasses formed from the other chalcogen elements. One of the principle differences between oxide and chalcogenide glasses is their bandgap energy, for example $SiO_2$ has a band gap around 10eV, yet chalcogenides have bandgaps between 1eV and 3eV.[31] Chalcogenide glasses display several potentially useful photoinduced effects including photodarkening,[31] photobleaching,[32] photocrystallisation,[31] photopolymerisation,[33] photodissolution of metals,[31] photocompaction[34] and photoinduced anisotropy.[31] Photodarkening written waveguides in a chalcogenide thin film have been shown to exhibit strong self phase modulation.[35]

Chalcogenide glasses transmit to longer wavelengths in the IR than silica and fluoride glasses. Chalcogenide glasses based on sulphur, selenium and tellurium typically transmit up to around 10, 15 and 20 μm respectively.[36] This long wavelength transparency enables chalcogenide glasses to be utilised for several applications including thermal imaging, night vision, CO and $CO_2$ laser power delivery, radiometry and remote chemical analysis.[37] Chalcogenide fibres are well suited for chemical sensing since most molecular species vibrate in the IR region. Chalcogenide fibre based reflectance probes have been used to detect contaminants in soil and distinguish various tissues and organs in bio-medical samples.[36] Chalcogenide glasses often exhibit a low phonon energy, this allows the observation of certain transition in rare earth dopant that are not observed in silica. For example it is virtually impossible to measure 1.3 μm fluorescence from the $^1G_4 \rightarrow {}^3H_5$ transition of $Pr^{3+}$ in silica-based glass, but it has been observed in tellurite-based glass [38] and gallium lanthanum sulphide (GLS) glass.[39] The low phonon energy of chalcogenides can be thought of as resulting from the relatively large mass of their constituent atoms and the relatively weak bonds between them. Chalcogenide glasses have a nonlinear refractive index around two orders of magnitude higher than silica. This makes them suitable for ultra-fast switching in telecommunication systems. An efficient optical Kerr shutter with a ps response time has been demonstrated in 48 cm of $As_2S_3$ fibre.[40] Table 1.1 compares some important optical properties of silica with ZBLAN, $As_2S_3$ and GLS glass. The table shows that the chalcogenides $As_2S_3$ and GLS have a high linear and nonlinear refractive index, a low phonon energy and longwave IR transmission compared to silica and fluoride glass.

TABLE 1.1 Optical properties of silica, ZBLAN, $As_2S_3$ and GLS glass.

| Optical property | Silica[41] | ZBLAN[42] | $As_2S_3$[34] | GLS[34, 43] |
|---|---|---|---|---|
| Refractive index (at 700 nm) | 1.44 | 1.48 | 2.56 | 2.48 |
| Nonlinear index ($10^{-20}$ $m^2$ $W^{-1}$) | 2.5 | ~2 | 250 | 300 |
| Transmission window (μm) | 0.16-2.0 | 0.22-4.0 | 0.7-10 | 0.5-10 |
| Phonon energy ($cm^{-1}$) | 1150 | 600 | 360 | 425 |
| Zero dispersion wavelength (μm) | 1.3 | 1.6 | 5.5 | 4 |
| dn/dT ($10^{-5}$ $K^{-1}$) | 1.2 | -1.5 | - | 10 |
| Fibre attenuation at 1.5 μm (dB $km^{-1}$) | 0.19 | 0.24 | 220 | 1500 |
| Fibre attenuation at 5 μm (dB $km^{-1}$) | NA | NA | 0.2 | 0.3 |

## 1.2 Gallium lanthanum sulphide (GLS) glass

The glass forming ability of gallium sulphide and lanthanum sulphide was discovered in 1976 by Loireau-Lozac'h *et al.*[44] GLS glasses have a wide region of glass formation centred about the $70Ga_2S_3 : 30La_2S_3$ composition and can readily accept other modifiers into their structure.[24] This means that GLS can be compositionally adjusted to give a wide variety of optical and physical responses. For example the addition of CsCl increases the thermal stability region of GLS[45] and the addition of $La_2F_6$ improves thermal stability, increases visible transmission and decreases OH impurity levels.[46] It is necessary however to add a small percentage, typically 2% by weight, of lanthanum oxide to form a glass. Without this oxide, whether added intentionally or as an impurity in the precursors, crystallisation of the glass is a problem and glass formation is hindered.[16] GLS has a high refractive index of ~2.4, a transmission window of ~0.5-10 μm and a low maximum phonon energy of ~425 cm$^{-1}$.[34] GLS glasses have a high dn/dT and low thermal conductivity, causing strong thermal lensing, thus they are not suitable for bulk lasers. However, the high glass transition temperature of GLS makes it resistant to thermal damage, it has good chemical durability and its glass components are non-toxic.[24]

Because of its high lanthanum content GLS has excellent rare-earth solubility. A high solubility of the ion is not required for the glass to support a lasing ion, but dispersion of the ions in the glass matrix is required to alleviate cross quenching.[47] This property motivated much of the original interest in GLS in the quest for a rare-earth host for solid state lasers. Laser action at 1075 nm has been demonstrated in UV laser written channel waveguides in neodymium-doped GLS.[48] Other active area of research into GLS include its acousto-optics properties, IR lens moulding, 2.9 μm Er:YAG laser power delivery for dentistry applications, nonlinear micro resonators and electrical and optical data storage utilising the change in resistivity and reflectance of GLS in its crystalline and vitreous phase respectively.

## 1.3 Transition metal dopants

Solid state lasers that use transition metals as the active ion have a long history and can in fact be traced back to the first demonstration of laser action – the chromium doped ruby laser. Figure 1.2 shows the tuning range of lasers based on various first row transition metals in crystalline hosts. The figure show an almost continuous coverage of laser wavelengths from 600-4500 nm that is available from lasers based on nickel, vanadium, titanium, cobalt and iron active ions; which illustrates the huge potential of these elements for active optical devices. Apart from being of considerable academic interest, the demonstration of laser action from one of these elements in a glass host would have important implications for other optical devices in that it could lead to a broadband gain medium that could be incorporated into existing fibre and planar optical devices.

To date there has been no demonstration of a first row transition metal laser that uses glass as a host. The high maximum phonon energy of silica makes it one of the more unlikely candidates for the host material. Chalcogenide glasses have low maximum phonon energies due to the relatively large atomic mass of the constituent atoms. In particular GLS has a maximum phonon energy of 425 cm$^{-1}$[34] This low maximum phonon energy allows emission from transition metal dopants which are weakly or not at all observed in silica. This is because its higher maximum phonon energy increases the probability of non radiative decay from excited states. Of particular interest is vanadium doped GLS (V:GLS) which exhibits a broad infrared emission peaking at 1500 ~nm with a FWHM of ~500 nm.[49] Close examination of figure 1.1 indicates that if an optical amplifier could be fabricated from V:GLS it could have a gain bandwidth that covers the entire low loss window of silica.

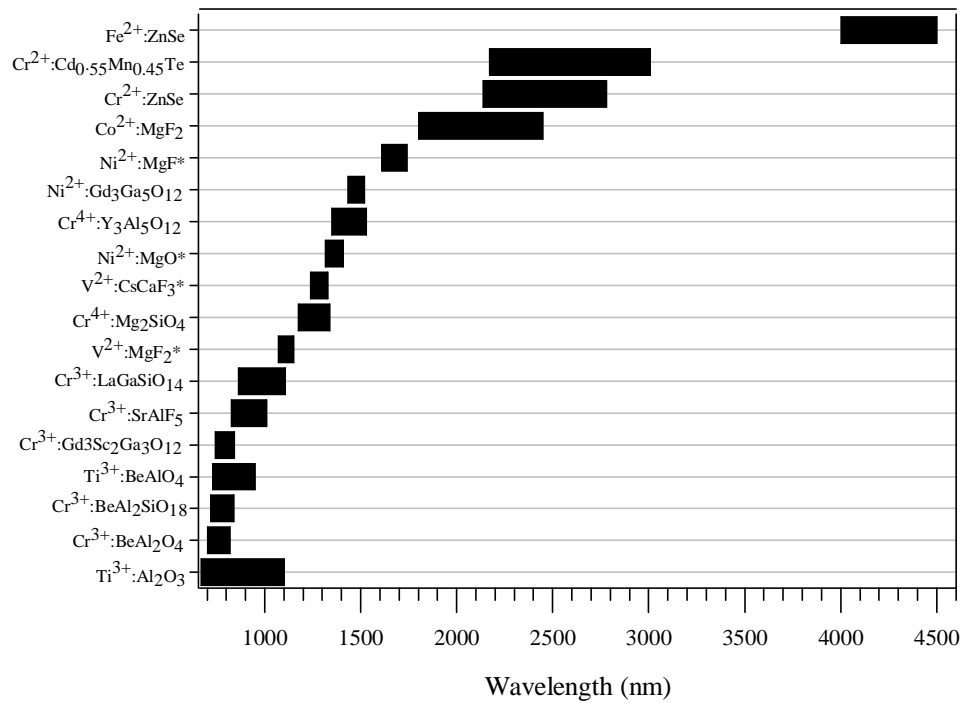

FIGURE 1.2 Overview of the tuning range of selected transition metal doped ion lasers, * these lasers operate at low temperatures. [2-13].

## 1.4 Waveguide technology

Most practical optical devices require a waveguide of some form to confine and direct light signals. Single mode guidance is highly preferable as it avoids dispersion from and interactions between different spatial modes. Optical fibre geometry is frequently used, however, the fundamental structure of fibres essentially limits device construction to one dimension. Waveguides written onto a surface allows the realisation of two dimensional optical devices. Waveguides written below a surface, such as femtosecond laser direct written waveguides (see chapter 6), allows three dimensional optical devices to be constructed. Despite a lot of effort it has not been possible to fabricate low loss single mode GLS fibre to date. Single mode UV written waveguides in GLS have been demonstrated,[34] these are however limited to two dimensional structures and the waveguides are extremely fragile. Femtosecond laser writing is particularly attractive because as well as having rapid processing times, waveguiding structures can be formed below the surface of the glass enabling 3-D structures to be fabricated. From experimental studies of writing waveguides using a fs laser in various glass and crystalline materials it has been suggested that achieving a refractive index change without any physical damage is restricted to glasses.[50] The formation of sub diffraction limited structures is feasible using a focused fs laser beam because of the nonlinear process involved in material modification. The fabrication of buried fs laser written waveguides in GLS has been demonstrated[51] and they show promise for the development of optical devices based on high quality waveguide structures in GLS.

## 1.5 Scope of the thesis

This thesis focuses on two different, but complementary, aspects of the modification of GLS. Firstly the addition of transition metal ions as dopants is examined and their potential for use as active optical materials is explored. It is also argued that the spectroscopic analysis of transition metal ions is a useful tool for evaluating the local environment of their host. Secondly fs laser modification of GLS is investigated as a method for waveguide formation. The observation of spectral broadening indicates that these waveguides may have applications for nonlinear optical devices. The change in direction of this thesis, from investigating transition metal dopants to fs laser written waveguides, was compelled by the loss of ORC glass fabrication and characterisation facilities in a fire. Because of this certain glass samples could not be fabricated, notably vanadium doped GLSO and bismuth doped GLS, and certain characterisations, notably quantum efficiency, could not be completed on all samples.

In this thesis the following are presented for the first time:
- Calculation of the crystal field parameters for a transition metal ion in GLS using the Tanabe-Sugano model
- Calculation of the lifetime distribution in a transition metal doped chalcogenide glass using the continuous lifetime distribution model.
- Quantum efficiency measurement of a transition metal ion in GLS
- X-ray photoelectron spectroscopy measurement of a dopant ion in GLS
- Electron paramagnetic resonance measurement of a dopant ion in GLS
- The emission and emission lifetime of titanium doped GLS.
- Optical characterisation of bismuth doped GLS.
- Characterisation of fs laser written waveguides in GLS
- Broadening of an ultra-short pulse in a GLS waveguide.
- Loss measurement of a GLS waveguide using the Fabry-Perot technique.
- Index change profile measurement using quantitative phase microscopy in a GLS waveguide.

The thesis is structured into seven chapters, including this introduction, chapter 1.

- Chapter 2 provides sufficient background related to the spectroscopy of transition metal ions for the understanding of chapters 3, 4 and 5.
- Chapter 3 details the melting procedures for the fabrication of transition metal doped GLS and all of the spectroscopic techniques used in the analysis of transition metal doped GLS
- Chapter 4 presents a rigorous optical characterisation of vanadium doped GLS glass. The emission lifetime and its non-exponential decay characteristics are investigated in detail. Absorptions from three spin-allowed transition and one spin-forbidden transition were identified. The energy of these transitions was used to identify the oxidation state and coordination number of the vanadium ion. X-ray photoelectron spectroscopy was used to identify that vanadium exists in a broad range of oxidation states in GLS.

- Chapter 5 details the spectroscopic properties of titanium, nickel and bismuth doped GLS. Arguments based on the number of observed absorption peaks and comparisons with dopants in other hosts were used to identify the oxidation state of these dopants.

- Chapter 6 is somewhat self-contained and describes the fabrication and characterisation of buried waveguides written into GLS glass using 800 nm focused fs laser pulses. The spectral broadening of 1550 nm fs laser pulse coupled into these waveguides is also reported.

- Chapter 7 draws conclusions and identifies topics that might provide the basis for further studies.

# Chapter 2

# Background

## 2.1 Introduction

The scope of this chapter is to provide sufficient background for the understanding of subsequent chapters. It begins with an introduction to absorption and lifetime measurements, then an overview of crystal field theory which describes how the energy levels of the d orbital split in the presence of a ligand or crystal field is given. Group theory is then introduced to describe how the labelling of the energy levels of a transition metal ion is arrived at. Next the single configurational coordinate model is introduced this is followed by spectral broadening mechanisms and finally some important structural properties of GLS are described.

## 2.2 Spectroscopy basics

### 2.2.1 Absorption measurements

In absorption spectroscopy the experimenter observes what frequencies of radiation are absorbed from incident radiation as it passes through a sample. If light of frequency ν is absorbed, it signifies that an absorbing species of the sample has undergone a transition from a state of energy $E_1$ to a state of energy $E_2$ and that equation 2.1 is satisfied.[52]

$$h\upsilon = E_2 - E_1 \tag{2.1}$$

Consider the reduction of intensity that occurs when light of intensity I passes through a slab, with infinitesimal thickness dz, of the sample. The loss of intensity dI is proportional to the thickness dz and the intensity of the incident light I and is given by:

$$dI = -\alpha I dz \tag{2.2}$$

Where $\alpha$ is the absorption coefficient, which depends both on the absorbing species and the frequency of the incident light, and commonly has units of $cm^{-1}$. Integrating both sides of equation 2.2 gives I as a function of z: $\ln(I) = -\alpha z + C$. For a sample of thickness l the difference between the incident intensity $I_0$ at z = 0 and the intensity $I_T$ that emerges from the sample at z = l is given by $\ln(I_0) - \ln(I_T) = (-\alpha 0 + C) - (-\alpha l + C) = \alpha l$, this can be expressed as equation 2.3 which is otherwise known as the Beer-Lambert law.[19, 52]

$$I_T = I_0 \exp(-\alpha l) \tag{2.3}$$

If the concentration of absorbing species c is taken into account then equation 2.2 becomes:

$$dI = -\sigma I c dz \tag{2.4}$$

If c is expressed as a number per unit volume then σ is the absorption cross section and has units of area. Data obtained from the spectrophotometers used in this study is in units of absorbance A:

$$A = \log_{10} \frac{I_0}{I_T} \tag{2.5}$$

The absorption coefficient $a$ was then calculated from $a = A/l$. Note the decadic form $a$ of absorption coefficient is used where $a = \alpha / \ln 10 = \alpha/2.303$.

## 2.2.2 Excited state absorption

Excited state absorption (ESA) occurs when the energy sate $E_1$, described in section 2.2.1, is not the lowest energy level of the absorbing species. In this case absorption occurs with the promotion of an electron in an excited state higher than the initial excited state. ESA needs to be addressed when considering a material as a gain medium for a laser or optical amplifier because ESA can induce parasitic loss of pump or laser radiation which increases the pump power threshold. ESA is a problem for broadband gain media in particular but it is also likely to be relevant for laser ions with multiple electronic levels, such as erbium or thulium. ESA is usually measured using the pump-probe technique[53-55] in which the transmission of a weak probe beam is measured with and without the presence of a strong pump beam.

## 2.2.3 Lifetime measurements

In lifetime measurements the experimenter observes the emission intensity from a sample as a function of time after an initial excitation pulse, which ends at t = 0. Consider the ground state (level 1) and excited state (level 2) of an absorbing species. After an initial excitation pulse the population density in level 2 is expressed as $N_2$ (number per unit volume). Assuming that no quenching or interaction between excited species occurs then $N_2$, as a function of time, will decay with an exponential decay rate. The change in population density $N_2$, as the population is transferred to level 1, can be expressed as:[56]

$$\frac{dN_2}{dt} = -A_{21}N_2 \tag{2.6}$$

Where $A_{21}$ is the rate at which the population is transferred from level 2 to level 1. $A_{21}$ has units of 1/time, and is referred to as the radiative transition rate. The solution to equation 2.6 is:

$$N_2 = N_2^0 \exp(-A_{21}t) \tag{2.7}$$

Where $N_2^0$ is the initial population density in level 2 at t = 0. Defining a time $\tau_2$ as the time taken for the population $N_2$ to decay to 1/e of its original value and considering that the observed emission intensity I(t) is proportional to the population $N_2$, equation 2.7 can be expressed as:

$$I(t) = I_0 \exp(-t/\tau_2) \tag{2.8}$$

Where $I_0$ is the emission intensity at t = 0. The time $\tau_2$ is referred to as the lifetime of level 2.

## 2.3 Crystal field theory

This relatively simple theory was first proposed by Bethe[57] and Van Vleck[58] but is still useful for visualising the electrostatic interaction between the orbitals of a central metal ion and the surrounding ligand field. In crystal field theory the positive ions are regarded as point charges and neutral molecules as dipoles with their negative ends directed towards the metal. Covalent bonding is completely neglected.[15] The arrangement of the orbitals in a transition metal (TM) is illustrated in figure 2.1, adapted from.[2]

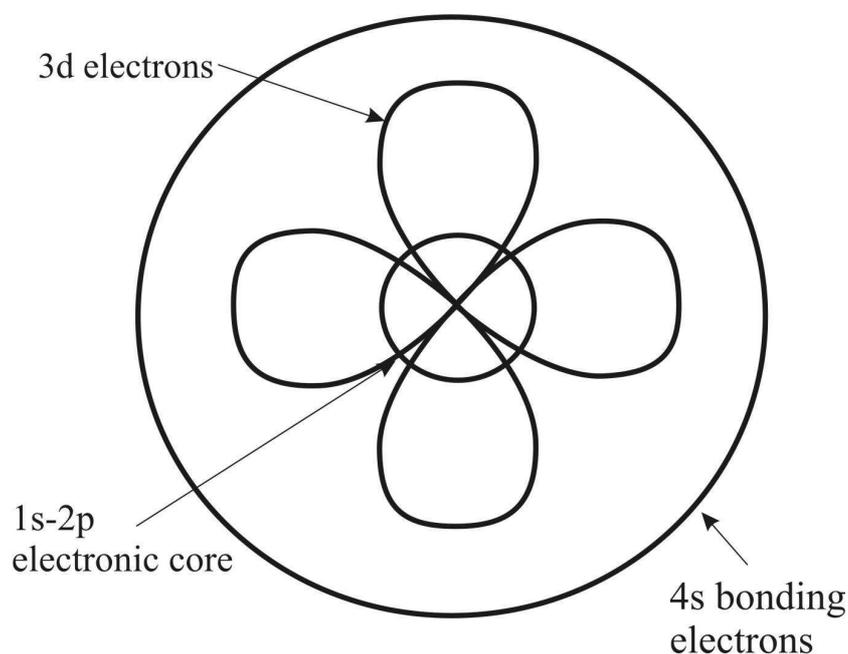

FIGURE 2.1 Orbital arrangement in a transition metal ion.

The 4s electrons are used to form chemical bonds leaving 3d electrons exposed to the electric field of neighbouring atoms, this field is also called the crystal or ligand field. Therefore the 3d electrons are strongly affected by both the strength of the neighbouring atoms electric field and their arrangement around the transition metal ion (coordination). The angular dependence of the d orbital wave function consists of five orthogonal sets of independent orbitals as illustrated in figure 2.2. The five orbitals are degenerate, in other words they have the same energy in the absence of an external electrostatic field.

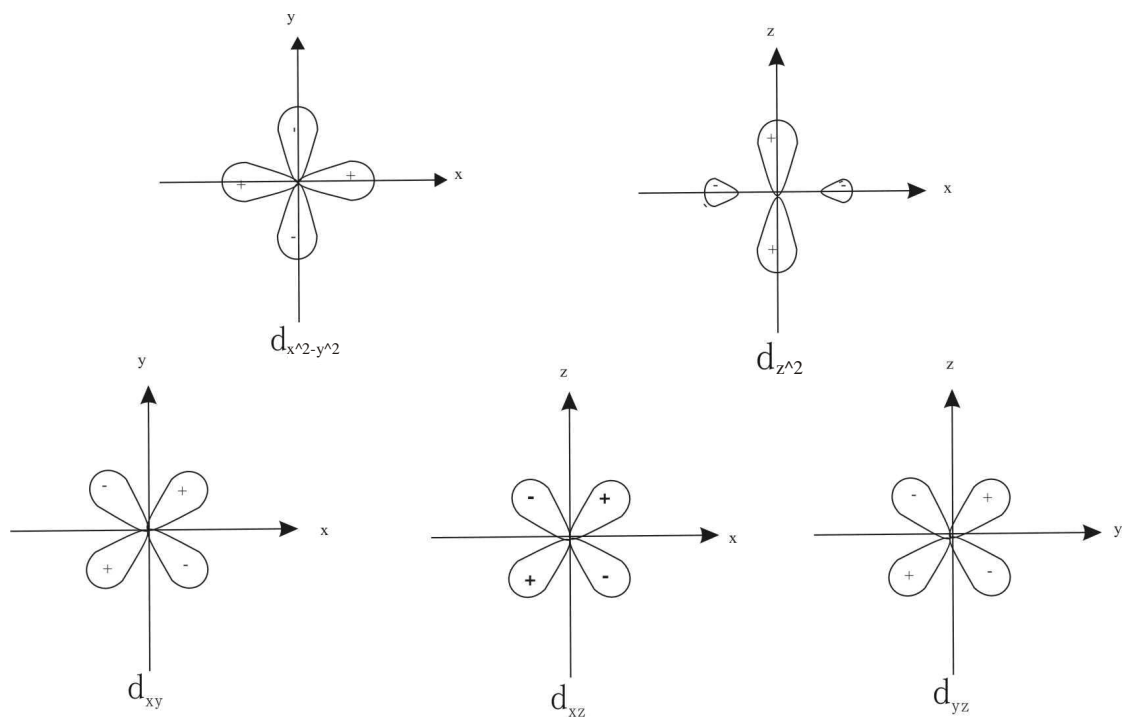

FIGURE 2.2 Shape of the five degenerate d orbitals, after[14].

When the central metal ion is surrounded by six ligands, with an axis of symmetry, the ion is said to be in octahedral coordination, this case is illustrated in figure 2.3

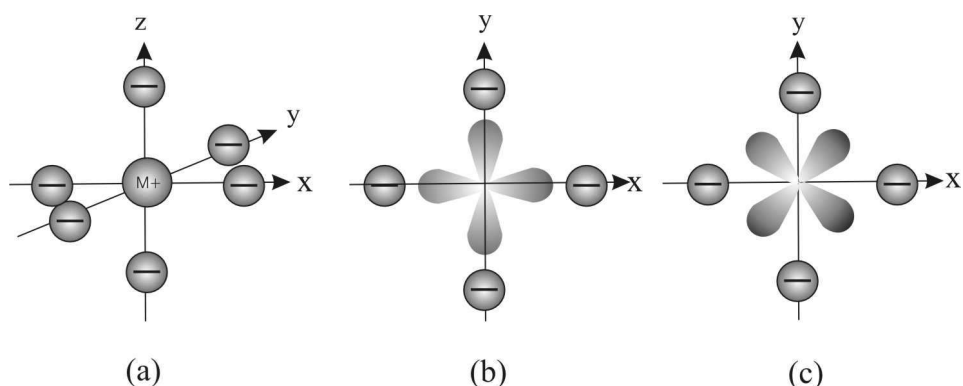

FIGURE 2.3 (a) The arrangement of ligands around an ion in octahedral coordination. (b) $d_{x^2-y^2}$ orbital. (c) $d_{xy}$ orbital, after[15].

It can be seen from figure 2.3 that electrons in the $d_{x^2-y^2}$ orbital experience greater repulsion from the negatively charged ligands than electrons in the $d_{xy}$ orbital, this has the effect of destabilising the $d_{x^2-y^2}$ relative to its energy in the absence of ligands. The $d_{yz}$ and $d_{zx}$ orbitals have the same spatial orientation relative to ligands in the xz and yz planes of the $d_{xy}$ orbital and therefore have the same energy. The $d_{z^2}$ orbital is destabilised to the same extent as the $d_{x^2-y^2}$ orbital. Therefore, in octahedral coordination, the five d orbitals that were originally the same energy are split into two sets, one triply degenerate set of $d_{xy}$,$d_{yz}$ and $d_{xz}$ (denoted $t_{2g}$) and another less stable doubly degenerate set of $d_{x^2-y^2}$ and $d_{z^2}$ (denoted $e_g$). Figure 2.4 illustrates this splitting. The energy difference between the $t_{2g}$ and $e_g$ levels is denoted 10 Dq and is called the crystal field splitting parameter.

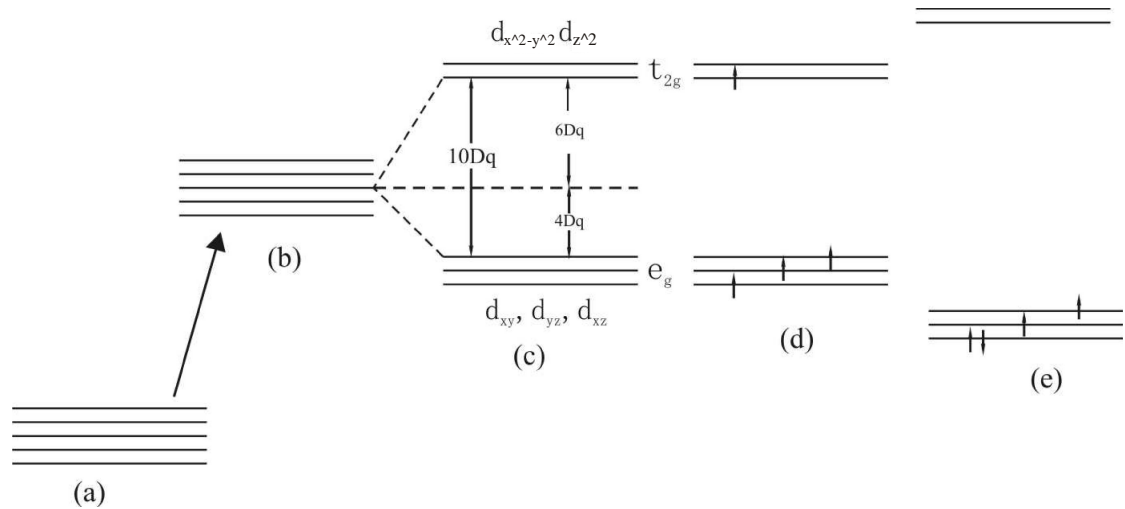

FIGURE 2.4 Splitting of d orbitals in an octahedral ligand field. (a) Free ion. (b) Ion in hypothetical spherically symmetric field. (c) Ion in an octahedral field. (d) Occupation of d orbitals by electrons for $d^4$ configuration in a weak field (e) Occupation of d orbitals by electrons for $d^4$ configuration in a strong field.

When the central metal ion is surrounded by four ligands, with an axis of symmetry, the ion is said to be in tetrahedral coordination, this case is illustrated in figure 2.5 which shows that the triply degenerate set of $d_{xy}$, $d_{yz}$ and $d_{xz}$ orbitals (referred to as $t_2$ orbitals) are closer to lines connecting the ligands than the doubly degenerate set of $d_{x^2-y^2}$ and $d_{z^2}$ orbitals (referred to as e orbitals). Hence the $t_2$ orbitals experience a stronger repulsion than the e orbitals and the order of energy levels is inverted relative to that for the octahedral environment, this is illustrated in figure 2.6. The reason why the $t_2$ and e orbitals take the value of 6 or 4 Dq above or below the barycentre is simply related to the fact that the $t_2$ and e orbitals are triply and doubly degenerate respectively. Therefore the triply degenerate orbital will contribute 3/5 of the total splitting and the double degenerate orbital will contribute 2/5 of the total splitting.

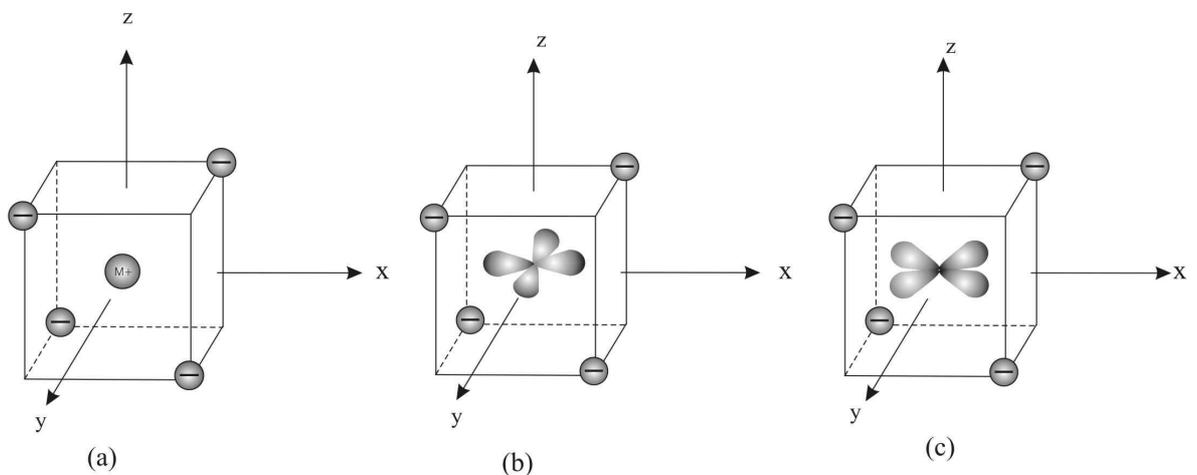

FIGURE 2.5(a) The arrangement of ligands around an ion in tetrahedral coordination. (b) $d_{x^2-y^2}$ orbital. (c) $d_{xy}$ orbital, after[15].

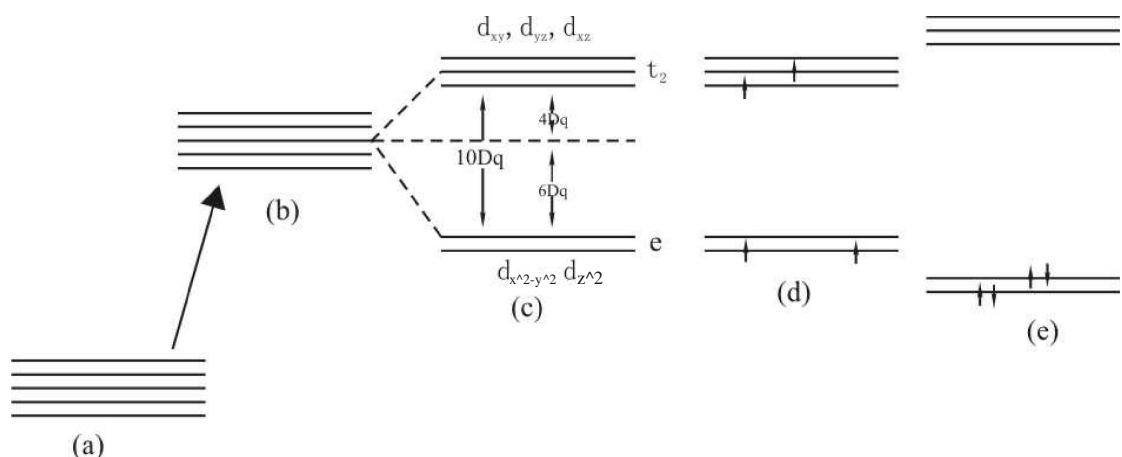

FIGURE 2.6 Splitting of d orbitals in an tetrahedral ligand field. (a) Free ion. (b) Ion in hypothetical spherically symmetric field. (c) Ion in an tetrahedral field. (d) Occupation of d orbitals by electrons for $d^4$ configuration in a weak field (e) Occupation of d orbitals by electrons for $d^4$ configuration in a strong field.

In the situation where the TM ion has multiple 3d electrons group theory can give a more accurate representation of how the crystal field splits the 3d shell. In group theory the symmetry operations, such as rotations and reflections, form mathematical groups which can then be decomposed into the irreducible representations for that symmetry group. A very useful method of displaying the dependence of energy states upon the strength of the crystal field was developed by Tanabe and Sugano[59]. In these Tanabe-Sugano (TS) diagrams, the nomenclature for the various levels corresponds to the irreducible representations for the symmetry group of the ion in question. The TS diagrams in section 4.16 illustrate how the degenerate levels of the 3d orbital split in the presence of an increasing crystal field strength. A detailed description of TS diagrams is given in section 4.16.1. In TS diagrams the crystal field strength is denoted Dq/B where Dq is the crystal field parameter and can be thought of as a measure of the overlap between electrons in the 3d orbital and the orbitals of neighbouring atoms. The mutual repulsion contribution of the energy levels is represented by the Racah parameter B. After the crystal field the next strongest interaction to cause splitting of the energy levels in transition metals is the spin orbit interaction, this is however very weak in comparison to the splitting caused by the crystal field. Because of this the designation of energy levels in transition metals typically has the following form: $^{(2S+1)}A$. Where S is the total spin quantum number and A is a letter associated with the coordination of the active ion. However in the case of rare earth ions the opposite is true because the optically active 4f electrons are shielded from the crystal field by 5s and 5p electrons causing the crystal field splitting of the energy levels to be much weaker than the spin orbit interaction. Hence the designation of energy levels in rare earth metals typically has the following form: $^{(2S+1)}L_J$. Where S is the total spin quantum number, L is the total orbital angular momentum and J is the total angular momentum. If the value of Dq/B is known for a particular ion then the position of its energy levels can be read from its TS diagram by drawing a vertical line at the correct value of Dq/B and reading on the Y axis where the line intersects the energy levels.

## 2.4 Group theory

Group theory is an extremely useful technique for interpretation of the optical spectra of ions in transparent materials. In this work it is applicable in the determination of the number of energy levels of a transition metal ion, labelling these energy levels in a proper way, determining their degeneracy and establishing selection rules for transitions between these energy levels.

Consider for example the octahedrally coordinated ion in figure 2.3 (a). If a rotation of 90° is made around the z-axis, the system is invariant from before the rotation operation. This rotation is called a $C_4$ (001) symmetry operation, where (001) denotes the rotational axis and (4) refers to the $2\pi/4$ rotation angle. There are 24 rotation operations possible which leave the octahedron invariant. There are also 24 reflection transformations which leave the octahedron invariant, however each of these symmetry reflections can be achieved by applying both a symmetry rotation and an inversion to the octahedron.[60] This gives a larger group of symmetry operations containing 48 elements. These are 24 rotation operations and 24 rotation plus inversion operations. This set of symmetry operations is referred to as the $O_h$ point symmetry group. There are 32 possible point symmetry groups denoted by the Schoenflies symbols, for example the point symmetry group for an ideal tetrahedron is labelled $T_d$. The elements which bear a relationship to each other comprise a group. A set of symmetry operation elements constitute a group if they can be multiplied together under the following rules.[60]

1. The set is closed under group multiplication. If A and B are elements in the set, then the product AB is also a member of the set.
2. The associative law holds: A(BC) = (AB)C
3. A unit element e exists, such that eR = Re = R, for any element R
4. For any element R there is an inverse element $R^{-1}$ which is also an element of the set. The inverse element has the property $RR^{-1} = R^{-1}R = e$.

Elements A and B are said to be in the same class if there exists an element R of the group such that $A = RBR^{-1}$. For example the 90° rotation of the octahedron constitutes the class $C_4$. The six possible operations of class $C_4$ lead to the class $6C_4$. The various rotation reflection and inversion operations of the $O_h$ group belong to the following ten different symmetry classes: E, $8C_3$, $6C_2$, $6C_4$, $3C'_2$, i. $6S_4$, $8S_6$, $3\sigma_h$ and $6\sigma_d$.[19]

The rotation operation $C_4$ (001) transforms the coordinates (x, y, z) into (y, -x, z). This transformation can be written as a matrix equation.[19]

$$C_4(001)(x, y, z) = (y, -x, z) = (x, y, z)\begin{bmatrix} 0 & -1 & 0 \\ 1 & 0 & 0 \\ 0 & 0 & 1 \end{bmatrix} \qquad (2.9)$$

Thus the effect of the 48 symmetry operations of the $O_h$ group over the functions (x, y, z) can be represented by 48 matrices. This set of 48 matrices constitutes a representation, and the basic functions x, y, and z are called basis functions. Each set of orthonormal basis functions $\varphi_i$ generates a representation $\Gamma$ such that

$$R\phi_i = \sum_j \phi_j \Gamma^{ji}(R) \qquad (2.10)$$

Where R is a symmetry operation and $\Gamma^{ji}(R)$ are components of the matrix. The representation that involve the lowest dimension matrices capable of representing the group are called irreducible representations.[19] The number of inequivalent irreducible representations equals the number of classes. The reduction of a representation to its irreducible representations is performed using a character table. To construct a character table for the $O_h$ group a suitable set of basis functions are used, these are the orbital wavefunctions s, p, d… of the ion. The sum of the diagonal elements of the matrices that constitute a representation are called their characters. For example the character of the matrix in equation 2.9 is 1. The character represents the number of orbitals that remain unchanged by a particular symmetry operation. All of the information needed for a symmetry operation is contained in the character its associated matrix. Thus the character table for a point group contains all the information needed to describe it. A character table of the $O_h$ group using orbital wavefunctions as basis functions is given in table 2.1.

TABLE 2.1 A character table of the $O_h$ group, after[19].

| $O_h$ | E | $8C_3$ | $6C_2$ | $6C_4$ | $3C'_2$ | i. | $6S_4$ | $8S_6$ | $3\sigma_h$ | $6\sigma_d$ | Basis function |
|-------|---|--------|--------|--------|---------|----|--------|--------|-------------|-------------|----------------|
| $A_{1g}$ | 1 | 1 | 1 | 1 | 1 | 1 | 1 | 1 | 1 | 1 | s |
| $A_{2g}$ | 1 | 1 | -1 | -1 | 1 | 1 | -1 | 1 | 1 | -1 | |
| $E_g$ | 2 | -1 | 0 | 0 | 2 | 2 | 0 | -1 | 2 | 0 | $(d_z{}^2, d_{x^2-y^2})$ |
| $T_{1g}$ | 3 | 0 | -1 | 1 | -1 | 3 | 1 | 0 | -1 | -1 | |
| $T_{2g}$ | 3 | 0 | 1 | -1 | -1 | 3 | -1 | 0 | -1 | 1 | $(d_{xz}, d_{yz}, d_{xy})$ |
| $A_{1u}$ | 1 | 1 | 1 | 1 | 1 | -1 | -1 | -1 | -1 | -1 | |
| $A_{2u}$ | 1 | 1 | -1 | -1 | 1 | -1 | 1 | -1 | -1 | 1 | $f_{xyz}$ |
| $E_u$ | 2 | -1 | 0 | 0 | 2 | -2 | 0 | 1 | -2 | 0 | |
| $T_{1u}$ | 3 | 0 | -1 | 1 | -1 | -3 | -1 | 0 | 1 | 1 | $(p_x, p_y, p_z)$ $(f_x{}^3, f_y{}^3, f_z{}^3)$ |
| $T_{2u}$ | 3 | 0 | 1 | -1 | -1 | -3 | 1 | 0 | 1 | -1 | $(f_x(y^2 z^2),$ $f_y(z^2 x^2), f_z(x^2 y^2))$ |

The irreducible representations which correspond to the various energy levels of an ion with $O_h$ symmetry are given in the left column of table 2.1 according to Mulliken notation. The dimension of the matrices for each irreducible representation is given by the corresponding character in class E. The subscripts u and g indicate whether an irreducible representation is anti-symmetric (u) or symmetric (g) in respect to the inversion operation i.

## 2.5 The single configurational coordinate model

The single configurational coordinate model (SCCM) is a very useful model for analysing and interpreting the transitions within transition metal ions. Consider an optically active ion (A) in a transparent host material consisting of ions (B). The A ion will be surrounded by a number of B ions belonging to the host material. This environment is dynamic because the A and B ions form part of a vibrating lattice. Also consider that the optically active ion A is coupled to the vibrating lattice. This means that neighbouring B ions can vibrate about some average point and this affects the electronic states of the A ion.[19] The SCCM is dependent on two main approximations:

- The ions move very slowly in comparison to the valence electrons, this approximation is reasonable because the nuclei are much heavier than electrons and therefore move on a much slower timescale.
- The movement of the ligand B ions is considered as a single symmetrical 'breathing' mode. In this case only one nuclear coordinate, which corresponds to the distance A-B, is needed to describe the position of all the ligands. This coordinate is called the configurational coordinate Q,

The potential energy curves for the ground state (electronic state a) and an excited state (electronic state b) for the one-coordinate dynamic centre A are represented diagrammatically in the SCCM in figure 2.7. These potential energy curves are approximated by parabolas according to the harmonic oscillator approximation. In this approximation the B ions pulsate in harmonic oscillation around the equilibrium positions.[19] The horizontal lines on each potential energy curve represent the allowed vibration modes or phonon levels. For the harmonic oscillator of electronic state a at frequency $\omega$, the permitted phonon energies $E_n$ are given by

$$E_n = \left( n + \frac{1}{2} \right) \hbar \omega \qquad (2.11)$$

Where n = 0, 1, 2… and so on. Similarly for electronic state b, which may have a different harmonic oscillator frequency, the allowed phonon levels are characterised by m = 0, 1, 2… and so on. The probability distribution in each of these phonon levels is given by the square of its oscillator function. These probability distributions are represented very approximately by the red curves on each of the phonon levels. These distributions show that in the lowest phonon level the probability distribution is centred around the equilibrium position and in higher order phonon levels the maximum amplitude probability occurs where the phonon levels cross the potential energy curves. This has a strong influence on determining the line shapes of absorption and emission spectra. The Frank-Condon principle states that electronic transitions are most likely to occur when two vibrational wavefunctions overlap and that they are very fast in comparison to the motion of the lattice. This implies that electronic transitions can be represented by vertical lines, as in figure 2.7.

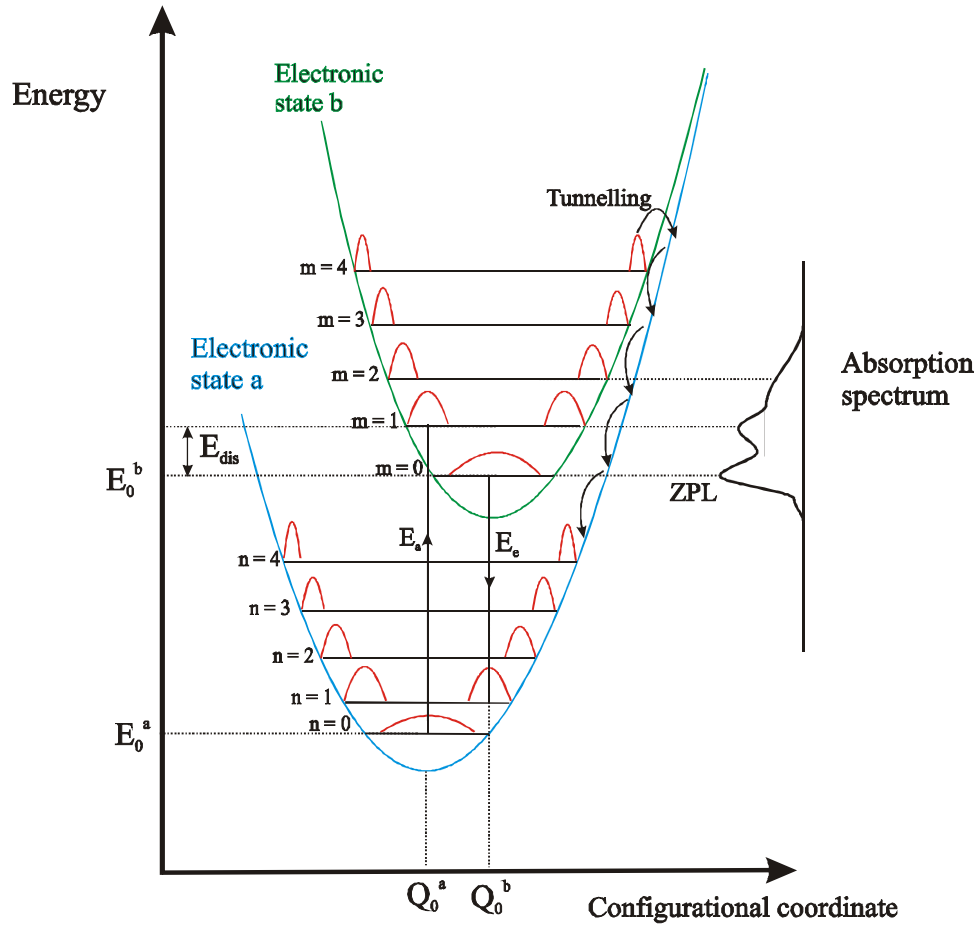

FIGURE 2.7 The single configurational coordinate model, showing how phonon assisted absorption gives rise to absorption line shapes and the mechanisms for phonon assisted non-radiative decay.

The peak in the absorption band occurs at an energy where the overlap between the probability distribution in the phonon levels is at a maximum, which is illustrated in figure 2.7 by the transition $E_a$. Similarly the peak in the emission band corresponds to the transition $E_e$. The difference between the absorption and emission peaks ($E_a - E_e$) is known as the Stokes shift (SS), i.e. SS = $E_a - E_e$. It should be noted that the equilibrium position coordinates $Q_0^a$ and $Q_0^b$ are different for the electronic states a and b. This reflects the difference in electron-phonon coupling between the two states. The dimensionless Huang-Rhys parameter (S) quantifies this difference in electron-phonon coupling and is given by

$$S = \frac{E_{dis}}{\hbar\omega} \qquad (2.12)$$

Where $E_{dis}$ is defined in figure 2.7 and $\hbar\omega$ is the energy of the breathing mode vibration. The Huang-Rhys parameter is related to the Stokes shift by

$$SS = E_a - E_e = (2S - 1)\hbar\omega \qquad (2.13)$$

The absorption band shape, induced from phenomena illustrated in the SCCM in figure 2.7, is due to overlapping occurring between the vibrational m states and the n = 0 phonon level, which would occur at very low temperatures (~0 K). The transitions n = 0 ↔ m = 0 are termed zero phonon lines (ZPL) as they occur without the participation of phonons. ZPL are characterised by relatively narrow line-widths which, disregarding the effect of the host, is the natural linewidth discussed in section 2.6.1. These transitions can commonly be observed in the low temperature absorption and emission spectra of transition metal doped crystals, as in $V^{2+}$ doped ZnSe for example.[61] However they are rarely observed in transition metal doped glasses because of the greater inhomogeneous broadening in these hosts. It can be seen from figure 2.7 that for sufficient S no ZPL will be observed.

Once in an excited state, the ion A can reach its ground state through the emission of a photon (radiative decay) or through the emission of phonons (non-radiative decay). Non-radiative decay can be accounted for by the SCCM, illustrated in figure 2.7. For sufficiently large S, excitation from the ground state results in the population of higher order phonon modes in the excited state. These higher order phonon modes can coincide with the crossing of the potential energy curves of excited state *a* and *b* and therefore the system can relax through the phonon levels of excited state a. If the populated higher order phonon modes coincides with the proximity of the potential energy curves of excited state a and b then the same process may occur by tunnelling.

## 2.6 Broadening mechanisms

The SCCM shows how various absorption and emission lines are generated from the overlap between probability distributions in phonon levels of the ground and excited state. In a real system these absorption and emission line are usually broadened further. The mechanisms responsible for this broadening can be categorised as either homogeneous or inhomogeneous.

## 2.6.1 Homogeneous broadening

Homogeneous broadening is an increase in absorption and emission linewidth caused by phenomena that influence each ion equally. The most fundamental of the homogeneous broadening mechanisms is the natural, or minimum, linewidth. This arises from the Heisenberg uncertainty principle which states that the uncertainty in determining the energy width $\Delta E$ of an energy level that has a minimum uncertainty in its lifetime $\Delta t$, is obtained from the relationship $\Delta E = \hbar / \Delta t$.[56]

Another type of homogeneous broadening is caused by collisions between phonons and optically active ions. These collisions can cause a decrease in decay lifetime when the phonons "knock" an electron from the excited state before it has the opportunity to radiate spontaneously.[56] Dephasing collisions between phonons and optically active ions can also occur; these collisions interrupt the phase of radiating ions without increasing their population decay rate. Consequently temporal coherence is reduced and the emission linewidth is broadened.[23]

## 2.6.2 Inhomogeneous broadening

Inhomogeneous broadening arises from the range of local environments experienced by different ions. This range of local field environments is the manifestation of a variety of crystal field strengths, coordination number, symmetry and proximity to defects. Crystal hosts have a relatively ordered structure; therefore inhomogeneous broadening is relatively weak and arises principally from defects and strains in the crystal. In glass hosts inhomogeneous broadening is relatively strong which causes the absorption and emission spectra of ions in glass hosts to have characteristic broad linewidths. This strong inhomogeneous broadening is related to a fundamental characteristic of glasses, this is their disordered structure. Inhomogeneous broadening results in the superposition of a range of homogeneously broadened lines which generates the observed absorption and emission spectra of an ion.

## 2.7 Selection rules

Since electric dipole processes dominate over magnetic dipole transition strength, magnetic dipole transitions are neglected in this discussion. For electric dipole interaction the rules for allowed transition are:[60]

- No change in spin, i.e. $\Delta S=0$
- The change in angular momentum $\Delta L$ is $\pm 1$
- The change in total momentum $\Delta J$ is $0$, $\pm 1$, but not $J = 0 \rightarrow J = 0$
- No change in parity

Radiative transitions in the transition metal ions involve electrons in the same $3d^n$ configuration, electric dipole transitions are not allowed between these equal parity states and mixing of the higher lying 4p odd parity states is required for electric dipole transitions to occur. The mixing of different parity states is permitted when inversion symmetry of the crystal field is broken, either by an asymmetric distribution of ligands around the ion or a distortion of the symmetry by odd parity phonons. Electric dipole transitions, due to mixing by odd-parity phonons, are called vibronic transitions and are more common in transition metals than rare earths because of the stronger electron-lattice coupling.[62]

## 2.8 Structure of GLS

This section gives a brief general description of the structure of glass, some structural phenomenon relevant to chalcogenide glasses and some structural properties of GLS glass relevant to the discussion in section 4.7.

## 2.8.1 General structure of glass

Glasses can be thought of as having a disordered structure since unlike crystals they do not have long range order. However, there is usually short range order present in that each constituent ion has a specific number of ligands. This description of glasses forms the basis of the continuous random network model, first proposed by Zachariasen in 1932.[63] Constituents of glass are generally divided into two broad categories: network formers and network modifiers. Network formers can be thought of as the backbone of the glass structure through an interconnecting network of polyhedra. For example, $SiO_4$, $ZrF_4$ and $GaS_4$ are the network forming polyhedra for silicate fluorozirconate and GLS glass, respectively. Network modifiers take up the interstitial space between the network forming polyhedra, breaking up the periodicity and preventing crystallisation. For example Al, Ba and $La_2S_3$ are network modifiers in silicate fluorozirconate and GLS glass, respectively.[62]

## 2.8.2 Chalcogenide glass

Chalcogenide glasses can be either stoichiometric or non-stoichiometric. The structure of arsenic chalcogenides can be characterised in terms of the correlation between hetropolar (arsenic-chalcogen) and homopolar (arsenic-arsenic, chalcogen-chalcogen) chemical bonds. For example in the stoichiometric $As_2Se_3$ glass the concentration of homopolar bonds is 10-35%.[64]

## 2.8.3 GLS glass

A study of the structure of bulk GLS glass using extended x-ray fine structure spectroscopy (EXAFS) has been presented by Benazeth *et al*.[65] The Ga-S distance was reported to be 2.26 Å which is characteristic of a covalent bond and therefore $GaS_4$ units were identified as the glass forming units. Comparisons between the Raman and IR spectra of crystalline and amorphous GLS also indicate the presence of $GaS_4$ structural units.[66, 67] In contrast, the crystal $Ga_2S_3$ presents many crystallographic sites and dispersed Ga-S distances.[65] The structure of the crystal $Ga_2S_3$ is shown in figure 2.8. Three sulphur atoms are bound to three gallium atoms. Two of these sulphur atoms ($S_1$) are each engaged in two covalent bonds and one dative bond ($S_2$) with three surrounding gallium atoms. The remaining sulphur atom ($S_3$) is the usual bridging atom as it is bound to two gallium atoms.[34]

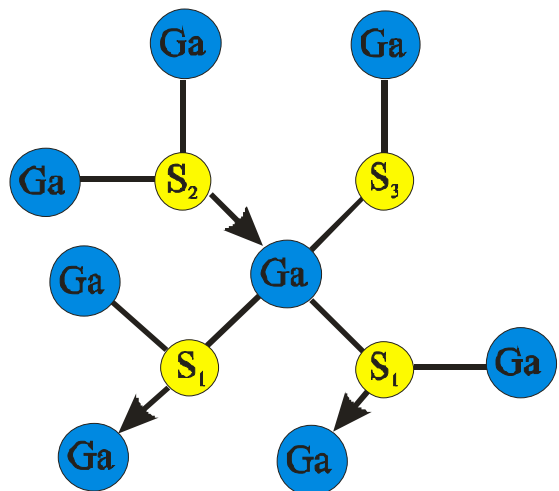

FIGURE 2.8 The covalent gallium environment of the crystalline $Ga_2S_3$, after [65]

Such an environment of sulphur atoms, where most of them present coordination numbers greater than two, is not usually known in glassy sulphides. And indeed, it is impossible experimentally to obtain an amorphous structure from pure $Ga_2S_3$.[65] Because of this a network modifying agent is required for glass formation. The main network modifier in GLS is $La^{3+}$[68] which is 8 fold coordinated to sulphur with an undetermined symmetry.[69] $GaS_4$ tetrahedra are formed from the reaction of $Ga_2S_3$ crystal with sulphur anions ($S^{2-}$) brought by the addition of $La_2S_3$.[16, 65] These sulphur ions break the Ga→S dative bonds, characteristic of the crystalline phase, which forms some $GaS_4$ tetrahedra with a negative charge (figure 2.9 (a)). These negative ionic cavities form some reception sites for $La^{3+}$ ions, which act as charge compensators for these negative charges.[16, 68]. GLS also contains a small quantity of Ga-related tetrahedra, containing at least one threefold coordinated oxygen atom linked to three tetrahedra. Thus in GLS there is both an oxide and a sulphide environment for the $La^{3+}$ ion.

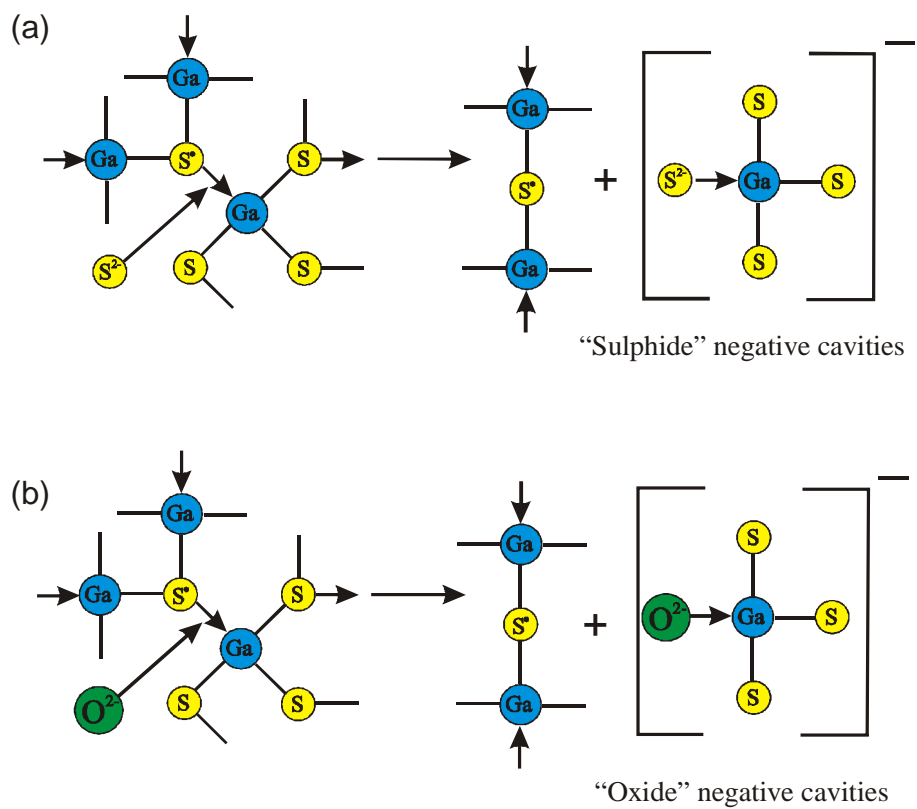

FIGURE 2.9 Formation of sulphide negative cavities (a) and oxide negative cavities (b), after[16].

When $La_2S_3$ is substituted by $La_2O_3$ to form GLSO it is expected that the reaction will be that same as in figure 2.9 (a) but with $O^{2-}$ replacing $S^{2-}$ anions, as illustrated in figure 2.9 (b). Thus in GLSO reception sites for the $La^{3+}$ ion are predominantly oxide in nature.[16] However, it is proposed that a small quantity of sulphide sites exist in GLSO in order to explain the two lifetime components observed in titanium doped GLSO, see section 5.2.4. This leads to a model the GLS system in which a covalent network of $GaS_4$ tetrahedra are inter-dispersed by essentially ionic La-S channels.[65, 68, 69]

# Chapter 3

# Glass melting and spectroscopic techniques

## 3.1 Introduction

The first part of this chapter details the melting procedures for the fabrication of transition metal doped GLS, in particular, vanadium doped GLS. Other authors, notably Mairaj[34] and Brady,[70] have carried out a very exhaustive analysis of the raw material purification and glass melting procedures for GLS. The fabrication of transition metal doped samples examined in this work used many of the techniques developed by Mairaj and Brady without further enhancement; these techniques are therefore only briefly summarised. Detailed in this chapter are melting procedures developed for the fabrication of low doping concentration transition metal doped GLS glass. The second part of this chapter details all of the spectroscopic techniques used in the analysis of transition metal doped GLS in sufficient detail for the reader to understand, critique and repeat them. A brief introduction to each spectroscopic technique is given. Absorption, Raman, XPS and EPR measurements were taken on commercially manufactured equipment since they were available with the required specifications. The lack of commercially manufactured equipment with the required specifications for photoluminescence, excitation, lifetime and quantum efficiency measurements meant they were taken on in house equipment. The in house equipment may not have been able to compete with commercially manufactured versions, were they available, in repeatability but they did allow a degree of functionality and specialisation not possible with a manufactured system.

## 3.2 Glass melting procedures

### 3.2.1 Batching and melting details of transition metal doped GLS samples

Raw material purity is particularly important for the synthesis of GLS glass, as ion concentrations as low as 1ppm can result in strong absorption signatures.[70] Hydroxyl impurities are also detrimental to the optical transmission of GLS and the quantum efficiency of doped samples; to that end water needs to be eliminated both from the raw material and the synthesis process as much as possible. The precursor materials required for the production of GLS glass are: gallium sulphide $Ga_xS_y$ (this comes in three phases GaS, $Ga_2S_3$ and $Ga_4S_5$) lanthanum sulphide $La_2S_3$, and lanthanum oxide $La_2O_3$. These melt materials are not available from any supplier at the desired purity, however their precursors (gallium metal, lanthanum halides and gaseous hydrogen sulphide) are. So gallium and lanthanum sulphides were synthesised in-house from gallium metal (9N purity) and lanthanum fluoride (5N purity) precursors in a flowing $H_2S$ gas system. It may be possible to use chlorides as precursors but researches who developed the GLS fabrication process[34, 70] chose to use gallium metal and lanthanum fluoride because they were the highest purity precursors available. The flowing system sweeps the by-products of the reaction out of the hot zone, preventing them from reacting, and by Le Chatelier's principle shifting the position of equilibrium to (as much as 100%) favour the sulphurised product.[34] This principle can be illustrated by imagining the reaction between gallium and hydrogen sulphide occurring in a sealed container, in this case gallium and hydrogen sulphide will react to give gallium sulphide and hydrogen but gallium sulphide and hydrogen will also react to give gallium and hydrogen sulphide, this reaction can be written: $Ga + H_2S \leftrightarrow GaS_x + H_2$ Both these reactions will occur until some equilibrium position occurs, now if the hydrogen is swept away there will be none to react with the gallium sulphide so the reaction will be: $Ga + H_2S \rightarrow GaS_{x} + H_2\uparrow$. So the equilibrium position of the reaction will now be (almost) 100% in favour of the sulphurised product. Lanthanum sulphide and gallium sulphide powders were processed at 1150 °C and 965 °C respectively. Before sulphurisation lanthanum fluoride was purified and dehydrated in a dry-argon purged furnace at 1250 °C for 36 hours to reduce OH⁻ and transition metal impurities. The lanthanum oxide and transition metal sulphides and oxides were purchased commercially and used without further purification.

Batching of the melt components was carried out in a dry nitrogen purged glovebox. The nitrogen source was cryogenic grade and was filtered by a molecular sieve and Millipore[TM] particle filter (0.5 µm). Moisture levels were routinely measured with the aid of a dewpoint meter and were < 1 ppm. Melt components were batched into vitreous carbon crucibles using plastic spatulas; the spatulas were changed for each material to minimise contamination. Melt components were weighed using a scale with a resolution of 0.001g. Batches were then transferred to the furnace using a custom built closed atmosphere transfer pod. All glass melt processes are carried out in a dry argon (moisture < 1 ppm) purged, silica lined furnace (Lenton LTF 6/50/610). This method was chosen in favour of the sealed ampoule method because volatile impurities such as OH⁻ are carried downstream away from the melt and because of safety concerns of the ampoules exploding. The precursors were melted at 1150°C for around 24 hours with an initial ramp rate of 20°C min⁻¹ and a constant argon flow of 200 ml min⁻¹. The melt is rapidly quenched to form a glass by pushing the crucible holder into a silica water jacket. The quenching process is designed to prevent crystallisation of the glass by rapidly increasing the viscosity of the glass through rapid temperature drop, hence arresting the nucleation and growth of crystals.

It was found that several attempts at melting glass failed due to the melt components not fully reacting when gallium sulphide was placed in the crucible first. Glass melts were more successful when gallium sulphide was placed in the crucible last.

TABLE 3.1. Melting point and density of various melt components[20].

| Melt component | Melting point(°C) | Density (kg m⁻³) |
|---|---|---|
| $Ga_2S_3$ | 1090 | 3700 |
| GaS | 965 | 3860 |
| $La_2S_3$ | 2110 | 4900 |
| $La_2O_3$ | 2315 | 6500 |

This is believed to be because gallium sulphide has the lowest melting temperature (as shown in table 3.1) of the melt components; so once melted the gallium sulphide will be able trickle through the other melt components and react with them more effectively if it is placed in last. The dopant was placed in-between the melt components to minimise the amount of dopant that could stick to the side of the crucible or react with the flowing argon atmosphere of the furnace. Hence melt components were added to the crucible in the following sequence: lanthanum oxide, lanthanum sulphide, transition metal dopant then gallium sulphide. The transition metals and their compounds, that were used to dope GLS glass for this study, are detailed in table 3.2

TABLE 3.2. Transition metals dopants and their compounds.

| Transition metal dopant | Compound | Supplier | Purity (%) |
|---|---|---|---|
| vanadium | $V_2S_3$, $V_2O_5$ | Cerac | 99.8, 99.9 |
| chromium | $Cr_2S_3$ | Cerac | 99 |
| titanium | $TiS_2$, $Ti_2S_3$ | Johnson Matthey | 99.95 |
| nickel | $NiS$ | Cerac | 99.9 |
| iron | $Fe_2O_3$ | Merck | 99 |
| cobalt | $CoS_3$, $CoS_2$ | Cerac | 99.5 |
| bismuth | $Bi_2S_3$ | Cerac | 99.9 |

## 3.2.2 Batching and melting details of vanadium doped GLS samples

Because vanadium doped GLS has been studied in more detail than other transition metal dopants in this study, the fabrication process for vanadium doped GLS samples is detailed in this section.

Initial samples of vanadium doped GLS were batched from the melt components in the molar percentages specified in table 3.3 the concentration of vanadium ions is also given.

TABLE 3.3. Molar percentages of melt components for initial vanadium doped GLS melts.

| Melt code | GaxSy (% molar) | $La_2S_3$ (% molar) | $La_2O_3$ (% molar) | $V_2S_3$ (% molar) | $V^{x+}$ (% molar) |
|---|---|---|---|---|---|
| LD1175-C | 64.55 | 30.38 | 5.054 | 0.0083 | 0.015 |
| LD1257-1 | 71.15 | 24.00 | 4.73 | 0.054 | 0.108 |
| LD1257-2 | 69.62 | 23.85 | 6.01 | 0.259 | 0.518 |
| LD1257-3 | 69.72 | 23.19 | 6.05 | 0.519 | 1.038 |

Given the 0.001 gram resolution of the weighing scales, which gave a molar % concentration resolution of around 0.015% for a 20 gram melt, a method was devised for giving greater control of the amount of dopant at low concentrations. To accomplish this a high doping concentration glass of around 2.5 % molar vanadium was melted first, this is referred to as the dopant glass. The dopant glass in the appropriate proportions was then re-melted with the melt components to produce low doping concentration glass. Two vanadium compounds with different oxidation states were used; $V_2S_3$ with vanadium in a 3+ oxidation state and $V_2O_5$ with vanadium in a 5+ oxidation state. This was done in order to determine if the initial oxidation state of the vanadium dopant would affect the optical properties of the glass. Copper and silver co-dopants have been shown to enhance the emission of vanadium doped CdS, ZnS CdSe and ZnSe by up to a factor of 10, this was attributed by the authors to copper and silver helping to incorporate vanadium in the trivalent state.[71] Copper and silver co-dopants were used with V:GLS in order to determine if this effect would be observed in the GLS system.

Tables 3.4-3.6 show that composition of the dopant glasses and the glasses they were used to produce. Vanadium ion concentration is given as $V^{x+}$ because as detailed in chapter 4, although the optically active ion is thought to be $V^{2+}$, vanadium exists in several oxidation states in the GLS system.

TABLE 3.4. Molar percentages of melt components for "dopant glasses".

| Melt code | GaxSy (% molar) | $La_2S_3$ (% molar) | $La_2O_3$ (% molar) | $V_2S_3$ (% molar) | $V_2O_5$ (% molar) | $V^{x+}$ (% molar) |
|---|---|---|---|---|---|---|
| LD1283-1 | 69.98 | 21.25 | 5.93 | 0 | 1.42 | 2.84 |
| LD1283-2 | 70.17 | 21.31 | 5.95 | 1.29 | 0 | 2.58 |
| LD1283-3 | 75.49 | 0 | 21.87 | 0 | 1.32 | 2.64 |
| LD1283-4 | 75.65 | 0 | 21.92 | 1.21 | 0 | 2.43 |

TABLE 3.5. Molar percentages of melt components for vanadium doped GLS glass doped with "dopant glasses" and co-doped with copper and silver.

| Melt code | GaxSy (% molar) | $La_2S_3$ (% molar) | $La_2O_3$ (% molar) | Dopant glass (% molar) | CuS (% molar) | $Ag_2S$ (% molar) | $V^{x+}$ (% molar) |
|---|---|---|---|---|---|---|---|
| LD1285-2 | 67.49 | 22.99 | 5.78 | LD1283-2 (3.64) | 0 | 0 | 0.0977 |
| LD1285-3 | 67.63 | 23.10 | 5.78 | LD1283-1 (3.39) | 0 | 0 | 0.0983 |
| LD1285-4 | 67.48 | 22.92 | 5.78 | LD1283-2 (3.63) | 0 | 0.05 | 0.0979 |
| LD1285-5 | 67.52 | 22.92 | 5.79 | LD1283-2 (3.63) | 0.05 | 0 | 0.0978 |

TABLE 3.6. Molar percentages of melt components for vanadium doped GLSO glass doped with "dopant glasses".

| Melt code | GaxSy (% molar) | $La_2S_3$ (% molar) | $La_2O_3$ (% molar) | Dopant glass (% molar) | $V^{x+}$ (% molar) |
|---|---|---|---|---|---|
| LD1284-1 | 74.619 | 0 | 21.652 | LD1283-3 (3.62) | 0.0970 |
| LD1284-3 | 74.384 | 0 | 21.59 | LD1283-4 (3.93) | 0.0963 |
| LD1284-4 | 76.871 | 0 | 22.299 | LD1283-4 (0.81) | 0.0199 |
| LD1284-5 | 76.915 | 0 | 22.318 | LD1283-3 (0.74) | 0.0199 |

## 3.3 Spectroscopic techniques

### 3.3.1 Absorption spectroscopy

When the frequency of light incident on a solid is resonant with an electronic transition between two energy levels intrinsic to the material an increase in absorption is observed. Hence an absorption spectroscopy can be used to probe the energy level structure of a material

Absorption spectra were taken on a Varian Cary 500 spectrophotometer which operates over a range of 175-3300nm with a resolution of ±0.1nm in the UV-VIS region and ±0.4nm in the infrared region. A basic schematic of the optics in the Cary 500 is shown in figure 3.1; the inset shows the chopper blade. One section of the chopper is mirrored and reflects the light beam, one section of the chopper is cut-out and allows the light beam to pass, and one section is matt black. This allows light to be directed alternately to the sample and reference beams. The matt black section of the chopper allows the grating to move to the next wavelength and enables correction for dark current. The system then compares the intensity of monochromatic light transmitted through the sample with the reference beam, 100% transmission and zero corrections are taken to account for system response and dark noise. Samples were held vertically in a V groove against an aperture 3mm in diameter, the reference beam also passed through an aperture 3mm in diameter.

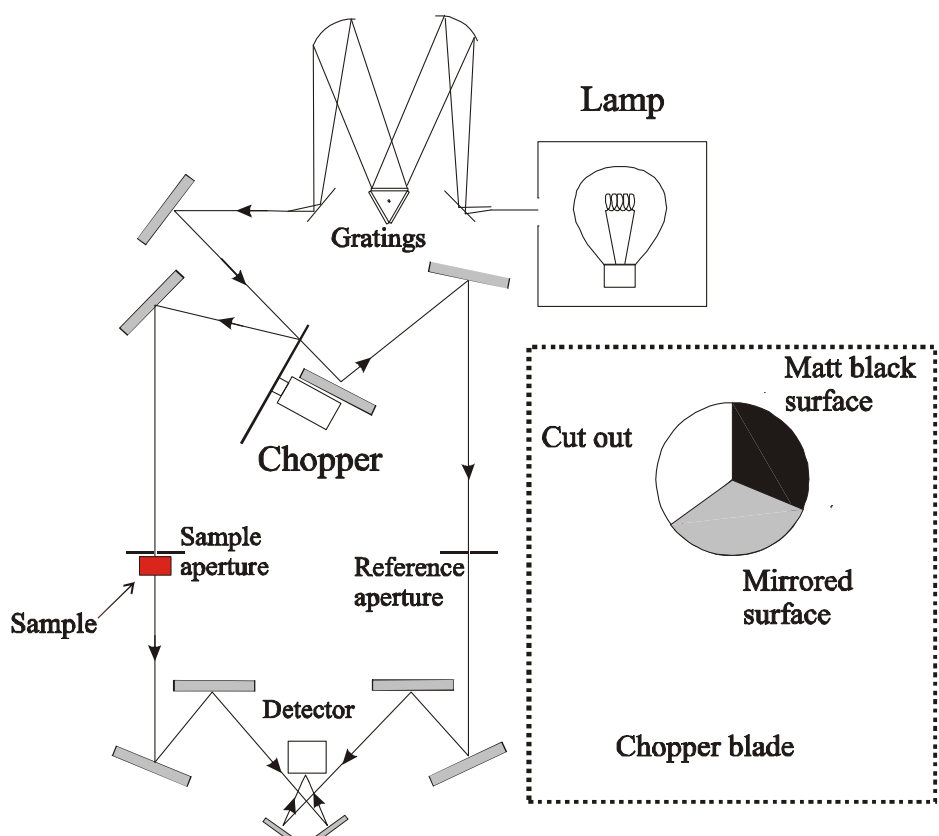

FIGURE 3.1 Basic schematic of the optics in the Varian Cary 500 spectrophotometer.

The Cary 500 uses two gratings to produce monochromatic light: a 1200 line/mm grating which is used when scanning 175 - 800 nm and a 300 line/mm grating which is used when scanning 800 – 3300 nm. Two lamps are available: a deuterium lamp (below 350 nm) and a tungsten halogen lamp (above 350 nm). The detectors used are a photomultiplier tube (below 800 nm) and a thermoelectric cooled PbS detector (above 800 nm). The change of grating and detector at 800 nm causes a small but sudden increase in transmission at 800 nm that was not always fully corrected for by the background scan. To overcome this, the increase in transmission observed was subtracted from wavelengths > 800 nm.

Samples that were cut and polished into thick and thin samples with thicknesses of $l_1$ and $l_2$ allowed reflection corrected absorption coefficient spectra to be calculated using equation 3.1.

$$a_{rc}(\lambda) = \frac{A_{l_1}(\lambda) - A_{l_2}(\lambda)}{l_1 - l_2}$$

(3.1)

Where $a_{rc}(\lambda)$ is the reflection corrected absorption coefficient spectrum and $A_{l_i}(\lambda)$ is the absorbance spectra of a sample of thickness $l_i$.

### 3.3.2 Photoluminescence spectroscopy

Similarly to absorption, photoluminescence can be used to probe the energy level structure of a material, however in the case of photoluminescence light is absorbed by the material (usually at one wavelength) exciting electrons (usually from the ground state) into higher energy levels. The electrons then relax back down to the ground state in a spin-allowed transition, called fluorescence, or in a spin-forbidden transition, called phosphorescence. Photoluminescence spectra were taken with the equipment setup illustrated in figure 3.2.

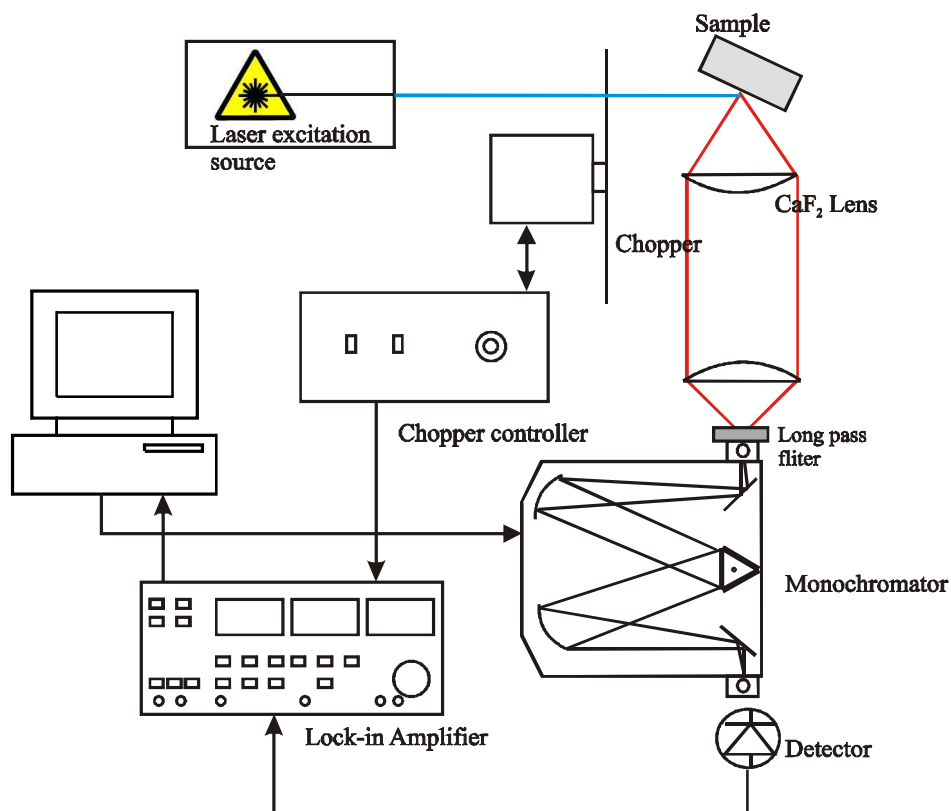

FIGURE 3.2. Photoluminescence spectroscopy equipment setup.

Various laser excitation sources were used to obtain photoluminescence spectra. These were: an Amoco Laser Company ALC D500 1064nm Nd:YAG laser, Thorlabs 830 808 and 658 nm diode lasers, 633 nm HeNe laser and the 514 nm line from a Spectraphysics Ar ion laser. The laser beam was then chopped with a Scitec optical chopper and a reference signal is sent to a Stanford research systems SR830 DSP lock-in amplifier in the form of a square wave reference signal. The laser beam then hits the sample at a shallow angle, such that most of the reflected laser light does not hit the $CaF_2$ lenses which are used to collimate and focus the fluorescence into a Bentham TMc300 monochromator. The TMc300 incorporates three gratings mounted on a single motorised turret, the gratings used were a ruled 1200 groove/mm, 600 groove/mm and a 150 groove/mm with useable wavelength ranges of 0.35-1.6μm, 1-2.7 μm and 2.5-6 μm respectively. The TMc300 has an aperture ratio of f/4.1, a resolution maximum of 0.1nm and a dispersion of 2.7nm/mm. The exit beam from the monochromator is picked up by a EG&G optoelectronics J10D liquid nitrogen cooled InSb detector which sends a reading to the lock-in amplifier via a preamp or by a Newport 818-IG InGaS detector. A long pass filter with an appropriate cut off wavelength was used to cut out the laser light but allow photoluminescence into the monochromator.

The system was corrected for the wavelength dependent response of the gratings, detectors and other system components by passing a Bentham IL6 quartz halogen white light source of known spectral luminance along the beam path taken by the luminescence; this gave a spectrum of the white light response of the system ($R_{wl}(\lambda)$). Care was taken to place the white light source as close as possible to the source of the photoluminescence, i.e. the sample, since it was found that if the beam path of the white light source was much larger than that of the photoluminescence features in the photoluminescence spectra caused by atmospheric absorption would be "over corrected", as illustrated in figure 3.3.

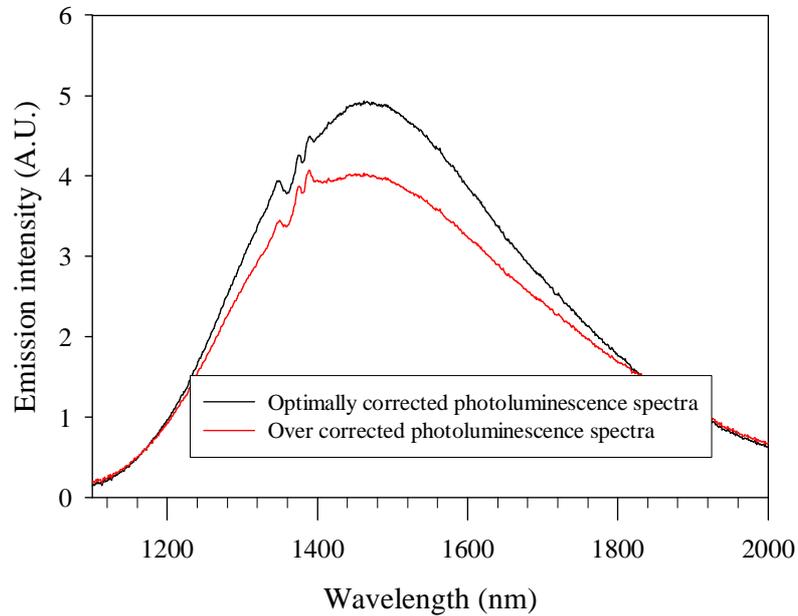

FIGURE 3.3. Optimally and over corrected photoluminescence spectra of vanadium doped GLS.

The intensity of the white light source was attenuated, using an aperture, so that it was as close as possible to the intensity of the spectra it was correcting for while still having a reasonable signal to noise ratio. This was done in order to minimise any deviation from linearity which was <0.5% for the Newport 818 IG and <1% for the JD10 InSb detectors; it also kept the detectors away from their saturation region where the response becomes nonlinear.

The spectrum of the white light source was obtained by approximating it to a black body emitter at its specified colour temperature of 3200 K. The correction spectrum (C($\lambda$)) was then calculated using equation 3.2.

$$C(\lambda) = \frac{2\pi hc^2 R_{wl}(\lambda)}{\lambda^5 e^{\left(\frac{hc}{k\lambda T}\right)}}$$

(3.2)

Where h is the Plank's constant, k is the Boltzmann constant, c is the speed of light, T is the temperature, which was set at 3200 K, and $R_{wl}(\lambda)$ is the white light response of the system. The correction curves, for the various system configurations used, are illustrated in figure 3.4.

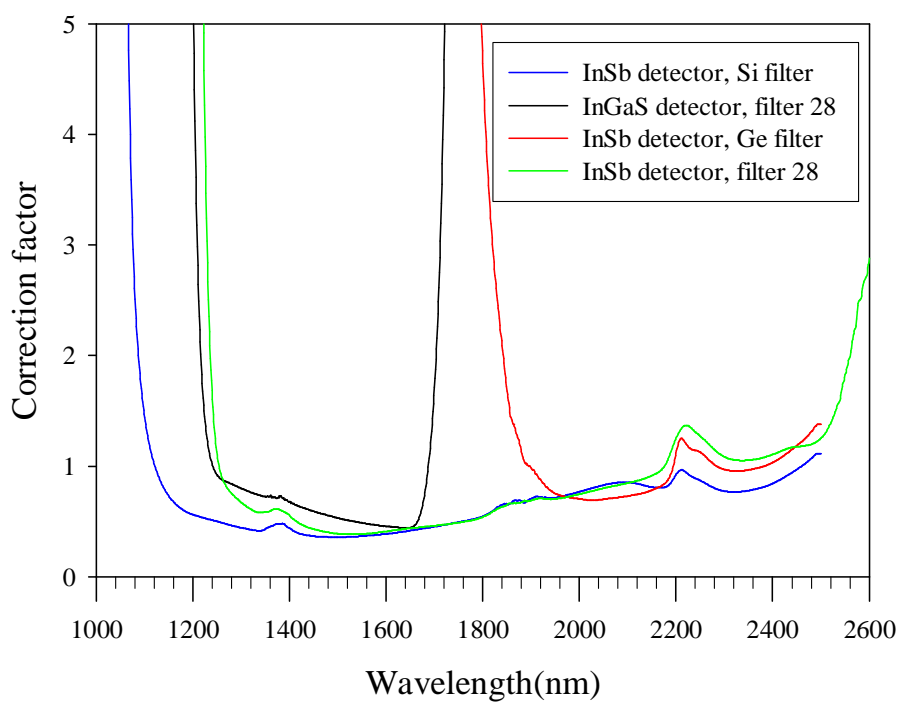

FIGURE 3.4. Correction spectra for various system configurations (all with 600 line/mm grating); filter 28 is a 1200 nm long pass filter.

### 3.3.3 Photoluminescence Excitation Spectroscopy

Normally, photoluminescence (PL) spectra are obtained at a fixed laser excitation energy. The photoluminescence excitation (PLE) spectrum is, however, obtained by detecting at a fixed emission energy and varying the energy of the exciting light source. In this way the absorption of the exciting light can be probed, but the signal detected is dependent on absorption which leads to emission. PLE spectra can elucidate absorbing transitions that are obscured in absorption spectra because the detected radiation can be at an energy which is transmitted by the sample and weakly fluorescing absorptions such as a band-edge are not detected. PLE spectra also indicate which excitation energies will give rise to photoluminescence at the energy being detected in the PLE spectra.

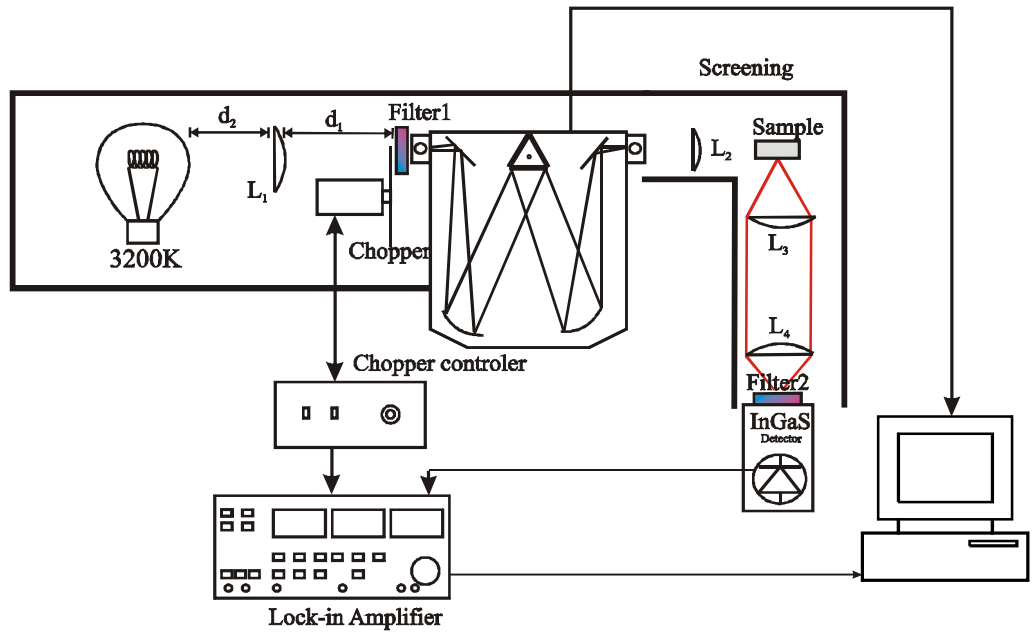

FIGURE 3.5. Photoluminescence Excitation spectroscopy equipment setup.

Photoluminescence excitation spectra were obtained using the setup shown in figure 3.5. White light from a Philips 250 W tungsten halogen bulb was focused by a 40 mm diameter silica lens ($L_1$) into an Acton Spectrapro 300i monochromator which dispersed the light to give a variable energy exciting light source. The exciting light was focused onto the edge of the sample at 90º by $L_2$ and fluorescence was collimated and focused by CaF$_2$ lenses $L_3$ and $L_4$ onto a Newport 818IG InGaS detector. Filter 2 was a silicon or 1400 nm long pass filter that was placed in front of the InGaS detector to give an effective detection range of 1000 – 1700 nm or 1400 – 1700 nm respectively. Filter 1 was a 715 nm long pass colour glass filters that was placed in front of the monochromator when scanning wavelengths longer than 750 nm to cut out second order light; from absorption measurements the response of the filter was taken to be flat. When taking initial measurements it was found that the monochromator transmitted light in the detection range, 1000 – 1700 nm or 1400 – 1700 nm, when scanning the excitation range of 400 – 1400 nm. This caused a strong intrinsic background signal in the measurement that could not be corrected for. The origin of this background signal is believed to be imperfections in the optics of the monochromator and reflections from internal surfaces which were detectable because of the high intensity of light incident on the monochromator. In order to minimise this background signal the input cone of light entering the monochromator was matched to its aperture ratio or F/#; this meant that distances $d_1$ and $d_2$ became the critical design parameters for minimising this background signal. The monochromator has an acceptance cone which is determined by the area of the grating and the gratings path length from the entrance slit. The acceptance cone is usually described by an F/# which is defined as the path length from the entrance slit to the grating divided by the diameter of an "equivalent circle" with the same area as the square grating. The half angle ($\theta_{1/2}$) of the acceptance cone of a monochromator can be calculated from its F/# using equation 3.3.

$$\theta_{1/2} = \sin^{-1} \frac{1}{2nF/\#}$$
(3.3)

Where n is the order of diffraction. The Acton Spectrapro 300i has a F/# of 4 from which a $\theta_{1/2}$ of 7.2° was calculated. Adjusting $d_1$ to give $\theta_{1/2}$ = 7.2° still resulted in interference. In order to optimise the system experimentally, the setup shown in figure 3.6 was arranged. The monochromator was fixed at 590 nm this wavelength was detected by a Newport 818-SL silicon photo-diode from a reflection off a microscope slide, to give the intensity of the excitation ($I_{ex}$). The background signal ($I_{back}$) was detected by a Newport 818IG InGaS detector fronted by a silicon filter; $d_1$ and $d_2$ were then varied to maximise $I_{ex}/I_{back}$. The optimal setting for $d_1$ and $d_2$ was found to be 250 and 350 mm respectively; this corresponded to $\theta_{1/2}$ and F/# of 4.6° and 6.3 respectively. The reason why the interference was minimised when the acceptance cone is under-filled may be because the specified F/# for the monochromator is calculated for an equivalent circle which would have sections of light missing the grating.

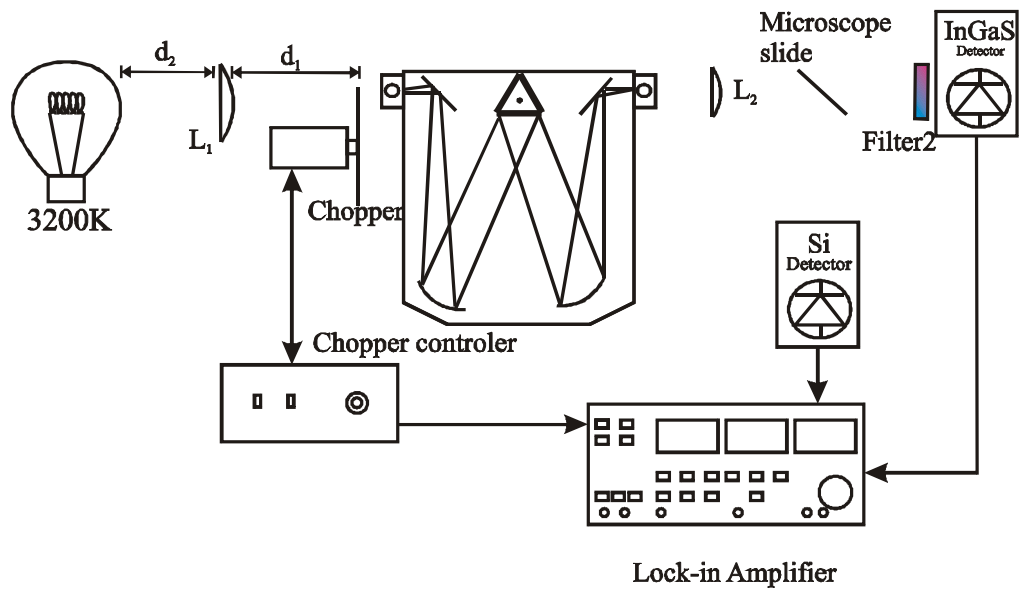

FIGURE 3.6. Experimental setup for optimisation of system interference.

Another source of interference was found to be scattered chopped white light that was detected by the detector which was in free space. To minimise this, black card screening was erected around the optics.

The PLE system was corrected for the varying intensity of exciting light, due to varying grating response and spectral output of the white light source, by obtaining the white light response ($(R_{wl}(\lambda)$) for each grating with Newport 818-SL and 818-IG detectors. The correction curves for each grating were then calculated using the calibration report supplied with the detectors. The calibration reports gave the responsivity of the detectors in steps of 10 nm. The spectral responsivity data was entered onto a computer and smoothed with a running average algorithm, with a sampling proportion of 0.05, to give detector responsivity spectra ($R_{det}(\lambda)$) in steps of 1 nm. The correction spectra ($C(\lambda)$) for each grating were then calculated using equation 3.4. The grating correction spectra are given in figure 3.7.

$$C(\lambda) = \frac{R_{wl}(\lambda)}{R_{det}(\lambda)} \qquad (3.4)$$

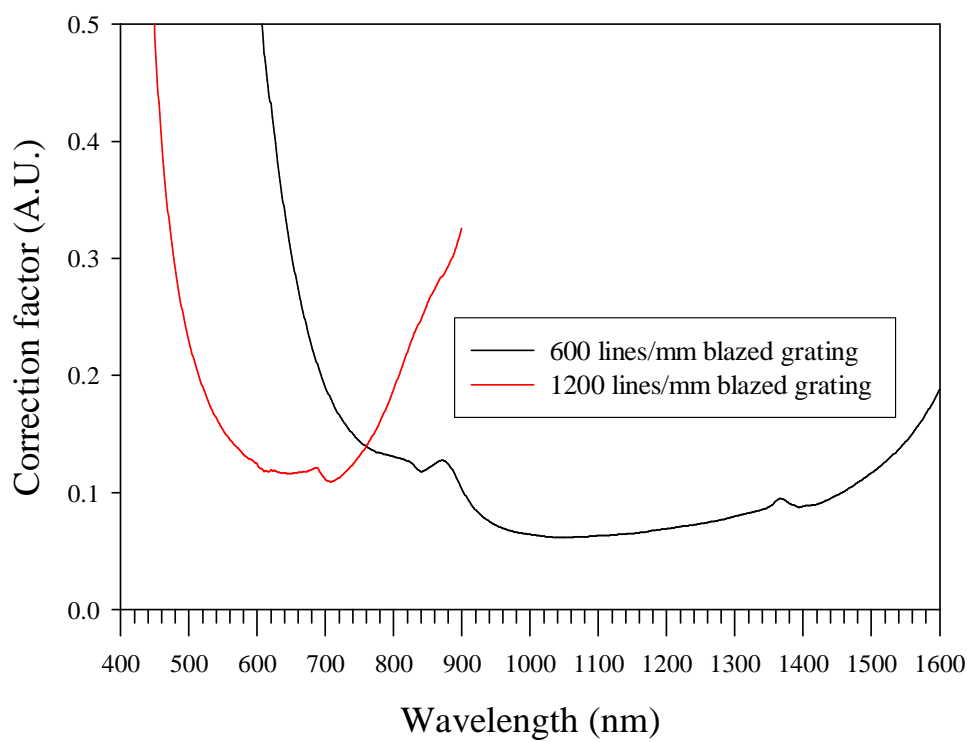

FIGURE 3.7. Correction spectra for gratings used in PLE measurements.

In order to test if the PLE system would give valid results, a sample of neodymium doped GLS was scanned. The results are shown in figure 3.8, along with the absorption of neodymium GLS; as expected for a valid scan the PLE peaks match the absorption peaks. Note that an absorption band centred at 540 nm, that could not be clearly identified in the absorption spectrum because of its proximity to the band-edge, has been elucidated in the PLE spectra. This is because the PLE system detects emission from incident light at a wavelength that is transmitted by the glass. A very weak absorption band at 700 nm is also much clearer in the PLE spectrum.

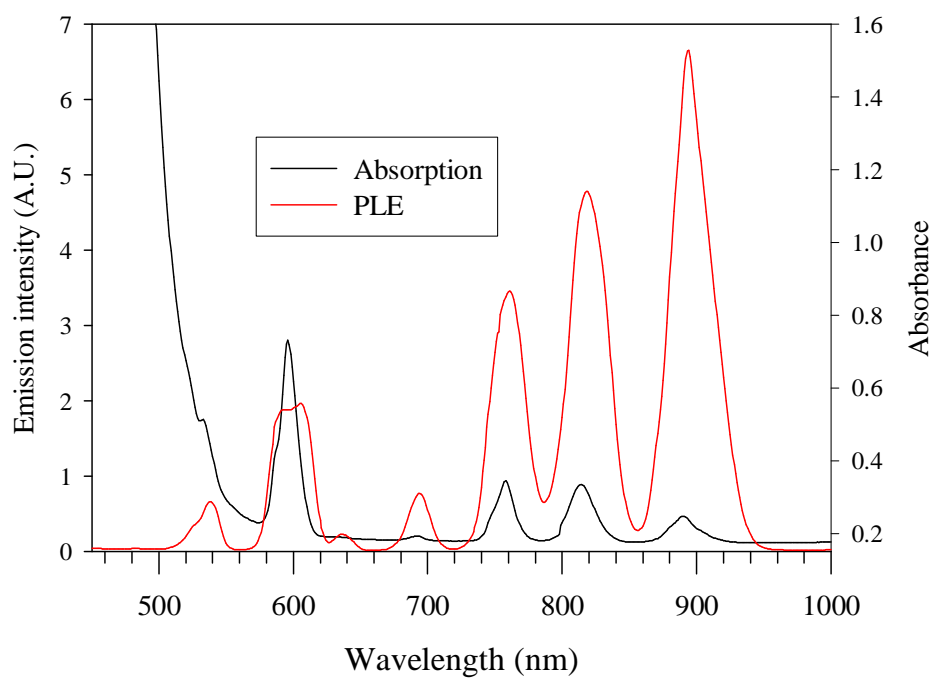

FIGURE 3.8. Absorption and PLE spectra of neodymium doped GLS.

### 3.3.4 Temporally resolved fluorescence lifetime

Fluorescence lifetime is defined as the time taken for the intensity of fluorescence to decay to $1/e \approx 0.368$ of its initial value. Fluorescence lifetime measurements were taken using the setup illustrated in figure 3.9.

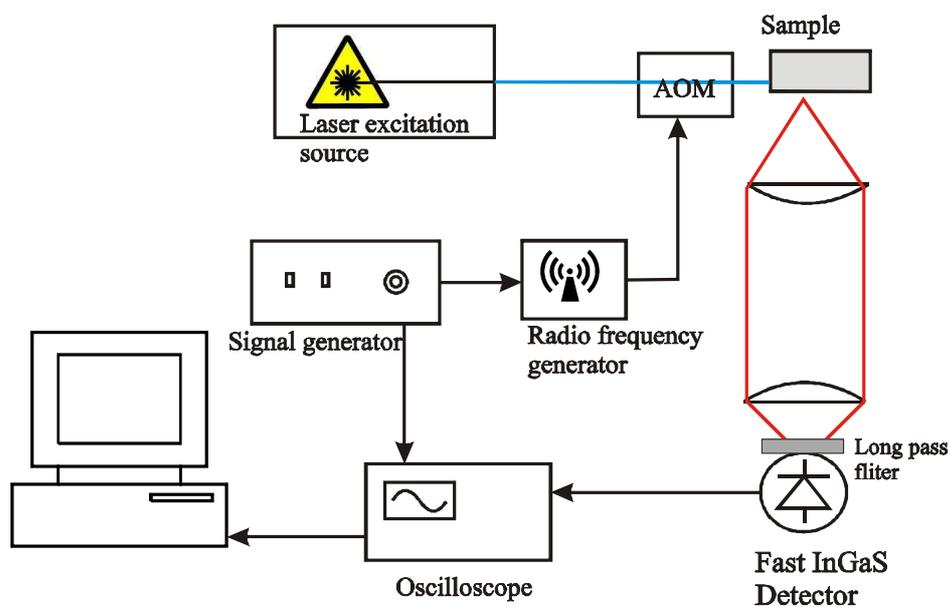

FIGURE 3.9. Fluorescence lifetime equipment setup.

Temporally resolved fluorescence lifetime (TRFL) measurements were obtained by exciting transition metal doped GLS samples with various laser excitation sources. The excitation sources were modulated by a Gooch and Housego M080-1F-GH2 acousto-optic modulator (AOM). The modulation signal was generated by a Thurlby Thandar TG230 2MHz sweep/function generator which then activated a Gooch and Housego A103 radio frequency generator. The excitation beam was attenuated, to give an incident power on the sample of around 10mW, in order to minimise heating. The fluorescence was detected with a New Focus 2053 InGaS detector which was set to a gain that corresponded to a 3dB bandwidth of 3MHz. The data was captured by a Picosope ADC-212 virtual oscilloscope with a 12 bit intensity resolution and a maximum temporal resolution of 700 ns, the signal was averaged for around 2 minutes to improve signal to noise ratio in the measurement. Lifetime measurements were taken several times at different alignments in order to give an estimate of random experimental error.

### 3.3.5 Frequency resolved fluorescence lifetime

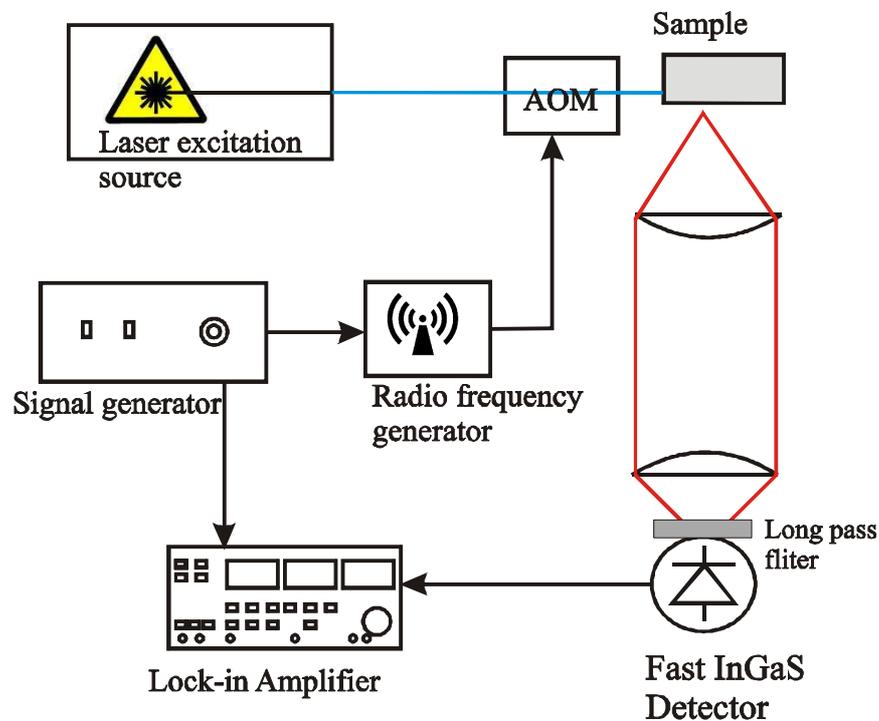

FIGURE 3.10. Frequency resolved fluorescence lifetime equipment setup.

An alternative technique to using an oscilloscope to make fluorescence lifetime measurements is frequency resolved fluorescence lifetime measurements (FRFL). For this the intensity of the fluorescence is measured by standard phase sensitive detection with a lock in amplifier and the modulation frequency of the excitation source is varied. The principle behind this technique is that once the modulation frequency matches the fluorescence lifetime of the active ion, the intensity measured by the lock-in amplifier will start to fall. It can be shown[72, 73] that once the fluorescence power has fallen by 3dB the modulation frequency ($\upsilon_m$) will be related to the fluorescence lifetime ($\tau_f$) of the active ion by equation 3.5.

$$\upsilon_m = \frac{1}{2\pi\tau_f} \tag{3.5}$$

Figure 3.10 shows the experimental setup used to take these measurements. The laser excitation was modulated by an acousto optic modulator (AOM). The modulation frequency was varied manually on the AOM controller and the value of the fluorescence intensity and the modulation frequency were read from the display on the lock-in amplifier and recorded in a log book. This technique was used in order to verify the lifetime measurements taken using an oscilloscope. The advantage of this technique is that the phase sensitive detection allows measurement of signals far weaker than can be detected with a transient system. The disadvantage is that the transient decay profile cannot be deduced since there is no solution to the Fourier transform of a stretched exponential which was found to describe the decay profiles in this work and is given in more detail in chapter 4.

### 3.3.6 Raman spectroscopy

The Raman effect occurs as a result of inelastic photon scattering by a substance. When light (usually at one wavelength) is incident on a medium most of it is scattered at the same wavelength in an elastic process called Rayleigh scattering. Raman spectra consist of Stokes and anti-Stokes lines which are symmetrical about the Rayleigh line. Stokes Raman scattering occurs as a result of photon absorption to a virtual state, followed by relaxation to a higher order phonon level. In anti-Stokes Raman scattering a photon is absorbed from a higher order phonon level to a virtual state followed by depopulation to the ground state.

Raman spectra were taken with a Renishaw Ramanscope with a 10mW 633 nm HeNe laser and a 50× objective lens. A basic schematic of the optics in the Renishaw Ramanscope is shown in figure 3.11. Light from a 633 nm laser is reflected by a holographic 633 nm line reject filter onto the sample via a 50X microscope objective. Light is then reflected from the sample back onto the line reject filters which allow Stokes and anti-Stokes shifted light into the spectrometer but rejects Rayleigh scattered light. The spectrometer consists of a diffraction grating and CCD camera which allows continuous and static scans to be taken. The slit width of the spatial filter in front of the diffraction grating and the width of the detection elements on the CCD camera dictated the resolution of the system. In a continuous scan the grating rotates allowing the measurement of a range of wavelengths, at a resolution of 4 cm$^{-1}$. In a static scan the grating remains stationary and the array of detectors in the CCD camera allows a scan, with a low resolution and wavelength range, to be taken in around 1 s.

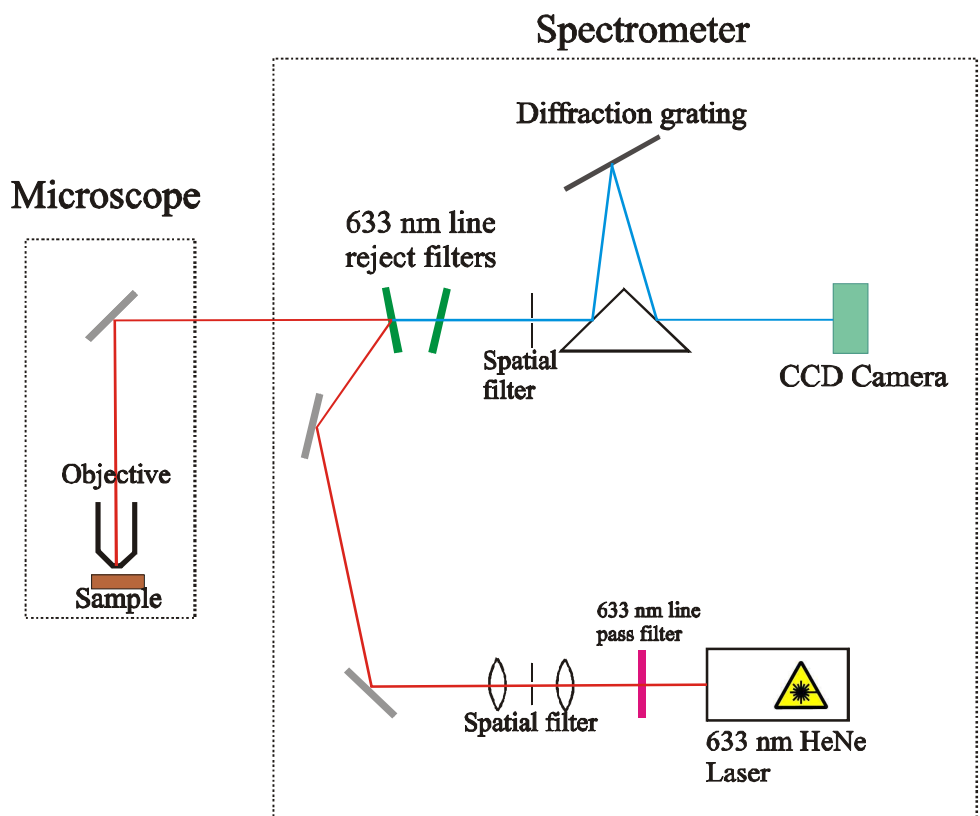

FIGURE 3.11 Schematic representation of micro Raman system.

## 3.3.7 Quantum efficiency

The quantum efficiency (QE) was measured by taking spectra of both the fluorescence and the excitation laser using standard phase sensitive detection, with a lock-in amplifier and monochromator. The important difference compared to standard fluorescence measurements, was that the sample was placed inside an integrating sphere. An integrating sphere is a hollow sphere coated on the inside with a material that reflects diffusely and has a high reflectivity. The effect of this is that any light entering, or produced within, the sphere becomes distributed isotropically, therefore the flux received at an aperture in the sphere is proportional to the total amount of light entering or produced within the sphere. The integrating sphere used was a Labsphere model FM-040-SF which has an interior coating of a highly reflective white paint ("Spectraflect") and was around 15cm in diameter. A custom built sample holder was produced consisting of a small crocodile clip attached to the end of a brass thread, this went through a threaded and polished aluminium plate which was clamped to one of the ports of the integrating sphere. This sample holder allowed fine adjustment of the samples' position in the integrating sphere by turning the brass thread. Because of the sample holders' small surface area and relatively high reflectivity, it was not thought to significantly degrade the performance of the integrating sphere. The baffle was also moved from its original position on the inside surface of the sphere to one where it covered the exit aperture, so that there was no direct line of sight from the sample to the detector. The original baffle position was found to give anomalous results, thought to result from laser light being scattered from the sample directly into the exit aperture. The setup used is illustrated in Figure 3.12.

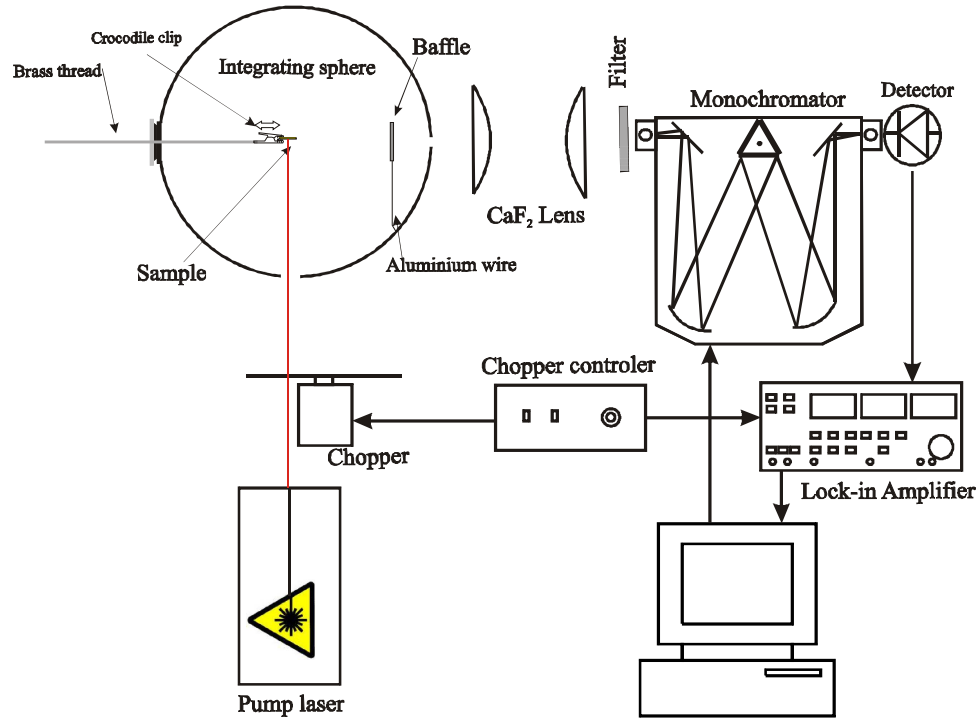

FIGURE 3.12. Quantum efficiency measurement setup.

A "photons out / photons in" method, similar to that described by,[74, 75] for calculating the quantum efficiency was used. This consisted of comparing the area under the fluorescence spectra to that of the laser line with and without the sample in place. Spectra were first corrected for the spectral response of the detection system, by passing a halogen white light source of know spectral luminance through the entrance port of the integrating sphere the correction spectra ($C(\lambda)$) was then calculated using the method described in section 3.3.2. The spectra were also corrected for the photon energy since a higher photon flux is required at longer wavelengths to produce the same irradiance per unit area than at shorter wavelengths. So the number of photons (n) detected at a particular wavelength is proportional to the measured irradiance $I(\lambda)$ multiplied by the wavelength ($\lambda$), $n=a\lambda I(\lambda)$, where a is a constant of proportionality. The number of photons absorbed was taken to be proportional to the difference between the area under the corrected laser line spectra with the sample present ($I_{sample}(\lambda)$) and without the sample present ($I_{sphere}(\lambda)$). The number of photons emitted was taken to be proportional to the area under the corrected emission spectra ($I_{PL}(\lambda)$). Hence the quantum efficiency ($\eta_{QE}$) was calculated from:

$$\eta_{QE} = \frac{\int \lambda I_{PL}(\lambda) C(\lambda) d\lambda}{\int \lambda I_{sphere}(\lambda) C(\lambda) d\lambda - \int \lambda I_{sample}(\lambda) C(\lambda) d\lambda} \qquad (3.6)$$

Spectra of the emission were taken in steps of 1nm and spectra of the laser line were taken in steps of 0.2nm, to improve the accuracy of the area measurement. The correction spectra used to correct the laser line spectra were corrected for the response of the filter used to cut out second order light, when the white light response was taken, since this filter was not present when the laser line spectra were taken. QE measurements are used in section 4.12.

## 3.3.8 X-ray Photoelectron Spectroscopy

Determination of the oxidation state of an active ion dopant is an important part of the characterisation of a material being considered for optical device applications because it determines which energy levels exist within this ion. Knowledge of the oxidation state is therefore needed when modelling the radiative and non-radiative transitions that occur in an optical material.

X-ray photoelectron spectroscopy (XPS) can be used to determine the oxidation state of an active ion dopant. XPS involves exposing a sample to a monochromatic X-ray source which liberates core electrons at energies directly proportional to their binding energy in a phenomenon originally described as the *photoelectric effect*. If the sample is irradiated with photons of frequency $\upsilon$, the energetics of the process are defined by the Einstein relation,[76]

$$h\nu = I_k + E_k + \phi \qquad (3.7)$$

$I_k$ is the binding energy of the $k$th species of electron in the material, $E_k$ is the kinetic energy at which the electron is ejected and $\phi$ is the workfunction of the material. An energy spectrum of these electrons, called photoelectrons, shows peaks at energies which can be ascribed to the elemental constituents of the sample. Small shifts in the peaks can be used to identify the oxidation state of the elements, since higher oxidation states increase the relative nuclear attractive force on the core electrons and hence increases their binding energy.

Photoelectrons travelling through a solid material have a relatively high probability of experiencing inelastic collisions with locally bound electrons, which results in energy loss. The attenuation of the photoelectron flux through inelastic scattering can be described as follows. If $I_0(x)$ is the photoelectron flux (at a particular electron kinetic energy E) originating at a depth x below the surface of the solid, the flux I(x) emerging at the surface is given by:

$$I(x) = I_0(x)e^{-(x/\lambda_a)} \qquad (3.8)$$

Where $\lambda_a$ is referred to as the attenuation length of an electron and represents the depth from which 1/e photoelectrons produced there can escape. The attenuation length for most solid materials is around 20Å.[77] XPS is therefore a highly surface sensitive technique.

A sample of 1% molar vanadium doped gallium lanthanum sulphide glass, which had dimensions of 0.5x5x5mm and was highly polished on both faces, was placed into a vacuum chamber as illustrated in figure 3.13.

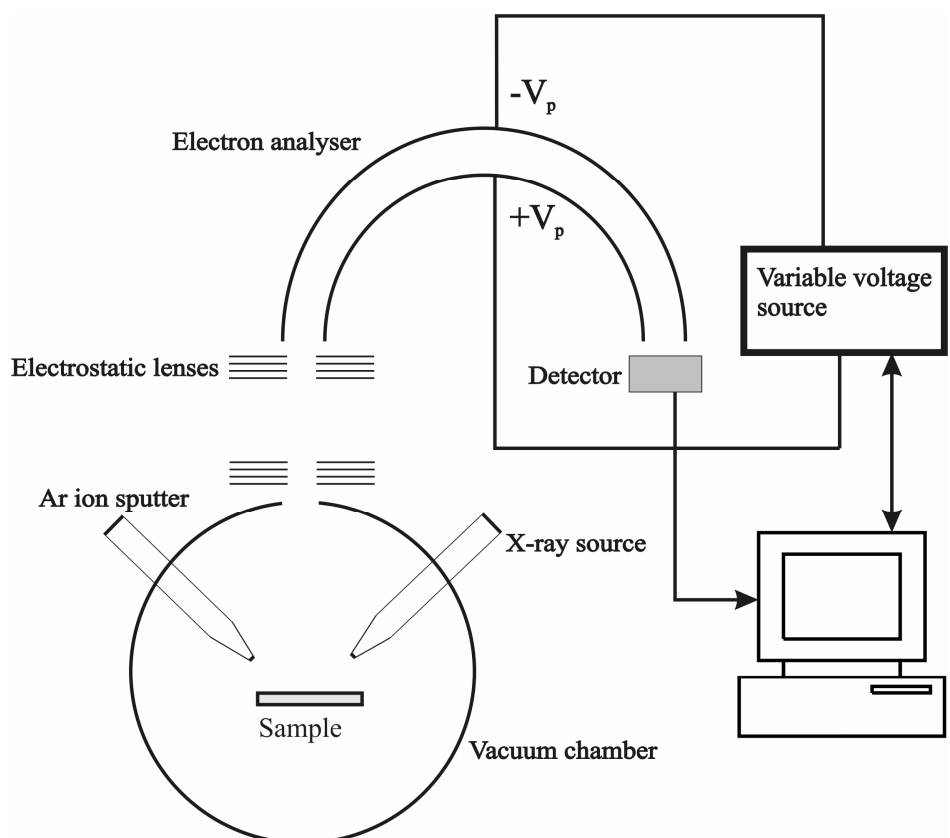

FIGURE 3.13. X-ray photoelectron spectra equipment schematic.

The vacuum chamber was evacuated to around $10^{-9}$ mbar and the sample was exposed to X-ray radiation from an Mg Kα anode source centred at 1253.6 eV. The FWHM of the anode source limited the resolution of the photoelectron spectra to 1eV. Higher resolution can be obtained using a monochromated X-ray source, however this comes at the expense of counts, due to the larger source-sample separation. An initial photoelectron spectrum was taken and no vanadium signature was detected but around 75% carbon was detected. This was believed to be a surface effect caused by polishing. The sample was then sputtered for 30 minutes with a 5 keV Ar ion sputterer in order to remove any surface contamination, photoelectron spectra indicated no vanadium to be present and a high carbon signature. Further photoelectron spectra taken after 1 and 2 hours of sputtering indicated that no vanadium was present and a high carbon signature. It is believed that the effects of polishing penetrated the sample further than can be practically removed by sputtering. So a sample was then fractured by tapping sharply with a razor blade. It was noted that fractures were conchoidal, typical of fractures in brittle materials that have no natural planes of separation. After several attempts a roughly flat face of around 5x5mm was obtained, this was placed immediately into the vacuum chamber and evacuated in order to minimise any atmospheric reaction. The sample introduction chamber was pumped for ~12 hours to allow for the sample out-gassing. A weak vanadium signature was found, in order to obtain good signal to noise, a spectrum was collected over an 8 hour period. XPS measurements are used in section 4.13. XPS measurements and analysis where carried out under the supervision of Dr. N. Blanchard using facilities of the Advanced Technology Institute, University of Surrey.

### 3.3.9 Electron paramagnetic resonance

All atoms have an inherent magnetism because electron spin contributes a magnetic moment. In most atoms the magnetic moment of paired electrons with opposite spin cancel each other out, but in atoms containing an unpaired electron the cancellation is incomplete, materials containing these atoms are classified as paramagnetic. Electron Paramagnetic Resonance (EPR) is a technique which is used to detect the presence of unpaired electrons in chemical species. The various oxidation states of a transition metal ion often give a unique EPR fingerprint which can be used as a qualitative identification of the oxidation states present in a particular glass system.

In the presence of a high intensity magnetic field (B) the spin axis of an unpaired electron will precess around the field direction at the Larmor frequency, as illustrated in figure 3.14.

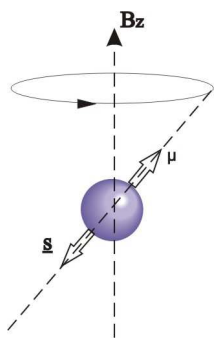

FIGURE 3.14. Precessing electron spin.

The electron spin can now be orientated either parallel or anti parallel to the applied magnetic field. This creates distinct energy levels for the unpaired electron as shown in figure 3.15. If a microwave field at the Larmor frequency is applied to a paramagnetic sample, then the spins can change their direction relative to the magnetic field. This results in the absorption of microwave radiation with energy hυ, which can be measured. The energy separation between the two levels (ΔE) at a magnetic field strength B is expressed as ΔE = hυ = gμB, where g is the gyromagnetic ratio which is defined at the ratio between the electrons magnetic dipole moment and its angular momentum and μ is the Bohr magneton.

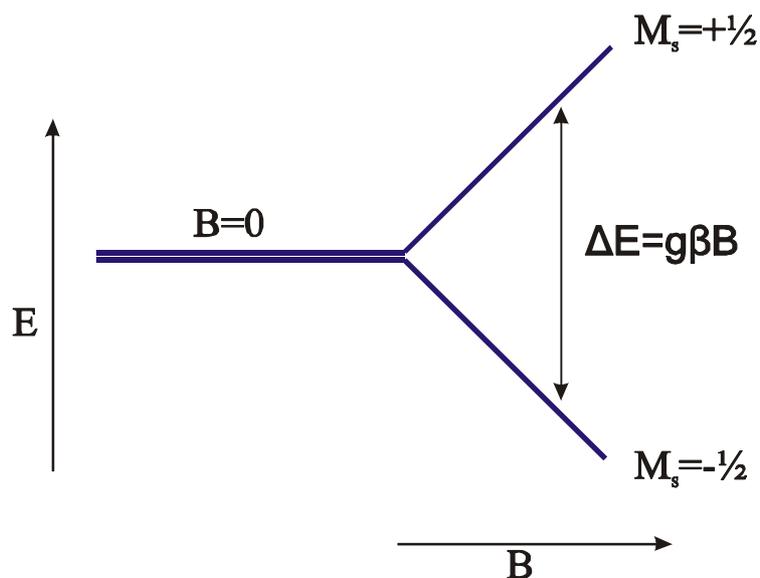

FIGURE 3.15. Energy-level diagram for two spin states as a function of applied field B.

Figure 3.16 shows the block diagram for a typical EPR spectrometer. The klystron generates narrow band microwaves at a frequency which is largely determined by the strength of the magnet. The microwaves are then adjusted to the required intensity using the attenuator. The microwaves then enter the circulator and are routed toward the cavity where the sample is mounted. The microwaves are then reflected back from the cavity (less when power is being absorbed) and are routed to the diode detector. Any power reflected from the diode is absorbed completely by the load. In order to take EPR spectra the intensity of the magnetic field is swept from low to high and plotted against absorbed microwave power.

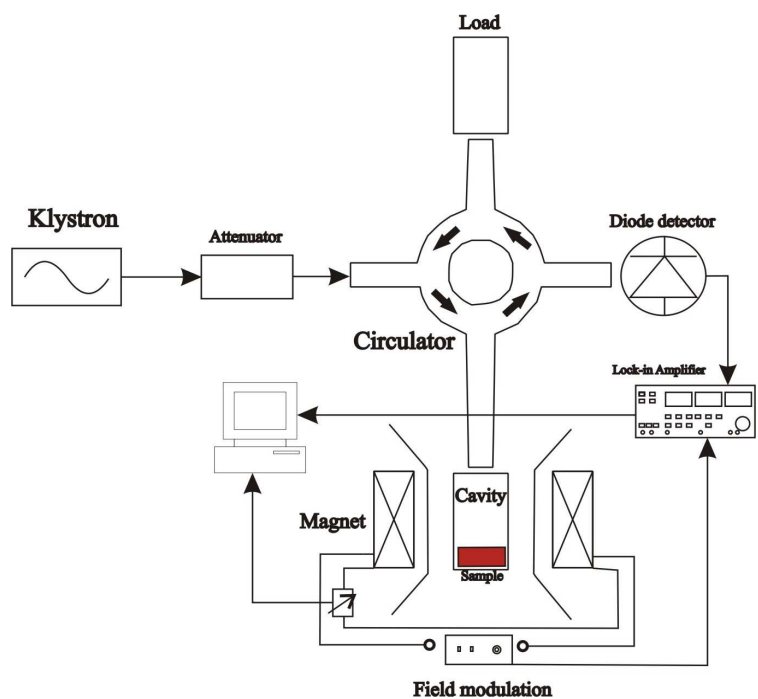

FIGURE 3.16. Schematic block diagram for a typical EPR experimental setup.

In practice this technique alone gives a very poor signal to noise ratio. However the signal to noise ratio can be greatly improved by introducing a small amplitude magnetic field modulation superimposed on the large d.c. magnetic field by small magnetic coils. This allows the use of phase sensitive detection which selects the a.c. component of the diode current. The detected a.c. signal is proportional to the change in sample absorption which amounts to detection of the first derivative of the absorption curve. EPR measurements of vanadium doped GLS were taken on two different systems the first operated with H-fields of around 0.3 Tesla, corresponding to microwave frequencies of about 10GHz (X band). The second was a "high field EPR" instrument operating with H-fields of up to 8 Tesla, corresponding to microwave frequencies of 80-300 GHz. The high field instrument had a sensitivity and resolution which is between 100 and 100,000 times better than the X-band system.[78] The samples required no special preparation but needed to have dimensions of 1-5mm diameter, hence all samples measured were irregular shaped pieces with a diameter of around 3mm. EPR measurements are used in section 4.14. EPR measurements where carried out using facilities of the School of Physics and Astronomy, University of St-Andrews, by  Dr. Hassane El Mkami.

## 3.3.10 Summary of spectroscopic techniques

Table 3.7 gives a summary of all the spectroscopic techniques used in this work. Including the application they were used for, the range of the controlled variable and its resolution.

TABLE 3.7. Summary of spectroscopic techniques used in this work.

| Spectroscopic technique | Application in this work | Range of controlled variable used in this work | Resolution of controlled variable used in this work |
|---|---|---|---|
| Absorption | Determination of energy levels | 400 to 3300 nm | ~0.4 nm |
| Photoluminescence (PL) | Calculation of Stokes shift. Determining the possible wavelength range of an active optical device. | 800 to 1800 nm | ~5 nm |
| Photoluminescence excitation (PLE) | Elucidating absorbing transitions obscured in absorption spectra. Determining the optimal excitation wavelength | 400 to 1800 nm | ~5nm |
| Temporally resolved fluorescence lifetime (TRFL) | Determination of lifetime. Determination of decay profile. | 0 to 4 ms | 700 ns |
| Frequency resolved fluorescence lifetime (FRFL) | Compliment to TRFL. Detection of signals too weak for TRFL | 10 to 100,000 Hz | 4 significant figures |
| Raman spectroscopy | Determining structural modifications | 30 to 2000 $cm^{-1}$ | ~4 $cm^{-1}$ |
| Quantum efficiency (QE) | Determining suitability for active optical devices. Calculation of emission cross-section | 1000 to 1750 nm | ~5 nm |
| X-ray photoelectron spectroscopy (XPS) | Determining oxidation state of the vanadium ion in GLS | 505 to 545 eV | 1 eV |
| Electron paramagnetic resonance (EPR) | Compliment to XPS | 0 to 10 kG | Unspecified |

# Chapter 4

# Vanadium doped chalcogenide glass

## 4.1 Introduction

Transition metals have long been used as the active ion in solid state lasers. In fact, laser action was first demonstrated in 1960 in a system which used $Cr^{3+}$ doped $Al_2O_3$ (ruby) as the active medium.[79] Since then a number of transition metal ions have been used in tuneable laser sources, the most successful of which has been the $Ti^{3+}$ doped $Al_2O_3$ (Ti:Sapphire) laser which is tuneable from 690-1100 nm[19] and has been used to generate ultra short laser pulses with durations of 8 fs.[80] Transition metal ions in glasses are usually regarded as a nuisance because they can exhibit a strong and broad absorption, even at sub ppm concentrations, which can seriously degrade the performance of low loss fibre and active optical devices. Transition metal doped glasses are not often considered for active optical devices because they usually have low quantum efficiencies and short lifetimes which results in a high pump threshold. Because of this most publications reporting the spectroscopy of transition metal doped glasses concentrate on absorption in order to study optical loss mechanisms in the glass, such as in silica fibres[81] and fluoride fibre.[82, 83] The only demonstration of lasing in a non rare-earth metal doped glass know to the author is in a bismuth doped aluminosilicate fibre laser.[84]

Transition metal doped GLS including vanadium, chromium, nickel, iron, copper and cobalt has previously been studied by Petrovich who investigated the effect of transition metal ion impurities on the infra red absorption of GLS.[24] The absorption of titanium, vanadium, chromium, nickel, iron, copper and cobalt doped GLS has been reported by Brady.[70] Chromium doped chalcogenide glasses has been studied by Haythornthwaite who investigated these glasses as potential materials for broadband amplification.[85] Optical characterisation of transition metal doped chalcogenides, including V:GLS, was carried out by Aronson.[23] Aronson suggested that vanadium was in a 3+ oxidation state and tetrahedrally coordinated based on comparisons of the optical properties of V:GLS to that of vanadium in other hosts. With the benefit of further measurement and analysis, Aronson's assignment is revised to octahedral $V^{2+}$ in this work. There have been relatively few publications reporting the spectroscopy of other vanadium doped glasses. Some of these include the absorption of vanadium in zirconium fluoride glass,[82] silica glass[81] and sodium silicate glass.[86] Even rarer are reports of emission from vanadium doped glass, one of the few examples being vanadium doped phosphate glass.[87] Most of the comparisons made in this chapter are to vanadium doped crystals. Laser action at low temperatures has been reported in $V^{2+}$ doped $CsCaF_3$[6] and $V^{2+}$ doped $MgF_2$.[7]

Unlike rare earth ions the optically active orbitals of transition metals are not shielded from the surrounding glass ligands. Because of this the optical properties of transition metal ions in glass is strongly affected by the local bonding environment experienced by the ion, including the ligands nature, distance from the ion, coordination and symmetry. This fundamental difference with rare earths makes transition metals less suitable for active optical devices in certain respects. However this means that the optical characterisation on transition metals can be used to deduce more information about the local bonding environment in the glass; which, negating optical device applications, justifies characterisation of transition metal doped glass.

This chapter presents a rigorous optical characterisation of vanadium doped GLS glass (V:GLS) in the form of absorption, photoluminescence (PL), photoluminescence excitation (PLE), emission lifetime, quantum efficiency, X-ray photoelectron spectroscopy (XPS) and electron paramagnetic resonance (EPR) measurements. As described later in the chapter vanadium exists as a broad range of oxidation states in GLS with $V^{2+}$ being the only optically active oxidation state. The proportion of the vanadium content that exists as $V^{2+}$ is unknown. Because of this two concentrations are reported for each sample. In the first three figures (4.1-4.3) the batched vanadium content is quoted which is calculated from the weight of elemental vanadium dopant incorporated with the melt components. All other concentrations quoted are a relative vanadium content which is calculated from the intensity of the vanadium absorption normalised to the highest doping concentration used. The suffixes to the vanadium dopant used in this chapter refer to the oxidation state of the vanadium dopant, not the oxidation state in which it exists in the glass system. These are V(V) for $V_2O_5$ and V(III) for $V_2S_3$. Where analyses in terms of Gaussians are used spectral units are give in energy.

## 4.2 Absorption measurements

Absorption spectroscopy is arguably the most fundamental optical characterisation technique. Here it is used to identify the wavelength that transitions between the ground state and excited states of the vanadium ion occur.

Absorption measurements were taken with a Varian Cary 500 spectrophotometer, which is described in section 3.3.1. The reference aperture was left blank. The range of the measurements was 400 to 3300 nm. Figure 4.1 shows the absorption spectra of GLS doped with varying concentrations of vanadium. All have a peak at 1100 nm with relative intensities consistent with the doping concentration. The red shift of the band-edge with increasing dopant concentration indicates that there is a strong absorption by vanadium within the GLS band-edge region around 500nm. There is also a shoulder visible around 700 nm indicating a third vanadium absorption band in this region. In addition there is also evidence of a very weak shoulder at around 1000nm which could be attributed to a spin forbidden transition.

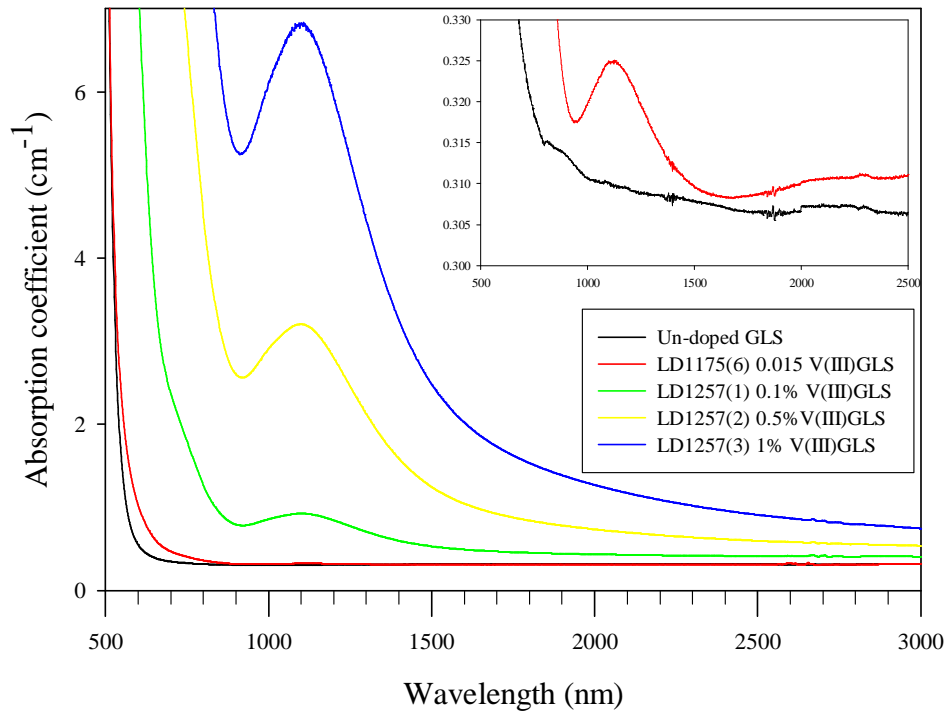

Figure 4.1 Absorption spectra of 0.015% , 0.1%, 0.5% and 1% molar vanadium doped GLS and undoped GLS in 5mm thick slabs. The inset shows a close-up of the lowest doping concentration. Batched concentrations are quoted here.

Figure 4.2 shows the absorption spectra of vanadium doped GLS produced with the "doping glass" method discussed in chapter 3. These samples were also cut and polished into thick and thin slabs which allowed reflection corrected absorption spectra to be calculated with the method described in section 3.3.1.   As discussed in section 4.13 vanadium is believed to exist in a mixed oxidation state $V^{5+}/V^{4+}/V^{3+}/V^{2+}/V^{+}$ with $V^{2+}$ being the only optically active ion. The different starting vanadium oxidation states show no difference in the form and position of the absorption bands, indicating that the optically active vanadium ion is the same when the starting vanadium oxidation state is +3 or +5. There is however an inconsistency in the intensity of the absorption and the concentration of vanadium (calculated from the vanadium weighed during batching) between both the absorption spectra of samples in figure 4.2 and those in figure 4.1. It is proposed that this occurs because the proportion of vanadium that is incorporated as the optically active $V^{2+}$ ion is sensitive to the starting vanadium oxidation state, small variations in melting conditions and small compositional variations, as observed in transition metal doped phosphate glass.[88]

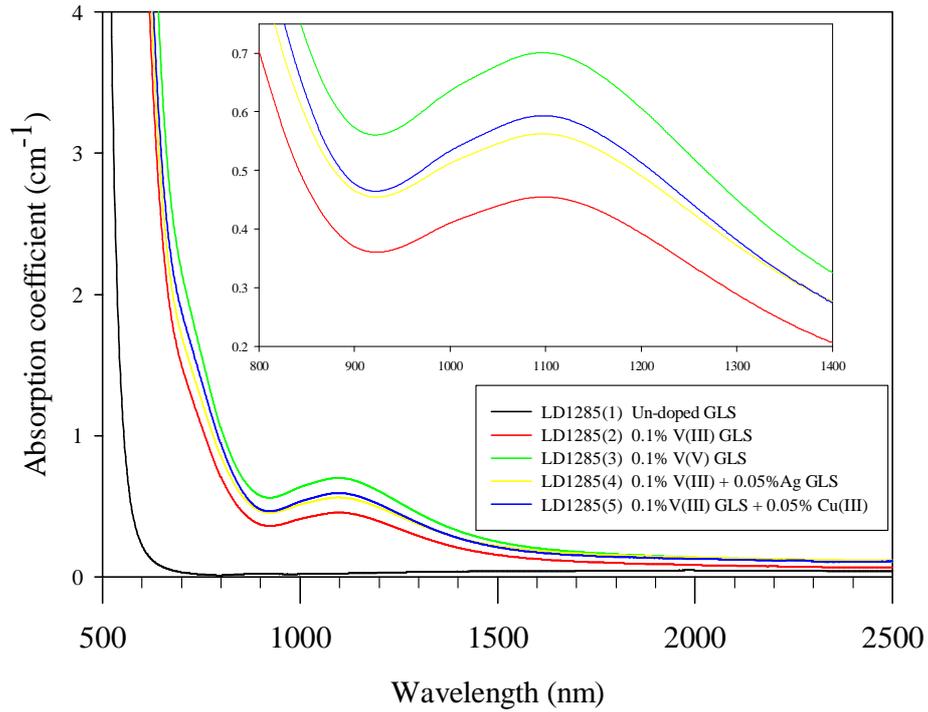

Figure 4.2 Reflection corrected absorption spectra of 0.1 % vanadium doped GLS with the vanadium dopant in a +3 and +5 oxidation state before melting. Silver and copper co-dopants are also shown. Batched concentrations are quoted.

Because the concentration of optically active $V^{2+}$ ions cannot be measured accurately from the concentration of vanadium in the glass melt components, a relative concentration is calculated from the intensity of the absorption peak at 1100nm and normalised to the intensity of the absorption peak at 1100nm for 1% V:GLS, the background loss of GLS is also accounted for. This is expressed in equation 4.1.

$$[V^{2+}] = \frac{a_{V:GLS}(1100) - a_{GLS}(1100)}{a_{1\% V:GLS}(1100) - a_{GLS}(1100)} \qquad (4.1)$$

Here $[V^{2+}]$ is the relative concentration of optically active $V^{2+}$ ions, $a_{V:GLS}$, $a_{1\% V:GLS}$ and $a_{GLS}$ is the absorption coefficient at 1100nm for the relevant sample, 1.038% V:GLS and un-doped GLS respectively. Results are shown in table 4.1

Table 4.1 Relative V$^{2+}$ ion concentration for vanadium doped GLS samples.

| Melt code | Batched vanadium concentration (% molar) | Relative V$^{2+}$ concentration (% molar) |
|-----------|------------------------------------------|-------------------------------------------|
| LD1285-2 | 0.0977 | 0.0616 |
| LD1285-3 | 0.0983 | 0.0955 |
| LD1285-4 | 0.0979 | 0.0713 |
| LD1285-5 | 0.0978 | 0.0821 |
| LD1257-1 | 0.108 | 0.0944 |
| LD1257-2 | 0.518 | 0.4443 |
| LD1257-3 | 1.038 | 1.038 |
| LD1175-6 | 0.015 | 0.0023 |
| LD1284-1 | 0.0970 | 0.0608 |
| LD1284-3 | 0.0963 | 0.0489 |
| LD1284-4 | 0.0199 | 0.0087 |
| LD1284-5 | 0.0199 | 0.0242 |

In figure 4.2 the addition of silver and copper co-dopants produces no difference in the form and position of the absorption bands. There is a small increase in the intensity of the absorption, indicating that copper and silver co-dopants are helping to incorporate vanadium in the divalent state, but this could be accounted for by the variability of the proportion of vanadium that actually becomes V$^{2+}$ as was discussed earlier.

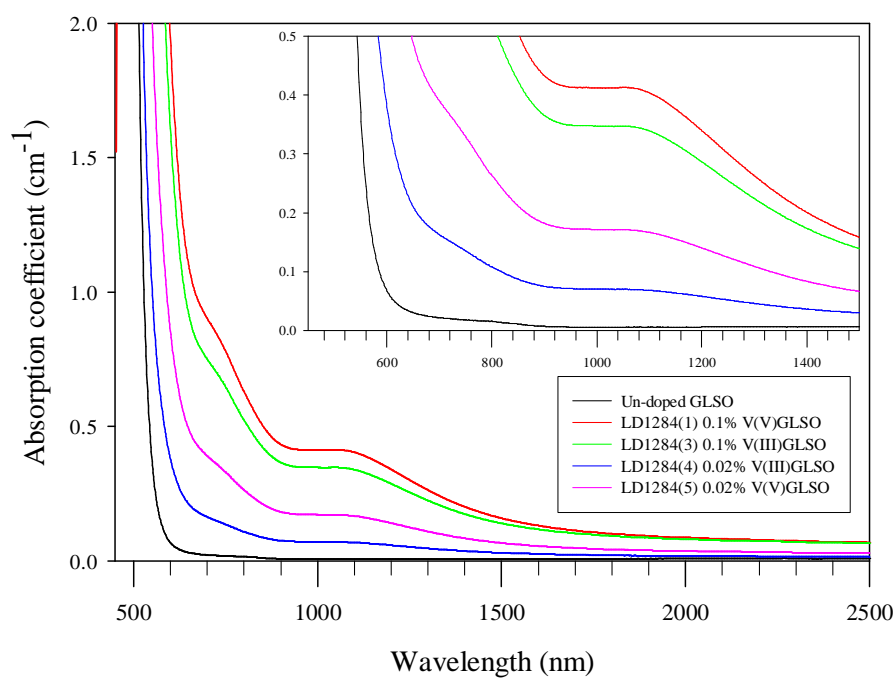

Figure 4.3 Reflection corrected absorption spectra of 0.1 and 0.02 % vanadium doped GLSO with the vanadium dopant in a +3 and +5 oxidation state before melting. Batched concentrations are quoted here.

Figure 4.3 shows the reflection corrected absorption of vanadium doped GLSO produced with the "doping glass" method. A notable difference compared to the V:GLS absorption in figure 4.2, is the blue shift in the electronic absorption edge for the un-doped sample. There is still a peak at around 1100nm but its position is less defined as it is more of a shoulder on the absorption at around 700nm. Possible explanations for this are a broadening of one or both of the absorption bands at 700 and 1100 nm and/or an increase in the relative intensity of the absorption band at 700 nm to that at 1100nm. As in V:GLS, doping with V(III) or V(V) show no difference in the form and position of the absorption bands. Similarly to V:GLS there is an inconsistency in the intensity of the absorption and the batched concentration of vanadium. It is also noted that the absorption of the optically active $V^{2+}$ is consistently higher when GLS and GLSO is doped with $V_2O_5$ with vanadium in a +5 oxidation state than with $V_2S_3$ with vanadium in a +3 oxidation state. This may be because the $V^{2+}$ is believed to be preferentially incorporated into a high efficiency oxygen related site and the oxygen from $V_2O_5$ could aid in the formation of these sites. It can also be seen that the largest change in absorption is in LD1175-6 which was fabricated at a very low doping concentration without the doping glass method and highlights the difficulty of accurate fabrication of low doping concentration glass with the usual method. From this point on, all vanadium concentrations quoted refer to the relative $V^{2+}$ concentration given in table 4.1.

## 4.3 Derivative absorption spectroscopy

The mathematical differentiation of spectroscopic data is often used as a resolution enhancement technique, to facilitate the detection and location of poorly resolved spectral components including peaks which appear only as shoulders as well as the isolation of small peaks from an interfering large background absorption.[89] All spectral features that are attributed to optical transitions in transition metals are assumed to be composed of a sum of Gaussian peaks. The standard Gaussian curve function for an absorption band peaking at $x_0$ with absorbance A is:

$$A = A_0 e^{\left[-4\ln 2\left(\frac{x-x_0}{w}\right)^2\right]} \qquad \text{0}^{\text{th}} \text{ order} \qquad (4.2)$$

Where $A_0$ is the maximum band height at $x_0$ and w is the full width at half maximum (FWHM)

Differentiation gives:

$$\frac{dA}{dx} = -\frac{8\ln 2 A_0 (x-x_0)}{w^2} . e^{\left[-4\ln 2\left(\frac{x-x_0}{w}\right)^2\right]} \qquad \text{1}^{\text{st}} \text{ order} \qquad (4.3)$$

$$\frac{d^2 A}{dx^2} = \left(\frac{16\ln 2 (x-x_0)^2 A_0}{w^4} - . \frac{8\ln 2 A_0}{w^2}\right) . e^{\left[-4\ln 2\left(\frac{x-x_0}{w}\right)^2\right]} \qquad \text{2}^{\text{nd}} \text{ order} \qquad (4.4)$$

$$\frac{d^3 A}{dx^3} = \left( \frac{12 \ln 2\, A_0\, (x - x_0)}{w^4} - \frac{\left( A_0\, (512 \ln 2)^3\, (x - x_0)^2 \right)}{w^6} \right) . e^{\left[ -4 \ln 2 \left( \frac{x - x_0}{w} \right)^2 \right]} \ 3^{rd} \ \text{order} \ (4.5)$$

The points of inflection on a Gaussian (or any) curve are given when the first derivative is at an extremum and when the second derivative equals zero.[90] By setting equation 4.3 equal to zero it can be shown[91] that the ratio of the FWHM to the width of the Gaussian band between the points of inflection ($\sigma$) is given by equation 4.6:

$$\frac{FWHM}{\sigma} = \sqrt{2 \cdot \ln 2} = 1.177 \tag{4.6}$$

Higher order derivatives discriminate strongly in favour of narrower bands[92] however the signal-to-noise ratio (SNR) is degraded with increasing differentiation order.[89, 93] Relative signal to noise ratios for unsmoothed derivatives are given in table 4.2. Setting $x=x_0$ for equations 4.2-4.5 gives the peak amplitudes in table 4.2, this shows that second derivatives are inversely proportional to the square of the FWHM.

TABLE 4.2 Peak amplitude and relative signal-to-noise ratio for Gaussian peaks and some derivatives.

| Order | Peak amplitude at $x=x_0$ | Relative SNR |
|---|---|---|
| $0^{th}$ order (A) | $A_0$ | 1 |
| $1^{st}$ order $\left( \dfrac{dA}{dx} \right)$ | 0 | $2.02/w$ |
| $2^{nd}$ order $\left( \dfrac{d^2 A}{dx^2} \right)$ | $-\dfrac{8 \ln 2 A_0}{w^2}$ | $3.26/w^2$ |
| $3^{rd}$ order $\left( \dfrac{d^3 A}{dx^3} \right)$ | 0 | $8.1/w^3$ |

In order to improve the signal-to-noise ratio, "running average" smoothing filters (RASF) can be applied. The optimum number of passes for a RASF is believed to be n+1 where n is the derivation order.[89] The relative SNR of the smoothed $n^{th}$ derivative is given by:

$$\frac{(SNR)_n}{(SNR)_0} = \frac{\alpha_n C_n N^{n+0.5}}{W^n} = \frac{\alpha_n C_n (rW)^{n+0.5}}{W^n} = \alpha_n C_n\, r^{n+0.5} \sqrt{W} \tag{4.7}$$

Where $(SNR)_0$ is the SNR of the unsmoothed $0^{th}$ derivative. Smoothing with a RASF reduces noise by a factor of $N^{n+05}$, where N is the number of data points in the smoothing operation, but reduces the signal by a factor of $\alpha_n$, where r is the sampling proportion and $C_n$ is a constant which depends on the derivative order and band shape. SNR will reach a maximum for sampling proportions of 1,[89] however large sampling proportions cause severe peak height attenuation, As a trade-off between SNR and signal attenuation a sampling proportion of 0.05 was used for this work.

After trials with various smoothing filters it was found that a negative exponential with a sampling proportion of 0.05 and a polynomial degree of 3, gave a better SNR than a running average. Figure 4.4 shows the first derivative of the absorption coefficient of 0.0955% V:GLS. The absorption data was first smoothed, then the first derivative was taken and then it was smoothed again. The zero crossing gives a peak position at 9100 cm$^{-1}$, which is consistent with the absorption peak measured in figure 4.1. The extrema occur at 7790 cm$^{-1}$ (1284 nm) and another which is partially obscured by the weak shoulder at around 10200 cm$^{-1}$ (980 nm). This gives a width between the points of inflection ($\sigma$) of 2410 cm$^{-1}$ (304 nm) which, using the factor in equation 4.6, gives a FWHM of 2836 cm$^{-1}$ (358 nm). The very weak shoulder at around 10000 cm$^{-1}$ is much clearer as is the shoulder at around 13300cm$^{-1}$.

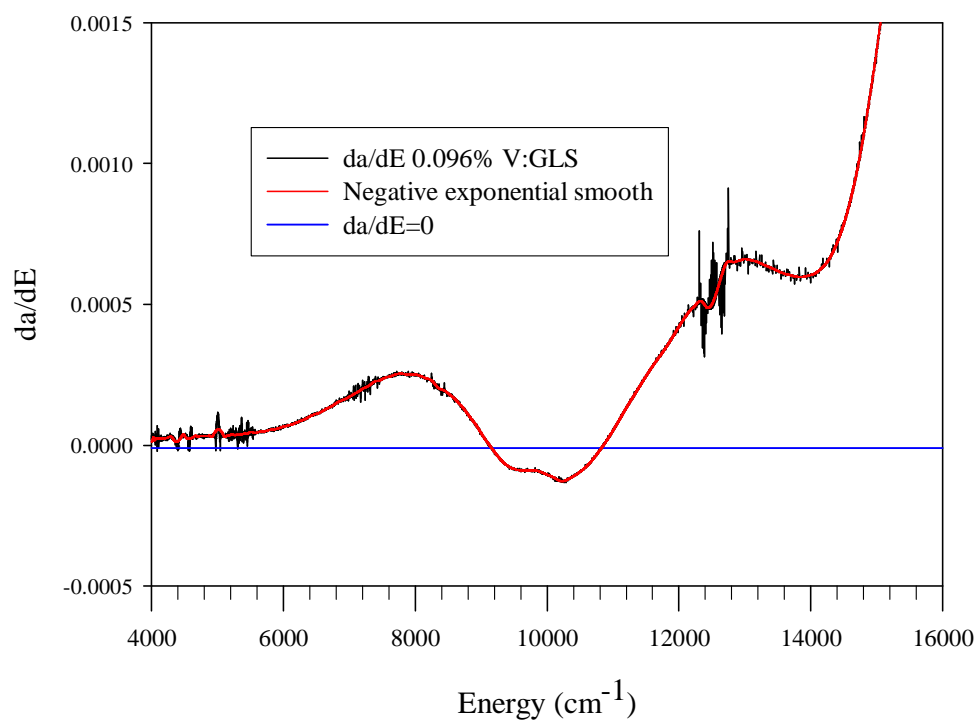

FIGURE 4.4 First derivative of the absorption coefficient of 0.0955% V:GLS, smoothed with a negative exponential smoothing filter with a sampling proportion of 0.05 and a polynomial degree of 3.

The feature around 12500 cm$^{-1}$ (800 nm) is believed to be a result of the detector and grating change in the spectrophotometer used to take the absorption measurements, as discussed in section 3.3.1.

Figure 4.5 shows the first derivative of the smoothed first derivative of the smoothed absorption data, i.e. the second derivative. As shown in table 4.2 absorption peaks correspond the where d$^2$a/dE$^2$ < 0. Figure 4.5 shows negative peaks at 9044, 9983 and 13362 cm$^{-1}$ which correspond to 1105, 1002 and 748 nm respectivley. Comparing figures 4.2, 4.4 and 4.5 clearly demonstrates that the weak absorption around 10000 cm$^{-1}$ increases in intensity relative to the other absorption bands as the order of derivation increases. This indicates that it is much narrower than the other absorption bands.

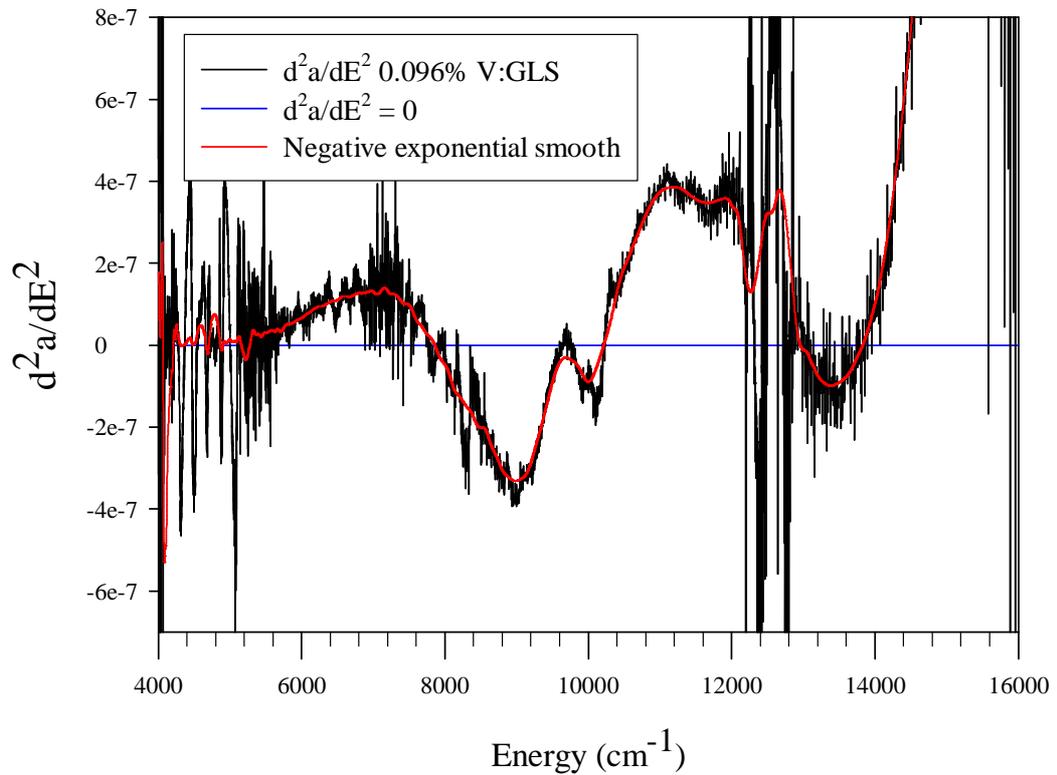

FIGURE 4.5 Second derivative of the absorption coefficient of 0.0955% V:GLS smoothed with a negative exponential smoothing filter with a sampling proportion of 0.05 and a polynomial degree of 3.

The second derivative of the absorption spectrum of undoped GLS did not show any features, apart from one at 800 nm related to the grating change-over discussed earlier. Figure 4.6 shows the second derivative of the absorption spectrum of 0.061% V:GLSO. The absorption peaks occur at 9300, 10200 and 13600 cm$^{-1}$ which are at a higher energy than in V:GLS and is consistent with a higher crystal field strength.

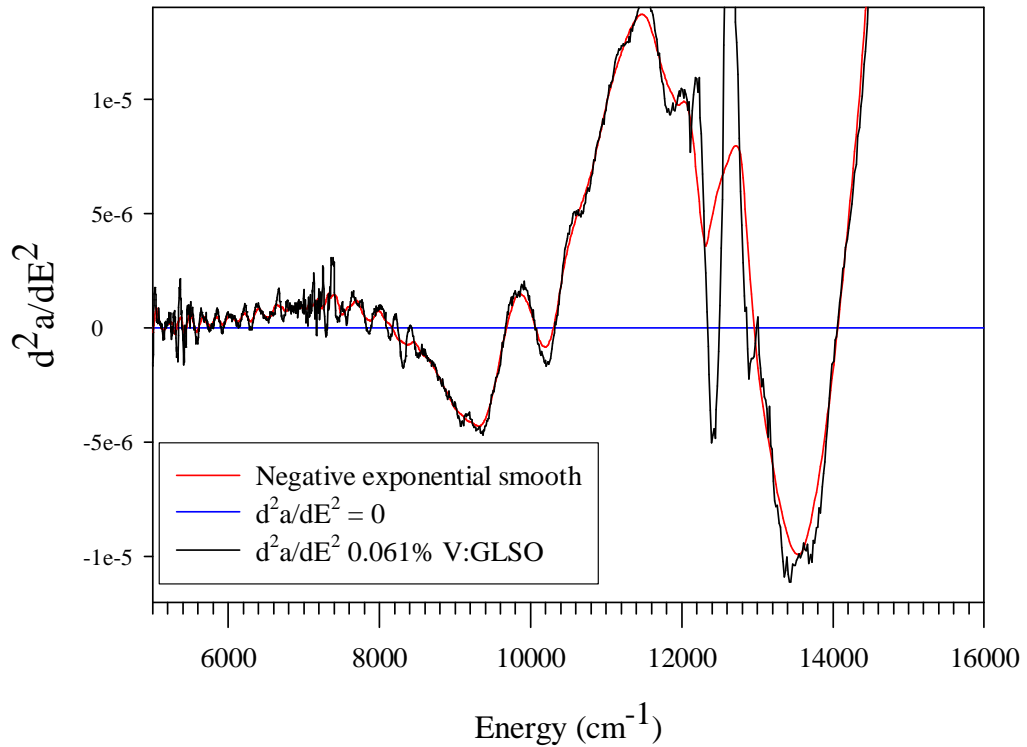

FIGURE 4.6 Second derivative of the absorption coefficient of 0.061% V:GLSO smoothed with a negative exponential smoothing filter with a sampling proportion of 0.05 and a polynomial degree of 3.

## 4.4 Photoluminescence of vanadium doped GLS

The photoluminescence (PL) of V:GLS reveals the Stokes shift, discussed in section 2.5 and 4.11, and the wavelength range over which an active device based on V:GLS could be used. The measurements were taken using the experimental setup described in section 3.3.2.

### 4.4.1 Photoluminescence spectra

Figures 4.7 to 4.10 show the PL spectra of 0.0023, 0.0944, 0.4443 and 1.038% molar vanadium doped GLS excited at wavelengths of 514, 808 and 1064 nm and temperatures of 77 and 300 K. The excitation wavelengths were chosen as the closest available laser sources that could excite into the three main absorption bands of V:GLS which were identified in section 4.2 and 4.5. The low temperature PL spectra were taken to give an indication of the relative contribution of homogeneous and inhomogeneous broadening. It was not possible to obtain spectra for all combinations of excitation wavelength and temperature in the highly doped samples as emission was significantly weaker than at lower concentrations. This is attributed to re-absorption of the emission and concentration quenching of the excitation.

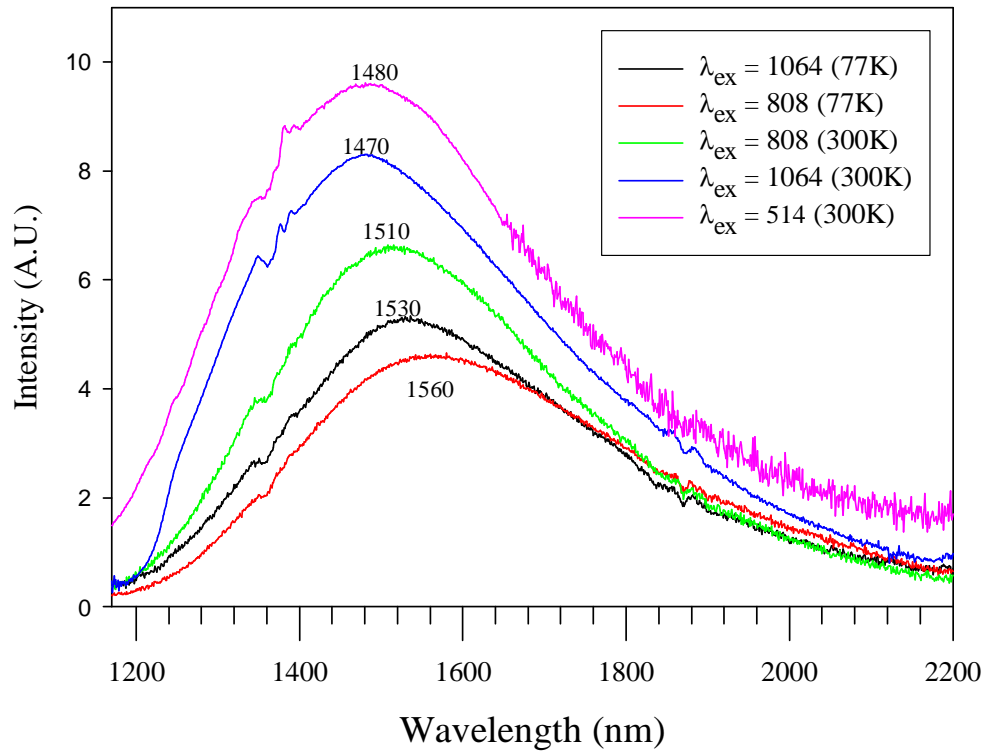

FIGURE 4.7 Photoluminescence spectra of 0.002% vanadium doped GLS excited with various laser excitation sources at 514, 808 and 1064 nm at temperatures of 77 and 300K. The peak wavelength is given for each spectrum.

The peak positions of the photoluminescence spectra in figures 4.7 to 4.10 are summarised in table 4.3. It can be seen that there is a trend for the emission peak to move towards shorter wavelengths as the temperature increases. This is the opposite trend that would be expected from the SCCM model. However it can be explained by considering that non-radiative decay is expected to preferentially affect vanadium ions in low crystal field sites which emit at relatively long wavelengths because non-radiative decay involves fewer phonons in these sites. Ions in these sites will be quenched non-radiatively to a greater extent at high temperatures than at low temperatures, leaving ions in high crystal field sites which emit at shorter wavelengths to dominate the radiative decay. There is a trend for the emission peak to shift to longer wavelengths with increasing vanadium concentration. This is attributed to increased re-absorption of the fluorescence on its high energy side with increasing concentration. There is also a trend for the 808 nm excited emission peak to shift to longer wavelengths from the 1064 nm excited emission peak. This is explained by examining the PLE spectrum of V:GLS in figure 4.12 where it can be seen that 1064 nm is on the high energy side of the first excitation peak whereas 808 nm is on the low energy side of the second excitation peak. The FWHM of the photoluminescence spectra, in figures 4.7 to 4.10, was around 460 nm and differences at the various vanadium concentrations, pump wavelengths and temperatures were less than the estimated 20 nm resolution of the measurements. This indicates that the broadness of the V:GLS emission is almost entirely caused by an inhomogeneous mechanism, such as a range of crystal field strengths. A description of various broadening mechanisms is given in section 2.6.

TABLE 4.3. Photoluminescence peak positions (nm) for varying concentration, excitation wavelength and temperature.

| Vanadium Concentration (% molar) | Excitation wavelength/Temperature | | | | | |
|---|---|---|---|---|---|---|
| | 1064 nm | | 808 nm | | 514 nm | |
| | 300 K | 77K | 300 K | 77K | 300 K | 77K |
| 0.0023 | 1470 | 1530 | 1510 | 1560 | 1480 | - |
| 0.0944 | 1460 | 1536 | 1495 | 1525 | - | - |
| 0.4443 | 1500 | - | 1530 | - | - | - |
| 1.038 | 1520 | 1490 | 1540 | - | - | - |

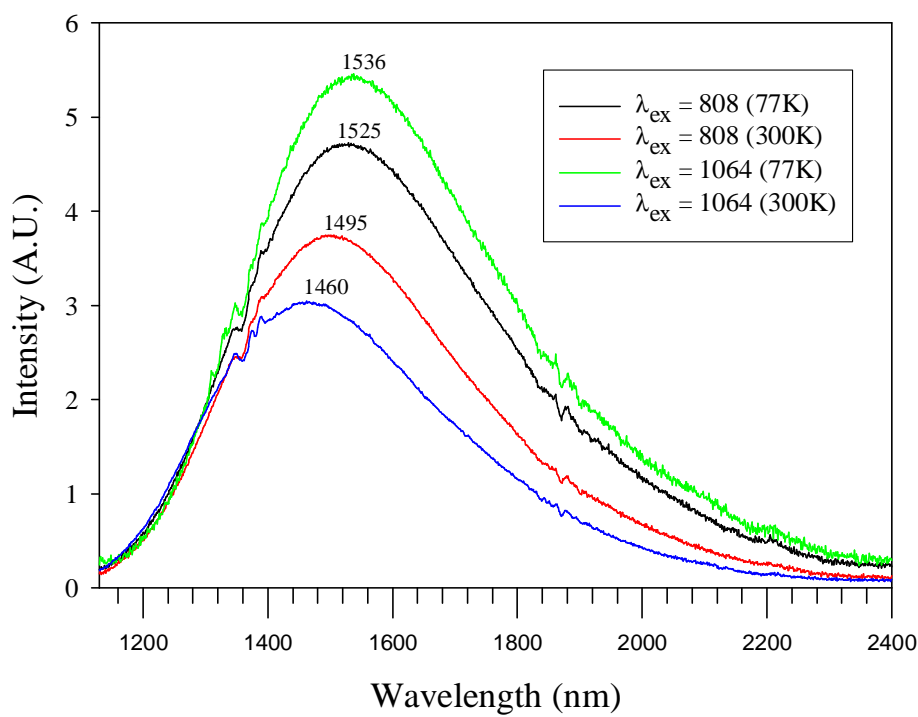

FIGURE 4.8 Photoluminescence spectra of 0.09% vanadium doped GLS excited with various laser excitation sources at 808 and 1064 nm, at temperatures of 77 and 300K. The peak wavelength is given for each spectrum.

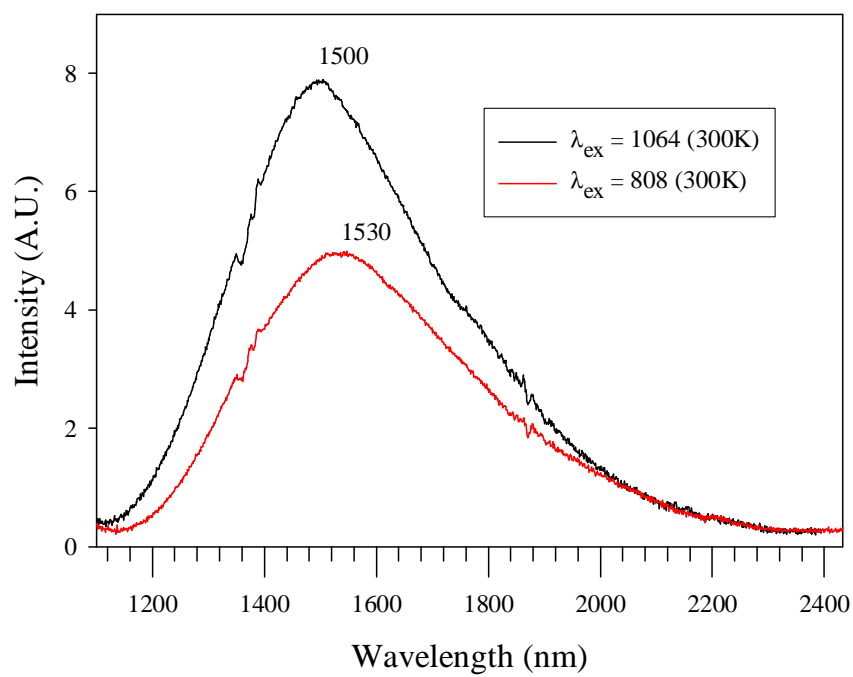

FIGURE 4.9 Photoluminescence spectra of 0.44% vanadium doped GLS excited with various laser excitation sources at 808 and 1064 nm at temperatures of 77 and 300K. The peak wavelength is given for each spectrum.

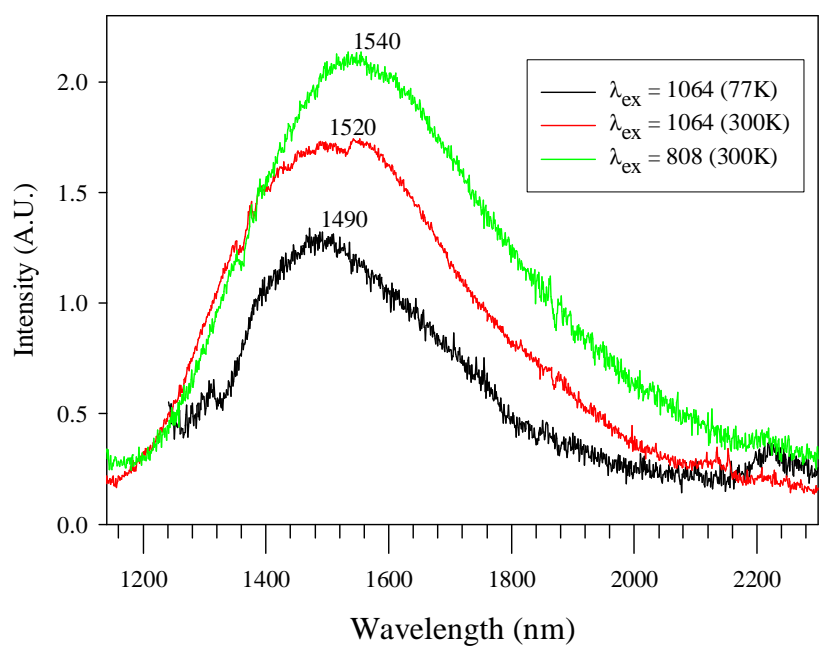

FIGURE 4.10 Photoluminescence spectra of 1% vanadium doped GLS excited with various laser excitation sources at 808 and 1064 nm at temperatures of 77 and 300K. The peak wavelength is given for each spectrum.

## 4.4.2 Discussion of photoluminescence spectra

Exciting V:GLS at 514, 808 and 1064 nm roughly equates to exciting into each of the three main absorption bands identified for V:GLS in figure 4.12. The PL from excitation at these wavelengths show similar characteristic spectra, peaking at ~1500 nm, with a full width at half maximum FWHM of ~500 nm. This indicates that the three absorption bands belong to the same oxidation state, rather than two or more oxidation states which is commonly observed in transition metal doped glasses[86, 94, 95] and crystals.[96] The broadness of the PL spectra indicates that the vanadium ion is in a low crystal field site. A result of this is that the lowest energy level with a spin allowed transition to the ground state (which is strongly dependent on crystal field strength) is the lowest energy level. Conversely, in a strong crystal field site the lowest spin forbidden level (which is almost independent of crystal field strength) is the lowest energy level. In this case, characteristic narrow *R*-line emission should be observed, as in $V^{3+}$ doped phosphate glass[87] and $V^{3+}$ doped corundum.[97]

For comparison the PL of 1064 nm excitation of tetrahedrally coordinated $V^{3+}$ doped $LiGaO_2$, $LiAlO_2$ and $SrAl_2O_4$ peaks at 1650, 1730 and 1800 nm respectively. The origin of this emission has been assigned to the $^3T_2(^3F) \rightarrow ^3A_2(^3F)$ transition.[98] The low temperature (T=4.2K) photoluminescence spectrum of vanadium doped ZnTe, excited at 529 nm, displays three structured emission bands centred at 3195, 2793 and 2247 nm which have been attributed to internal relaxations of the three charge states $V^+$, $V^{2+}$ and $V^{3+}$ respectively; the vanadium ions are tetrahedrally coordinated and have been assigned to the $^5E(^5D) \rightarrow ^5T_2(^5D)$, $^4T_2(^4F) \rightarrow ^4T_1(^4F)$ and $^3T_2(^3F) \rightarrow ^3A_2(^3F)$ transition respectively.[99] Similarly for vanadium doped ZnS three structured emission bands centred at 2700, 2083 and 1785 nm were observed and attributed to tetrahedral $V^+$, $V^{2+}$ and $V^{3+}$ respectively and assigned to the $^5E(^5D) \rightarrow ^5T_2(^5D)$, $^4T_2(^4F) \rightarrow ^4T_1(^4F)$ and $^3T_2(^3F) \rightarrow ^3A_2(^3F)$ transition respectively.[100] In vanadium doped CdTe the same respective charge states and transitions occur at 3597, 2898 and 2439 nm.[101] These comparisons and others are detailed in table 4.4 which indicates that the PL peak observed in V:GLS is similar to that observed in $V^{2+}$ and $V^{3+}$. Table 4.4 also give the refractive index of ZnS, ZnTe and CdTe which illustrates an interesting trend; the peak emission wavelength of vanadium in 1+,2+ and 3+ oxidation states is directly proportional to the refractive index of the host. This is believed to be related to the degree of covalency of the bonds between the host and vanadium dopant. The nephelauxetic effect, which may be used as a measure of the covalency, is caused by the lowering of the excited state of the ion by the surrounding crystal field.[102]

TABLE 4.4 Summary of photoluminescence from vanadium in 1+,2+ and 3+ oxidation states in various hosts.

| Ion | Host | Refractive index ($\lambda$=2 $\mu$m) | Excitation wavelength (nm) | Emission peak (nm) | Coordination | Transition | Reference |
|---|---|---|---|---|---|---|---|
| V$^+$ | ZnS | 2.26 | - | 2700 | tetrahedral | $^5E(^5D)\rightarrow$ $^5T_2(^5D)$ | [100] |
| V$^+$ | ZnTe | 2.65 | 529 | 3195 | tetrahedral | $^5E(^5D)\rightarrow$ $^5T_2(^5D)$ | [99] |
| V$^+$ | CdTe | 2.71 | 1064 | 3597 | tetrahedral | $^5E(^5D)\rightarrow$ $^5T_2(^5D)$ | [101] |
| V$^{2+}$ | GLS | 2.4 | 1064 | 1500 | octahedral | $^4T_2(^4F)\rightarrow$ $^4A_2(^4F)$ | This work |
| V$^{2+}$ | GLSO | 2.3 | 1064 | 1500 | octahedral | $^4T_2(^4F)\rightarrow$ $^4A_2(^4F)$ | This work |
| V$^{2+}$ | ZnS | 2.26 | - | 2083 | tetrahedral | $^4T_2(^4F)\rightarrow$ $^4T_1(^4F)$ | [100] |
| V$^{2+}$ | ZnTe | 2.65 | 529 | 2793 | tetrahedral | $^4T_2(^4F)\rightarrow$ $^4T_1(^4F)$ | [99] |
| V$^{2+}$ | CdTe | 2.71 | 1064 | 2898 | tetrahedral | $^4T_2(^4F)\rightarrow$ $^4T_1(^4F)$ | [101] |
| V$^{2+}$ | KMgF$_3$ | - | - | 1102 | octahedral | $^4T_2(^4F)\rightarrow$ $^4A_2(^4F)$ | [7] |
| V$^{2+}$ | KMnF$_3$ | - | - | 1141 | octahedral | $^4T_2(^4F)\rightarrow$ $^4A_2(^4F)$ | [7] |
| V$^{2+}$ | RbMnF$_3$ | - | - | 1193 | octahedral | $^4T_2(^4F)\rightarrow$ $^4A_2(^4F)$ | [7] |
| V$^{2+}$ | CsCaF$_3$ | - | - | 1351 | octahedral | $^4T_2(^4F)\rightarrow$ $^4A_2(^4F)$ | [7] |
| V$^{2+}$ | CsCdCl$_3$ | - | - | 1639 | octahedral | $^4T_2(^4F)\rightarrow$ $^4A_2(^4F)$ | [7] |
| V$^{2+}$ | NaCl | - | - | 1754 | octahedral | $^4T_2(^4F)\rightarrow$ $^4A_2(^4F)$ | [7] |
| V$^{2+}$ | MgF$_2$ | - | 514 | 1140 | octahedral | $^4T_2(^4F)\rightarrow$ $^4A_2(^4F)$ | [103] |
| V$^{3+}$ | LiGaO$_2$ | - | 1064 | 1650 | tetrahedral | $^3T_2(^3F)\rightarrow$ $^3A_2(^3F)$ | [98] |
| V$^{3+}$ | LiAlO$_2$ | - | 1064 | 1730 | tetrahedral | $^3T_2(^3F)\rightarrow$ $^3A_2(^3F)$ | [98] |
| V$^{3+}$ | SrAl$_2$O$_4$ | - | 1064 | 1800 | tetrahedral | $^3T_2(^3F)\rightarrow$ $^3A_2(^3F)$ | [98] |
| V$^{3+}$ | ZnS | 2.26 | - | 1785 | tetrahedral | $^3T_2(^3F)\rightarrow$ $^3A_2(^3F)$ | [100] |
| V$^{3+}$ | ZnTe | 2.65 | 529 | 2247 | tetrahedral | $^3T_2(^3F)\rightarrow$ $^3A_2(^3F)$ | [99] |
| V$^{3+}$ | CdTe | 2.71 | 1064 | 2439 | tetrahedral | $^3T_2(^3F)\rightarrow$ $^3A_2(^3F)$ | [101] |

## 4.5 Photoluminescence excitation of vanadium doped GLS

The photoluminescence excitation (PLE) spectra of V:GLS, for reasons discussed in chapter 3, clarifies the identification of the two highest energy absorption bands that could not be fully determined in absorption measurements because of their proximity to the band-edge absorption of GLS. The spectra were taken using the setup described in section 3.3.3.

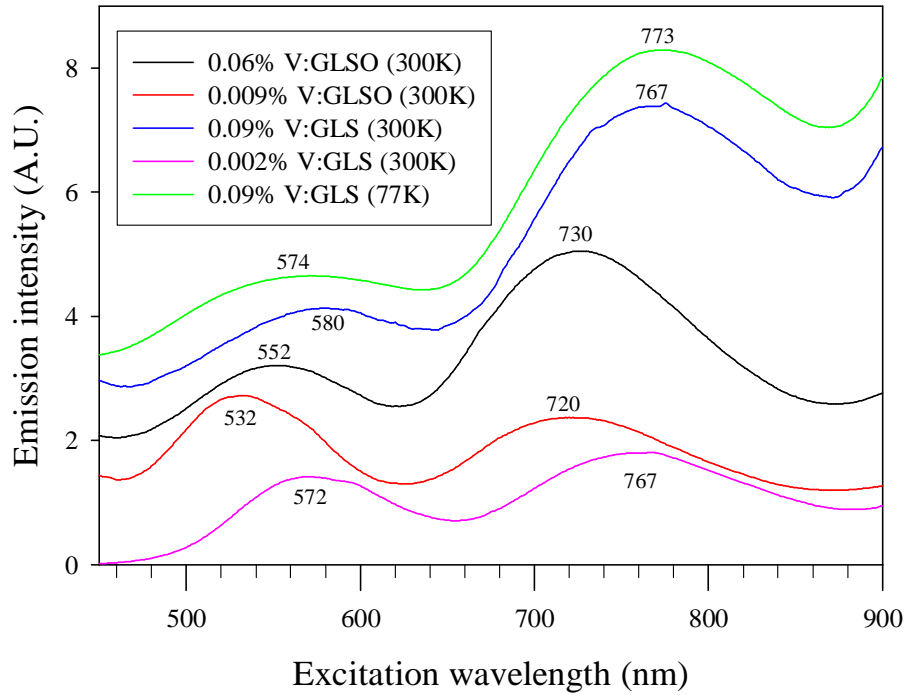

FIGURE 4.11 PLE spectra detecting emission at 1000-1700 nm of various concentrations of vanadium doped GLS and GLSO at temperatures of 300 and 77K. Peak positions are given. Concentrations quoted are relative. Spectra are offset on the y-axis for clarity.

Figure 4.11 shows PLE spectra taken by detecting emission at 1000-1700 nm. This gave greater coverage of the emission band than detecting at 1400-1700nm as in figure 4.12 which increased the signal strength and is believed to give a more representative spectrum as data is collected from a greater proportion of emitting ions.

Figure 4.11 shows that 0.06% V:GLSO has peaks at 552 and 730 nm whereas 0.009% V:GLSO has peaks at 532 and 720 nm. This red shift of the excitation peaks with increasing concentration, especially for the high energy peak, can be reconciled by considering the overlap between the absorption band centred at 1100 nm and the emission band centred at around 1500 nm. As concentration increases, a greater proportion of the high energy side of the emission band is reabsorbed by the absorption band centred at 1100 nm which leads to a red shift in the fluorescence, as detailed in table 4.3. This red shift of the fluorescence with increasing concentration means that when exciting at longer wavelengths a greater proportion ions in low crystal field strength sites, that fluoresce at longer wavelengths, will be excited. This will shift the PLE peak to longer wavelengths with increasing concentration since there would be less re-absorption of fluorescence. It is also observed that there is a red shift of the PLE peaks in vanadium doped GLS in comparison to vanadium doped GLSO. This can be explained by considering the overlap of the GLS band-edge with the vanadium absorption band around 500-600 nm which would cause the high energy side of the absorption band to be suppressed by competing absorption of the band-edge. The red shift of the band-edge absorption of GLS, in comparison to that of GLSO, means that PLE peaks will be shifted to longer wavelengths in V:GLS .

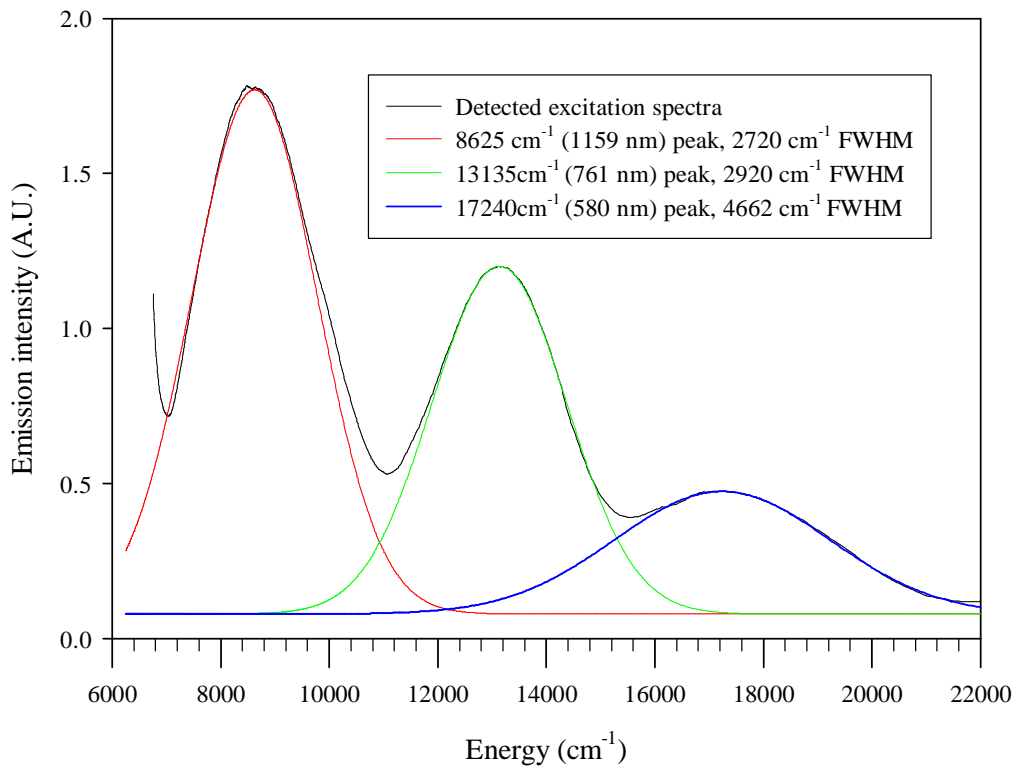

FIGURE 4.12 PLE spectra detecting emission at 1400-1700 nm of 0.09% vanadium doped GLS fitted with three Gaussians.

Figure 4.12 shows the PLE spectra of 0.09% V:GLS where emission was detected at 1400-1700 nm, this allowed the lowest energy absorption to be covered. It can be seen from the Gaussian fits to the PLE spectra that the three main peaks are very Gaussian in nature. The lowest energy peak occurs at 1159 nm which is red shifted from the absorption peak at 1100 nm this can be explained by the same process described for the red shift of the highest energy peak in V:GLSO. Similarly to the absorption spectra in figure 4.2, a weak shoulder is also visible at around 10000 cm$^{-1}$ (1000 nm) which is again interpreted as the result of a spin forbidden transition.

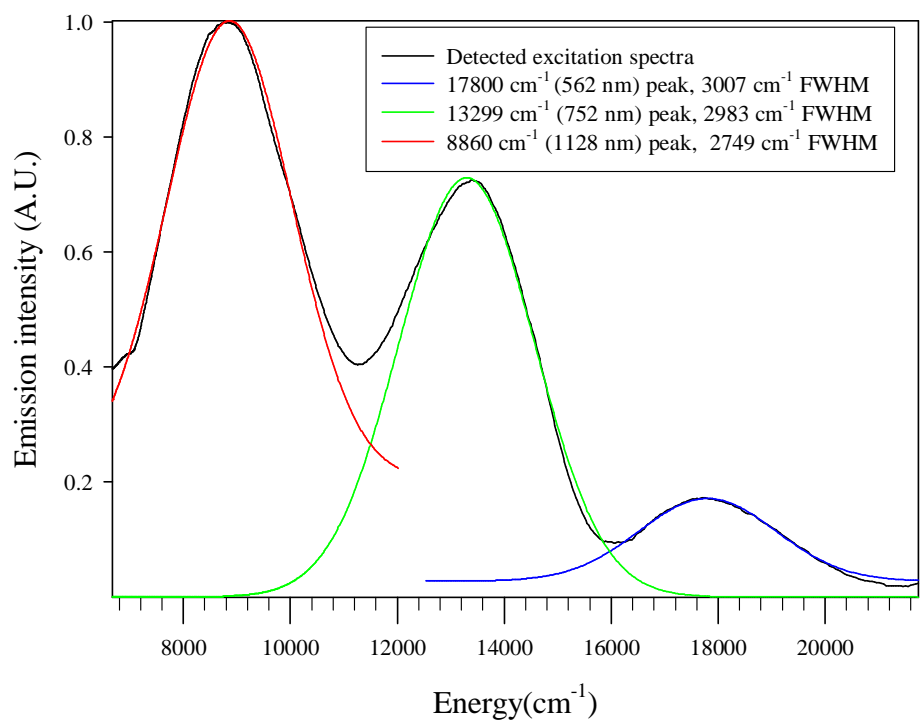

FIGURE 4.13 PLE spectra detecting emission at 1400-1700 nm of 0.06% vanadium doped GLSO fitted with three Gaussians.

The PLE peaks in V:GLSO in figure 4.13 are all shifted to higher energy than for V:GLS indicating that the vanadium ion experiences a higher crystal field strength in GLSO. Both re-absorption of fluorescence and the overlap of the GLS band-edge with vanadium absorptions are thought to cause red shifting of the PLE peaks therefore the lowest energy peaks observed can be considered an upper limit of the true position of the PLE peaks. The broad Gaussian nature of the PLE bands indicate they originate from spin allowed transitions.

## 4.6 Fluorescence Lifetime

Fluorescence lifetime measurements are one of the simplest and most reliable optical characterisation techniques.[104] However, in the case of non-exponential decay (as observed in V:GLS) interpretation can be far from trivial. In this section the fluorescence decay of V:GLS is analysed using stretched exponential, bi exponential, average lifetime, frequency resolved lifetime and continuous lifetime distribution models. From the analysis solid conclusions are drawn as to the local environment of the vanadium ion.

### 4.6.1 Introduction to the stretched exponential function

Many relaxation processes in complex condensed systems such as glasses have long been known to follow the Kohlraush-Williams-Watts (KWW) function,[105] which is also currently know as the stretched exponential function and is given in equation 4.8

$$I(t) = y_0 + I_0 \exp\left(-\left(\frac{t}{\tau}\right)^{\beta}\right) \tag{4.8}$$

Where $\tau$ is a characteristic relaxation time, $\beta$ is the stretch factor (ranging between 0 and 1) and $y_0$ is the offset. The closer $\beta$ is to 0 the more the function deviates from a single exponential. Stretched exponential relaxation was first described by Kohlraush in 1847 to model the decay of the residual charge on a glass Leyden jar. Since then the stretched exponential function has been shown to fit many other relaxation processes in amorphous materials such as nuclear relaxation,[105] magnetic susceptibility relaxation,[105] fluorescence decay[106] and photoinduced dichroism.[107] Stretched exponential behaviour is rarely observed in crystalline solids; with some exceptions.[108] There have been several mutually exclusive microscopic explanations for the observed stretched exponential relaxation in glasses, this can be seen as being due to the use of different models of the fundamental structure of glasses. The models for stretched exponential relaxation behaviour in glasses generally fall into two categories: spatially heterogeneous dynamics and temporally heterogeneous dynamics. The spatially heterogeneous dynamic model assumes that the relaxation of a single excited ion follows a single exponential law and that the system remains homogeneous over the time taken for the relaxation to occur. Validation of this model has been allowed through the development of fluorescence microscopy which allows the study of single molecules in complex condensed environments.[109] It has been shown that that the fluorescence decay of a single molecule in varying local environments leads to a stretched-exponential decay as a result of the presence of a continuous distribution of lifetimes.[110, 111] The temporally heterogeneous dynamics model assumes that the system remains homogeneous in space but that random sinks, in disordered material such as glass, capture excitations and become progressively depleted with time. This causes the decay rate itself to slow with the progress of time, stretching the decay. These two processes may occur at the same time in a given relaxation process.

## 4.6.2 Experimental and analysis techniques

Temporally resolved fluorescence lifetime (TRFL) measurements were obtained using the experimental setup described in 3.3.4 and a 1064 nm laser source. An estimate of random experimental error was obtained by taking the measurements several times at different alignments.

Regression analysis was implemented using the Marquardt-Levenberg algorithm,[112] given in equation 4.9. This algorithm seeks the values of the parameters that minimize the sum of the squared (SS) differences between the values of the observed and predicted values of the dependent variable.

$$SS = \sum_{i=1}^{n} (y_i - \hat{y}_i)^2 \qquad (4.9)$$

Where $y_i$ is the observed and $\hat{y}_i$ is the predicted value of the dependent variable, the index i refers to the $i^{th}$ data point and n is the total number of data points. The goodness of the fit was measured using the coefficient of determination ($R^2$), given in equation 4.10.

$$R^2 = \frac{\sum_{i=1}^{n} (\hat{y}_i - \overline{y})^2}{\sum_{i=1}^{n} (y_i - \overline{y})^2} \qquad (4.10)$$

Where $\overline{y}$ is the mean value of the observed dependant variable. The coefficient of determination indicates how much of the total variation in the dependent variable can be accounted for by the regressor function. If $R^2 = 1$ then this indicates that the fitted model explains all variability in the observed dependant variable, while $R^2 = 0$ indicates no linear relationship between the dependant variable and regressors.

The fitting procedure was found to return exactly the same results if the fitting of the decay was carried out up to a time just before the detection limit was reached, as in figure 4.14, or after it was reached, as in figure 4.15. This indicates that the offset in the regression model properly accounted for the detection limit of the system. The linearity of the detector used to collect the decay data was around 0.5%. The decay curves of erbium doped GLS,[113] taken with a similar detector over a similar dynamic range, showed no deviation from single exponential behavior; as would be expected from a rare earth ion. This indicates that the non-exponential decay observed in transition metal doped GLS is a physical phenomenon and is not related to a deviation from linearity in the detector.

## 4.7 Time resolved fluorescence decay data for vanadium doped GLS

Figure 4.14 shows the fluorescence decay of 0.002% V:GLS, fitted with a stretched exponential. The best fit to the experimental data was with a lifetime of 33 μs and a stretch factor (β) of 0.8. Visual inspection indicates an excellent fit to the experimental data, this is confirmed with an $R^2$ of 0.9996.

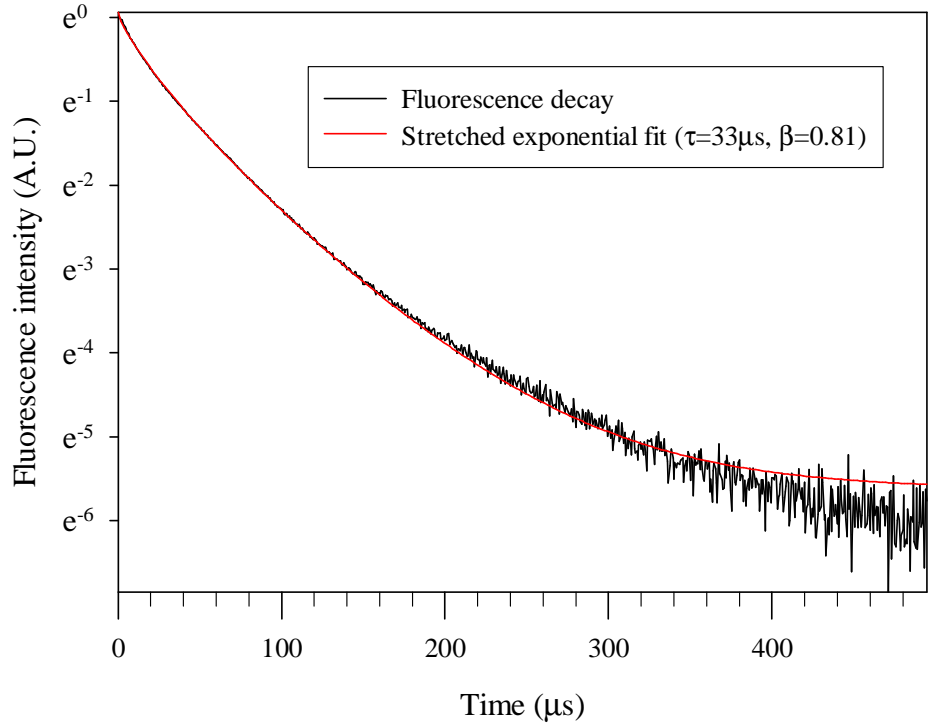

FIGURE 4.14 Fluorescence decay of 0.002% vanadium doped GLS fitted with a stretched exponential. The lifetime was 33 μs and the stretch factor was 0.81.

Figure 4.15 shows the fluorescence decay of 0.44% V:GLS fitted with a stretched and double exponential function. The double exponential function is given in equation 4.11.

$$I(t) = y_0 + I_1 \exp{-\left(\frac{t}{\tau_1}\right)} + I_2 \exp{-\left(\frac{t}{\tau_2}\right)} \qquad (4.11)$$

Where $\tau_1$ and $\tau_2$ are the two characteristic lifetimes; $I_1$ and $I_2$ are their respective coefficients. Inspection reveals that the stretched exponential does not describe the data as well as it does at lower concentrations and the double exponential function is a better fit for higher concentrations. The $R^2$ for the double and stretched functions are 0.9969 and 0.9898 respectively. The lifetimes for the double exponential fit are 6 and 29 μs; this is significant as it indicates that the lifetime observed at low concentrations is still present at high concentrations.

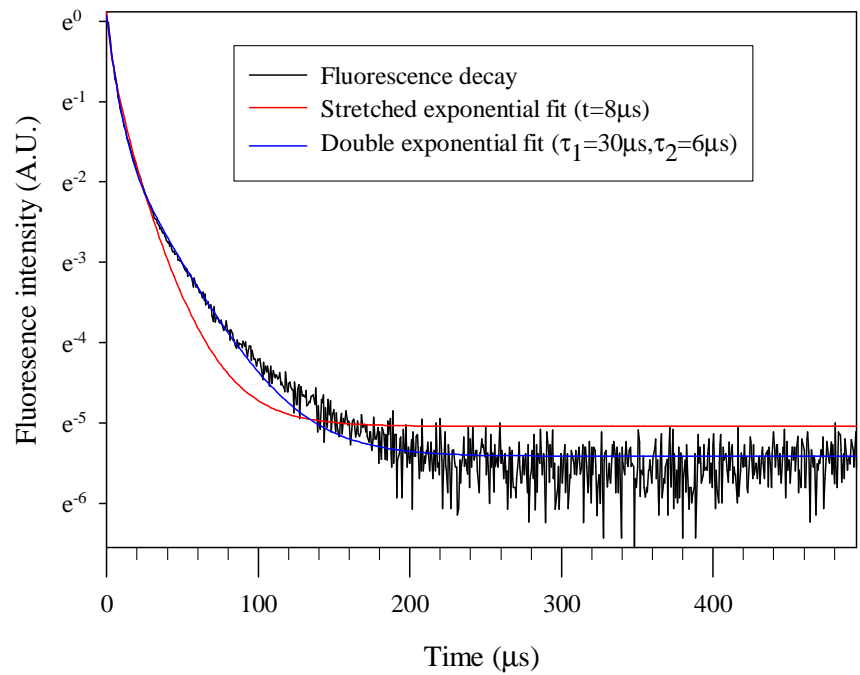

FIGURE 4.15 Fluorescence decay of 0.444% vanadium doped GLS fitted with a stretched and double exponential.

Figure 4.16 shows the residuals for the double and stretched exponential fits to the fluorescence decay of 0.44% V:GLS. The initial positive peak, in both residuals, is attributed to a slight timing uncertainty between the AOM switching off and the oscilloscope triggering. The next negative peak can also be attributed to this as the model readjusts to its initial displacement from the data. Subsequent residual is attributed to further inadequacies of the model. For the double exponential these are small and the residual approaches zero at around 80 µs. For the stretched exponential there is a relatively large and broad residual peak at around 150 µs and zero is not approached until about 300 µs. These differences are consistent with the hypothesis that the fluorescence decay has a bi exponential form at this concentration.

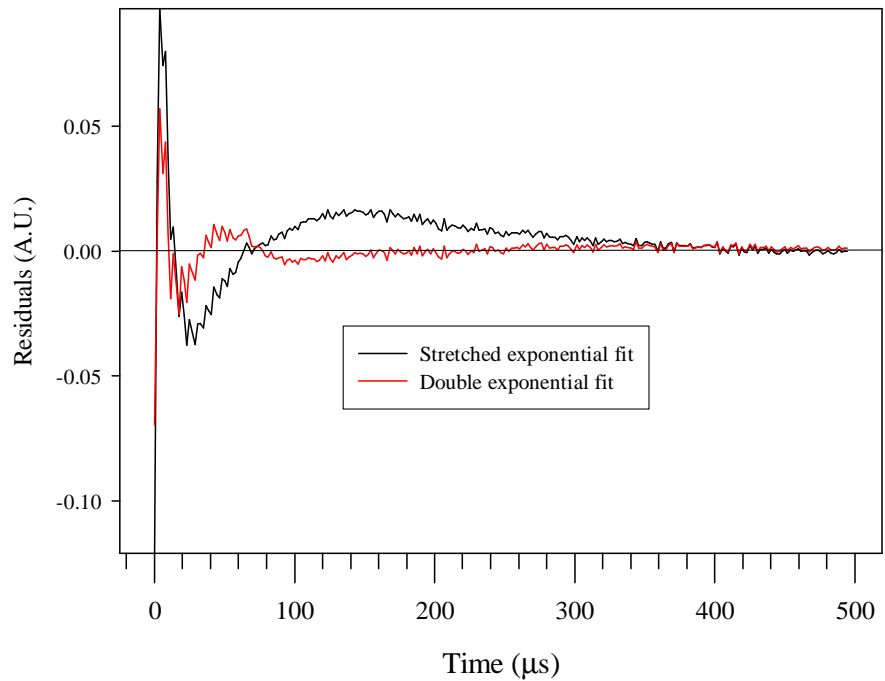

FIGURE 4.16 Comparison of the residuals of stretched and double exponential fits to the fluorescence decay of 0.444% vanadium doped GLS.

Figure 4.17 shows the fluorescence decay of 1.038% V:GLS fitted with a stretched and double exponential function. Similarly to 0.44% inspection reveals the stretched exponential does not describe the data as well as at lower concentrations and the double exponential function is a better fit. The $R^2$ for the double and stretched functions are 0.9945 and 0.9839 respectively. The lifetimes for the double exponential fit are 4 and 28 μs.

This observation of a slow lifetime component at 0.44% and 1.04% doping concentrations  which is very similar to the single stretched exponential lifetime observed at lower doping concentrations, is thought to have phenomenological significance and indicates that a characteristic lifetime component of ~ 30  μs is present at all concentrations measured.  It is noted that since we are modelling the decay in our glass with the stretched exponential model we should fit this decay with a double stretched exponential. However in order to minimise the number of regressor variables, so that a fair comparison can be made between the two models, a double exponential is thought to be a good approximation. This is because a single exponential function can be fitted to a stretched exponential (β = 0.8) with $R^2$ = 0.9839.

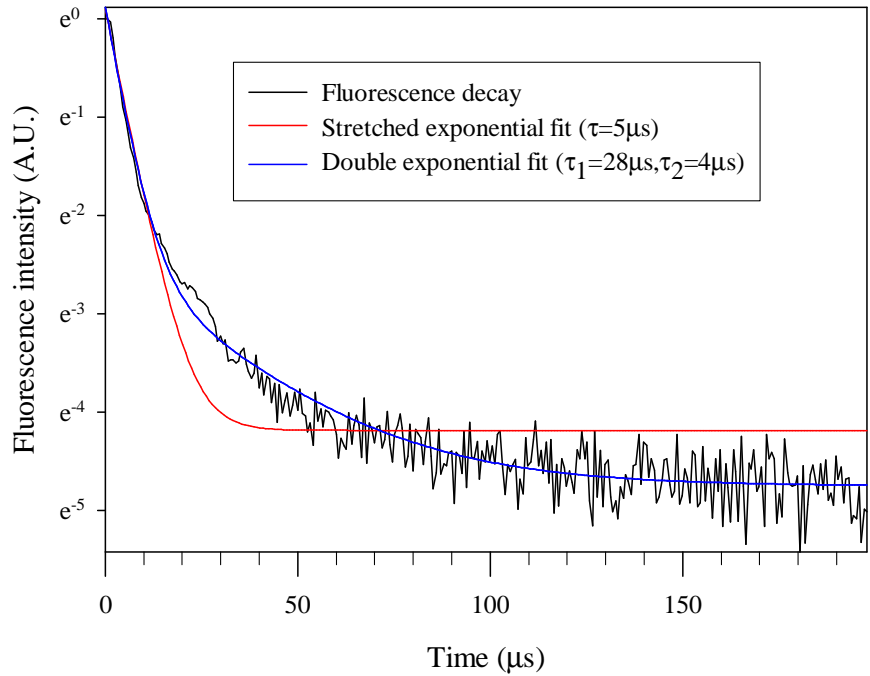

FIGURE 4.17 Fluorescence decay of 1.038% vanadium doped
GLS fitted with a stretched and double exponential.

When exciting at 830 nm, the lifetime of 0.09% V:GLS was 35 µs, compared to 31 µs when exciting at 1064 nm; this slight increase in lifetime can be explained by the preferential excitation of ions in higher crystal field sites. Lifetimes were also measured for vanadium doped GLSO samples although samples with a doping concentration higher than ~0.1% were not fabricated because of the loss of ORC facilities. There are two known mechanisms whereby concentration quenching can occur: cross relaxation and concomitant transfer. In cross relaxation an excited ion and an unexcited ion interact and some of the energy of the excited ion is transferred to the unexcited ion whereby both ions end up in an intermediate energy level, from which they both decay non-radiatively. In concomitant transfer the excitation is transferred from one ion to a neighbouring ion such that the second ion ends up in the same excited state, the excitation is transferred very rapidly between the ions, which greatly increases the probability that an ion close to a defect in the glass will be excited and decay non-radiatively. The theory of concentration quenching predicts the decay rate W will increase linearly with the square of ion concentration C: [114]

$$W = W_0 + UC^2 \qquad (4.12)$$

Where $W_0$ is the decay rate in the absence of concentration quenching and U is the energy transfer parameter. Figure 4.18 shows the lifetimes of V:GLS and V:GLSO fitted with a stretched exponential up to concentrations where no deviation from stretched exponential behavior was observed. The figure shows that V:GLSO has a longer lifetime than V:GLS, for the same doping concentration and that both have a slight negative dependence of lifetime with doping concentration, with this dependence being greater in V:GLSO. The fit of equation 4.12 shows that for vanadium doped GLSO $W_0$ and U are 0.0236 µs$^{-1}$ and 1.3221 µs$^{-1}$%molar$^{-2}$ respectively, whereas for vanadium doped GLSO $W_0$ and U are 0.0302 $\mu s^{-1}$ and 0.3979 $\mu s^{-1}\%$ molar$^{-2}$ respectively. Since there is not known to be an intermediate energy level between the emitting energy level and the ground state of V:GLS concomitant transfer is proposed as the concentration quenching process occurring in the V:GLS system.

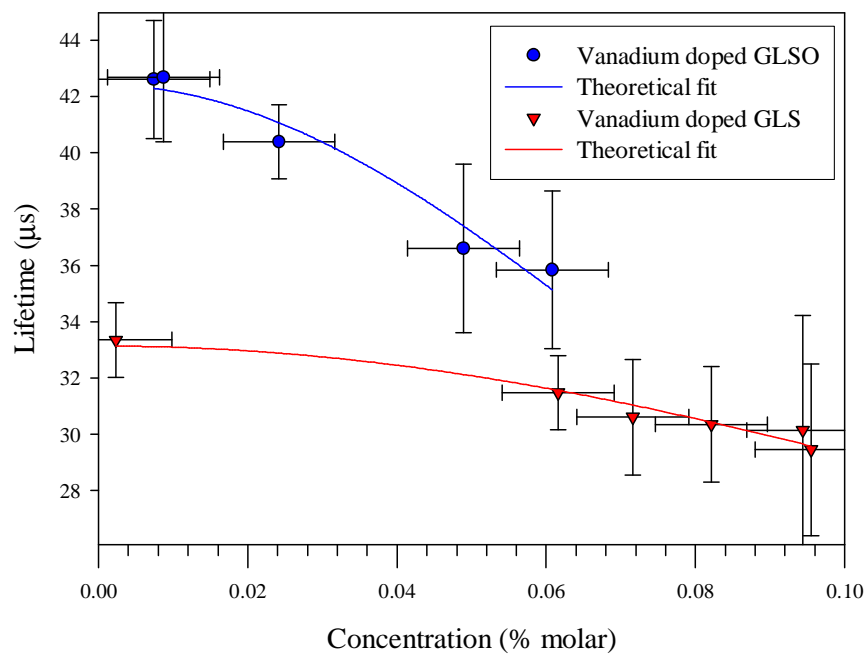

FIGURE 4.18 Lifetimes of vanadium doped GLS and GLSO as a function of doping concentration for fluorescence decays which could be fitted to the stretched exponential function. These lifetime have been fitted with equation 4.12.

The coefficient of determination of an unconstrained double exponential can be misinterpreted if the data follows stretched exponential behaviour. This is because certain combinations of the parameters of a double exponential function can give the appearance of stretched exponential behaviour. This is illustrated by fitting a double exponential to an artificially generated stretched exponential which often has an $R^2$ very close to 1 i.e. an almost perfect fit. Investigating the mathematical relevance of this is beyond the scope of this work. To overcome this problem it was assumed that if bi-exponential behaviour was present it should have a long lifetime component of around 30 $\mu s$ and a short lifetime component of around 5 $\mu s$. This assumption was implemented by fitting equation 4.11 to the decay data with the following constraints: $6 > \tau_1 > 4$, $31 > \tau_2 > 29$. These ranges are somewhat arbitrary and the object of the constraints are to show the conformity to bi-exponential behaviour with lifetime components of around 30 and 5 $\mu s$ at concentrations greater than ~0.1%.

Figure 4.19 shows the $R^2$ of stretched and constrained double exponential fits as a function of vanadium concentration, to allow easier interpretation of the data a log scale has been used. Figure 4.19 indicates that the stretched exponential gives an almost perfect fit for vanadium concentrations up to 0.1%. Above ~0.1% concentration the fluorescence decay starts to deviate from stretched exponential behaviour which is manifested as a decrease in $R^2$. The $R^2$ of the constrained double exponential shows a relatively poor fit up to a concentration of ~0.1% where the decay is then better described by a bi-exponential. If higher doping concentrations of V:GLSO had been fabricated it is believed they would follow the same trend observed for Ti:GLSO (section 5.2.4) where the $R^2$ of the stretched exponential did not decrease significantly up to a concentration of ~1%.

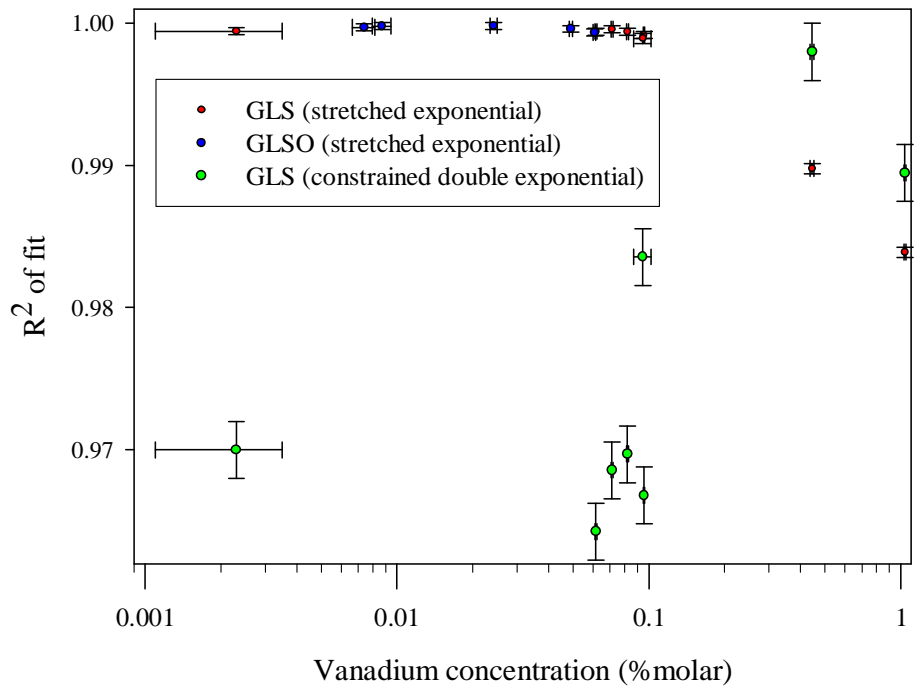

FIGURE 4.19 $R^2$ of stretched and constrained double exponential fits as a function of vanadium concentration.

Figure 4.20 shows that the stretch factor decreases (i.e. increased stretching) with increasing doping concentration. This behaviour is consistent with the temporally heterogeneous dynamics model of stretched exponential behaviour and can be explained by the increased proximity of fluorescing ions to defects in the glass as the concentration is increased. However, projecting the stretch factor to zero concentration indicates an initial stretch factor of ~0.8 for lowest possible concentrations. This could either be caused by a high concentration of defects in the glass or a continuous distribution of lifetimes (spatially heterogeneous dynamics).

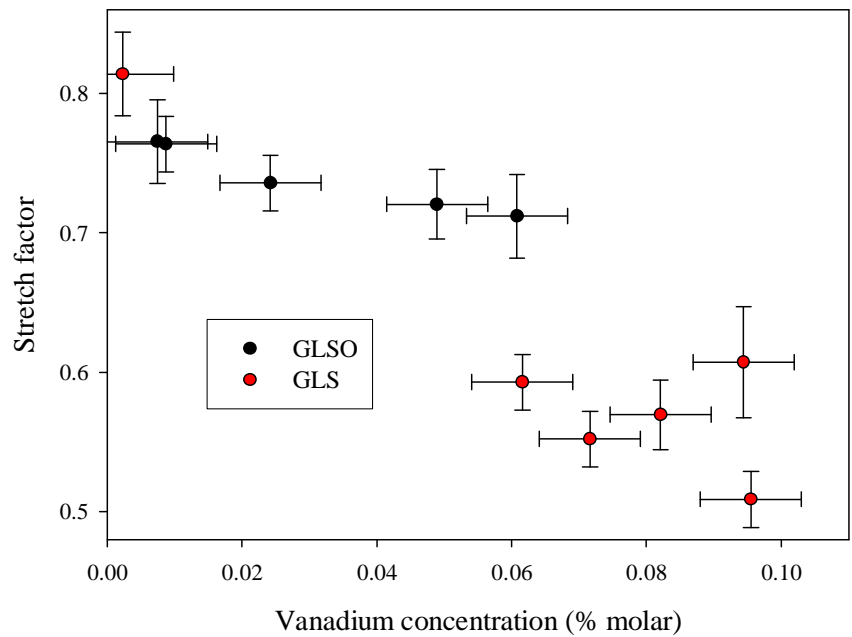

FIGURE 4.20 stretch factor as a function of vanadium doping
concentration in GLS and GLSO.

The finding that there is a deviation from stretched exponential behaviour in
concentrations above 0.1% in GLS but not in GLSO (see Ti:GLSO lifetime
measurements, section 5.2.4) and that lifetimes are longer in GLSO indicates that
another effect is taking place and that it is something to do with the oxygen content of
the glass. GLS contains ~0.5% (molar) oxygen whereas GLSO contains ~15% (molar)
oxygen. We therefore propose that two reception sites for transition metals exist in
GLS glass; a high efficiency oxide site and a low efficiency sulphide site. In GLS the
transition metal ion preferentially fills the high efficiency oxide sites until, at a
concentration of ~0.1%, they become saturated and the low efficiency sulphide sites
starts to be filled; this explains the deviation from stretched exponential behaviour at
concentrations > 0.1% and the appearance of  characteristic fast and slow lifetime
components. The peak position and shape of the absorption bands of V and Ti:GLS do
not change significantly as concentration increases from 0.01 to 1%, however one
would expect a noticeable red shift in absorption going from an oxide coordinated
transition metal to a sulphide coordinated transition metal. So the oxide site probably
doesn't have oxygen directly bonded to the transition metal ion. The observation that
the 0.1% threshold is five time less than the oxygen content of GLS indicates that the
oxide site contains around five oxygen atoms.

To put these findings in a structural context please refer to the discussion of oxide and
sulphides site in GLS in section 2.8.3. Dopant ions in glasses are generally expected to
enter substitutionally for network modifier cations.[60] The main network modifier in
GLS is $La^{3+}$[68] which is 8 fold coordinated to sulphur with an undetermined
symmetry.[69] We therefore propose that V and Ti substitute for $La^{3+}$ and are sulphide
coordinated; however in the high efficiency oxide sites one or more of the sulphur atoms is part of an oxide negative cavity whereas in the low efficiency sulphide sites none, one or more of these sulphur atoms is part of a sulphide negative cavity. Two dopant reception sites have been reported for dysprosium doped GLS (Dy:GLS),[16] key differences with this work are that the lifetimes in Dy:GLSO and Dy:GLS were 100 µs and 2.54 ms respectively. This meant that the oxide site was low efficiency and the sulphide site was high efficiency. Separate absorption bands were also identified for the two sites.

Dysprosium has the same outer electron structure as La; so if Dy substitutes for La the 8-fold coordination and any symmetry that may already exist would be expected to be maintained. However as a consequence of the outer electron structure and the relative size of V and Ti ions, tetrahedral and octahedral coordination are the most likely coordinations to occur. Analysis in this chapter indicates that vanadium is in a 2+ oxidation state and is octahedrally coordinated. So the addition of a transition metal to GLS will change the coordination of the La site it substitutes for from 8 to 6 or 4 (6 in the case of V:GLS) and bring about symmetry that may not have already existed. The oxide site being high efficiency for V:GLS and Ti:GLS, but low efficiency for Dy:GLS can be explained because O is more electronegative than S so the oxide site will have a slightly higher crystal field strength than the sulphide site. The separation of the lowest energy levels in transition metals is strongly influenced by crystal field strength so there will be a greater separation of the two lowest energy levels and therefore a lower probability of non-radiative decay. Energy levels in rare earths on the other hand are influenced very little by crystal field strength. The oxide site is expected to have a higher phonon energy than a sulphide site[16] this would increase the probability of non-radiative decay for both transition metal and rare-earth dopants. If the decrease in non-radiative decay caused by the increase in crystal field strength is greater than the increase in non-radiative decay caused by the increase in phonon energy for a transition metal in an oxide site then this would explain why the oxide site is high efficiency for transition metals and low efficiency for rare-earths.

## 4.8 Average Lifetime

Because the fluorescence decay in this study has been modelled with both stretched and double exponential function, comparisons between the lifetimes are difficult. Because of this the average lifetime is used for comparison in this section. The average lifetime is the statistical average lifetime as used in population analysis, and can be thought of as the summation over all time of the number of members (excited ions in this case) lost in a time interval $\Delta t$ multiplied by their age at loss; this quantity is then averaged over the total population.[104] Equation 4.13 gives the average lifetime in integral form[85, 104, 115]

$$\tau_{av} = \frac{\int_{t=0}^{t=t_{lim}} tI(t)\,dt}{\int_{t=0}^{t=t_{lim}} I(t)\,dt} \qquad (4.13)$$

Where I(t) is the detected emission decay data and $t_{lim}$ is the time when the detection limit is reached. Figure 4.21 is a graphical representation of tI(t) and I(t) which should aid the reader in understanding how the average was calculated, tI(t) was divided by 10 so that it was scaled to I(t).

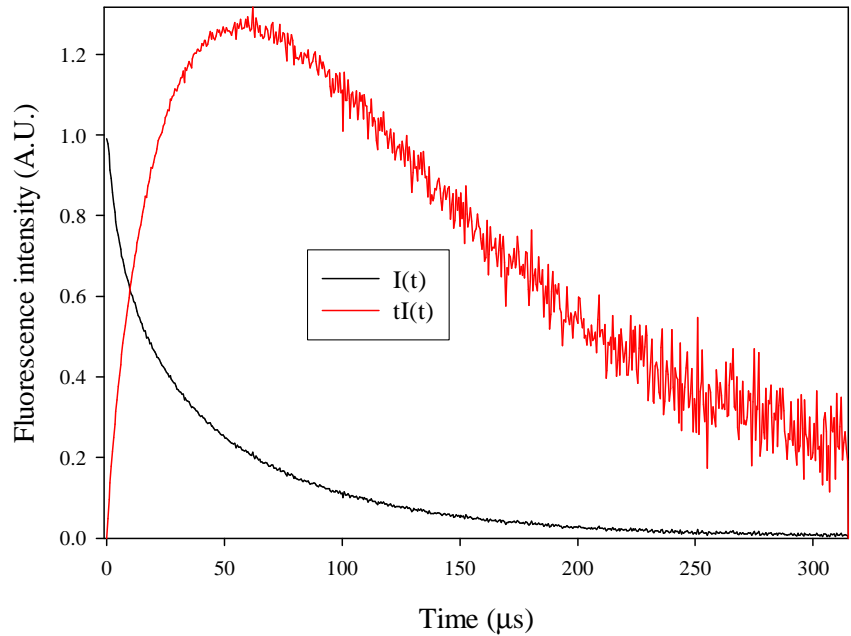

FIGURE 4.21 Graphical representation of I(t) and tI(t) for 0.096% vanadium doped GLS.

As shown in figure 4.21 I(t) and tI(t) were not integrated past a time when the signal detection limit was reached, since at this point I(t) and tI(t) are almost entirely comprised of noise; this increases $\int tI(t)$ to a much greater extent than $\int I(t)$ which increases $\tau_{av}$. The detection limit was set to the time at which I(t) was 1% of its initial value.

Figure 4.22 shows the average lifetime of V:GLS as a function of doping concentration. The first point to note about the average lifetime is that it is nearly double the lifetime calculated with the stretched exponential model. This is because stretching of an exponential effectively causes y=0 to be approached at a slower rate than in a single exponential and will therefore preferentially increase $\int tI(t)$ to a much greater extent than $\int I(t)$ which will increase the average lifetime in comparison to the lifetime calculated using the stretched exponential model. This is not significant because the average lifetime is used here as a statistical comparison tool.

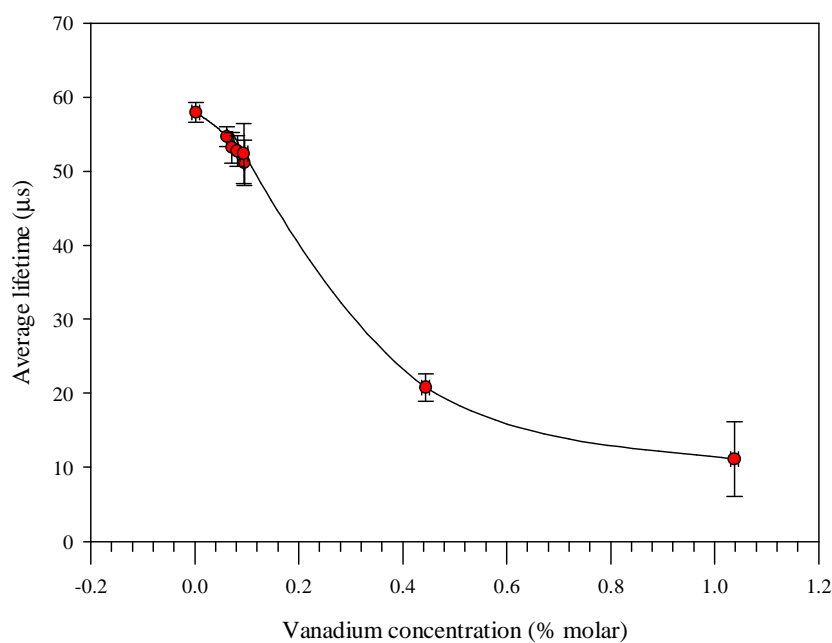

FIGURE 4.22 Average lifetime of vanadium doped GLS as a function of doping concentration. The lines are a guide for the eye.

The average lifetime as a function of doping concentration in figure 4.22 shows how the rate of decrease of average lifetime slows at concentrations > 0.5 %, this is attributed to saturation of defect centres in the glass.

## 4.9 Frequency resolved lifetime measurements of vanadium doped GLS

Frequency resolved lifetime measurements were made using the setup described in section 3.3.5 using 1064 nm excitation. Frequency resolved lifetime, otherwise known as frequency resolved spectroscopy (FRS), has been proposed as a method of avoiding misinterpretation of temporally resolved lifetime measurements, otherwise known as temporally resolved spectroscopy (TRS).[72] While TRS and FRS are equivalent in systems displaying exponential decay with a single lifetime, in more complicated systems this equivalence breaks down. For example in a bi-exponential system, variation of the excitation pulse width in TRS will vary the relative contributions of the fast and slow components, however in FRS the two components are present in a proportion relative to their intrinsic populations.[72] FRS also avoids problems that can occur with TRS in systems with second order recombination kinetics.[72] Another advantage of FRS over TRS is that a lock-in amplifier can be used to take the measurement which allows much weaker signals to be detected.

The FRS measurements presented here are intended as a reference and back-up to the TRS measurements. FRS measurements are typically taken with the in-phase and phase quadrature signal from a lock-in amplifier. Phase quadrature FRS (QFRS) measurements can reveal clearer information about distributions of lifetimes than in-phase FRS measurements[116-118] but are beyond the scope of this work. A useful analogy for understanding how fluorescence lifetime can be calculated from FRS measurements is to think of the fluorescing material as acting as a single RC time-constant filter with a cut-off frequency ($\upsilon_c$) inversely proportional to the lifetime of the material. Hence the lifetime ($\tau$) can be calculated from an in phase FRS measurement using equation 4.14.

$$\tau = \frac{1}{2\pi\upsilon_c} \qquad (4.14)$$

Where the cut-off frequency $\upsilon_c$ is defined as the frequency in the stop-band at which the output power is half the output power in the pass-band ($P_{pass}$).

The in phase FRS measurements for 0.0023%, 0.0944% and 1.038% V:GLS are shown in figure 4.23 along with the respective $\upsilon_c$ used to calculate the lifetime. The lifetimes calculated from FRS ($\tau_{FRS}$) and TRS ($\tau_{TRS}$) measurements are given in table 4.5.

TABLE 4.5 Lifetimes of various concentrations of V:GLS, calculated by FRS and TRS. † TRS lifetime calculated from stretched exponential fit (excited at 1064 nm), ‡ TRS lifetime calculated from average lifetime (excited at 1064 nm). †† 633 nm excitation.

| Vanadium concentration (% molar) | $\upsilon_c$(KHz) | $\tau_{FRS}$ (μs) | $\tau_{TRS}$ (μs) |
|---|---|---|---|
| 0.0023 | 4.8429 | 32.86 | 33.34† |
| 0.0944 | 5.0441 | 31.55 | 30.15† |
| 1.038 | 16.5015 | 9.64 | 11.10‡ |
| 0.0944 | 5.014 | 31.86†† | - |

The lifetimes calculated by FRS are in excellent agreement with TRS lifetimes calculated using the stretched exponential function; this validates the use of the stretched exponential function in calculating the fluorescence lifetime. The FRS lifetime measurement for 1.038% V:GLS is in reasonably good agreement with the average lifetime. The lifetime measured using a 1 mW 633 nm laser excitation source is in excellent agreement with the lifetime measured at 1064 nm in table 4.5. TRS measurement was not possible at 633 nm excitation because of the relatively low emission intensity at this wavelength (see V:GLS PLE figure 4.12) and the low power available; this highlights the advantage of FRS over TRS measurements. The lifetimes of 0.0944% V:GLS measured at wavelengths of 633, 830 and 1064 nm were 32, 35 and 30 μs respectively, see table 4.5 and section 4.7. Excitation at these wavelengths roughly equates to excitation into each of the three observed absorption bands and the observation of the same characteristic lifetime indicates that the three absorption bands all belong to the same oxidation state of vanadium.

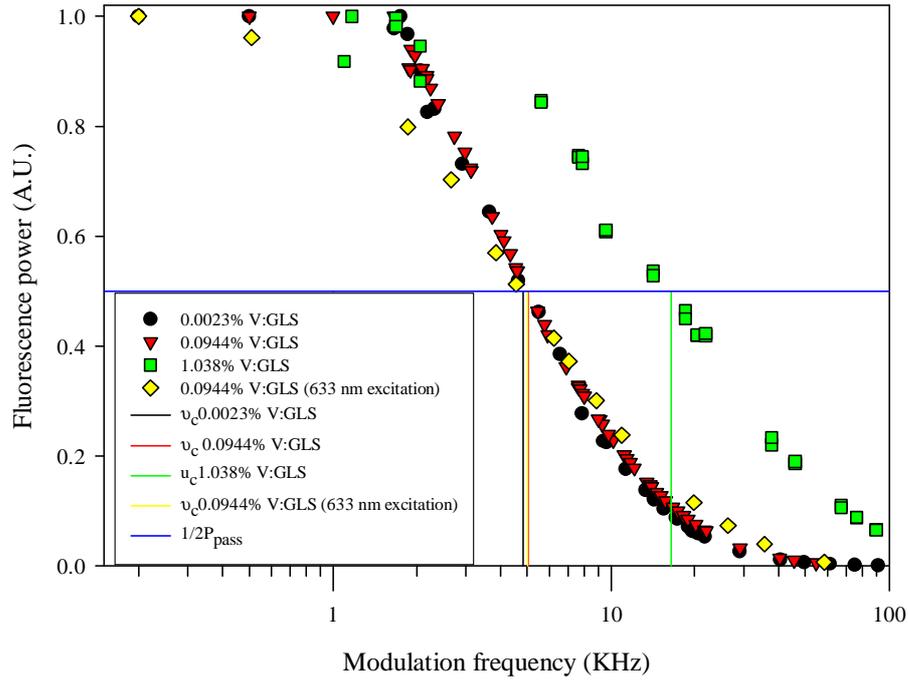

FIGURE 4.23 In phase FRS measurement of V:GLS at various concentrations. Excitation was at 1064 nm unless stated otherwise.

## 4.10 Continuous lifetime distribution analysis of vanadium doped GLS

So far the analysis of fluorescence decay data as a function of vanadium concentration using stretched and double exponential fits has indicated that at low concentrations there is a continuous distribution of lifetimes around 30 μs; which, in part at least, leads to a stretching of the decay. At higher concentration there are two decay constants of ~5 and 30 μs. Another method for recovering decay constants is to use an approach originally used in molecular physics and biophysics[119-123] and later applied to chromium doped aluminosilicate and gahnite glass[124-127] in which the distribution of the luminescence decays is approximated by a continuous function of decay constants, A(τ). Thus the luminescence decay is given by

$$I(t) = \int \frac{A(\tau)}{\tau} e^{-t/\tau} d\tau \qquad (4.15)$$

For ease of calculation equation 4.15 can be approximated by the discrete representation

$$I(t) = \sum_i \frac{A_i}{\tau_i} e^{-t/\tau_i} \qquad (4.16)$$

If a logarithmic scale is used for τ equation 4.15 becomes

$$I(t) = \int A(\tau) e^{-t/\tau} d(\ln \tau) \qquad (4.17)$$

And equation 4.16 becomes

$$I(t) = \sum_i A_i e^{-t/\tau_i} \qquad (4.18)$$

The coefficients $A_i$ enumerate the contributions of those sites that have a decay constant $\tau_i$ and were calculated from the decay data by regression analysis using the Marquardt-Levenberg algorithm given in equation 4.9. Equation 4.18 was implemented in MATLAB with the assistance of Dr. Giampaolo D'Alessandro, School of Mathematics, University of Southampton; the code is given in appendix A. The lifetime values, $\tau_i$, were spaced logarithmically between two specified lifetimes. The number of $\tau_i$ values was also specified. In previous work 125 $\tau_i$ values were shown to give unambiguous results;[125] 120 $\tau_i$ values were chosen for this work. Because such a large number of parameters were being fitted the reliability of the process was tested. This was done by creating a computer generated lifetime distribution consisting of two Gaussians with 120 logarithmically spaced lifetime values. The exponential decay for this artificial lifetime distribution was then computed and finally the continuous lifetime distribution model was used to recover the lifetime distribution as show in figure 4.24. The figure show that the model unambiguously identifies two lifetime distributions and accurately recovers their centre of gravity, there is however some discrepancy with the relative intensity and shape of the lifetime distributions.

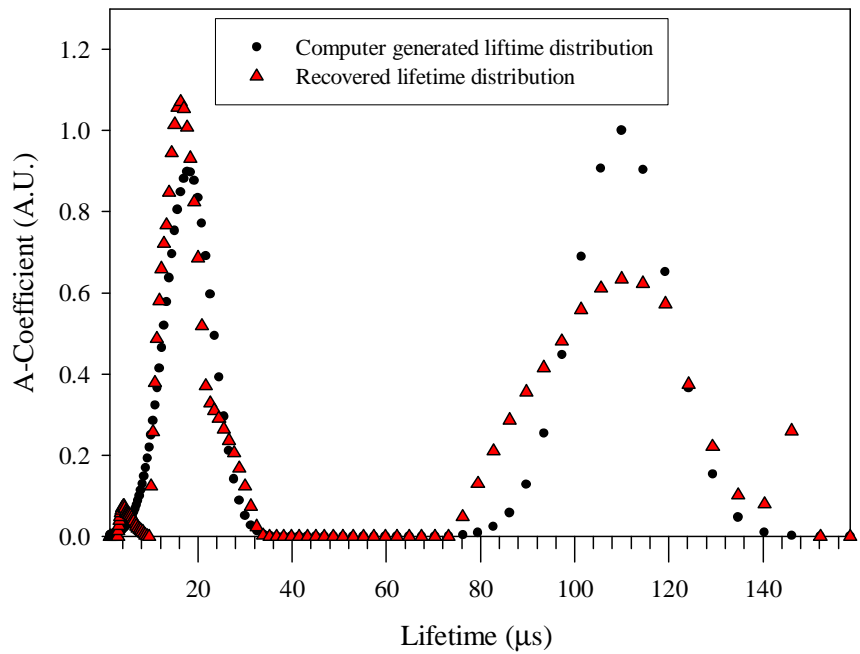

FIGURE 4.24 Test of continuous lifetime analysis technique using a computer generated continuous lifetime distribution.

The decay for the artificial lifetime distribution and the fit to the continuous lifetime distribution model are shown in figure 4.25 and indicate an excellent fit to the artificial decay.

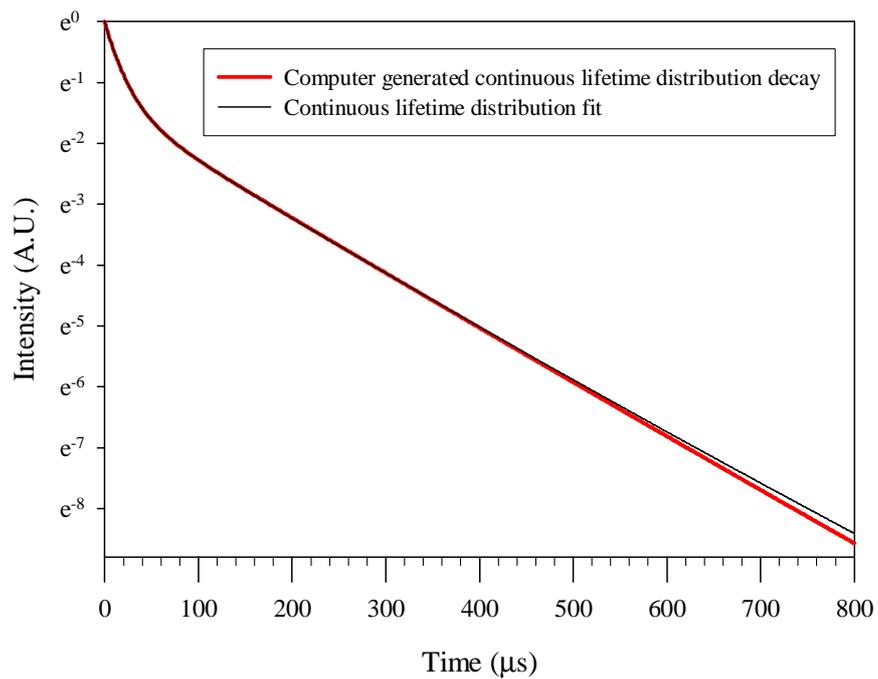

FIGURE 4.25. Exponential decay for the artificial lifetime distribution fitted to the continuous lifetime distribution model.

The choice of the range of decay constants was found to be extremely important for the quality of the fits, because of this the ranges were based on lifetime measurements in section 4.7. Figure 4.26 shows the fluorescence decays of 1.038% and 0.0955% fitted with the continuous lifetime distribution model, the figure indicates an excellent fit to the observed decay.

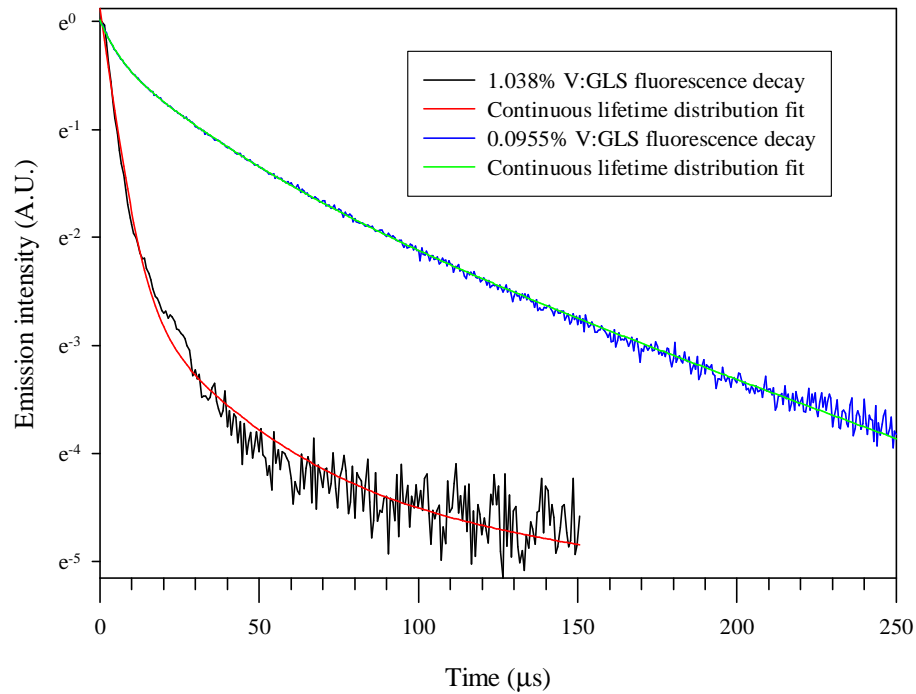

FIGURE 4.26 Some fluorescence decays of V:GLS fitted with a
continuous lifetime distribution.

Figure 4.27 shows how the distribution of lifetimes varies with vanadium concentration. The figure shows that at the lowest concentration there is a single distribution of lifetimes with a centre of gravity around 35 µs, which is in good agreement with the lifetime calculated from stretched exponential fit of 33.4 µs. As the concentration increases, two distribution peaks become apparent, one centred around 30 µs (peak 1) and another around 5 µs (peak 2). It is also clear that, as the concentration increases, peak 2 becomes more intense in comparison to peak 1. The distribution of lifetimes also appears to be narrower in peak 2. Since the centre of gravity of the lifetime distribution peaks can not be identified precisely from figure 4.27 they are given in table 4.6 along with lifetimes calculated from double and stretched exponential fits. Table 4.6 shows that there is excellent agreement between the lifetime distribution peaks and the stretched and double exponential peaks. The lifetime distributions also verify the assignment of double exponential behaviour at concentrations > ~0.1. However the lifetime distribution for 0.0955% V:GLS, which was found to be better described by a stretched than a double exponential, shows that a lifetime component ~ 5 µs is present although this component is small in comparison to the distribution ~ 30 µs.

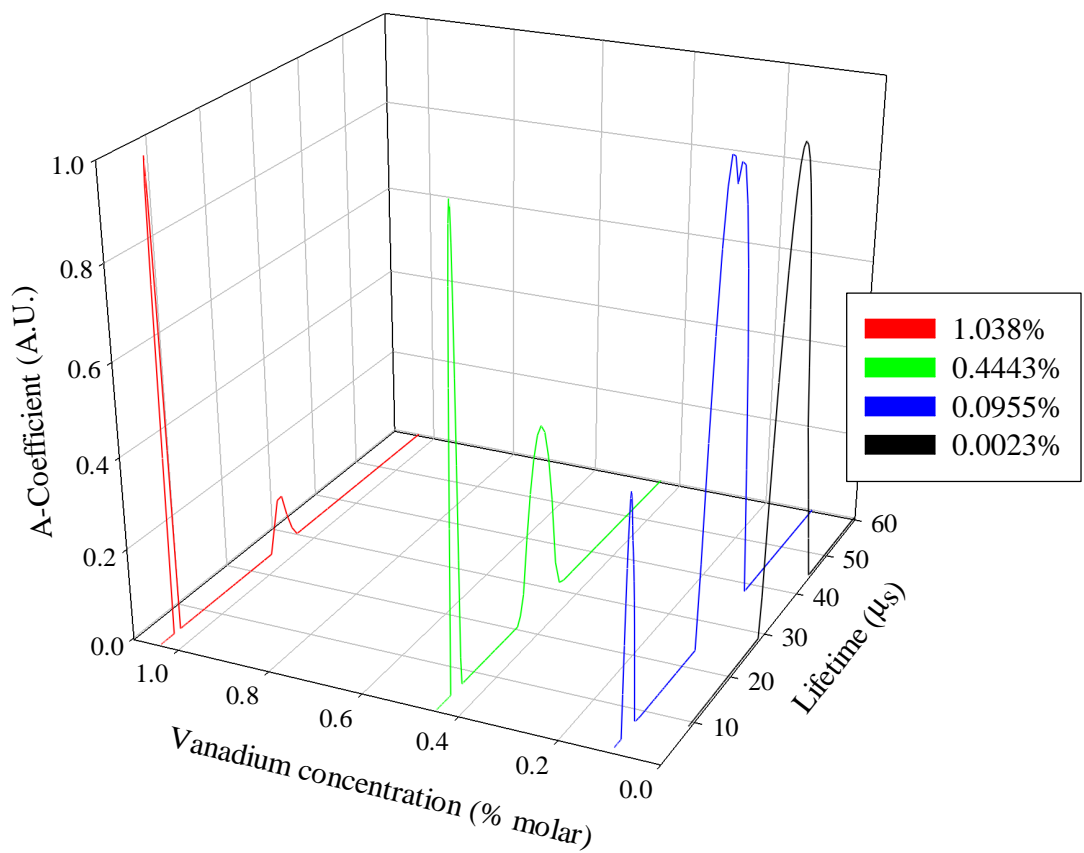

Figure 4.27 Lifetime distribution in V:GLS at various vanadium concentrations.

TABLE 4.6 Lifetimes identified by continuous lifetime distribution, double exponential and stretched exponential fits to the fluorescence decay of V:GLS at various vanadium concentrations.

| Vanadium concentration (%molar) | Continuous lifetime distribution peak 1 ($\mu$s) | Continuous lifetime distribution peak 2 ($\mu$s) | Double exponential fit $\tau_1$ ($\mu$s) | Double exponential fit $\tau_2$ ($\mu$s) | Stretched exponential fit ($\mu$s) |
|---|---|---|---|---|---|
| 0.0023 | 34.9 | - | - | - | 33.34 |
| 0.0944 | 30.0 | 4.8 | - | - | 30.15 |
| 0.4443 | 26.5 | 5.5 | 28.8 | 5.9 | - |
| 1.038 | 26.3 | 4.4 | 28.3 | 4.4 | - |

.

## 4.11 Temperature dependence of emission lifetime

### 4.11.1 Introduction

The temperature dependence of emission lifetime of 0.0023% V:GLS is modelled using a unified model of the temperature quenching of narrow-line and broad-band emissions developed by Struck and Fonger.[128] The Struck-Fonger model for non-radiative temperature dependence accounts for the presence of multiple activation energies in initial vibrational states which is thought to be an improvement on earlier single activation energy models.

### 4.11.2 Determination of quantum efficiency

Transitions from the excited state of a ion can occur either radiatively with the emission of a photon or non-radiatively through multi-phonon decay. The total decay rate W of a transition is the sum of the radiative decay rate $W_r$ and the non-radiative decay rate $W_{nr}$:

$$W = W_r + W_{nr} = \frac{1}{\tau} = \frac{1}{\tau_r} + \frac{1}{\tau_{nr}} \qquad (4.19)$$

In emission lifetime measurement it is the total emission lifetime $\tau$ that is measured, $\tau_r$ is the radiative lifetime and $\tau_{nr}$ is the non-radiative lifetime. The quantum efficiency $\eta$ can be calculated from the ratio of the radiative decay rate $W_r$ and the total decay rate W:[3]

$$\eta = \frac{W_r}{W} = \frac{W - W_{nr}}{W} = \frac{\tau}{\tau_r} = \frac{\tau}{\tau - \tau_{nr}} \qquad (4.20)$$

The total decay rate can be measured relatively simply, therefore the radiative decay rate $W_r$ or the non-radiative decay rate $W_{nr}$ must be calculated in order to obtain the quantum efficiency.

### 4.11.3 Struck- Fonger fit

The temperature dependence of the radiative decay rate $W_r(T)$ can be expressed by a coth law:[60]

$$W_r(T) = R_{vib} \coth\left(\frac{E_{vib}}{2kT}\right) \qquad (4.21)$$

Where $R_{vib}$ is the radiative decay rate at 0K, $E_{vib}$ is the energy of the acentric (odd parity) phonon. The acentric phonon differs from the totally symmetric ("breathing mode") phonon in that it can force an electric-dipole transition probability due to mixing with higher lying levels of different parity.

The temperature dependence of the non-radiative decay rate $W_{nr}(T)$ can be expressed by the Stirling approximation of the model of Struck and Fonger:[128]

$$W_{nr} = R_{nr} e^{-S\langle 2m+1 \rangle} \frac{e^{p^*}}{\sqrt{2\pi p^*}} \left[ \frac{2S\langle 1+m \rangle}{p+p^*} \right]^p \qquad (4.22)$$

Where

$$p^* = \sqrt{p^2 + 4S^2 \langle 1+m \rangle \langle m \rangle},$$

$$\langle m \rangle = \frac{1}{e^{\hbar\omega/kT} - 1},$$

Where S is the Huang-Rhys parameter, which describes the amount of overlap between the ground and excited state parabola, p is the number of phonons bridging the energy gap, $R_{nr}$ is the non-radiative decay constant and hω is the energy of the effective symmetric phonon. This approximation is valid for $P^*>1$. At 0K <m> → 0, therefore $W_{nr}$ at 0K is given by:

$$W_{nr}(0) = R_{nr} \frac{e^{p-S}}{\sqrt{2\pi p}} \left( \frac{S}{p} \right)^p \qquad (4.23)$$

Hence equation (4.22) can now be expressed in a form which is used to fit to the experimental data:

$$W_{nr}(T) = W_{nr}(0) \left[ \frac{p}{p^*} \right]^{1/2} \left[ \frac{2p\langle 1+m \rangle}{p+p^*} \right]^p e^{(p^*-p-2mS)} \qquad (4.24)$$

The total decay rate is thus given by:

$$W = W_r + W_{nr} = W_r(R_{vib}, E_{vib}) + W_{nr}(R_{nr}, \hbar\omega, S, p)$$

$$= R_{vib} \coth\left( \frac{E_{vib}}{2kT} \right) + R_{nr} \frac{e^{p-S}}{\sqrt{2\pi p}} \left( \frac{S}{p} \right)^p \left[ \frac{p}{p^*} \right]^{1/2} \left[ \frac{2p\langle 1+m \rangle}{p+p^*} \right]^p e^{(p^*-p-2mS)} \qquad (4.25)$$

This equation can now be fitted to the measured experimental data (the total decay rate W as a function of temperature) in order to find fit parameters to calculate the radiative decay rate $W_r$ and hence calculate the quantum efficiency.

## 4.11.4 Parameter estimation

In order to achieve a valid fit to the experimental data, initial estimates of the start values of the fit parameters need to be made as accurately as possible. The Huang-Rhys parameter, S, and the acentric phonon energy, $E_{vib}$, can be estimated from the temperature dependence of emission bandwidths Γ(T):[60]

$$\Gamma(T) = 2.36\hbar\omega\sqrt{S} \left( \coth\left( \frac{E_{vib}}{2kT} \right) \right)^{1/2} \qquad (4.26)$$

Temperature dependent emission bandwidth measurements were taken using the photoluminescence spectroscopy setup described in section 3.3.2 but with the 0.0023% V:GLS sample enclosed in a Leybold AG helium gas closed cycle cryostat. When the cryostat reached the desired temperature it was left at that temperature for 40 minutes to allow for any thermal inertia between the temperature sensor and the sample. The excitation source was a CW 1064nm Nd:YAG laser with approximately 0.5 Watt output power. Emission was detected with an EG&G optoelectronics J10D liquid nitrogen cooled InSb detector. Figure 4.28 shows the spectra plotted in energy taken at temperatures varying from 6.6 K to 293 K. These spectra where fitted to the 4 parameter Gaussian in equation 4.2.

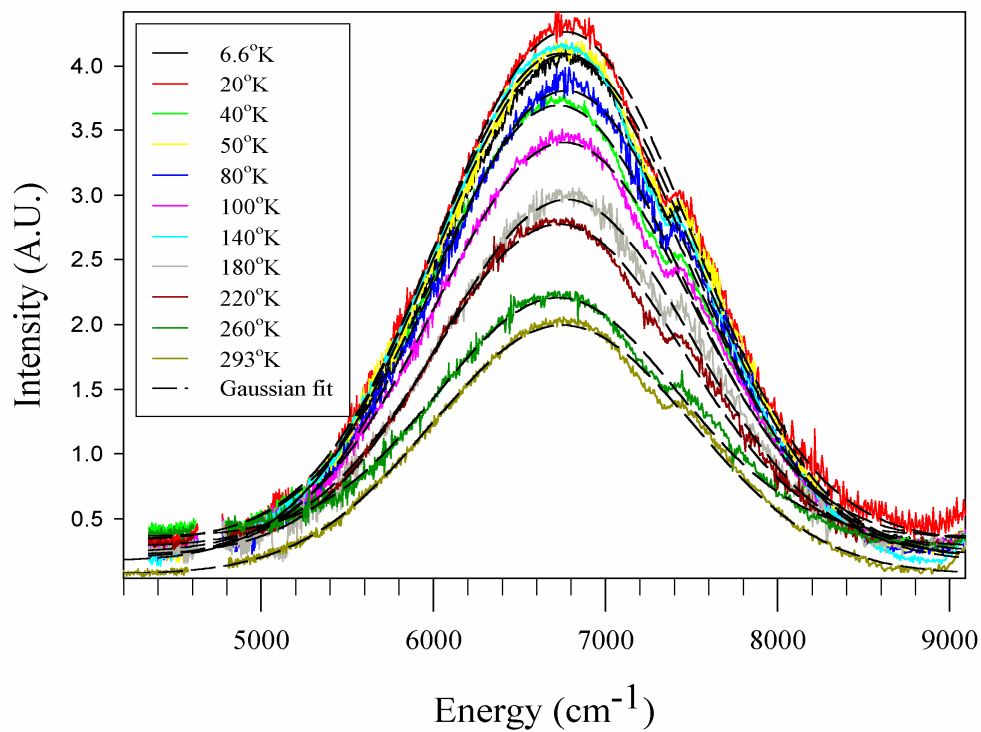

FIGURE 4.28 Emission spectra of 0.0023% V:GLS at various temperatures fitted with a 4 parameter Gaussian.

The variation in FWHM of the Gaussian curves in figure 4.28 was less than the estimated resolution of approximately 60 cm$^{-1}$. This indicates that emission broadening due to an inhomogeneous broadening mechanism, such as the range of crystal field strengths (Dq) that can be created in different glass sites,[129] is much stronger than that caused by the coupling of vibrational modes. The S and E$_{vib}$ parameters cannot therefore be estimated using this method. The resolution was estimated to be approximately 60 cm$^{-1}$ by averaging the difference between bandwidth measurements taken at the same temperature. The accuracy of this experiment was hampered by noise from the detector caused by vibration of the closed cycle cryostat. Several methods of isolating the detector from the vibration were tried but none succeeded in eliminating the vibration.

The Huang-Rhys parameter S can also be calculated from the emission bandwidth at 0K:[60]

$$\Gamma(0) = 2.36\hbar\omega S^{1/2} \qquad (4.27)$$

The bandwidth measured at 6.6K was measured to be 1740 cm$^{-1}$ and is a good approximation for the 0K bandwidth ($\Gamma(0)$). The maximum phonon energy, $\hbar\omega$, has been measured previously[34] to be 425 cm$^{-1}$. This gives an estimate for the S parameter of 2.95.

If the ground and excited state parabola are identical the Huang-Rhys parameter S can also be calculated from the difference between the emission and absorption band peaks (known as the Stokes shift (SS)) and the maximum phonon energy $\hbar\omega$:[60] SS=(2S-1) $\hbar\omega$. A description of the Huang-Rhys parameter and Stokes shift in terms of the single configurational coordinate model is given in section 2.5. The 300K absorption peak at 9091 cm$^{-1}$ and an emission peak at 6725 cm$^{-1}$ gives a Stokes shift of 2366 cm$^{-1}$. This gives an estimate of the S parameter of 3.28, the difference to the 0K emission bandwidth calculation may be because the ground and excited state parabola are not identical. The number of phonons bridging the energy gap $E_{gap}$ can be calculated from p=$E_{gap}$/$\hbar\omega$. $E_{gap}$ for V:GLS is 6725 cm-1 which gives p=15.82.

Assuming that $E_{vib}$ is equal to the maximum phonon energy $\hbar\omega$, the radiative rate at 0 K can be estimated:[130]

$$R_{vib}(0K) = \frac{1}{\tau_r(0K)} = \frac{\eta(T)}{\tau(T)\coth\left(\dfrac{E_{vib}}{2kT}\right)} \qquad (4.28)$$

Then using the quantum efficiency for V:GLS at 300 K of 0.04 calculated in section 4.12 and the lifetime at 300 K of 33 µs, $R_{vib}(0K)$=$W_r(0K)$ was estimated to be 936 s$^{-1}$. Approximating W(0K) to 1/$\tau$(6.5K) and using equation 4.19, $W_{nr}(0K) \approx$ 18290 s$^{-1}$. Using equation 4.23 with S=2.95, $W_{nr}(0K)$=8290 s$^{-1}$ and p=15.82 the non-radiative decay constant $R_{nr}$ is 9x10$^{10}$ s$^{-1}$.

### 4.11.5 Temperature dependent lifetime measurements

Temperature dependent lifetime measurements were taken using the fluorescence lifetime setup described in section 3.3.4, except with the sample enclosed in a Leybold AG helium gas closed cycle cryostat. Fluorescence was detected by a new focus 2034 extended InGaS detector. The excitation source was a 1064nm Nd:YAG laser with the output power attenuated to approximately 10 mW. When the cryostat reached the desired temperature it was left at that temperature for 40 minutes to allow for any thermal inertia between the temperature sensor and the sample. The emission decay curves, for various temperatures between 6.6 K and 293 K are shown in figure 4.29 together with the stretched exponential fits that were used to calculate the 1/e lifetimes.

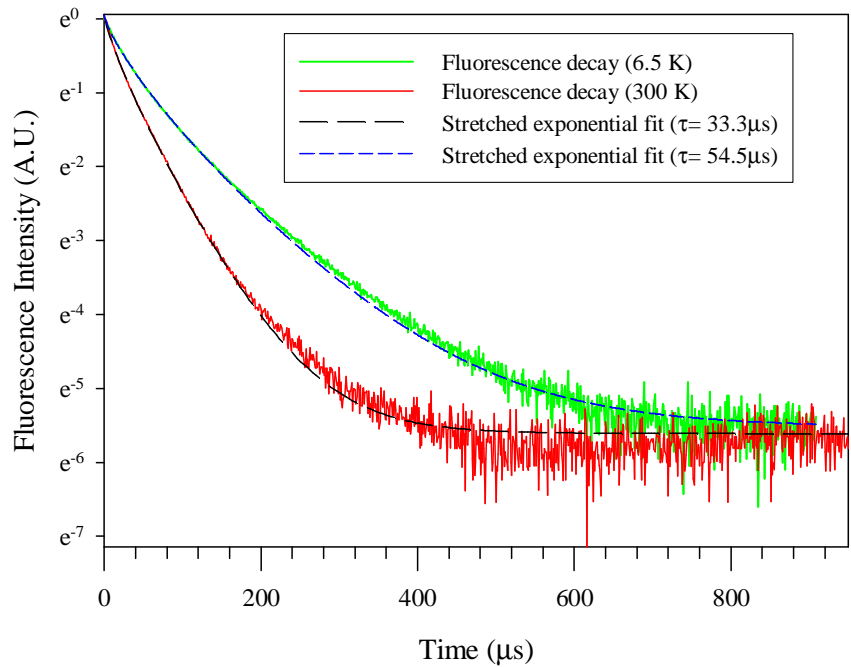

FIGURE 4.29 Emission decay of 0.0023% V:GLS at 6.5 and
300 K together with stretched exponential fits.

The experimental data was fitted to equation 4.25 with the parameters estimated
previously as initial parameters. All of the parameters were fitted to the data except $\hbar\omega$
which has previously been determined experimentally and was set as a constant. The
initial and fitted parameters are given in table 4.7. Figure 4.30 shows the experimentally
determined decay rates, the Struck-Fonger fit and the non-radiative decay rate, $W_{nr}(T)$,
calculated from the fitted parameters $R_{nr}$, $\hbar\omega$, $S$ and $p$ with equation 4.24. The figure
indicates that the Struck-Fonger model describes the temperature dependent decay rate
of V:GLS very well from 300-210 K but then there is a small deviation that can not be
accounted for by the model, this may be related to the disordered nature of the glass
host. It can also be seen that at 0 K there are still strong non-radiative processes
occurring.

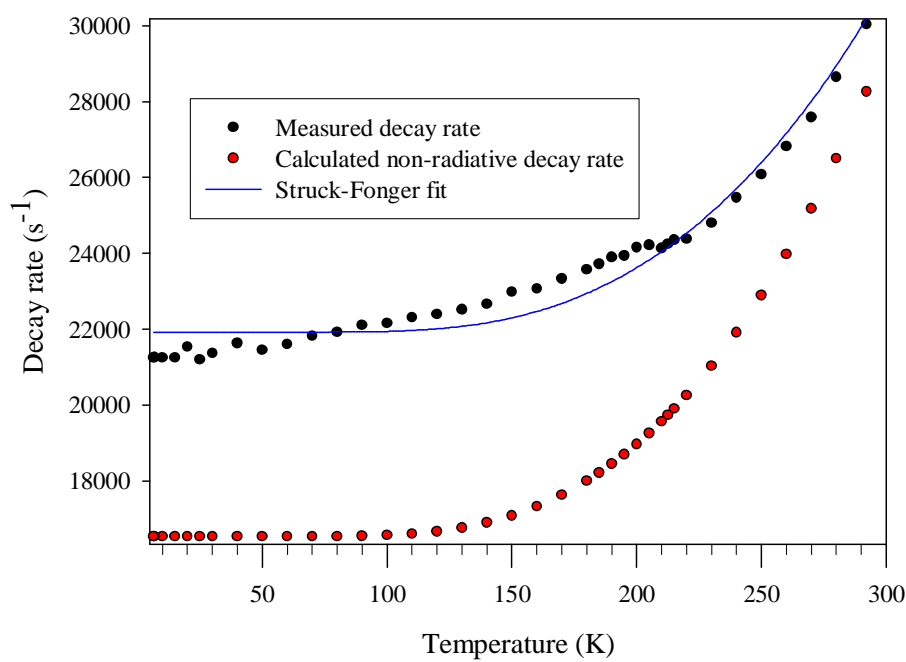

FIGURE 4.30 Experimental data for the total decay rate of 0.0023% vanadium doped GLS as a function of temperature fitted to the model of Struck and Fonger and the non-radiative decay rate as a function of temperature was calculated from the fit parameters.

In figure 4.31 the quantum efficiency (QE) as a function of temperature $\eta(T)$ was calculated from experimentally measured total decay rate as a function of temperature $W(T)$ and the non-radiative decay rate as a function of temperature $W_{nr}(T)$ with: $\eta(T)= W(T)-W_{nr}(T)/ W(T)$. The fit shows a 5.1% QE at room temperature which is in excellent agreement with the spectroscopically determined QE of 4.2%. Errors in the calculated QE were estimated from the coefficient of determination of the Struck-Fonger fit.

TABLE 4.7 Initial estimate and fit parameters for Struck-Fonger fit.

| | $R_{vib}$ (s$^{-1}$) | $E_{vib}$(cm$^{-1}$) | $R_{nr}$ (s$^{-1}$) | $\hbar\omega$ (cm$^{-1}$) | S | p |
|---|---|---|---|---|---|---|
| Initial estimate | 936 | 425 | $9\times10^{10}$ | 425cm$^{-1}$ | 2.95 | 15.82 |
| Struck-Fonger fit | 4549 | 1414 | $7.9\times10^{10}$ | - | 2.79 | 15.05 |

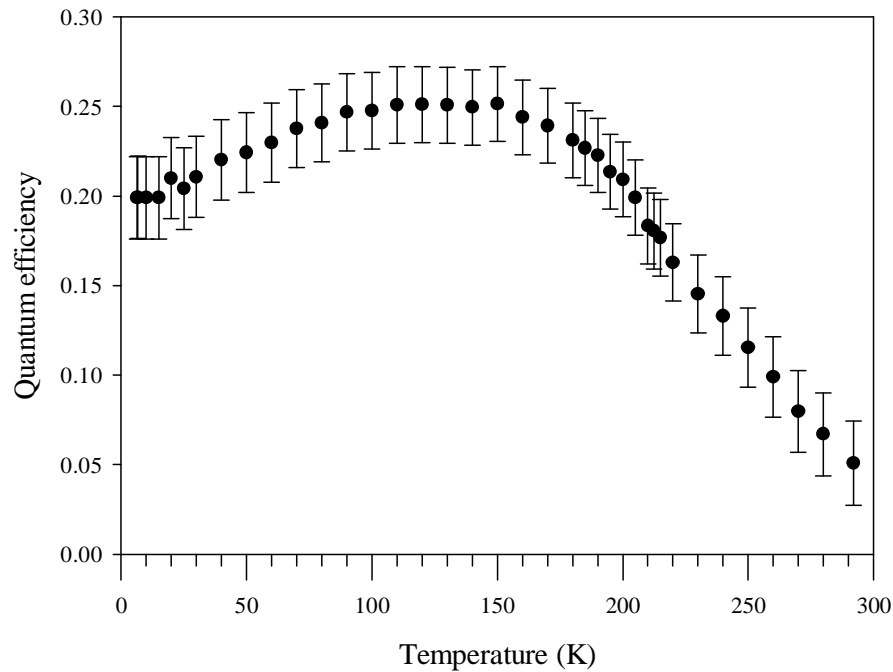

FIGURE 4.31 Temperature dependence of the quantum efficiency of 0.0023% V:GLS calculated from the Struck-Fonger model.

The temperature behaviour of the quantum efficiency is similar to that observed in $Cr^{4+}$ doped garnates[130] which showed and increase up to 130 K, caused by coupling of non-totally-symmetric phonons which forced electric dipole transition probability. Above 130 K non-radiative decay processes dominate and the quantum efficiency decreases.

## 4.12 Quantum efficiency measurements

Measurement of a material's fluorescence quantum efficiency (QE) is a key parameter in determining if it will be useful for various active devices such as optical amplifiers and lasers. It is particularly important for materials that have transition metals as the active ion because a key laser parameter, the radiative decay rate (otherwise known as the Einstein coefficient A), can not be determined from Judd-Ofelt analysis as it can be for rare earth dopants. If the quantum efficiency ($\eta_{QE}$) is known then the radiative decay rate ($W_r$) can be calculated from the, relativity easily measured, fluorescence lifetime ($\tau$) using the following relationship: $W_r = \eta_{QE}/\tau$.

Quantum efficiency measurements were taken using the method described in section 3.3.7. Figure 4.32 illustrates how the entire emission spectrum was found by fitting a Gaussian to the limited emission spectrum detected from the output of the integrating sphere.

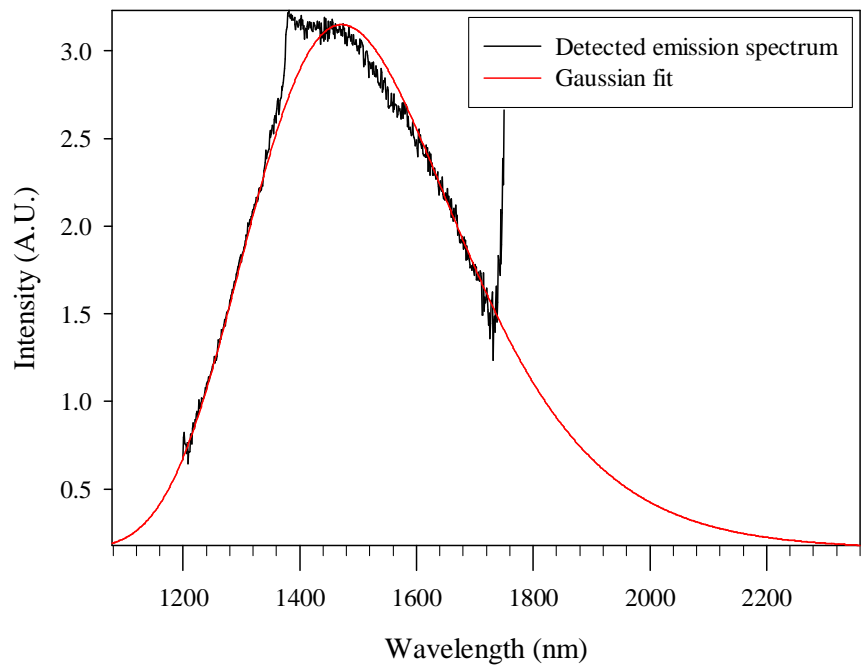

FIGURE 4.32 Emission spectrum of 0.0023% V:GLS taken with an integrating sphere and fitted to a Gaussian.

Appendix B gives the area under all the spectra taken, together with the calculated quantum efficiencies. The quantum efficiency for the vanadium doped GLS samples are illustrated in figure 4.33 including the experimental error bounds. The quantum efficiency as a function of vanadium doping concentration follows a similar trend to the lifetime i.e. decreasing with increasing doping concentration. However whereas the lifetime in V:GLSO was higher than V:GLS for the same vanadium concentration it appears that the quantum efficiency is the same or slightly lower in V:GLSO for the same doping concentration. This means that in V:GLSO the radiative rate is lower than in V:GLS since $W_r = W\eta$, where $\eta$ is the quantum efficiency. The decrease in QE with increasing concentration can be attributed to increased re-absorption of the emission and/or increased concentration quenching.

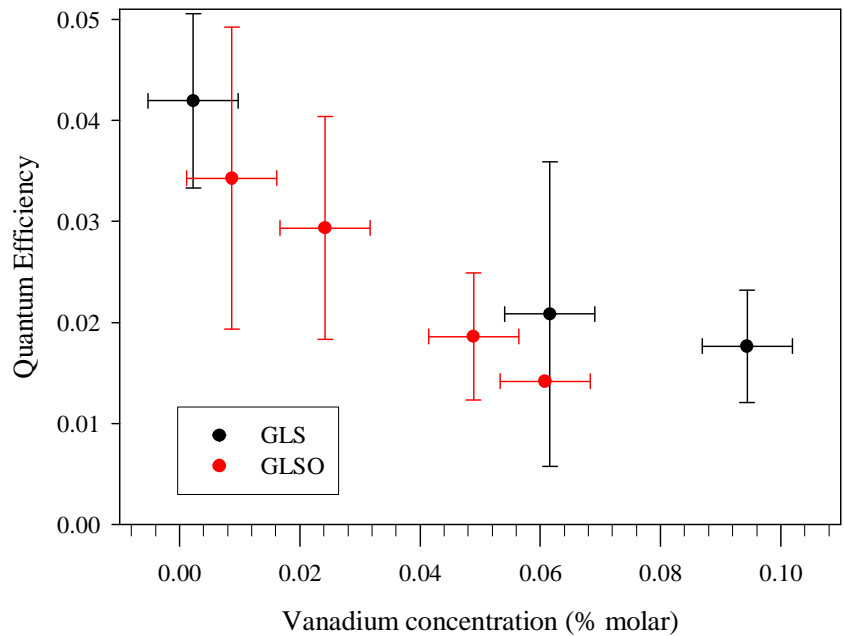

FIGURE 4.33 Quantum efficiency of vanadium doped GLS as a function of doping concentration measured with an integrating sphere.

The formula of McCumber,[131] in equation 4.29, is used to calculate the peak emission cross section.

$$\sigma_{em} = \sqrt{\frac{\ln 2}{\pi}} \frac{A}{4\pi c n^2} \frac{\lambda_0^4}{\Delta\lambda} \tag{4.29}$$

Where $\lambda_0$ is the peak fluorescence wavelength, $\Delta\lambda$ is the FWHM, n is the refractive index, c is the speed of light and A is the Einstein coefficient (also known as the radiative rate ($W_r$)) and is calculated from: $A = QE/\tau$. The peak emission cross-sections, calculated for various concentrations of V:GLS and V:GLS, are given in table 4.9.

TABLE 4.9 Details of vanadium samples with their respective quantum efficiency, peak emission wavelength ($\lambda_{max}$), emission bandwidths ($\Delta\lambda$), emission lifetimes ($\tau$), emission cross sections ($\sigma_{em}$) and $\sigma_{em}\tau$ products at room temperature.

| Vanadium concentration (%molar) | Host | Sample dimensions (mm) | QE (%) | $\lambda_{max}$ (nm) | $\Delta\lambda$ (nm) | $\tau$ ($\mu$s) | $\sigma_{em}$ ($10^{-21}$ cm$^2$) | $\sigma_{em}\tau$ ($10^{-26}$scm$^2$) |
|---|---|---|---|---|---|---|---|---|
| 0.0023 | GLS | 8x5x4 | 4.19 | 1470 | 484 | 33.35 | 2.721 | 9.08 |
| 0.0616 | GLS | 8x5x3 | 2.08 | 1508 | 447 | 31.47 | 1.716 | 5.40 |
| 0.0944 | GLS | 8x8x0.5 | 1.56 | 1536 | 462 | 30.15 | 1.579 | 4.76 |
| 0.0087 | GLSO | 8x8x0.5 | 3.43 | 1461 | 450 | 42.68 | 2.051 | 8.75 |
| 0.0242 | GLSO | 8x8x0.5 | 2.93 | 1466 | 448 | 40.39 | 1.886 | 7.62 |
| 0.0489 | GLSO | 8x8x0.5 | 1.86 | 1477 | 467 | 36.60 | 1.306 | 4.78 |
| 0.0608 | GLSO | 8x8x0.5 | 1.42 | 1482 | 447 | 35.84 | 1.078 | 3.86 |

The emission lifetime and cross section are important parameters for the characterisation of a laser material because the laser threshold is inversely proportional to $\sigma_{em}\tau$.[130] Because of this the most favourable $\sigma_{em}\tau$ found in this work for 0.0023% V:GLS is compared to that in other laser materials in table 4.10.

TABLE 4.10 Overview of the spectroscopic parameters for various laser materials compared to V:GLS. * These lasers only operate at low temperatures.

| Ion | Host | QE (%) | $\tau$ ($\mu$s) | $\sigma_{em}$ ($10^{-19}$ cm$^2$) | $\sigma_{em}\tau$ ($10^{-24}$scm$^2$) | Reference |
|------|---------|--------|--------|--------|--------|-----------|
| $V^{2+}$ | GLS | 4.19 | 33.35 | 0.02721 | 0.0908 | This work |
| $V^{2+}$ | MgF$_2$* | - | 40 | 0.045 | 0.18 | [103] |
| $V^{2+}$ | KMgF$_3$* | 87 | 1200 | 0.04 | 4.8 | [103] |
| $V^{2+}$ | CsCaF$_3$* | 40 | 1400 | - | - | [6] |
| Ti$^{3+}$ | Al$_2$O$_3$ | 100 | 3.1 | 4.5 | 1.40 | [132] |

Table 4.10 compares the spectroscopic parameters from this work with those of other $V^{2+}$ doped laser materials and the commercially successful Ti:Sapphire. Comparisons indicate that the lifetime of V:GLS is comparable or better than existing doped laser hosts. Though the QE and emission cross-section do not compare favourably, the ability to form optical fibres from V:GLS may overcome potential heat dissipation problems caused by the low QE (due to the large surface area to volume ratio of optical fibres). Additionally, the high pump beam confinement that can also be achieved in a fibre could compensate for the low emission cross section.

## 4.13 X-ray Photoelectron Spectroscopy

X-ray Photoelectron Spectra (XPS) were taken using the setup described in section 3.3.8 and is used here to determine the oxidation state of the vanadium ion. An XPS spectrum of 1% vanadium doped GLS is shown in figure 4.34 with a close up of the vanadium peak in figure 4.35. Photoelectron spectra of non conducting samples are known to have a shift in energy due to charging of the sample which can exert an attractive force on escaping photoelectrons and hence cause an unknown energy shift in the spectra.[77] This is usually corrected for from the position of the C1s peak of non-intrinsic carbon present in nearly all samples.[133] However the region of the C1s signal showed a complex structure, making it difficult to assign the peak of the adventitious carbon, so the spectrum was corrected using the O1s peaks. Next the elastic tail, or Shirley background, which is caused by electrons rebounding off ion sites, was removed. Then the spectrum was deconvoluted using a series of Gaussian-Lorentzian peaks into a best-fit of the measured spectrum.

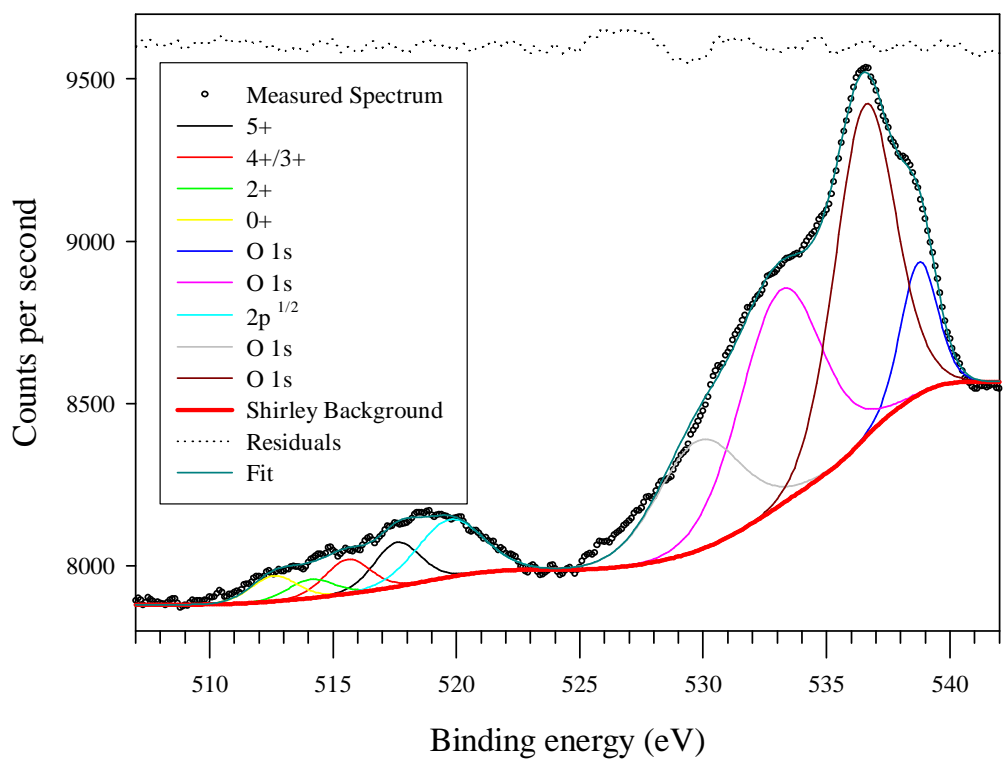

FIGURE 4.34 X-ray photoelectron spectra of 1% vanadium doped GLS.

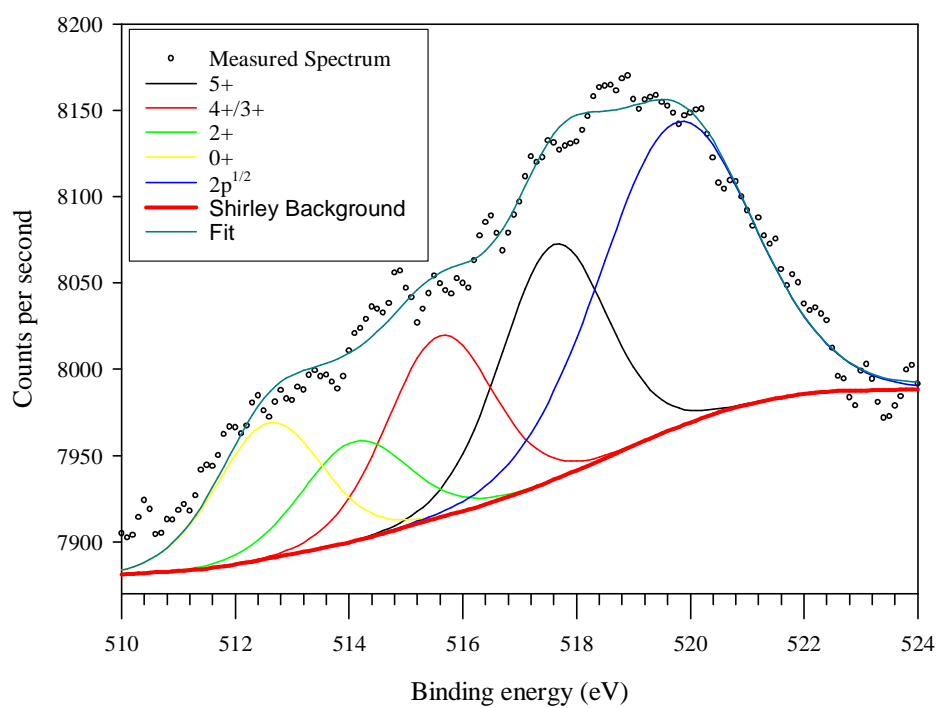

FIGURE 4.35 Close up of vanadium peak for X-ray photoelectron spectra of 1% vanadium doped GLS.

The spectra in figure 4.34 and 4.35 show a very broad vanadium peak suggesting the presence of mixed oxidation states $V^{5+}/V^{4+}/V^{3+}/V^{2+}/V^{0+}$, with $V^{5+}$ being the dominant species. A mixture of vanadium oxidation states has been observed in other glasses. Optical analysis indicates the presence of $V^{5+}$, $V^{4+}$ and $V^{3+}$ in vanadium doped flame-hydrolysed fused silica[81] and vanadium doped $Na_2O.2SiO_2$ glass.[86] The position of the peaks for the different vanadium oxidation states is consistent with XPS spectra previously taken of $V_2O_5$, $VO_2$, $V_2O_3$ and V metal, which have been attributed to $2p^{3/2}$ core electrons.[134] The peak at 520 eV is consistent with the $2p^{1/2}$ peak of vanadium.[134] Ar ion sputtering can alter the oxidation state of the vanadium; however the $V^{5+}$ would be expected to decrease with Ar ion sputtering, as the oxygen would be more likely to be sputtered than vanadium as it is a lighter element. XPS has been used to determine the oxidation state of chromium doped sodium silicate glass.[135] Which similarly shows a mixture of oxidation states present in the form of $Cr^{2+}$, $Cr^{3+}$ and $Cr^{6+}$.

The four O1s peaks give an indication of the structure of GLS glass. The peak at 530 eV is believed to be a non-bridging oxygen i.e $Ga-O^{2-}$ of the oxide negative cavities described in section 2.8.3 as it corresponds to the same binding energy of the non-bridging oxygen previously observed in XPS spectra of sodium silicate glass.[136] The other three peaks at 534, 537 and 539 eV are attributed to bridging bonds of La-O-La, Ga-O-Ga and S-O-S respectively.

The XPS spectra of vanadium doped GLS clearly indicate a broad range of vanadium oxidation states. However the resolution and signal strength of the measurement is not high enough to unambiguously identify each oxidation state and give an accurate compositional ratio. Because of this the absorption cross section can not be accurately calculated because the ratio of vanadium that remains in the optically active form is not known. A more sensitive XPS measurement, or a technique such as x-ray absorption near edge spectroscopy (XANES),[137] would be needed to find the ratios of each vanadium oxidation state.

XPS measurements and analysis where carried out under the supervision of Dr. N. Blanchard using facilities of the Advanced Technology Institute, University of Surrey.

## 4.14 Electron paramagnetic resonance.

The various oxidation states of a transition metal ion often give a unique electron paramagnetic resonance (EPR) fingerprint which can be used as a qualitative identification of the oxidation states present in a particular glass system. EPR measurements are used here to complement XPS measurements in the identification of the oxidation state of V:GLS.

Electron paramagnetic resonance (EPR) measurements were taken using the method described in section 3.3.9. The x-band EPR spectra of the $V^{4+}$ ion at low concentrations (<10%) is known to exhibit an absorption peak at around 3500 Gauss, with well resolved hyperfine lines due to the presence of the unpaired $3d^1$ electron.[138-143] The $V^{4+}$ ion almost exclusively exists as vanadyl ion ($VO^{2+}$) in a tetragonally compressed octahedron with $C_{4v}$ point symmetry.[144, 145] The broad peak at around 3500 in figures 4.36 and 4.37, is very similar to the unresolved hyperfine EPR lines of the $V^{4+}$ ion observed in vanadate glasses with $V_2O_5$ concentrations >10%[139, 141]. The disappearance of the hyperfine lines is attributed to spin-spin interactions, one such interaction occurs via a so called super-exchange of an electron, i.e. hopping of a mobile electron along a$V^{4+}$-O-$V^{5+}$ bond.[141, 146] The smallest concentration for the disappearance of the hyperfine EPR lines of $V^{4+}$ coupled by spin-spin interactions in tellurite and phosphate glasses was found to be $1.3 \times 10^{20} cm^{-3}$ or around 10% by weight.[141] However the disappearance of the $V^{4+}$ hyperfine EPR lines has been observed in vanadium doped silica with $V^{4+}$ concentrations of $< 1\%$[140] (similar to the concentrations in this study) and was attributed to clustering of the $V^{4+}$ ion. Disappearance of the $V^{4+}$ hyperfine EPR lines has also been observed in silver-vanadate-phosphate glasses with $V^{4+}$ concentrations of $\approx 10\%$ and was attributed to modifications of the glass structure caused by the modifier.[147] Variations in the ratio of glass modifier to glass network former in silver-vanadate-phosphate glasses have shown that this phenomenon can be attributed to an increasingly cross linked glass network.[146] Clustering may also be the explanation for the disappearance of the $V^{4+}$ hyperfine EPR lines in vanadium doped GLS at much lower concentrations than in tellurite and phosphate glasses, however there is some indication of hyperfine lines in the EPR spectra of 0.0023% vanadium doped GLS (figure 4.37) which indicates that spin-spin interactions of the $V^{4+}$ ion in GLS may occur at much lower concentrations than reported previously.[141]

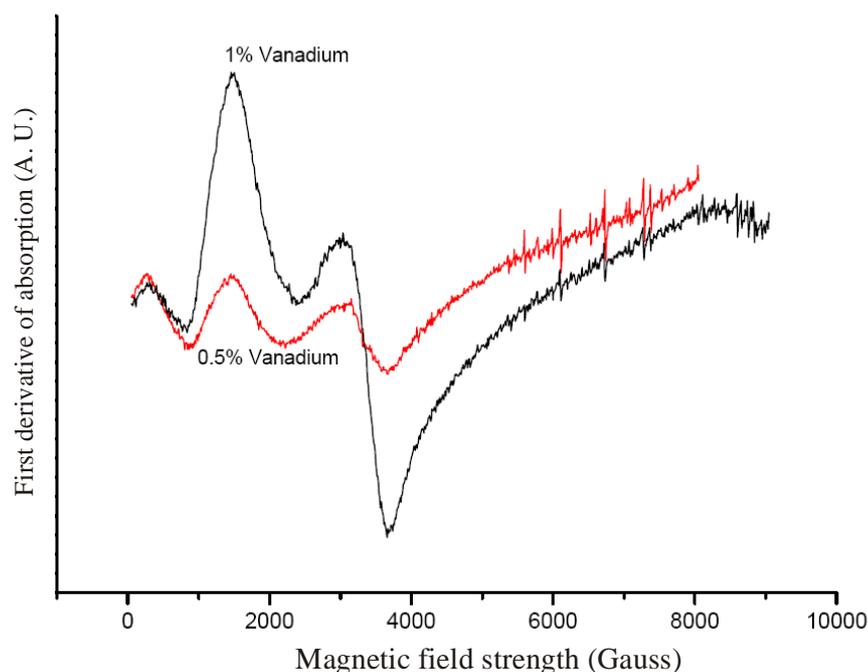

FIGURE 4.36 X-band EPR spectra (9.5 GHz) of 1% and 0.5% vanadium doped GLS at 300K.

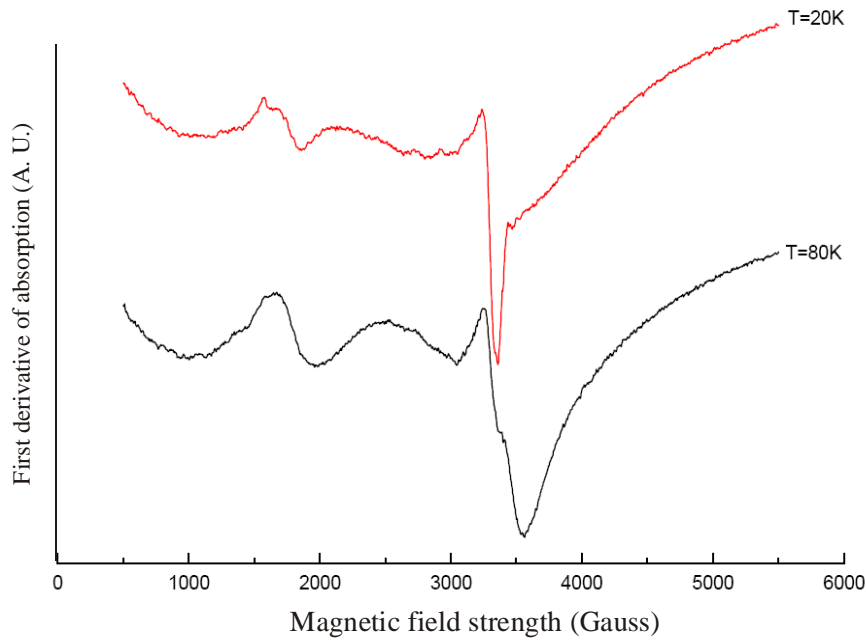

FIGURE 4.37 X-band EPR spectra (9.5 GHz) of 0.0023%
vanadium doped GLS at 20 and 80K.

The broad absorption peak at around 3500 Gauss in the X-band EPR spectra of
vanadium doped GLS is therefore attributed to unresolved hyperfine EPR lines of the
$V^{4+}$ ion caused by spin-spin interactions of the form $V^{4+}$-O-$V^{5+}$ or by strong cross-
linking in the GLS glass network. The electronic structure of $V^{5+}$ is $3p^6$ and as a result
has a total electronic spin equal to zero and cannot therefore be detected by EPR[147]
The high field EPR spectra (94GHz) of 1% vanadium doped GLS at 300K did not
display any feature that could be attributed to vanadium. EPR measurements where
carried out using facilities of the School of Physics and Astronomy, University of St-
Andrews, by Dr. Hassane El Mkami.

## 4.15 Determination of the oxidation state and coordination of V:GLS

Determination of the oxidation state of an active ion dopant is an important part of the
characterisation of a material being considered for optical device applications because it
determines which energy levels exist within this ion. Knowledge of the oxidation state
is therefore needed when modelling the radiative and non radiative transitions that occur
in an optical material.

XPS measurements of 1% $V_2S_3$ doped GLS, given in section 4.13, indicate a broad
range of oxidation states from $V^+$ to $V^{5+}$. It has been known for some time that at least
three vanadium valence states can coexist in a glass host.[86] Excitation spectra in
figure 4.12 show three distinct and very Gaussian peaks at 8625, 13135 and 17240 cm$^{-1}$,
Absorption spectra show a Gaussian peak at 9090 cm$^{-1}$, a shoulder at around 13000 cm$^{-1}$
and an absorption around 18000 to 20000 cm$^{-1}$, There is also evidence of a very weak
shoulder on both the excitation and absorption spectra at 10000cm$^{-1}$.

Only one of the possible oxidation states identified by XPS is believed to be responsible for the three peaks identified by excitation and absorption measurements for the following reasons. Firstly the same characteristic photoluminescence peaking at 1500 nm with a FWHM of 500 nm is observed when exciting into each band with 514, 808 and 1064nm laser sources, if one or more of the three absorption bands were caused by different oxidation states then it would be unlikely that they would generate very similar photoluminescence spectra. Secondly the lifetimes when exciting at different laser wavelengths are also very similar. The slight increase in lifetime from 30 μs (when exciting at 1064 nm) to 35 μs (when exciting at 830 nm) can be explained by the 830 nm pump source selectively exciting ions in higher crystal field sites which will have a higher efficiency and longer lifetime. The difference is not large enough to indicate that the 830 nm laser is exciting a different oxidation state to the 1064 nm laser. Thirdly the same characteristic absorption is observed when doping with vanadium in different initial oxidation states, $V_2S_3$ and $V_2O_5$, if more than one oxidation state contributed to the absorption bands one would expect the relative intensities to vary somewhat.

## 4.15.1 Treatment of each possible vanadium oxidation state

Vanadium 5+ has a $d^0$ electronic configuration and will not have any d-d optical transitions. However $d^0$ ions can contribute to optical transitions if there is a charge transfer process. Charge transfer transitions are usually high energy transitions which are predicted by molecular orbit theory but not by crystal field theory. These high energy transitions promote electrons that mainly belong to states of ligand ions to states that mainly belong to the transition metal ion.[19] Charge transfer transitions are displayed by the $Cr^{6+}$ ion in sodium silicate glass[148] which has absorption peaks at 270 and 370nm. Charge transfer bands of $Cr^{6+}$ in soda-lime-silicate glass have been identified at 370 nm[149] and in $Li_2B_4O_7$ glass at 370 nm [54]. Charge transfer bands are also displayed by the $V^{5+}$ ion in $Ca_2PO_4Cl$ crystal at 260 nm [150]. In $Na_2O.2SiO_2$ glass the transfer band $V^{5+}$ was measured up to 350 nm but was then obscured by the band-edge absorption of the host. [86] All of these charge transfer transitions occurred in or around the UV region and are summarised in table 4.11. Because the optical transitions observed for vanadium doped GLS occur at lower energies they are not attributed to a charge transfer transition. Therefore $V^{5+}$ is not believed to contribute to any observed optical transition of V:GLS. The absorption of $V^{5+}$, believed to be present in V:GLS, would probably be obscured by the band-edge absorption of GLS at ~500nm

TABLE 4.11 Summary of charge transfer transitions in $Cr^{6+}$ and $V^{5+}$.

| Ion | Host | Absorption peaks (nm) | Reference |
|------|------|------|------|
| $Cr^{6+}$ | sodium silicate glass | 270, 370 | [148] |
| $Cr^{6+}$ | soda-lime-silicate glass | 370 | [149] |
| $Cr^{6+}$ | $Li_2B4O_7$ glass | 370 | [54] |
| $V^{5+}$ | $Ca_2PO_4Cl$ crystal | 260 | [150] |
| $V^{5+}$ | $Na_2O.2SiO_2$ glass | <350 | [86] |
| $V^{5+}$ | CaYAlO$_4$ | 330, 430 | [151] |

Vanadium 4+ has a $d^1$ electronic configuration which means there is only one excited $E_g$ state[152] and therefore it has only one spin allowed transition ($^2T_2 \rightarrow \, ^2E_2$). Vanadium 4+ doped CaYAlO$_4$ displays a broad absorption band at 500 nm.[153] Two excitation peaks for $V^{4+}$ have been observed due to the Jan-Teller effect at 427 and 490 nm in $Al_2O_3$, at 419 and 486 nm in YAlO$_3$ and at 432 and 500 nm in yttrium aluminium garnet (YAG).[152] Based on these comparisons it is thought to be unlikely that $V^{4+}$ can account for the three broad absorption bands observed in V:GLS. The positions of the absorption peaks for $V^{4+}$ in various hosts, summarised in table 4.12, indicates that although $V^{4+}$ is thought to be present in V:GLS its absorption may be obscured by the band-edge of GLS.

TABLE 4.12 Summary of absorption transitions of $V^{4+}$.

| Host | Absorption peaks (nm) | Symmetry | Transition | Reference |
|------|------|------|------|------|
| CaYAlO$_4$ | 500 | Octahedral | $^2T_2 \rightarrow \, ^2E_2$ | [153] |
| $Al_2O_3$ | 427, 490 | - | - | [152] |
| YAlO$_3$ | 419, 486 | - | - | [152] |
| YAG | 432, 500 | - | - | [152] |

Vanadium 3+ has a $d^2$ electronic configuration and, from inspection of the Tanabe-Sugano diagram for a $d^2$ ion,[59, 60, 154] is expected to have three spin allowed ground state absorption transitions: $^3T_1(^3F) \rightarrow \, ^3T_2(^3F)$, $^3T_1(^3F) \rightarrow \, ^3A_2(^3F)$ and $^3T_1(^3F) \rightarrow \, ^3T_1(^3P)$ in octahedral coordination and $^3A_2(^3F) \rightarrow \, ^3T_1(^3P)$, $^3A_2(^3F) \rightarrow \, ^3T_1(^3F)$ and $^3A_2(^3F) \rightarrow \, ^3T_2(^3F)$ in tetrahedral coordination. Tetrahedral $V^{3+}$ in YAG has three absorption bands, centred at 600, 800 and 1320 nm which are attributed to spin allowed transitions from the $^3A_2(^3F)$ ground state to the $^3T_1(^3P)$, $^3T_1(^3F)$ and $^3T_2(^3F)$ levels respectively, a weak and narrow absorption at 1140 nm is attributed to the spin forbidden transition $^3A_2(^3F)$ to $^1E(^1D)$.[96, 155, 156] Likewise, tetrahedral $V^{3+}$ in LiAlO$_2$, LiGaO$_2$ and SrAl$_2$O$_4$ has three absorption bands, centred at ~550, 850 and 1350 nm which are attributed to transitions from $^3A_2(^3F)$ to $^3T_1(^3P)$, $^3T_1(^3F)$ and $^3T_2(^3F)$ levels respectively.[98] Optical transitions associated with octahedral $V^{3+}$ tend to occur at higher energies than those for tetrahedral $V^{3+}$. For example the $^3T_1(^3F)$ to $^3T_2(^3F)$ and $^3T_1(^3P)$ transitions of octahedral $V^{3+}$:YAG occur at 600 nm and 425 nm respectively,[96, 156, 157] at 707 nm and 440 nm in zirconium fluoride glass[82] and at 724 nm and 459 nm in phosphate glass.[87]

Octahedral $V^{3+}$ in $Na_2O.2SiO_2$ glass displays two absorption peaks, at 690 and 450 nm, and were assigned to the $^3T_1(^3F) \rightarrow ^3T_2(^3F)$ and $^3T_1(^3F) \rightarrow ^3A_2(^3F)$ transitions respectively.[86] The positions of absorption peaks for $V^{3+}$ in various hosts are summarised in table 4.13.

TABLE 4.13 Summary of absorption transitions of $V^{3+}$.

| Host | Absorption peaks (nm) | Symmetry | Transition | Reference |
|---|---|---|---|---|
| YAG | 600,800,1140,1320 | tetrahedral | $^3A_2(^3F) \rightarrow$ $^3T_1(^3P), ^3T_1(^3F),$ $^1E(^1D), ^3T_2(^3F)$ | [96, 155, 156] |
| YAG | 600,425 | octahedral | $^3T_1(^3F) \rightarrow ^3T_2(^3F), ^3T_1(^3P)$ | [96, 156, 157] |
| zirconium fluoride glass | 707,440 | octahedral | $^3T_1(^3F) \rightarrow ^3T_2(^3F), ^3T_1(^3P)$ | [82] |
| phosphate glass | 724,459 | octahedral | $^3T_1(^3F) \rightarrow ^3T_2(^3F), ^3T_1(^3P)$ | [87] |
| $Na_2O.2SiO_2$ | 690, 450 | octahedral | $^3T_1(^3F) \rightarrow ^3T_2(^3F), ^3T_1(^3P)$ | [86] |
| ZnTe | 2000,1333 | tetrahedral | $^3A_2(^3F) \rightarrow ^3T_1(^3P), ^3T_1(^3F)$ | [99] |
| ZnSe | 1736,1125,749 | tetrahedral | $^3A_2(^3F) \rightarrow ^3T_1(^3P), ^3T_1(^3F),$ $^3T_2(^3F)$ | [61] |

Vanadium 2+ has a $d^3$ electronic configuration and from inspection of the Tanabe-Sugano diagram for a $d^3$ ion[59, 60, 154] is expected to have three spin allowed ground state absorption transitions: $^4A_2(^4F) \rightarrow ^4T_2(^4F)$, $^4A_2(^4F) \rightarrow ^4T_1(^4F)$ and $^4A_2(^4F) \rightarrow ^4T_1(^4P)$ in octahedral coordination. In the case of a tetrahedrally coordinated $d^3$ ion there are also three spin allowed transitions: $^4T_1(^4F) \rightarrow ^4T_2(^4F)$, $^4T_1(^4F) \rightarrow ^4A_2(^4F)$ and $^4T_1(^4F) \rightarrow ^4T_1(^4P)$ in weak crystal fields (Dq/B <2.2). When Dq/B is > 2.2 the $^2E(^2G)$ level becomes the lowest energy level and there are four spin allowed transitions: $^2E(^2G) \rightarrow ^2T_1(^2G)$, $^2E(^2G) \rightarrow ^2T_2(^2G)$, $^2E(^2G) \rightarrow ^2A_1(^2G)$ and $^2E(^2G) \rightarrow ^2A_2(^2F)$. In octahedral $V^{2+}$ doped $MgF_2$ two spin allowed transition absorption bands, centred at 870 and 550 nm were observed and were attributed to spin allowed transitions from the $^4A_2(^4F)$ ground state to the $^4T_2(^4F)$ and $^4T_1(^4F)$ levels respectively.[8] In another reference, three spin allowed transition absorption bands centred for $V^{2+}$ doped $MgF_2$ where observed at 884, 550 and 366 nm and attributed to $^4A_2(^4F) \rightarrow ^4T_2(^4F)$, $^4A_2(^4F) \rightarrow ^4T_1(^4F)$ and $^4A_2(^4F) \rightarrow ^4T_1(^4P)$ transitions respectively, a spin forbidden transition was also observed at 787 nm and attributed to the $^4A_2(^4F) \rightarrow ^2E(^2G)$ transition.[103] The low temperature (T=10K) absorption spectra of octahedral $V^{2+}$ doped NaCl has spin allowed transition absorption bands, centred at 1222, 759 and 478 nm, which were attributed to the $^4A_2(^4F) \rightarrow ^4T_2(^4F)$, $^4A_2(^4F) \rightarrow ^4T_1(^4F)$ and $^4A_2(^4F) \rightarrow ^4T_1(^4P)$ transitions respectively.[7] The low temperature (T=10K) absorption spectra of octahedral $V^{2+}$ doped $RbMnF_3$ has spin allowed transition absorption bands centred at 976 and 615 nm which were attributed to the $^4A_2(^4F) \rightarrow ^4T_2(^4F)$, $^4A_2(^4F) \rightarrow ^4T_1(^4F)$ and transitions respectively.[7] The room temperature absorption spectra of octahedral $V^{2+}$ doped $CsCaF_3$ has spin allowed transition absorption bands centred at 1067, 662 and 424 nm which were attributed to the $^4A_2(^4F) \rightarrow ^4T_2(^4F)$, $^4A_2(^4F) \rightarrow ^4T_1(^4F)$ and $^4A_2(^4F) \rightarrow ^4T_1(^4P)$ transitions respectively a spin forbidden transition was also observed at 794 nm and attributed to the and $^4A_2(^4F) \rightarrow {}^2E(^2G)$ transition.[6]

The low temperature (T=4K) excitation spectra of tetrahedral $V^{2+}$ doped ZnTe has spin allowed transition absorption bands centred at 2247, 1389 and 1152 nm which were attributed to the $^4T_1(^4F) \rightarrow {}^4T_2(^4F)$, $^4T_1(^4F) \rightarrow {}^4A_2(^4F)$ and $^4T_1(^4F) \rightarrow {}^4T_1(^4P)$ transitions respectively; spin forbidden transitions were also observed at 990, 885 and 826 nm and attributed to the $^4A_2(^4F) \rightarrow {}^2E(^2H)$, $^4A_2(^4F) \rightarrow {}^2T_1(^2H)$ and $^4A_2(^4F) \rightarrow {}^2A_2(^2F)$ transitions.[99] The positions of absorption peaks for $V^{2+}$ in various hosts are summarised in table 4.14.

TABLE 4.14 Summary of absorption transitions of $V^{2+}$.

| Host | Absorption peaks (nm) | Symmetry | Transition | Reference |
|------|------|------|------|------|
| MgF$_2$ | 870,550 | octahedral | $^4A_2(^4F) \rightarrow {}^4T_2(^4F), {}^4T_1(^4F)$ | [8] |
| MgF$_2$ | 884,787,550,366 | octahedral | $^4A_2(^4F) \rightarrow {}^4T_2(^4F),$ $^2E(^2G), {}^4T_1(^4F), {}^4T_1(^4P)$ | [103] |
| KMgF$_3$ | 869,789,560,366 | octahedral | $^4A_2(^4F) \rightarrow {}^4T_2(^4F),$ $^2E(^2G), {}^4T_1(^4F), {}^4T_1(^4P)$ | [103] |
| NaCl | 1222,759,478 | octahedral | $^4A_2(^4F) \rightarrow {}^4T_2(^4F), {}^4T_1(^4F), {}^4T_1(^4P)$ | [7] |
| RbMnF$_3$ | 976,615 | octahedral | $^4A_2(^4F) \rightarrow {}^4T_2(^4F), {}^4T_1(^4F)$ | [7] |
| CsCaF$_3$ | 1067,794,662,424 | octahedral | $^4A_2(^4F) \rightarrow {}^4T_2(^4F),$ $^2E(^2G), {}^4T_1(^4F), {}^4T_1(^4P)$ | [6] |
| ZnTe | 2247,1389,1152 | tetrahedral | $^4A_2(^4F) \rightarrow {}^4T_2(^4F), {}^4A_2(^4F), {}^4T_1(^4P)$ | [99] |
| ZnTe | 990,885,826 | tetrahedral | $^4A_2(^4F) \rightarrow {}^2E(^2H), {}^2T_1(^2H), {}^2A_2(^2F)$ | [99] |
| ZnSe | 1957,1198,978 | tetrahedral | $^4A_2(^4F) \rightarrow {}^4T_2(^4F), {}^4A_2(^4F), {}^4T_1(^4P)$ | [61] |
| ZnSe | 838,739,667,617 | tetrahedral | $^4A_2(^4F) \rightarrow {}^2E(^2H), {}^2T_1(^2H), {}^2A_2(^2F),$ $^2T_1(^2F)$ | [61] |
| MgO | 719,502,344 | octahedral | $^4A_2(^4F) \rightarrow {}^4T_2(^4F), {}^4T_1(^4F), {}^4T_1(^4P)$ | [158] |

Vanadium 1+ has a $d^4$ electronic configuration and from inspection of the Tanabe-Sugano diagram for a $d^4$ ion[59, 60, 154] is expected to have one spin allowed ground state absorption transition in weak crystal fields (Dq/B <2.6): $^5T_2(^5D) \rightarrow {}^5E(^5D)$ in octahedral coordination. When Dq/B is > 2.6 the $^3T_1(^3H)$ level becomes the lowest energy level and there are four spin allowed transitions: $^3T_1(^3H) \rightarrow {}^3E(^3H)$, $^3T_1(^3H) \rightarrow {}^3T_2(^3H)$, $^3T_1(^3H) \rightarrow {}^3A_1(^3G)$ and $^3T_1(^3H) \rightarrow {}^3A_2(^3F)$. In the case of a tetrahedrally coordinated $d^4$ ion the spin allowed transition is $^5E(^5D) \rightarrow {}^5T_2(^5D)$ in weak crystal fields (Dq/B <2.0). When Dq/B is > 2.0 the $^1A_1(^1I)$ level becomes the lowest energy level and there are four spin allowed transitions: $^1A_1(^1I) \rightarrow {}^1T_1(^1I)$, $^1A_1(^1I) \rightarrow {}^1T_2(^1I)$, $^1A_1(^1I) \rightarrow {}^1E(^1I)$ and $^1A_1(^1I) \rightarrow {}^1A_2(^1F)$

In octahedral $V^+$ doped ZnTe one spin allowed transition absorption band is observed, centred at 2740, and is attributed to the $^5T_2(^5D) \rightarrow {}^5E(^5D)$ nm;[99] in octahedral $V^+$ doped ZnSe this transition occurred at 2604 nm.[61] Further spin forbidden transitions were observed and are detailed in table 4.15.

TABLE 4.15 Summary of absorption transitions of $V^+$.

| Host | Absorption peaks (nm) | Symmetry | Transition | Reference |
|------|----------------------|----------|------------|-----------|
| ZnTe | 2740 | octahedral | $^5T_2(^5D) \rightarrow ^5E(^5D)$ | [99] |
| ZnTe | 1379,1163,1000,930 | octahedral | $^5T_2(^5D) \rightarrow ^3T_1(^3H), ^3T_2(^3H), ^3T_2(^3F), ^3T_1(^3P)$ | [99] |
| ZnSe | 2604 | octahedral | $^5T_2(^5D) \rightarrow ^5E(^5D)$ | [61] |
| ZnSe | 1277,976,917,833 | octahedral | $^5T_2(^5D) \rightarrow ^3T_1(^3H), ^3T_2(^3H), ^3T_2(^3F), ^3T_1(^3P)$ | [61] |

These arguments are based on the Tanabe-Sugano model and comparisons with the absorption spectra of $V^{5+}$, $V^{4+}$, $V^{3+}$, $V^{2+}$ and $V^+$ in various hosts indicate that the three spin allowed transitions observed in V:GLS can only come from $V^{3+}$ or $V^{2+}$. In order to determine which of these ions is responsible for the observed absorption spectra and what its symmetry is the crystal field parameters: Dq, B and C will be calculated, using the Tanabe-Sugano model.[59, 60, 154]

## 4.16 Tanabe-Sugano analysis of V:GLS

### 4.16.1 Introduction

The Tanabe-Sugano model takes into account the interactions between two or more 3d electrons in the presence of a crystal field. The free ion states are shown on Tanabe-Sugano diagrams on the far left (Dq/B=0). The free ion states are governed by electron-electron interactions and so are labelled by $^{2S+1}L$ states (also called L-S terms) where S is the total spin and L is the total angular momentum.[19] The energy separation between the various $^{2S+1}L$ states is given in terms of the Racah parameters (A,B and C). These parameters describe the strength of the electrostatic interactions between multiple 3d electrons.[60] Tanabe and Sugano calculated the energy matrices for each state of $3d^2$ to $3d^5$ ions in an ideal octahedral crystal field,[59] These are reproduced in tables 4.16, 4.17, 4.19, 4.20 and in appendix C. Proof of their derivation has been published.[159] These energy matrices can then be used to calculate how the $^{2S+1}L$ free ion levels split up, and vary, as a function of the ratio between the crystal field strength and the inter-electronic interaction (measured in Dq/B). Represented graphically, these functions are called Tanabe-Sugano diagrams and they have been used since their introduction in 1954 to interpret the spectra of transition metal ions in a variety of crystalline and glass hosts. Tanabe-Sugano diagrams take advantage of the fact that C/B is almost independent of atomic number and the number of electrons and, for all first row transition metal elements C/B ≈ 4 to 5.[59]

As described in chapter 2, the d orbital splits into $t_2$ and e orbitals in the presence of a crystal field. The various states are represented as $t_2^n e^m$ (n+m=N) where N is the number of electrons in the d orbital. Tanabe and Sugano showed that it was unnecessary to calculate the energy matrices for N>5 (which becomes very laborious) because of the simple relationship between configurations in the state $t_2^n e^m$ and $t_2^{6-n} e^{4-m}$. This simple relationship results from the equivalence of electrons and holes. It has been shown,[154] that to obtain the full Hamiltonian energy matrices, (-4n+6m)Dq is added to the diagonal element in the state $t_2^n e^m$. For the state $t_2^{6-n} e^{4-m}$ this is $[-4(6-n)+6(4-m)]Dq = -(-4n+6m)Dq$. It is also unnecessary to calculate the energy matrices for a tetrahedral field because the energy matrices for a $d^n$ ion in a tetrahedral field are the same as a $d^{10-n}$ ion in a octahedral field.[60]

Cubic coordination can be thought of as comprising of two tetrahedral components. Hence the cubic crystal field interaction energy term has the same functional form as in a tetrahedral field but it is twice as large.[60] Because of their small ionic radii in proportion to rare earth ions, transition metals are usually found in tetrahedral or octahedral coordination where as rare earths are often found in dodecahedral coordination.[160] Because of the relatively small ionic radii of the $V^{2+}$ and $V^{3+}$ ion and the relatively large ionic radii of the $S^{2-}$ cation, cubic coordination in V:GLS is thought to be extremely unlikely. Low symmetry fields, such as tetragonal, cause a splitting of the energy terms. For example in tetrahedral $Cr^{4+}:Y_2SiO_5$ with $C_{3V}$ symmetry the $^3T_1(^3F)$ level splits into two components which were attributed to two closely spaced absorption peaks at 733 and 602 nm[161]. This sort of splitting is not evident in the absorption spectra of V:GLS so the data is analysed in terms of ideal (cubic symmetry) octahedral or tetrahedral coordination.

Each of the possible electronic configurations ($d^2$ or $d^3$) and coordination (tetrahedral or octahedral) is now analysed with the Tanabe-Sugano model.

## 4.16.2 Tetrahedral $d^2$ configuration

The energy matrix for the $^3T_1(^3F,^3P)$ state of the tetrahedral $d^2$ configuration is given in table 4.16.

TABLE 4.16 energy matrix for the $^3T_1(^3F,^3P)$ state, after [59].

| $^3T_1(^3F,^3P)$ | |
|---|---|
| $t_2^2$ | $t_2e$ |
| -5B | 6B |
| 6B | -10Dq+4B |

The eigenvalues of the matrix in table 4.16 give the diagonal terms of the diagonalized matrix which are the energy terms of the $^3T_1(^3F)$ and $^3T_1(^3P)$ states as a function of Dq and B.

$$E\left(^3T_1(^3F)\right) = \frac{1}{2}\left(-B + 5Dq - \sqrt{225B^2 - 180DqB + 100Dq^2}\right) \qquad (4.30)$$

$$E\left(^3T_1(^3P)\right) = \frac{1}{2}\left(-B + 5Dq + \sqrt{225B^2 - 180DqB + 100Dq^2}\right) \qquad (4.31)$$

Dividing 4.30 by B and arranging in terms of Dq/B, as is necessary for Tanabe-Sugano diagrams, gives:

$$E\left({}^3\mathrm{T_1}({}^3\mathrm{F})\right)/\mathrm{B} = -\frac{1}{2} + 5\mathrm{Dq/B} - \frac{1}{2}\sqrt{225 - 180Dq/B + 100(Dq/B)^2}\qquad(4.32)$$

Note that 4.32 is independent of C, in order to calculate C the energy term for a spin forbidden transition is needed. Table 4.17 gives the energy matrix for the ${}^1\mathrm{E}({}^1\mathrm{D},{}^1\mathrm{G})$ state.

Table 4.17 energy matrix for the ${}^1\mathrm{E}({}^1\mathrm{D},{}^1\mathrm{G})$ state, after [59].

| ${}^1\mathrm{E}({}^1\mathrm{D},{}^1\mathrm{G})$ | |
|---|---|
| $t_2{}^2$ | $e^2$ |
| B+2C | $-2\sqrt{3}$B |
| $-2\sqrt{3}$B | -20Dq+2C |

Diagonalising and dividing by B gives:

$$E\left({}^1\mathrm{E}({}^1\mathrm{G})\right)/\mathrm{B} = \frac{1}{2} + 2C/B - 10\mathrm{Dq/B} + \frac{1}{2}\sqrt{49 + 40Dq/B + 400(Dq/B)^2}\qquad(4.33)$$

and

$$E\left({}^1\mathrm{E}({}^1\mathrm{D})\right)/\mathrm{B} = \frac{1}{2} + 2C/B - 10\mathrm{Dq/B} - \frac{1}{2}\sqrt{49 + 40Dq/B + 400(Dq/B)^2}\qquad(4.34)$$

Figure 4.38 shows the energy terms of a tetrahedral $d^2$ ion, plotted as a function of Dq/B, note that the lowest energy level is less than zero and varies as a function of Dq/B. The energy matrices and energy terms for all the energy levels are given in appendix C.

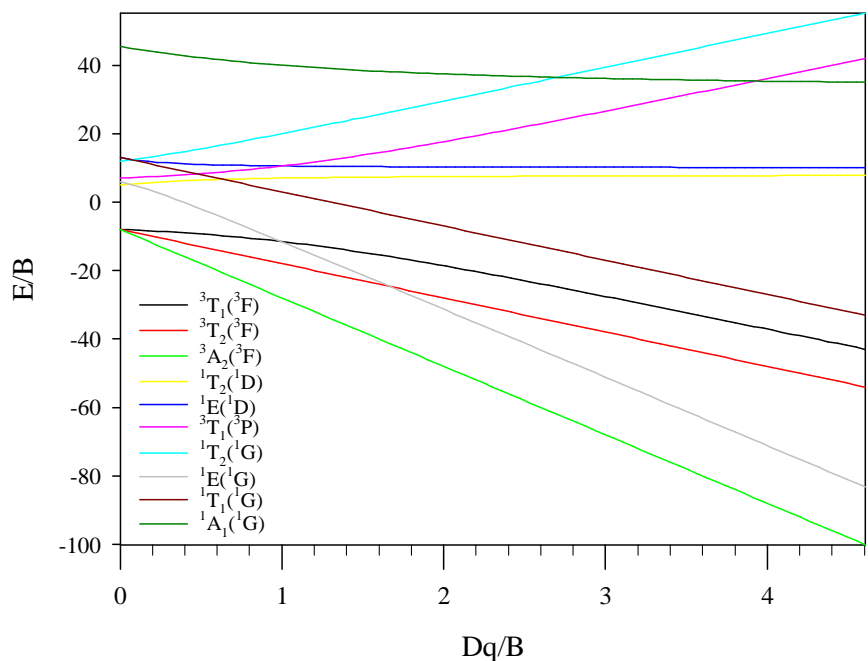

FIGURE 4.38 Energy terms of a tetrahedral $d^2$ ion plotted as a function of Dq/B.

In Tanabe-Sugano diagrams, the energy term of the lowest energy level is subtracted from the energy terms of all the energy levels, as has been done in equations 4.35 to 4.38 for the energy levels of interest i.e. the three spin allowed energy levels and the lowest spin forbidden energy level.

$$E\left(^3\text{T}_2(^3\text{F})\right)/\text{B} = 10Dq/B \qquad (4.35)$$

$$E\left(^3\text{T}_1(^3\text{F})\right)/\text{B} = 7.5 + 15Dq/\text{B} - \frac{1}{2}\sqrt{225 - 180DqB + 100Dq^2} \qquad (4.36)$$

$$E\left(^1\text{E}(^1\text{D})\right)/\text{B} = 8.5 + 2C/B + 10Dq/\text{B} - \frac{1}{2}\sqrt{49 + 40Dq/B + 400(Dq/B)^2} \quad (4.37)$$

$$E\left(^3\text{T}_1(^3\text{P})\right)/\text{B} = 7.5 + 15Dq/\text{B} + \frac{1}{2}\sqrt{225 - 180DqB + 100Dq^2} \qquad (4.38)$$

These energy terms are plotted in the Tanabe-Sugano diagram in figure 4.39.

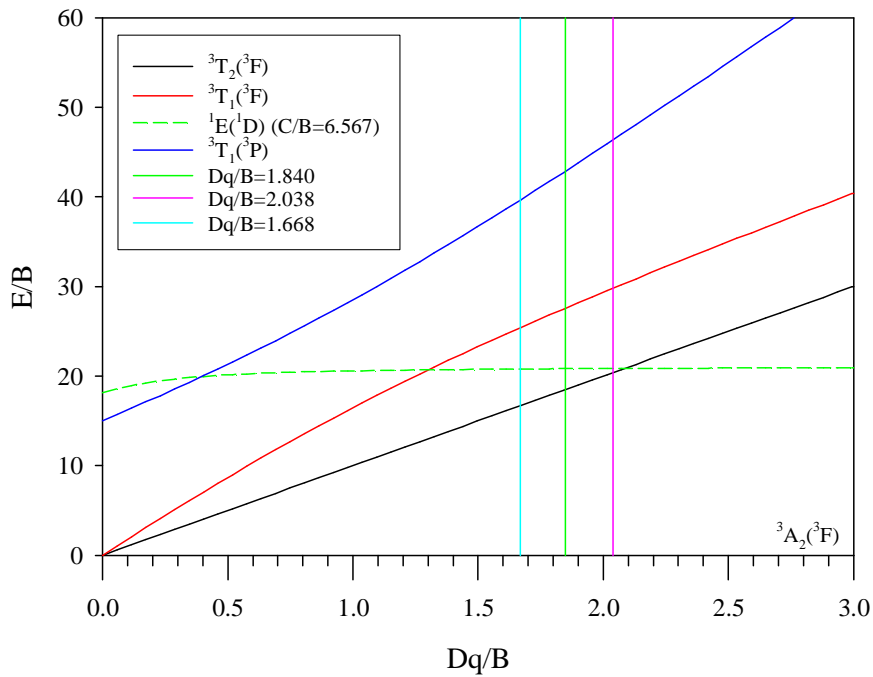

FIGURE 4.39 Tanabe-Sugano diagram of the tetrahedral $d^2$ configuration, the Dq/B value calculated is shown. The spin forbidden energy levels were calculated with C/B=6.5.

Dq is known (1/10 the energy of the lowest spin allowed absorption transition) so B is calculated from the experimentally determined energies of the $^3\text{T}_2(^3\text{F})$ and $^3\text{T}_1(^3\text{F})$ energy levels and then solving their energy terms simultaneously for B. The C/B ratio is calculated by rearranging equation 4.37 to make C/B the subject as in equation 4.39.

$$C/B = E\left({}^1\mathrm{E}({}^1\mathrm{D})\right)/2\mathrm{B} - 5\mathrm{Dq/B} - 4.5 + \frac{1}{4}\sqrt{49 + 40Dq/B + 400(Dq/B)^2} \qquad (4.39)$$

The values of Dq/B, B, and C/B, calculated from various absorption and PLE spectra in section 4.2 and 4.5, are given in table 4.18.

TABLE 4.18 Crystal field parameters calculated for a d² ion in tetrahedral coordination.

| Experimental data source | $E({}^3T_2({}^3F))$ (cm⁻¹) | $E({}^3T_1({}^3F))$ (cm⁻¹) | $E({}^1E({}^1D))$ (cm⁻¹) | Dq/B | B | C/B |
|---|---|---|---|---|---|---|
| Derivative absorption of V:GLS (figure 4.5) | 8965 | 13377 | 10100 | 1.848 | 485.12 | 6.567 |
| Derivative absorption of V:GLSO (figure 4.6) | 9300 | 13600 | 10200 | 2.038 | 456.32 | 7.319 |
| PLE of V:GLS (figure 4.12) | 8625 | 13135 | 10000 | 1.668 | 517.1 | 5.842 |
| PLE of V:GLSO (figure 4.13) | 8860 | 13299 | 10000 | 1.794 | 493.86 | 6.141 |

The calculated valued of Dq/B are shown on figure 4.39, all of which indicate a weak field site which is consistent with the emission absorption and lifetime measurements. The calculated C/B values are however slightly outside the allowed range of 4 to 5.

## 4.16.3 Octahedral d² configuration

The energy matrix for the ³T₁(³F,³P) state of the octahedral d² configuration is given in table 4.19

TABLE 4.19 energy matrix for the ³T₁(³F,³P) state, , after [59].

| ³T₁(³F,³P) | |
|---|---|
| $t_2^2$ | $t_2e$ |
| -5B | 6B |
| 6B | 10Dq+4B |

The eigenvalues of the matrix in table 4.19 give the diagonal terms of the diagonalized matrix which are the energy terms of the ³T₁(³F) and ³T₁(³P) states as a function of Dq and B.

$$E\left({}^3T_1({}^3F)\right) = \frac{1}{2}\left(-\mathrm{B} + 10\mathrm{Dq} - \sqrt{225B^2 + 180DqB + 100Dq^2}\right) \qquad (4.40)$$

$$E\left({}^3T_1({}^3P)\right) = \frac{1}{2}\left(-\mathrm{B} + 10\mathrm{Dq} + \sqrt{225B^2 + 180DqB + 100Dq^2}\right) \qquad (4.41)$$

Dividing equation 4.40 by B and arranging it in terms of Dq/B, as is necessary for Tanabe-Sugano diagrams, gives:

$$E\left(^3T_1(^3F)\right)/B = -\frac{1}{2} + 5Dq/B - \frac{1}{2}\sqrt{225 + 180Dq/B + 100(Dq/B)^2} \qquad (4.42)$$

Note that equation 4.42 is independent of C, in order to calculate C the energy term for a spin forbidden transition is needed. Table 4.20 gives the energy matrix for the $^1E(^1D,^1G)$ state.

TABLE 4.20 energy matrix for the $^1E(^1D,^1G)$ state, , after [59].

| $^1E(^1D,^1G)$ | |
|---|---|
| $t_2^2$ | $e^2$ |
| B+2C | $-2\sqrt{3}B$ |
| $-2\sqrt{3}B$ | 20Dq+2C |

Diagonalising and dividing by B gives:

$$E\left(^1E(^1D)\right)/B = \frac{1}{2} + 2C/B + 10Dq/B - \frac{1}{2}\sqrt{49 - 40Dq/B + 400(Dq/B)^2} \qquad (4.43)$$

and

$$E\left(^1E(^1G)\right)/B = \frac{1}{2} + 2C/B + 10Dq/B + \frac{1}{2}\sqrt{49 - 40Dq/B + 400(Dq/B)^2} \qquad (4.44)$$

Similarly, for the case of a $d^2$ tetrahedral ion, the energy terms of interest are given in equation 4.45 to 4.49 and plotted in the Tanabe-Sugano diagram in figure 4.40. The values of Dq/B, B and C/B were calculated using the same method as described for a $d^2$ tetrahedral ion. The energy term for the $^1E(^1D)$ level was used to calculate C/B, as shown in figure 4.40 this energy level is virtually indistinguishable from the $^1T_2(^1D)$ level.

$$E\left(^3T_2(^3F)\right)/B = -7\frac{1}{2} + 5Dq/B + \frac{1}{2}\sqrt{225 + 180Dq/B + 100(Dq/B)^2} \qquad (4.45)$$

$$E\left(^3A_2(^3F)\right)/B = -7\frac{1}{2} + 15Dq/B + \frac{1}{2}\sqrt{225 + 180Dq/B + 100(Dq/B)^2} \qquad (4.46)$$

$$E\left(^1E(^1D)\right)/B = 1 + 2C/B + 5Dq/B - \frac{1}{2}\sqrt{49 - 40Dq/B + 400(Dq/B)^2}$$
$$+ \frac{1}{2}\sqrt{225 + 180Dq/B + 100(Dq/B)^2} \qquad (4.47)$$

$$E\left({}^1T_2({}^1D)\right)/B = 1 + 2C/B - \frac{1}{2}\sqrt{49 - 20Dq/B + 100(Dq/B)^2}$$

$$+ \frac{1}{2}\sqrt{225 + 180Dq/B + 100(Dq/B)^2} \qquad (4.48)$$

$$E\left({}^3T_1({}^3P)\right)/B = \sqrt{225 + 180Dq/B + 100(Dq/B)^2} \qquad (4.49)$$

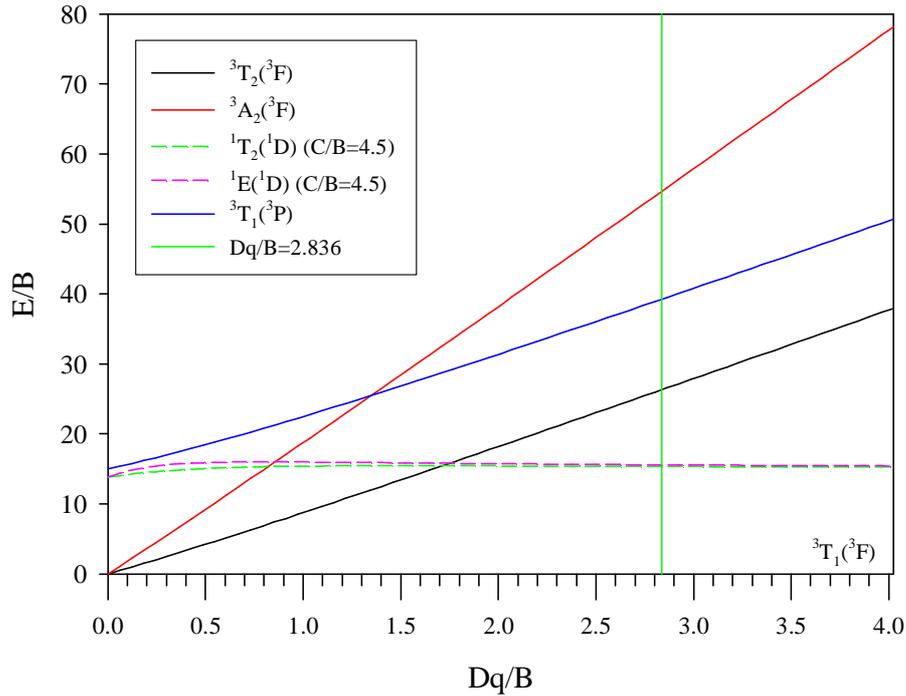

FIGURE 4.40 Tanabe-Sugano diagram of a octahedral $d^2$ ion, the Dq/B value calculated from absorption spectra is shown. The spin forbidden energy levels were calculated with C/B=4.5.

The values of Dq/B, B, and C/B, calculated from various absorption and PLE spectra are given in table 4.21. The calculated value of Dq/B is much larger than that for a $d^2$ ion in tetrahedral coordination and the C/B values are much larger that the allowed range of 4 to 5 and clearly are invalid. As can be seen in figure 4.40, the ${}^1E({}^1D)$ and ${}^1T_2({}^1D)$ levels are almost independent of crystal field strength therefore emission from these levels is characterised by narrow R-line emission, as in $V^{3+}$ doped phosphate glass[87] and $V^{3+}$ doped corundum[156], with a lifetime in the ms to s regime. When the Tanabe-Sugano diagram is plotted with a valid C/B of 4.5, in figure 4.40 it can be seen that the calculated Dq/B of 2.836 is in a strong field site, this is where the lowest energy transition is a spin forbidden transition. If this were the case we would expect to see narrow R-line emission, a long lifetime (in the ms to s regime) and have the characteristic weak and narrow spin forbidden absorption on the low energy side of the first spin allowed absorption. In V:GLS the emission is very broad (FWHM~500 nm) the lifetime is ~30μs  and  spin forbidden absorption is on the high energy side of the first spin allowed absorption peak. The octahedral $d^2$ ion is therefore discounted as a possible configuration for V:GLS, with a high degree of confidence.

TABLE 4.21 Crystal field parameters calculated for a $d^2$ ion in tetrahedral coordination.

| Experimental data source | $E(^3T_2(^3F))$ (cm$^{-1}$) | $E(^3T_1(^3P))$ (cm$^{-1}$) | $E(^1E(^1D))$ (cm$^{-1}$) | Dq/B | B | C/B |
|---|---|---|---|---|---|---|
| Derivative absorption of V:GLS (figure 4.5) | 8965 | 13377 | 10100 | 2.836 | 316.1 | 12.613 |
| Derivative absorption of V:GLSO (figure 4.6) | 9300 | 13600 | 10200 | 3.002 | 309.7 | 13.118 |
| PLE of V:GLS (figure 4.12) | 8625 | 13135 | 10000 | 2.686 | 322.2 | 12.143 |
| PLE of V:GLSO (figure 4.13) | 8860 | 13299 | 10000 | 2.790 | 321.1 | 12.205 |

## 4.16.4 Tetrahedral $d^3$ configuration

The energy matrices and energy terms of the tetrahedral $d^3$ configuration are given in appendix C. These are plotted in the Tanabe-Sugano diagram in figure 4.41. There was no solution found using the method previously described and when using the two lowest spin allowed energy terms in both high and low field regions. The tetrahedral $d^3$ ion is therefore discounted as a possible configuration for V:GLS.

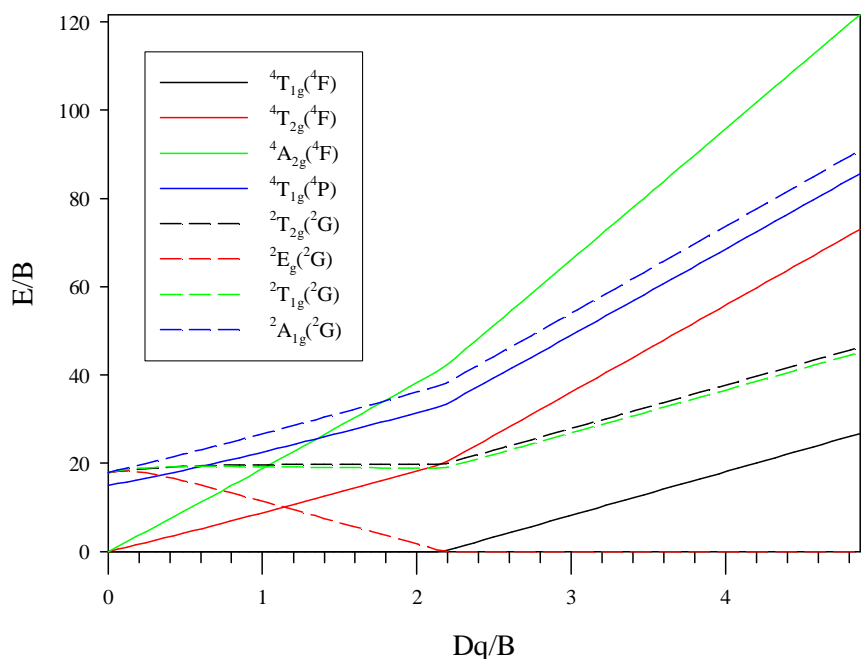

FIGURE 4.41 Tanabe-Sugano diagram of the tetrahedral $d^3$ configuration. The spin forbidden energy levels were calculated with C/B=4.63.

## 4.16.5 Octahedral d³ configuration

The energy terms of interest for the octahedral d³ configuration are given in equations 4.50 to 4.54

$$E\left(^4\mathrm{T}_2(^4\mathrm{F})\right)/\mathrm{B} = 10Dq\,/\,B \tag{4.50}$$

$$E\left(^4\mathrm{T}_1(^4\mathrm{F})\right)/\mathrm{B} = 7.5 + 15Dq/\mathrm{B} - \frac{1}{2}\sqrt{225 - 180DqB + 100Dq^2} \tag{4.51}$$

$$E\left(^4\mathrm{T}_1(^4\mathrm{P})\right)/\mathrm{B} = 7.5 + 15Dq/\mathrm{B} + \frac{1}{2}\sqrt{225 - 180DqB + 100Dq^2} \tag{4.52}$$

$$\frac{E\left(^2E(^2G)\right)}{B} = 1^{st}\,root\left(\det(M - \lambda I) = 0\right) - 12\frac{Dq}{B} - 15 \tag{4.53}$$

where

$$M = \begin{bmatrix} -6 + 3\dfrac{C}{B} - 12\dfrac{Dq}{B} & -6\sqrt{2} & -3\sqrt{2} & 0 \\[2mm] -6\sqrt{2} & 8 + 6\dfrac{C}{B} - 2\dfrac{Dq}{B} & 10 & \sqrt{3}(2 + \dfrac{C}{B}) \\[2mm] -3\sqrt{2} & 10 & -1 + 3\dfrac{C}{B} - 2\dfrac{Dq}{B} & 2\sqrt{3} \\[2mm] 0 & \sqrt{3}(2 + \dfrac{C}{B}) & 2\sqrt{3} & -8 + 4\dfrac{C}{B} + 18\dfrac{Dq}{B} \end{bmatrix}$$

$$\frac{E\left(^2T_1(^2G)\right)}{B} = 1^{st}\,root\left(\det(M - \lambda I) = 0\right) - 12\frac{Dq}{B} - 15 \tag{4.54}$$

where

$$M = \begin{bmatrix} -6 - 3\dfrac{C}{B} - 12\dfrac{Dq}{B} & -3 & 3 & 0 & -2\sqrt{3} \\[2mm] -3 & 3\dfrac{C}{B} - 2\dfrac{Dq}{B} & -3 & 3 & 3\sqrt{3} \\[2mm] 3 & -3 & -6 + 3\dfrac{C}{B} - 2\dfrac{Dq}{B} & -3 & -\sqrt{3} \\[2mm] 0 & 3 & -3 & -6 + 3\dfrac{C}{B} + 8\dfrac{Dq}{B} & 2\sqrt{3} \\[2mm] -2\sqrt{3} & 3\sqrt{3} & -\sqrt{3} & 2\sqrt{3} & -2 + 3\dfrac{C}{B} + 8\dfrac{Dq}{B} \end{bmatrix}$$

The energy terms for the ²E(²G) and ²T₂(²G) levels cannot be easily expressed explicitly, therefore they are defined in terms of the solution to the characteristic equation of the energy matrix of the ²E state, in which each matrix element has been divided by B to give the equation in terms of E/B. Where I is the identity matrix and the first root is the numerically smallest root in the zero crystal field (Dq=0) case. Comparing these equations to those for the tetrahedral $d^2$ configuration it can be seen that the energy terms for the three spin allowed energy levels are exactly the same for both configurations. The only difference between the energy terms of interest are for the spin forbidden energy levels. This now presents a problem for determining which of these configurations is most representative of V:GLS since only analysis of the spin forbidden terms can achieve this. The C/B ratio for this configuration was calculated from the energy term for the $^2E(^2G)$ multiplied by B to give it in terms of E($^2E(^2G)$). The values of Dq and E($^2E(^2G)$) were defined from experimentally determined values, B was calculated by solving equations 4.50 and 4.51 simultaneously for B. Then C was calculated by solving this equation for C. Because of the equations' complexity the equation solving facility of Mathematica software was used. As can been seen in the Tanabe-Sugano diagram for this configuration in figure 4.42 the $^2T_2(^2G)$ level lies close to the $^2E(^2G)$ level so the spin forbidden transition could be caused by a transition to either or both of these energy levels. Because of this, C/B was also calculated using the energy term for the $^2T_2(^2G)$ energy level. The values of Dq/B, B and the energy of the $^4T_1(^4P)$ level that was predicted by the Tanabe-Sugano model, calculated from various absorption and PLE spectra, are given in table 4.22.

TABLE 4.22 Dq and B crystal field parameters calculated for the octahedral $d^3$ configuration. The energy of the $^4T_1(^4P)$ level was calculated from the Tanabe-Sugano model.

| Experimental data source | E($^4T_2(^4F)$) (cm$^{-1}$) | E($^4T_1(^4F)$) (cm$^{-1}$) | E($^2E(^2G)$) (cm$^{-1}$) | E($^4T_1(^4P)$) (cm$^{-1}$) | Dq/B | B |
|---|---|---|---|---|---|---|
| Derivative absorption of V:GLS (figure 4.5) | 8965 | 13377 | 10100 | 20795 | 1.848 | 485.1 |
| Derivative absorption of V:GLSO (figure 4.6) | 9300 | 13600 | 10200 | 21146 | 2.038 | 456.3 |
| PLE of V:GLS (figure 4.12) | 8625 | 13135 | 10000 | 20499 | 1.668 | 517.1 |
| PLE of V:GLSO (figure 4.13) | 8860 | 13299 | 10000 | 20689 | 1.7940 | 493.9 |

Table 4.22 gives the energy of the $^4T_1(^4P)$ level that was calculated from its energy term equation 4.52. These are 20795 cm$^{-1}$(481 nm) from the absorption measurements of V:GLS, which is consistent with estimates of the centre of gravity for this absorption band in section 4.1 of ~500 nm. It was also noted that this estimate may be at a longer wavelength than the true centre of gravity for this absorption band because it was so heavily obscured by the band-edge absorption of GLS. The steeper gradient of the $^4T_1(^4P)$ level, in comparison the $^4T_2(^4F)$ and $^4T_1(^4F)$ levels is consistent with the greater bandwidth of the $^4T_1(^4P)$ level observed in the PLE spectra of V:GLS (figure 4.12). For V:GLSO the calculated energy of the $^4T_1(^4P)$ level is 21146 cm$^{-1}$(472 nm), the higher energy of this transition than for V:GLS is consistent with the higher energy of the experimentally determined transitions in V:GLSO. For the PLE measurements the lower energy calculated for the $^4T_1(^4P)$ level is consistent with the hypothesis that the absorption peaks of the PLE measurements were skewed to lower energy than with the absorption measurements because of a preferential excitation of ions in low field sites. The Dq/B of 1.668 calculated from PLE data, compared to 1.848 from absorption data, is also consistent with this hypothesis. It is also noted that Dq/B is higher in V:GLSO than in V:GLS which would be expected because oxygen is more electronegative than sulphur.

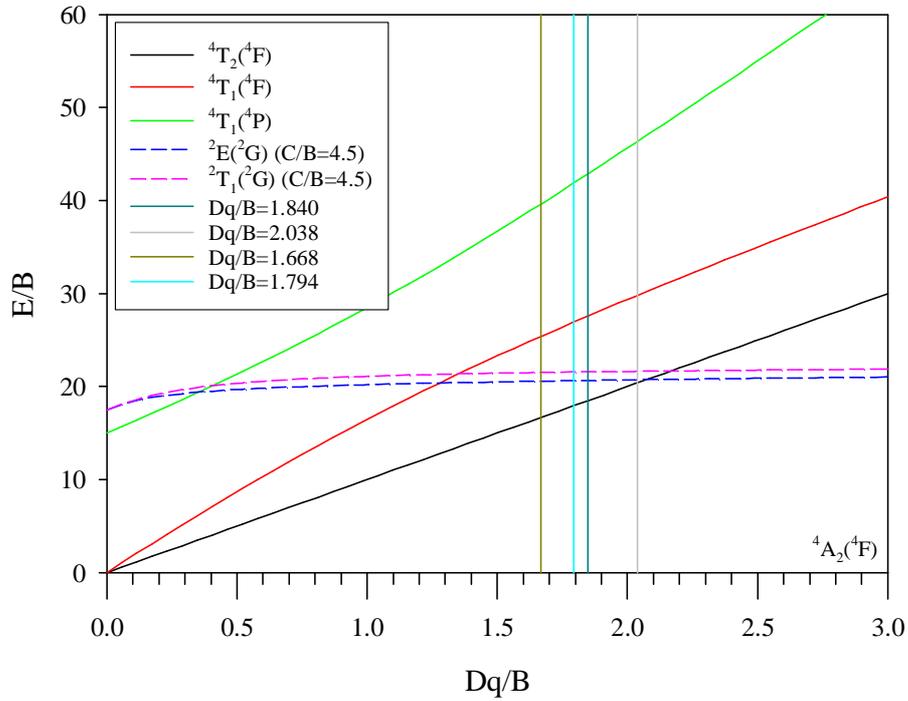

FIGURE 4.42 Tanabe-Sugano diagram of the octahedral $d^3$ configuration. The spin forbidden energy levels were calculated with C/B=4.5.

Because of the importance of the C/B parameter for this work it was checked using an approximate formula given by Rasheed,[162] for the calculation of C/B in the octahedral $d^3$ configuration (equation 4.55).

$$C/B = \left( \frac{E(^2E)}{B} + \frac{1.8B}{Dq} - 7.9 \right) \bigg/ 3.05 \qquad (4.55)$$

This equation was reported to be accurate to 5%;[162] a formula for the calculation of B was also given by Rasheed in 4.56

$$B = \frac{\frac{1}{3}\left(2E\left(^4T_2(^4F)\right) - E\left(^4T_1(^4F)\right)\right)\left(E\left(^4T_2(^4F)\right) - E\left(^4T_1(^4F)\right)\right)}{\left(9E\left(^4T_2(^4F)\right) - E\left(^4T_1(^4F)\right)\right)} \qquad (4.56)$$

TABLE 4.23 B and C/B crystal field parameters calculated for the octahedral $d^3$ configuration.

| Experimental data source | B | B (equation 4.56) | C/B ($^2E(^2G)$) | C/B ($^2T_1(^2G)$) | C/B (equation 4.55) |
|---|---|---|---|---|---|
| Derivative absorption of V:GLS (figure 4.5) | 485.12 | 485.21 | 4.553 | 4.247 | 4.554 |
| Derivative absorption of V:GLSO (figure 4.6) | 456.32 | 456.48 | 4.934 | 4.7367 | 5.025 |
| PLE of V:GLS (figure 4.12) | 517.09 | 517.67 | 4.164 | 3.777 | 4.097 |
| PLE of V:GLSO (figure 4.13) | 493.87 | 493.89 | 4.403 | 4.063 | 4.377 |

Table 4.23 gives the values of B calculated by solving the energy terms of the two lowest energy levels simultaneously for B and by using equation 4.56. Table 4.23 also gives C/B calculated from the energy term for the $^2E(^2G)$ and $^2T_2(^2G)$ energy levels and from equation 4.55. The values of B, calculated using the method detailed in this work and by equation 4.56 are almost identical. Calculation of B using equation 4.56 is the simplest method however this equation does not apply to all the configurations dealt with in this study. The values of C/B calculated using the energy term for the $^2E(^2G)$ energy level are all in the allowed range of 4 to 5 unlike those calculated for the tetrahedral $d^2$ configuration which were 5.8 to 6.6. The calculated values of Dq/B are shown in the Tanabe-Sugano diagram in figure 4.42 and all indicate a weak field site which is consistent with the emission absorption and lifetime measurements. For the Dq/B of 2.038 the C/B value is higher than that used to plot figure 4.42 while it appears in the figure to be at the cross over point of the spin forbidden and spin allowed transition, when plotted with the correct C/B it is in a weak field site. In tetrahedral $d^2$ configuration the $^3A_2(^3F) \rightarrow {}^3T_2(^3F)$ transition is expected to be significantly weaker than the other two spin allowed transitions because it is only magnetic dipole allowed,[163, 164] however this is not evident from the derivative absorption and PLE spectra of V:GLS in figure 4.5 and 4.12, this indicates that V:GLS, may not have a tetrahedral $d^2$ configuration.

The above arguments indicate that the octahedral $d^3$ configuration is more representative of V:GLS than the tetrahedral $d^2$ configuration. One of the values of C/B calculated using the energy term for the $^2T_2(^2G$ ) energy level is just outside the allowed range which indicates that the spin forbidden transition is more likely to be due to a transition to the $^2E(^2G)$ energy level. The C/B value calculated using equation 4.55 appears to be a good approximation to the method described in this work and it is much simpler to calculate. This indicated that equation 4.55 would be a more suitable method for calculating C/B when maximum precision is not required.

It is therefore proposed that the three spin allowed absorption bands identified in figure 4.12 at 1160, 760 and 580 nm are due to $^4A_2(^4F) \rightarrow {}^4T_2(^4F)$, $^4A_2(^4F) \rightarrow {}^4T_1(^4F)$ and $^4A_2(^4F) \rightarrow {}^4T_1(^4P)$ transitions respectively and the spin forbidden transition at 1000 nm identified in figure 4.5 is attributed to the $^4A_2(^4F) \rightarrow {}^2E(^2G)$. From the bandwidths of these absorption bands, calculated in figure 4.12, the energy level diagram for vanadium doped GLS is given in figure 4.43.

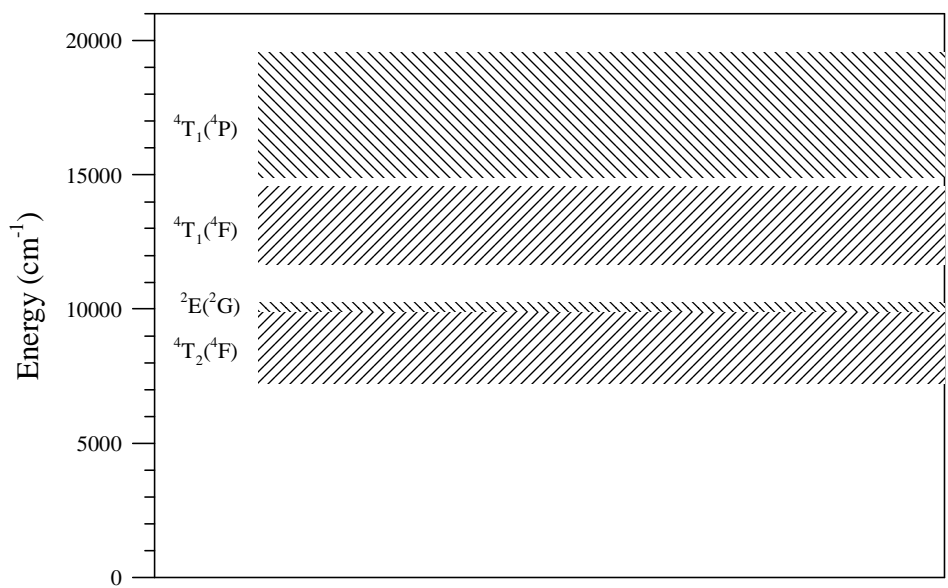

FIGURE 4.43 Energy level diagram of vanadium 2+ doped GLS

## 4.18 Conclusions

V:GLS was optically characterised to investigate its suitability as an active material for an optical device. Absorption measurements of V:GLS unambiguously identified one absorption band at 1100 nm, with evidence of a spin forbidden transition around 1000 nm, and two further higher energy absorption bands that could not be resolved. Derivative analysis of the absorption measurements clarified the identification of the spin forbidden transition and was able to resolve the second highest, but not the highest, energy absorption band at 750 nm. PLE measurements were able to resolve all three absorption bands, peaking at 1160, 760 and 580 nm. However there was a preferential detection of ions in low crystal field strength sites. XPS measurements indicated the presence of vanadium in a broad range of oxidation states from $V^+$ to $V^{5+}$. Excitation into each of the three absorption bands produced the same characteristic emission spectrum, peaking at 1500 nm with a FWHM of ~500nm. The decay lifetime and decay profile were also similar. This was a strong indication that only one of the vanadium oxidation states was responsible for the observed absorption bands. The quantum efficiency of 0.0023 % V:GLS was 4.2 %. Out of the possible vanadium oxidation states, only $V^{2+}$ and $V^{3+}$ is expected to exhibit three spin allowed transitions. Tanabe-Sugano analysis indicates that out of the possible configurations of coordination and oxidation state only tetrahedral $V^{3+}$ and octahedral $V^{2+}$ had a crystal field strength in the expected low field region. Out of these configurations only octahedral $V^{2+}$ had a C/B value in the expected range of 4-5. The configuration of the optically active vanadium ion in V:GLS is therefore proposed to be octahedral $V^{2+}$. The crystal field strengths (Dq/B) calculated from absorption measurements of V:GLS and V:GLSO are 1.84 and 2.04 respectively.

Lifetime measurements of V:GLS found that the decay was non exponential and at low concentrations could be modelled with the stretched exponential function. Analysis of the coefficient of determination of stretched and double exponential functions and results from a continuous lifetime distribution analysis of the emission decay, at various vanadium concentrations, indicated that at concentrations < 0.1% there was one lifetime component centred ~30 µs. At concentrations > 0.1 two lifetime components centred ~ 30 µs and 5 µs are present. This was argued to be caused by a preferentially filled, high efficiency, oxide site that gives rise to characteristic long lifetimes and a low efficiency sulphide site that gives rise to characteristic short lifetimes.

Comparisons of the $\sigma_{em}\tau$ product of V:GLS to that in other laser materials indicates the best possibility for demonstration laser action in V:GLS is in a fibre geometry. Modelling of laser action an a V:GLS fibre is not presented because the number of assumptions to be made about such a device is too great. Fabrication of a V:GLS fibre is suggested as further work.

Considering the track record of transition metal doped glasses as active optical devices the prospect of producing a commercially viable optical amplifier based on V:GLS is thought to be low, however the potential reward of producing a broadband optical amplifier centred at 1500 nm means that further development is justified on a risk/reward basis.

# Chapter 5

# Titanium, nickel and bismuth doped chalcogenide glass

## 5.1 Introduction

This chapter details the spectroscopic properties of titanium, nickel and bismuth doped GLS which, like vanadium doped GLS detailed in chapter 4, may have applications as active optical devices because they display strong infrared emission. Prior studies have been conducted on this topic but there are still some areas to investigate. It was found that the emission of Ti, Ni and Bi:GLS peaks at ~900 nm, which is less useful for telecommunication applications compared to the emission of V:GLS which peaked ~1500 nm. Reports of emission from Ti and Ni doped glasses are scarce in the literature. However, there are many reports of emission from Bi doped glasses[165-174] and laser action been demonstrated in a bismuth doped aluminosilicate fibre laser.[84]

Iron, cobalt and copper doped GLS were also investigated by the author however no emission in the range 800-1800 nm could be detected from these dopants so they are not included in this report. Chromium doped GLS was also studied, however this has been previously investigated in detail by Aronson.[23] The absorption of titanium, nickel, iron and cobalt doped GLS has been reported by Aronson[23] and Brady.[70] Petrovich[24] reported the absorption of nickel, iron and cobalt doped GLS. Aronson reported emission from Ni:GLS when exciting at 800 nm, which is at the long wavelength end of the Ni:GLS excitation range. Aronson did not attribute an oxidation state to Ti:GLS whereas Brady speculated it was in a 4+ oxidation state. Aronson, Brady and Petrovich all proposed Ni:GLS was in a 2+ oxidation state; in this chapter evidence is provided that Ni:GLS is in a 1+ oxidation. This is the first time the optical properties of bismuth doped GLS have been reported. Glasses were melted using the method detailed in section 3.2.1.

## 5.2 Titanium doped GLS

Titanium doped $Al_2O_3$ (Ti:Sapphire) has been used as a gain medium in room temperature tuneable lasers since laser action was first reported by Moulton in 1982.[4] The Ti:Sapphire laser is the most widely used near infrared tuneable laser source and is tuneable from 650 to 1100 nm, it is also used to generate ultra short laser pulses with durations as low as 8 fs.[80] Because of the success of the Ti:Sapphire crystal as a tuneable laser source little attention has been paid to titanium doped glass as an active medium and reports in the literature of photoluminescence from titanium doped glasses are extremely scarce. A titanium doped glass laser in a fibre geometry could potentially have several advantages over a Ti:Sapphire laser in that it could be more compact, have a higher alignment stability and robustness.

### 5.2.1 Absorption of titanium doped GLS

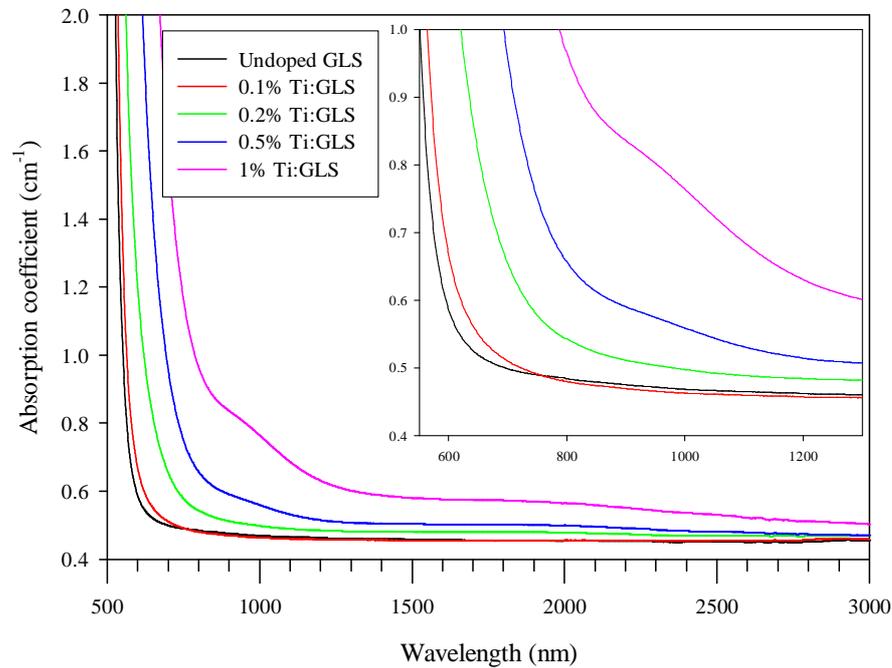

FIGURE 5.1 Absorption spectra of 0.1 to 1% molar titanium doped GLS and un-doped GLS in 3mm thick slabs.

Figure 5.1 shows the absorption spectra of GLS doped with varying concentrations of titanium. At concentrations up to 0.2% absorption from titanium is only visible as a red shift in the band-edge of GLS indicating absorption from titanium at ~ 500-600 nm. At concentrations of 0.5% and greater a shoulder at ~1000 nm is observed, there is also a weak and broad absorption centred at around 1800 nm. Figure 5.2 shows the absorption spectra of GLSO doped with varying concentrations of titanium. Similarly to Ti:GLS there is a red shift in the band-edge with increasing titanium concentration, however there is no evidence of a shoulder at ~1000 nm or a broad weak absorption at ~ 1800 nm as in Ti:GLS. The 1% Ti:GLSO sample had partially crystallised which meant that its base line absorption was higher than expected because of increased scattering.

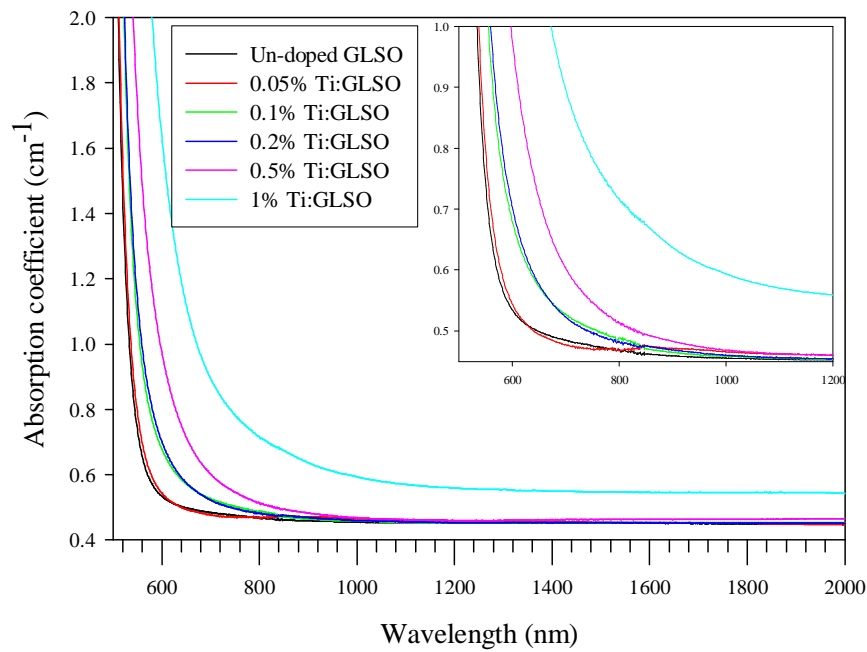

FIGURE 5.2 Absorption spectra of 0.05 to 1% molar titanium doped
GLSO and un-doped GLSO in 3mm thick slabs.

To confirm the observation of a shoulder at ~1000 nm or a broad weak absorption at ~ 1800 nm as in Ti:GLS but not in Ti:GLSO derivatives of the absorption spectra of 0.5% Ti:GLS and 0.5% Ti:GLSO were taken. As described in Section 4.3, absorption peaks correspond to where $d^2a/d\lambda^2 < 0$, where a is the absorption coefficient. The second derivative absorption spectra in figure 5.3 clearly show that there is an absorption peak at 980 nm in Ti:GLS but not in Ti:GLSO, absorption peaks at 615 and 585 nm are also identified for Ti:GLS and Ti:GLSO respectively. The second derivative absorption peak at 980 nm was much stronger in 1% Ti:GLS than in 0.5 % Ti:GLS but could not be identified at concentrations of 0.2% or lower.

The Ti$^{3+}$ ion has a single d electron, it is therefore expected to exhibit a single absorption band in ideal symmetry. In crystals of Al$_2$O$_3$, Ti$^{3+}$ ions have octahedral coordination with trigonal symmetry[175] and their absorption is characterised by a broad double humped absorption band extending from 400 to 600 nm. The main peak occurs at 490 nm with the shoulder at 550 nm, these are attributed to transitions from the $^2T_{2g}$ ground state to the Jahn-Teller split $^2E_g(E_{3/2})$ and $^2E_g(E_{1/2})$ excited states.[176] A weak residual IR absorption has been identified in Ti:Sapphire from around 650 to 1600 nm, peaking at 800 nm[176] and has been shown to be due to Ti$^{3+}$-Ti$^{4+}$ pairs.[177] It is now proposed that the absorption at ~600 nm in Ti:GLS and Ti:GLSO is due to the $^2T_{2g} \rightarrow ^2E_g$ transition of octahedral Ti$^{3+}$ and the absorption at 980 nm is due to Ti$^{3+}$-Ti$^{4+}$ pairs. The residual IR absorption coefficient of Ti$^{3+}$:Al$_2$O$_3$ has been shown to be proportional to the square of its blue-green absorption coefficient.[176, 177] If this relationship also exists in Ti:GLS then it would explain the inability to resolve the 980 nm absorption at low concentrations.

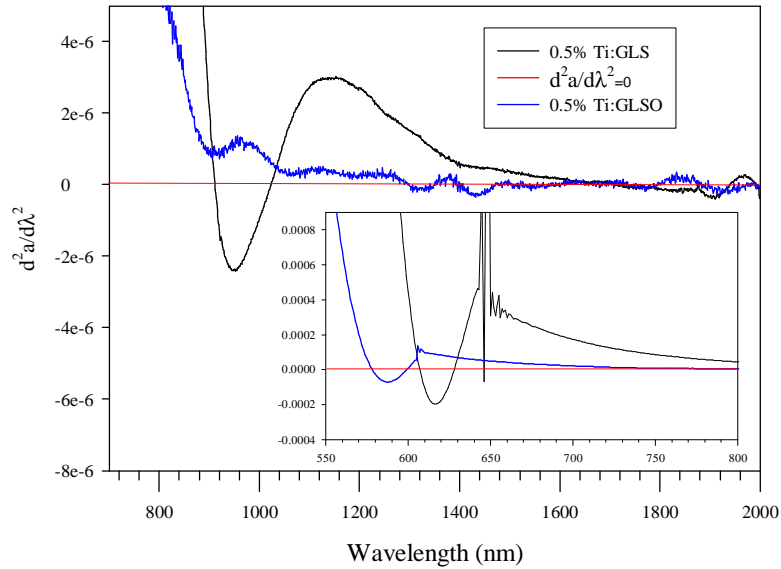

FIGURE 5.3 Second derivative of the absorption coefficient of 0.5%
Ti:GLS and 0.5% Ti:GLSO.

Because absorption due to $Ti^{3+}$-$Ti^{4+}$ pairs is detrimental to the performance of the
Ti:Sapphire laser, much effort has been made to minimise it. In Ti:Sapphire the valence
of Ti ions has been controlled by melting temperature and oxygen partial pressure[176,
178] In silica the concentration of $Ti^{3+}$ relative to $Ti^{4+}$ ([$Ti^{3+}$]/[ $Ti^{4+}$]) increased with
melting temperature and decreasing oxygen partial pressure and was maximised by
melting in deoxidized argon, i.e. a reducing atmosphere;[179] this is the same
atmosphere that GLS is melted in so  the oxygen partial pressure parameter is already
maximised for the minimisation of $Ti^{4+}$ concentration. In silicate, borate and phosphate
glasses the redox reaction of $TM^{m+} \leftrightarrow TM^{(m+1)+}$ is related to the glass basicity (B),[88,
179] TM is a transition metal ion and glass basicity is calculated in terms of the
coulombic force between the cation and oxygen ion of each glass component as in
equation 5.1.[179]

$$B_i = \frac{(r_i + 1.4)^2}{Z_i \times 2} \qquad (5.1)$$

Where $Z_i$ and $r_i$ are the valency and radius of the cation, the values of 2 and 1.4 are the
valency and radius of the oxygen ion respectively and $B_i$ is the basicity of glass
component i. The higher $La_2O_3$ content of GLSO compared to GLS is thought to cause
the formation of oxide negative cavities[16] whereby the oxygen coordination of
gallium is increased from 0 to 1, therefore GLSO should have a higher basicity than
GLS. In silicate glass [$Ti^{3+}$]/[ $Ti^{4+}$] is inversely proportional to B,  in borate glass
[$Ti^{3+}$]/[ $Ti^{4+}$] is proportional to B and in phosphate glass there is no dependence of
[$Ti^{3+}$]/[ $Ti^{4+}$] on B.[179] The relationship between [$Ti^{3+}$]/[ $Ti^{4+}$] and B in the GLS
system is not known however it is assumed that [$Ti^{3+}$]/[ $Ti^{4+}$] is proportional to B since
there is no absorption due to $Ti^{3+}$-$Ti^{4+}$ pairs in the more basic GLSO glass.

The absorption of titanium has been characterised in a variety of glasses.[179-187] In all of the glasses a broad double humped absorption band extending from 400 to 750 nm was observed and attributed to $Ti^{3+}$ in tetragonally distorted octahedron, except in silicate glass where a single absorption peak at 560 nm was observed and attributed to a continuous range of Jahn-Teller splittings. No absorption shoulder could be resolved in GLS and GLSO this is therefore attributed to the same effect. In some of the glasses an infrared absorption at ~800 nm was attributed to $Ti^{3+}$-$Ti^{4+}$ pairs. The peak absorptions for titanium in a variety of glasses and in sapphire are summarised in table 5.1

TABLE 5.1 Absorption details for titanium in a variety of glasses and in sapphire.

| Host | $Ti^{3+}$ Absorption peak (nm) | $Ti^{3+}$ Absorption shoulder (nm) | $Ti^{3+}$-$Ti^{4+}$ pair absorption peak (nm) | Reference |
|---|---|---|---|---|
| GLS glass | 615 | - | 980 | This work |
| GLSO glass | 585 | - | - | This work |
| Sodium phosphor aluminate glass | 560 | 697 | - | [188] |
| Silicate glass | 560 | | - | [186] |
| Lithium calcium phosphate glass | 510 | 680 | - | [185] |
| Lithium magnesium borate glass | 515 | 679 | - | [180] |
| Sodium silicate glass | 500 | 770 | - | [184] |
| Phosphate glass | 565 | 725 | - | |
| Barium borosilicate glass | 500 | 746 | - | [187] |
| Lithium silicate glass | 530 | 640 | 800 | [179] |
| Sodium phosphor aluminate glass | 580 | 650 | - | [179] |
| Fluorophosphate glass | 529 | 685 | - | [183] |
| $Al_2O_3$ crystal | 490 | 550 | 800 | [176] |

## 5.2.2 Photoluminescence of titanium doped GLS

Photoluminescence spectra were taken using the setup described in section 3.3.2. In Ti:$Al_2O_3$ emission peaks at ~750 nm when excited at 514 nm.[4] Out of all the titanium doped glasses in table 5.1 emission from d-d transitions in $Ti^{3+}$ was only reported in sodium phosphor aluminate glass[188] in the other glasses emission was not investigated or not detectable. In $Ti^{3+}$: sodium phosphor aluminate glass emission centred at 860 nm with a FWHM of 2020 cm$^{-1}$ was detected from excitation with a 633 nm He-Ne laser. This is close to the emission of Ti:GLS and Ti:GLSO in figure 5.4 where the emission peaked at 900 nm when excited at the same wavelength. This further backs up the hypothesis that titanium is in a 3+ oxidation state in GLS and GLSO. The broadness of the PL spectrum indicates that the titanium ion is in a low crystal field site. No emission in the range 1200-1800 nm was detected from Ti:GLS when exciting with a CW 500 mW 1064 nm laser source.

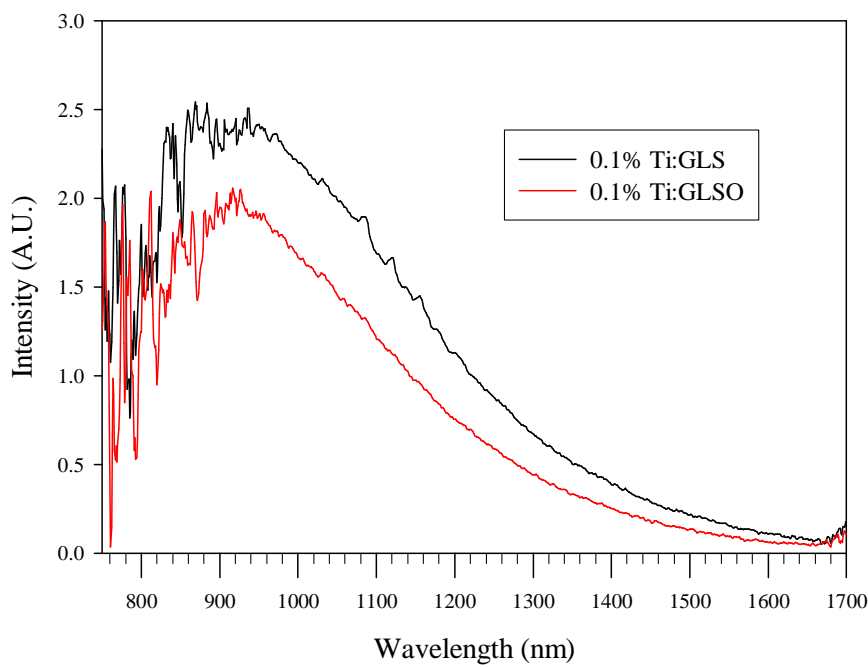

FIGURE 5.4 Photoluminescence spectra of 0.1% titanium doped GLS and GLSO excited with a 5mW 633 nm laser source.

The observation that emission from Ti:GLS peaks at 900 nm, which is at a higher energy than the weak absorptions at 980 and 1700 nm, implies that these absorptions cannot be due to the same oxidation state and coordination of titanium that produced the emission and backs up the hypothesis that they are due to $Ti^{3+}$-$Ti^{4+}$ pairs.

## 5.2.3 Photoluminescence excitation of titanium doped GLS

Photoluminescence excitation spectra were taken using the setup described in section 3.3.3. Figure 5.5 shows the excitation spectra of 0.1 % titanium doped GLS and GLSO, both of which show a single excitation peak at 580 nm which is in good agreement with the derivative absorption measurements. The excitation signal was relatively weak and the apparent increase in excitation in Ti:GLS at 800 nm is caused by correction for the system response. No excitation signal was detected up to 1000 nm, using a 600 line/mm grating to disperse the excitation source.

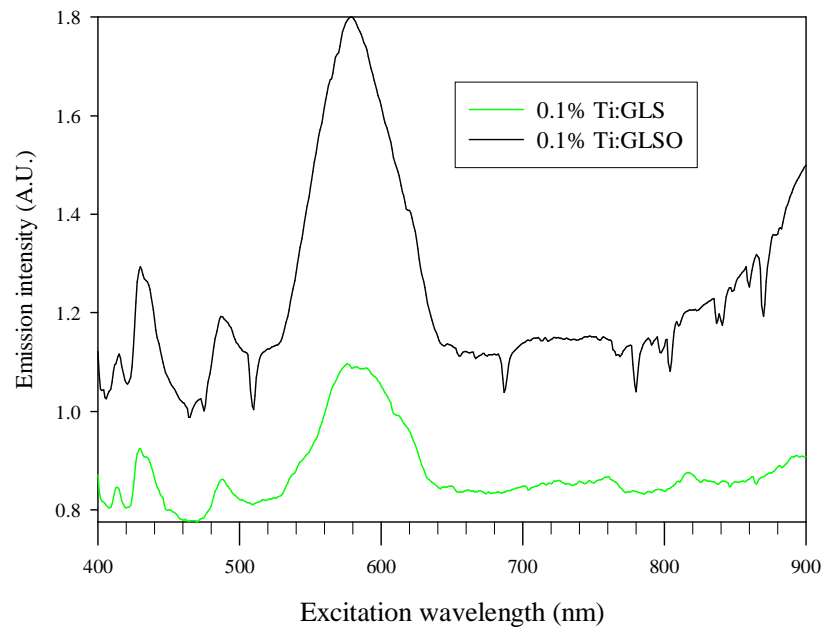

FIGURE 5.5 PLE spectra of 0.1% Ti:GLS and 0.1% Ti:GLSO. Emission was detected at 1000-1700 nm and the excitation source was dispersed with a 1200 line/mm blazed grating.

## 5.2.4 Fluorescence lifetime of titanium doped GLS

Similarly to vanadium doped GLS the lifetime of a range of concentrations is investigated in this section, using both the stretched and continuous lifetime distribution models. This data is used to infer the optimum concentration and composition of an active optical device based on Ti:GLS. Lifetime measurements were taken using the setup described in section 3.3.4.

### 5.2.4.1 Stretched and double exponential modelling

Figure 5.6 shows the fluorescence decay of 0.05% titanium doped GLS fitted with a stretched exponential. The best fit to the experimental data was with a lifetime of 67 µs and a stretch factor (β) of 0.5. Visual inspection indicates an excellent fit to the experimental data.

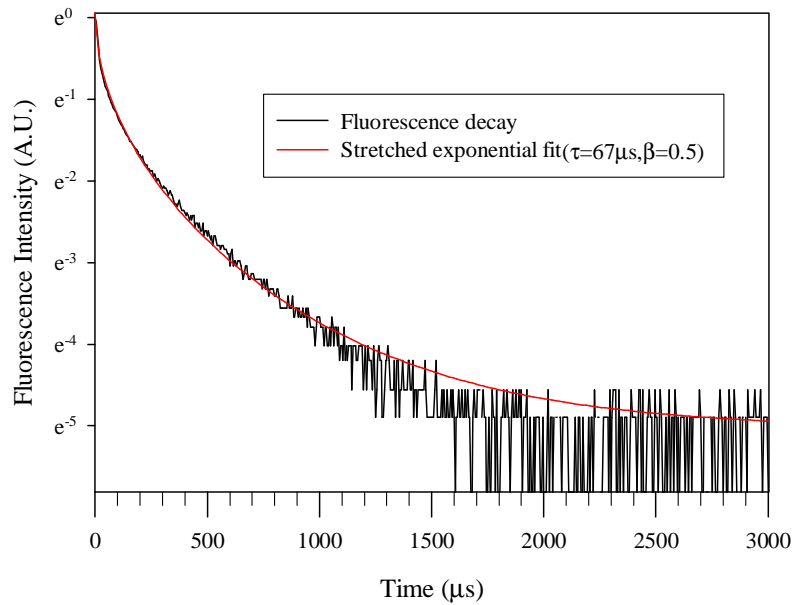

FIGURE 5.6 Fluorescence decay of 0.05% titanium doped GLS excited with a 10 mW 658 nm laser source fitted with a stretched exponential. The lifetime was 67 µs and the stretch factor was 0.5.

Figure 5.7 shows the fluorescence decay of 1% Ti:GLS, fitted with the stretched and double exponential. Similarly to 1% V:GLS the stretched exponential is no longer a good fit ($R^2$ = 0.9355) and the double exponential fit is better ($R^2$ = 0.9824). The lifetimes of the double exponential were 15 µs and 160 µs. The fact that the characteristic slow lifetime of the 67 µs is no longer observed at high concentrations when using the double exponential fit as it is in V:GLS is believed to be because the stretch factor is now 0.5 and a single exponential is no longer a good approximation. Fitting a single exponential to a stretched exponential with a stretch factor of 0.5 gives $R^2$ = 0.8632. To overcome this problem fluorescence intensity data from 0 to 100 µs was discarded, leaving just the slow component of the decay. This was fitted with a stretched exponential with β fixed at 0.5. This fit had the characteristic slow lifetime of ~67 µs. The same procedure for data 0 to 100 µs gave a lifetime of 15 µs.

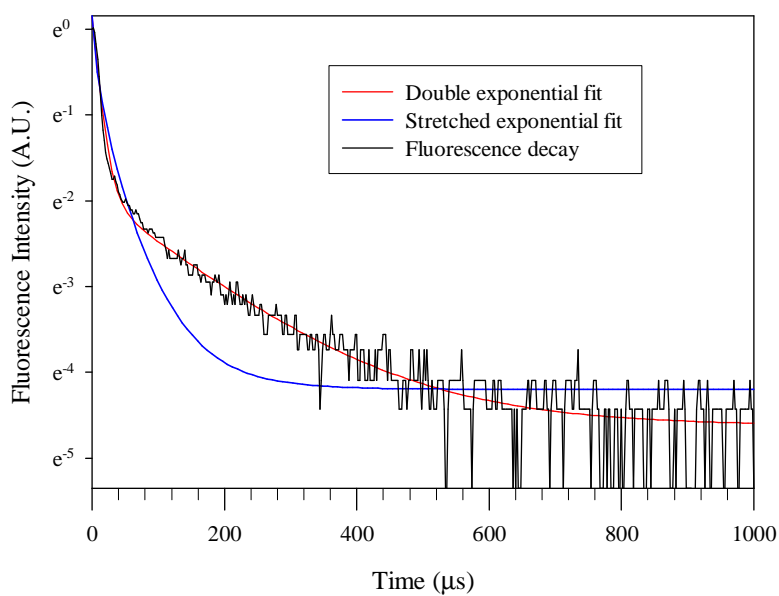

FIGURE 5.7 Fluorescence decay of 1% titanium doped GLS excited with a 10 mW 658 nm laser source fitted with a stretched and double exponential.

Figure 5.8 shows the emission decay of 0.05% titanium doped GLSO fitted with a stretched exponential with a lifetime of 97 μs and a stretch factor (β) of 0.5. Visual inspection indicates an excellent fit to the experimental data. The longer lifetime in the GLSO host follows the same trend observed in vanadium doped GLS and GLSO.

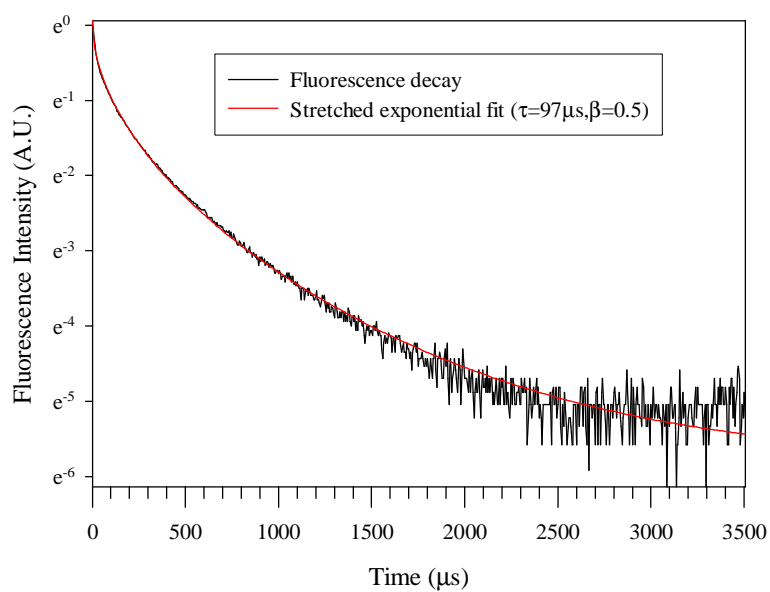

FIGURE 5.8 Fluorescence decay of 0.05% titanium doped GLSO excited with a 10 mW 658 nm laser source fitted with a stretched exponential. The lifetime was 97 μs and the stretch factor was 0.5.

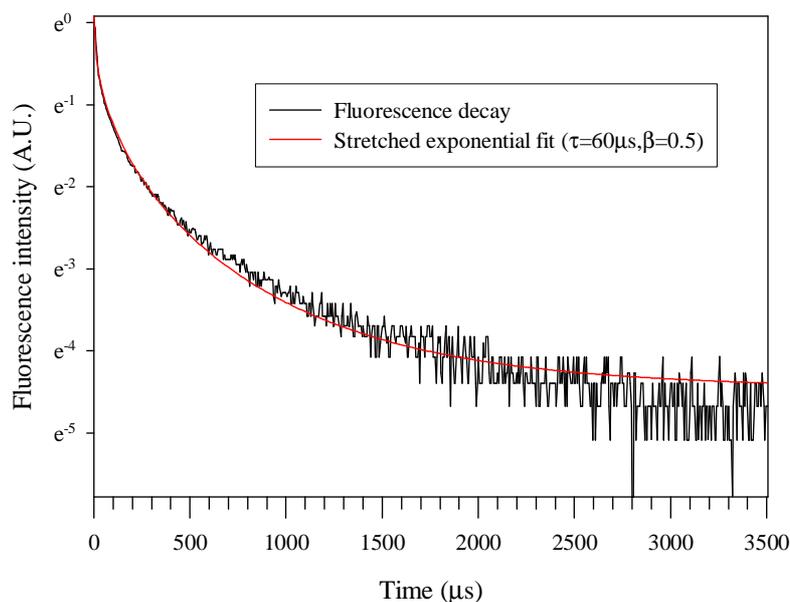

FIGURE 5.9 Fluorescence decay of 1% titanium doped GLSO excited with a 10 mW 658 nm laser source fitted with a stretched exponential. The lifetime was 60 μs and the stretch factor was 0.5.

Figure 5.9 shows the emission decay of 1% titanium doped GLSO fitted with a stretched exponential with a lifetime of 60 μs and a stretch factor (β) of 0.5. Unlike Ti:GLS the emission decay at this concentration is still well described by a stretched exponential.

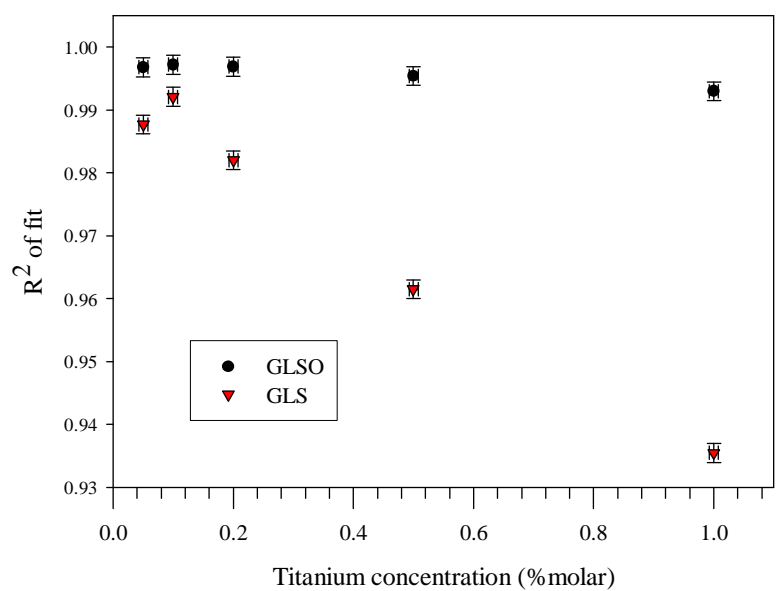

FIGURE 5.10 Coefficient of determination of stretched exponential fit as a function of titanium concentration in GLS and GLSO.

Figure 5.10 shows the $R^2$ of stretched exponential fits as a function of titanium concentration in GLS and GLSO. In Ti:GLS the $R^2$ appears to follow the same pattern as in V:GLS where above ~0.1% concentration the fluorescence decay starts to deviate from stretched exponential behaviour which is manifested as a decrease in $R^2$. In Ti:GLSO there is hardy any change in $R^2$ as titanium concentration increases. This can be explained in the same way that the data for V:GLS was explained, i.e. that two reception sites for transition metals exist in GLS glass; a high efficiency oxide site and a low efficiency sulphide site. In GLS the transition metal ion preferentially fills the high efficiency oxide sites until, at a concentration of ~0.1%, they become saturated and the low efficiency sulphide sites starts to be filled.

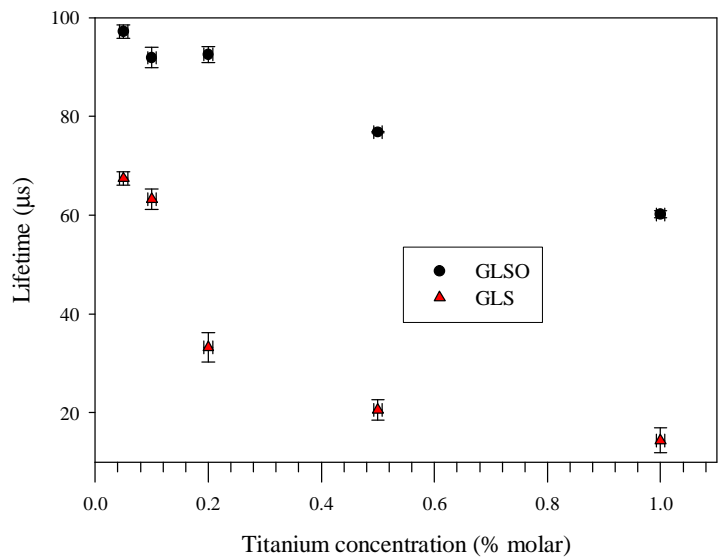

FIGURE 5.11 Lifetimes of titanium doped GLS and GLSO as a function of doping concentration. The emission decays were fitted with the stretched exponential model.

Figure 5.11 shows how the emission lifetime of titanium doped GLS and GLSO varies as a function of doping concentration. The figure shows that the lifetime is longer in Ti:GLSO than Ti:GLS, which indicates that GLSO is the most favourable host for an active optical device. The lifetime is still increasing slightly at the lowest concentration, 0.05 % molar titanium, investigated. This indicates that concentration quenching is still occurring at 0.05 %, therefore the optimum concentration for an active optical device based on Ti:GLSO may be lower than 0.05 %. In Ti:Al$_2$O$_3$, laser action has been demonstrated at titanium concentrations of ~ 0.03 to 0.15 % molar.[4] Figure 5.11 also shows that lifetimes decrease more rapidly as concentration increases in Ti:GLS than in Ti:GLSO. This can be explained by the oxide and sulphide site model. The higher oxygen content of GLSO means that in Ti:GLSO the high efficiency oxide sites don't get "used up" at higher concentrations, whereas in Ti:GLS the lower oxygen content means that as titanium concentration increases, the proportion of titanium ions in low efficiency sulphide sites increases. Therefore the lifetime will decrease more rapidly as concentration increases.

All the lifetimes in figure 5.11 were determined using the stretched exponential model. As shown in figure 5.10 there is a deviation from the stretched exponential model in Ti:GLS at concentrations > ~ 0.2 %, the lifetime at these concentrations is therefore an approximation. Because of the difficulties in comparing the lifetime of emissions where the decays follow different models, as discussed in section 4.8, this is believed to be a valid approximation.

### 5.2.4.2 Continuous lifetime distribution modelling

The fluorescence decays of titanium doped GLS were also analysed using the continuous lifetime distribution model, which is described in section 4.10. Figure 5.12 shows how the distribution of lifetimes varies with titanium concentration in GLS. The figure shows that there are two distribution peaks, one centred around 70 μs (peak S1) and another around 15 μs (peak S2). It is also clear that as the concentration increases peak S2 becomes more intense in comparison to peak S1. This effect was observed in V:GLS but not to such a great extent.

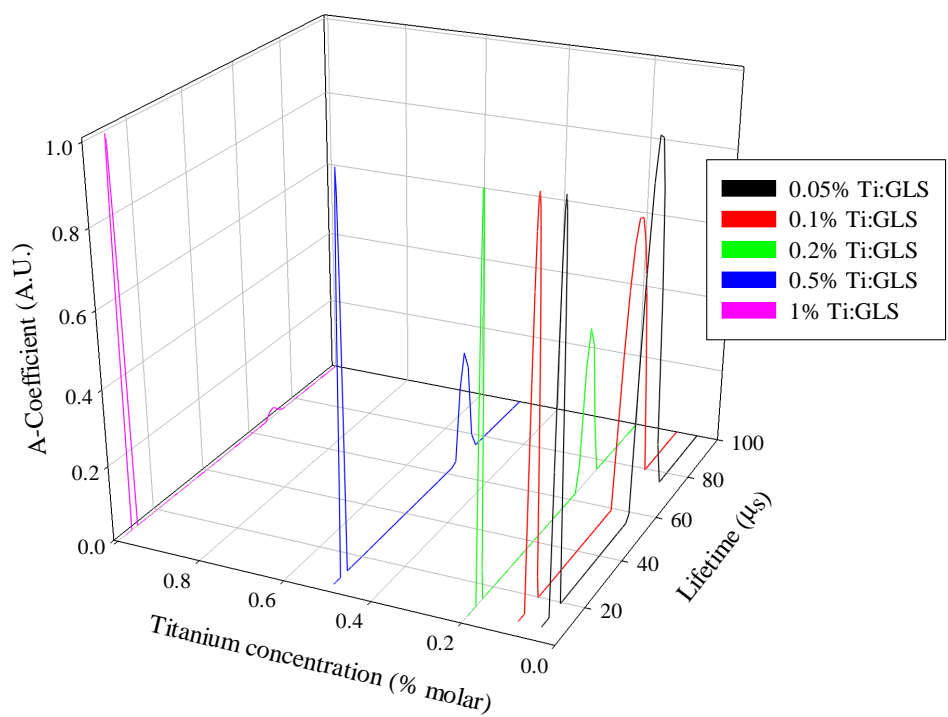

FIGURE 5.12 Lifetime distribution in Ti:GLS as a function of titanium concentration.

Figure 5.13 shows how the distribution of lifetimes varies with titanium concentration in GLSO. The figure shows that there are two distribution peaks, one centred around 100 μs (peak O1) and another around 18 μs (peak O2). As the concentration increases the relative intensities of peak O1 and O2 changes very little in comparison to Ti:GLS. A manifestation of this is that the overall lifetime will decrease more rapidly with increasing titanium concentration in Ti:GLS than Ti:GLSO. These results indicate that there is no preferential filling of high efficiency oxide sites down to a concentration of 0.05% (however this may not be the case at lower concentrations) and that in GLS the oxide sites become saturated at ~0.1% which means only low efficiency sulphide sites are filled. This explains why the relative intensity of the lifetime distribution centred ~70 µs decreases strongly with increasing concentration in Ti:GLS. The low dependence of the relative intensity of the lifetime distribution centred ~100 µs in Ti:GLSO is explained by the greater abundance of oxide sites in GLSO that do not become saturated, as they do in GLS.

It is noted that in all the decays, that were well described by the stretched exponential there were two distributions of lifetimes with similar intensities. In the decays that were better described by a double exponential, there were two distributions of lifetimes with dissimilar intensities. It therefore appears that, where two distributions of lifetimes are present, deviation from stretched exponential behaviour is manifest by a larger difference in the intensities of the two distributions of lifetimes.

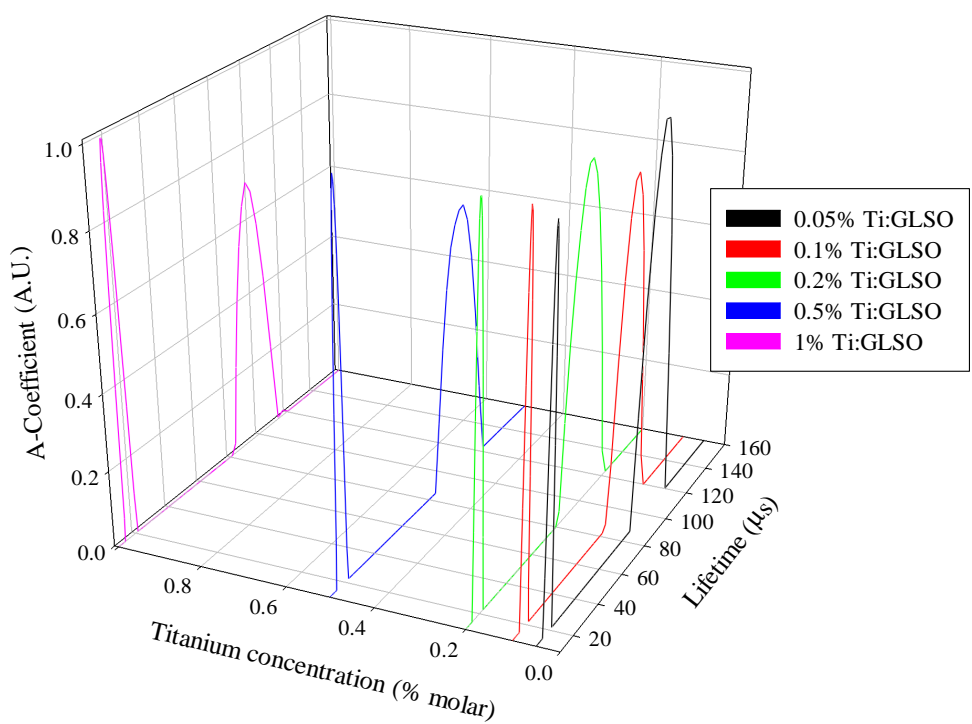

FIGURE 5.13 Lifetime distribution in Ti:GLSO as a function of titanium concentration.

There has been no report of the emission lifetime of titanium in a variety of glasses.[179-187] However the 97 µs emission lifetime of Ti:GLSO measured here compares very favourably with the lifetime of Ti:Sapphire of 3.1 µs.[4] The large difference in Ti$^{3+}$ lifetime between the oxide and sulphide sites in GLSO indicates that the lifetime of the Ti$^{3+}$ ion is highly sensitive to its host environment which may explain why the lifetime of Ti:GLSO is much longer than that of Ti:Sapphire.

## 5.3 Nickel doped GLS

Laser operation, tuneable from 1610 to 1740 nm with a CW output power of 20mW, has been demonstrated in $Ni^{2+}:MgF_2$ at 80 K.[5] Also in $Ni^{2+}:MgO$ CW output tuneable from 1316 to 1409 nm at 80 K[3] and in $Ni^{2+}:Gd_3Ga_5O_{12}$ tuneable from 1434 to 1520 nm at 100 K has been reported.[3]

### 5.3.1 Absorption of nickel doped GLS

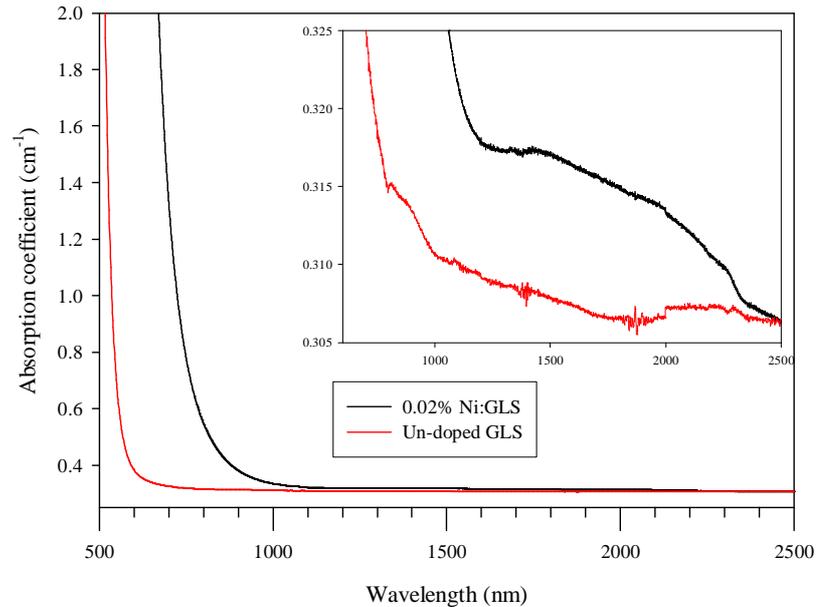

FIGURE 5.14 Absorption spectra of 0.02% (molar) nickel doped GLS and un-doped GLS in 5 mm thick slabs.

Figure 5.14 shows the absorption spectra of un-doped GLS and 0.02% nickel doped GLS. The absorption of Ni:GLS is characterised by a red-shift of ~ 300 nm in the band-edge indicating a nickel absorption in the region 500-800 nm. There is also a very weak absorption peaking at ~1500 nm. Similarly to the argument used for Ti:GLS the absorption bands at 500-800 nm and 1500 nm are not believed to originate from the same valence state and coordination of nickel because when exciting into the 500-800 nm absorption band, emission peaks at higher energy than the 1500 nm absorption band.

Nickel is most commonly observed in a 2+ oxidation state in various crystals and glasses,[3, 53, 55, 81, 83, 189-200] details of some of these are given in table 5.2. $Ni^{2+}$ and $Ni^{3+}$ have a $3d^8$ and $3d^7$ electronic structure respectively. As seen from the Tanabe-Sugano diagram in figure 4.39 and figure 4.42, $Ni^{2+}$ and $Ni^{3+}$ should have three spin allowed transitions which would be expected to produce three broad Gaussian absorption bands. Three absorption bands are observed for $Ni^{2+}$ in several glasses[81, 83, 189, 198] and crystals.[83, 190, 193] The excitation spectrum of Ni:GLS, in figure 5.16, confirms only one absorption band to be present, centred at 690 nm. It may be argued that this is the lowest energy absorption band of $Ni^{2+}$ or $Ni^{3+}$, since the higher energy absorptions of $Ni^{2+}$ or $Ni^{3+}$ are hidden by the band-edge absorption of GLS, but in glasses with similar band-edge absorptions to GLS two absorption bands from $Ni^{2+}$ are observed.[199] It is therefore thought unlikely that the absorption band of Ni:GLS centred at 690 nm is due to $Ni^{2+}$ or $Ni^{3+}$.

The $Ni^+$ ion has a $3d^9$ electronic structure, which is equivalent to a $3d^1$ electronic structure such as $Ti^{3+}$, but the energy terms of its energy levels have negative crystal field components. Therefore, like $Ti^{3+}$, $Ni^+$ should exhibit one absorption band. Reports in the literature of $Ni^+$ absorption are relatively scarce. In various hosts $Ni^+$ displays a single absorption band between 1867 and 2611 nm.[201-203] These are all at lower energy than the absorption band in Ni:GLS, however they are all in tetrahedral symmetry. In octahedral coordination, ions are expected to display a higher energy absorption transition than in tetrahedral symmetry. It is therefore proposed that the 690 nm absorption of Ni:GLS is due to $Ni^+$ in octahedral coordination. The weak absorption at ~1500 nm is attributed to small amounts of $Ni^+$ in tetrahedral coordination. The weak absorption at ~1500 nm is not attributed to ion pairs such as $Ni^+$-$Ni^{2+}$, as in Ti:GLS, because, unlike $Ti^{4+}$, $Ni^{2+}$ is expected to contribute d-d absorption transitions.

TABLE 5.2 Absorption details for nickel in a variety of glass and crystal hosts.

| Host | Ion | Absorption peaks (nm) | Transition | Coordination | Reference |
|---|---|---|---|---|---|
| GLS glass | $Ni^+$ | 690 | - | Octahedral | This work |
| GaP crystal | $Ni^+$ | 1867 | $^2T_2 \rightarrow ^2E$ | Tetrahedral | [201] |
| AgGaSe$_2$ crystal | $Ni^+$ | 2250 | $^2T_2 \rightarrow ^2E$ | Tetrahedral, tetragonal symmetry | [202] |
| CuAlS$_2$ crystal | $Ni^+$ | 2510 | $^2T_2 \rightarrow ^2E$ | Tetrahedral, tetragonal symmetry | [203] |
| CuGaS$_2$ crystal | $Ni^+$ | 2300 | $^2T_2 \rightarrow ^2E$ | Tetrahedral, tetragonal symmetry | [203] |
| AgGaS$_2$ crystal | $Ni^+$ | 2611 | $^2T_2 \rightarrow ^2E$ | Tetrahedral, tetragonal symmetry | [203] |
| LiF crystal | $Ni^{2+}$ | 1220, 712, 402 | $^3A_2(^3F) \rightarrow ^3T_2(^3F)$, $^3A_2(^3F) \rightarrow ^3T_1(^3F)$, $^3A_2(^3F) \rightarrow ^3T_1(^3P)$ | Octahedral | [193] |
| Silica glass | $Ni^{2+}$ | †650, †525, ‡410 | | ‡Octahedral, †Tetrahedral | [81] |
| ZBLAl fluoride glass | $Ni^{2+}$ | 1475, 847, 428 | $^3A_2(^3F) \rightarrow ^3T_2(^3F)$, $^3A_2(^3F) \rightarrow ^3T_1(^3F)$, $^3A_2(^3F) \rightarrow ^3T_1(^3P)$ | Octahedral | [83] |
| MgF$_2$ crystal | $Ni^{2+}$ | 1312, 768, 402 | $^3A_2(^3F) \rightarrow ^3T_2(^3F)$, $^3A_2(^3F) \rightarrow ^3T_1(^3F)$, $^3A_2(^3F) \rightarrow ^3T_1(^3P)$ | Octahedral | [83] |
| ZnNb$_2$O$_6$ crystal | $Ni^{2+}$ | 1385,848,460 | $^3A_2(^3F) \rightarrow ^3T_2(^3F)$, $^3A_2(^3F) \rightarrow ^3T_1(^3F)$, $^3A_2(^3F) \rightarrow ^3T_1(^3P)$ | Octahedral | [197] |
| Zinc-aluminosilicate glass | $Ni^{2+}$ | 1100, 600 | $^3A_2(^3F) \rightarrow ^3T_2(^3F)$, $^3A_2(^3F) \rightarrow ^3T_1(^3F)$ | Octahedral | [199] |
| Lithium gallium silicate glass | $Ni^{2+}$ | 1055, 627, 380 | $^3A_2(^3F) \rightarrow ^3T_2(^3F)$, $^3A_2(^3F) \rightarrow ^3T_1(^3F)$, $^3A_2(^3F) \rightarrow ^3T_1(^3P)$ | Octahedral | [198] |
| ZnSe crystal | $Ni^{2+}$ | 2000,1150,800 | $^3T_1(^3F) \rightarrow ^3T_2(^3F)$, $^3T_1(^3F) \rightarrow ^3A_2(^3F)$, $^3T_1(^3F) \rightarrow ^3T_1(^3P)$ | Tetrahedral | [190] |
| Borate glass | $Ni^{2+}$ | 1380,770,695 | $^3A_2(^3F) \rightarrow ^3T_2(^3F)$, $^3A_2(^3F) \rightarrow ^3T_1(^3F)$, $^3A_2(^3F) \rightarrow ^3T_1(^3P)$ | Octahedral | [189] |

### 5.3.2 Photoluminescence of nickel doped GLS

Figure 5.15 shows the photoluminescence spectrum of 0.02% nickel doped GLS, excited with a CW 5mW, 633 nm laser source which peaks at 910 nm with a FWHM of 330 nm. The broadness of the PL spectrum indicates that the nickel ion is in a low crystal field site. In $Ni^{2+}$ doped zinc aluminosilicate glass emission peaks at 1420 nm[199] and at 2350 nm in $CsCdCl_3$.[195] In $Ni^+$ doped GaP the emission peaks at 1870 nm.[201] Similarly to the absorption spectrum, the higher energy of the emission in Ni:GLS is explained by octahedral coordination.

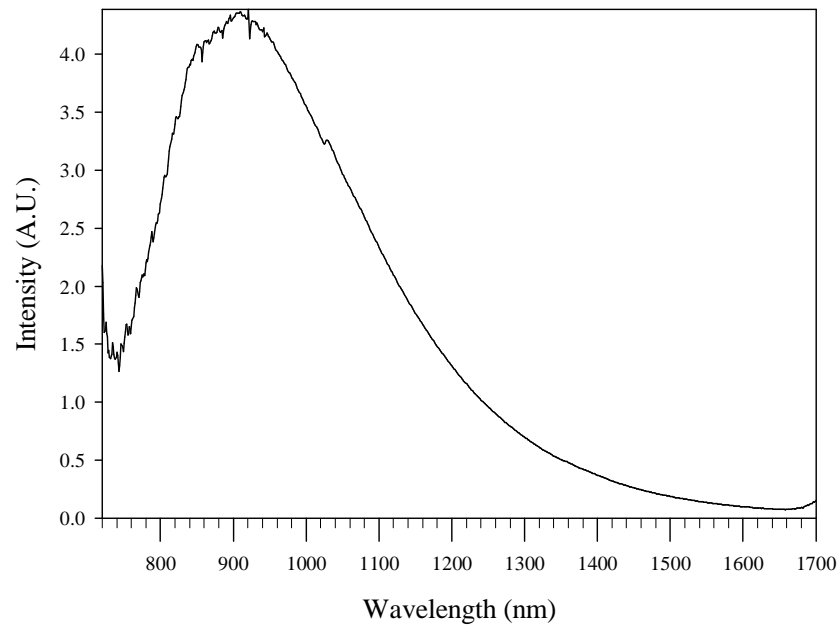

FIGURE 5.15 Photoluminescence spectrum of 0.02% nickel doped GLS excited with a 5mW, 633 nm laser source.

### 5.3.3 Photoluminescence excitation of nickel doped GLS

Figure 5.16 shows the excitation spectra of 0.02% nickel doped GLS indicating a single absorption band centred at 690 nm. The increase in the excitation spectrum at ~450 nm was caused by the system correction.

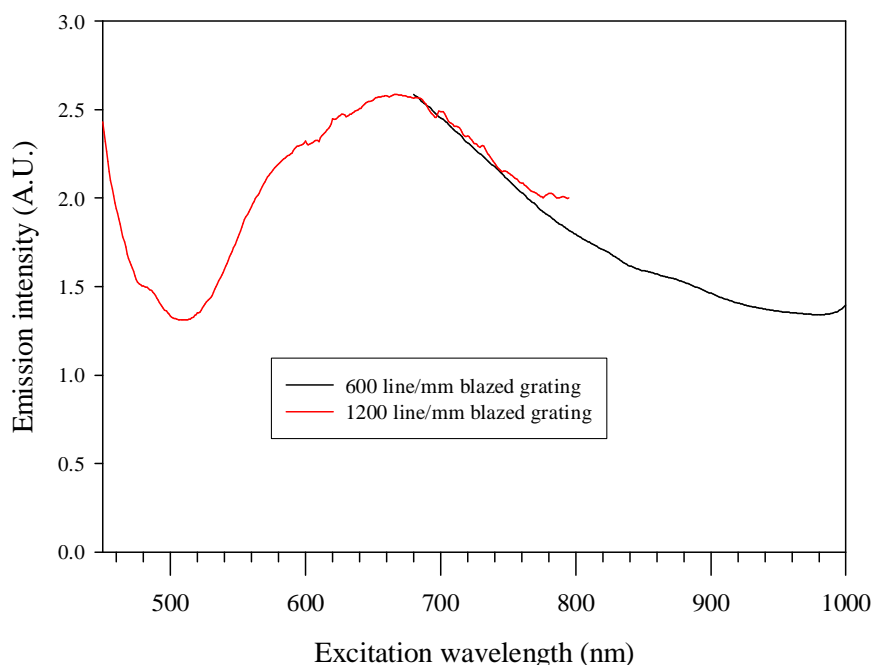

Figure 5.17 PLE spectra detecting emission at 1000-1700 nm of 0.02% nickel doped GLS.

### 5.3.4 Fluorescence lifetime of nickel doped GLS

Figure 5.17 shows the fluorescence decay of 0.01% nickel doped GLS and GLSO, fitted with a stretched exponentials. The best fit to the experimental data was with lifetimes of 28 and 70 µs for GLS and GLSO hosts respectively. Similarly to vanadium and titanium there is an increase in lifetime in the GLSO host except the effect is more pronounced with nickel. These lifetimes compare to 400 µs, 300 µs, 583 µs and 240 µs for $Ni^{2+}$ doped $MgAl_2O_4$,[55] $Mg_2SiO_4$,[200] zinc-aluminosilicate glass,[199] and $Li_2O$-$Ga_2O_3$-$SiO_2$ glass[198] respectively. Whereas the emission lifetime in $Ni^+$ doped ZnS is 25 µs[204] These comparisons indicate a better match with the lifetime of $Ni^+$.

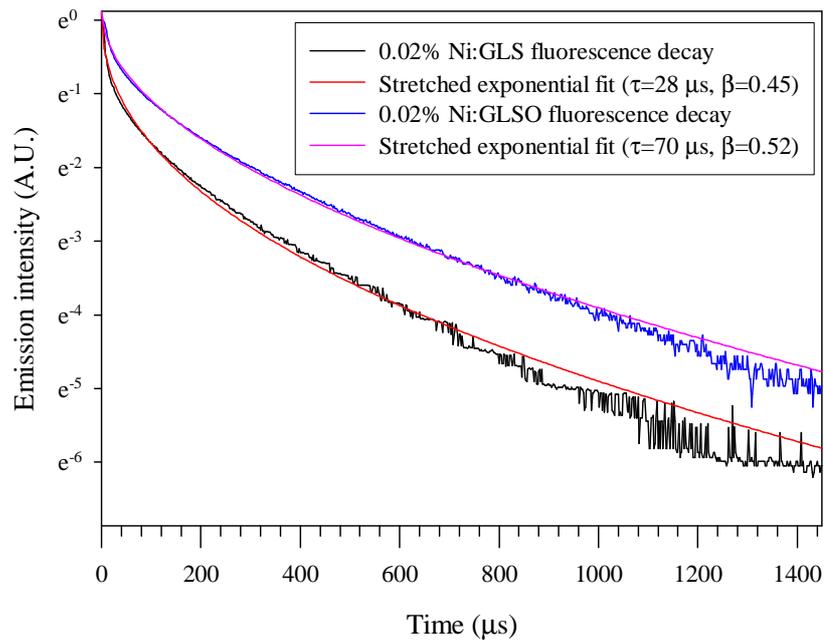

Figure 5.17 Fluorescence decay of 0.02% nickel doped GLS and GLSO exciting with a 10 mW 658 nm laser source fitted with stretched exponentials.

## 5.4 Bismuth doped GLS

Recently Fujimoto working at Osaka University, Japan, discovered a new broadband infrared emission from bismuth doped silica glass and demonstrated 1300 nm optimal amplification with 800 nm excitation.[166, 205] Lasing has also been demonstrated in a bismuth doped aluminosilicate fibre laser.[84] The optical properties of bismuth doped crystals[206-214] and glasses have been investigated previously.[165-174] In crystals assignment of oxidation states was generally made unambiguously. In bismuth doped glasses near infrared emission was often observed, however there is a lot of uncertainty as to the oxidation state of bismuth in glasses and most of the oxidation state assignments were made tentatively. The near infrared emission of bismuth doped glasses has been attributed to $Bi^+$,[167] $Bi^{5+}$,[166, 215] and Bi metal clusters.[172]

### 5.4.1 Absorption of bismuth doped GLS

Inspection of the polished sample (fabrication details in section 3.2) revealed dark patches in the glass indicating that the bismuth was not distributed evenly in the sample or had partly been incorporated as the black suboxide BiO.[173] Doping with lower concentrations may overcome this problem. The absorption spectrum of 1% bismuth doped GLS, in figure 5.18, shows a high baseline absorption compared to the un-doped sample. This is believed to be due to scattering from the dark patches in the glass. A weak shoulder can be observed at ~850 nm in the Bi:GLS absorption spectrum. Further identification of bismuth absorptions cannot be made because the Bi:GLS absorption cannot be directly related to the un-doped GLS absorption. The excitation spectrum of 1% Bi:GLS in figure 5.20 identifies two absorption bands centred at 665 and 850 nm.

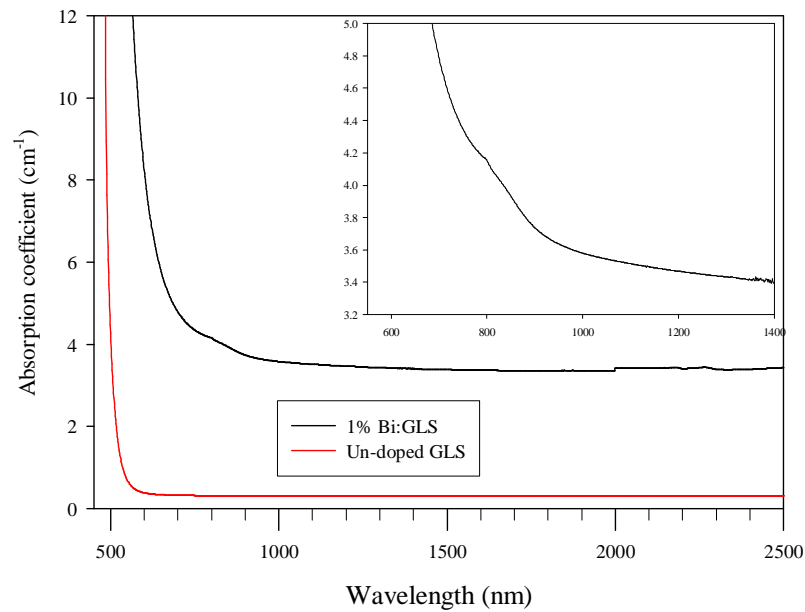

FIGURE 5.18 Absorption spectra of 1% (molar) bismuth doped GLS and un-doped GLS in 5 mm thick slabs.

The absorption peaks of various oxidation states of bismuth in various glasses and crystals are given in table 5.3. $Bi^{3+}$ is expected to exhibit two absorption bands, due to the $^1S_0{\rightarrow}^1P_1$ and $^1S_0{\rightarrow}^3P_1$ transitions.[206, 214] Examining the absorption peaks of $Bi^{3+}$ in table 5.3 indicates that $Bi^{3+}$ has two absorption peaks that do not have a large variation between different hosts, these two absorption peaks are located at approximately 250 and 350 nm. The electron configuration of the $Bi^{2+}$ ion is $6s^26p$ and the ground state is $^2P_{1/2}$.[214] In $Bi^{2+}$ doped crystals[207, 210, 216], shown in table 5.3, the first excited state ($^2P_{3/2}$) is split by the crystal field and the resulting crystal fields terms are denoted (1) and (2) in order of increasing energy. Examining the absorption peaks of $Bi^{2+}$ in table 5.3 indicates that $Bi^{2+}$ has two absorption peaks that do not have a large variation between different hosts, these two absorption peaks are located at approximately 450 and 600 nm. $Bi^{2+}$ can also display a third absorption peak at ~300 nm. Reconciling this with the absorption peaks of Bi:GLS at 665 and 850 nm, given in section 5.4.3, indicates that the absorption of Bi:GLS may not originate from $Bi^{3+}$ or $Bi^{2+}$.

TABLE 5.3 Absorption details for bismuth in a variety of glass and crystal hosts †Tentative assignment.

| Host | Ion | Absorption peaks (nm) | Transition | Reference |
|------|-----|----------------------|------------|-----------|
| Tantalum germanate glass | Bi clusters† | 508, 712, 800, 1000 | - | [172] |
| GLS glass | Bi$^+$† | 665, 850 | $^3P_0 \rightarrow ^1D_2$, $^3P_0 \rightarrow ^3P_2$ | This work |
| Boron barium aluminate glass | Bi$^+$† | 465, 700 | $^3P_0 \rightarrow ^1S_0$, $^3P_0 \rightarrow ^1D_2$ | [167] |
| Phosphor aluminate glass | Bi$^+$† | 460, 700 | $^3P_0 \rightarrow ^1S_0$, $^3P_0 \rightarrow ^1D_2$ | [168] |
| Germanium aluminate glass | Bi$^+$† | 500, 700, 800, 1000 | $^3P_0 \rightarrow ^1S_0$, $^3P_0 \rightarrow ^1D_2$, $^3P_0 \rightarrow ^3P_2$, $^3P_0 \rightarrow ^3P_1$ | [168] |
| SrB$_4$O$_7$ crystal | Bi$^{2+}$ | 312, 470, 575 | $^2P_{1/2} \rightarrow ^2S_{1/2}$, $^2P_{1/2} \rightarrow ^2P_{3/2}(2)$, $^2P_{1/2} \rightarrow ^2P_{3/2}(1)$ | [216] |
| BaSO$_4$ crystal | Bi$^{2+}$ | 455, 588 | $^2P_{1/2} \rightarrow ^2P_{3/2}(2)$, $^2P_{1/2} \rightarrow ^2P_{3/2}(1)$ | [207] |
| SrSO$_4$ crystal | Bi$^{2+}$ | 460, 575 | $^2P_{1/2} \rightarrow ^2P_{3/2}(2)$, $^2P_{1/2} \rightarrow ^2P_{3/2}(1)$ | [207] |
| BaBPO$_5$ crystal | Bi$^{2+}$ | 432, 622 | $^2P_{1/2} \rightarrow ^2P_{3/2}(2)$, $^2P_{1/2} \rightarrow ^2P_{3/2}(1)$ | [210] |
| LiScO$_2$ crystal | Bi$^{3+}$ | 253, 316 | $^1S_0 \rightarrow ^1P_1$, $^1S_0 \rightarrow ^3P_1$ | [211] |
| NaGdO$_2$ crystal | Bi$^{3+}$ | 253, 344 | $^1S_0 \rightarrow ^1P_1$, $^1S_0 \rightarrow ^3P_1$ | [211] |
| YOCl crystal | Bi$^{3+}$ | 270, 332 | $^1S_0 \rightarrow ^1P_1$, $^1S_0 \rightarrow ^3P_1$ | [206] |
| LaOCl crystal | Bi$^{3+}$ | 270, 333 | $^1S_0 \rightarrow ^1P_1$, $^1S_0 \rightarrow ^3P_1$ | [206] |
| YOF crystal | Bi$^{3+}$ | 200, 268 | $^1S_0 \rightarrow ^1P_1$, $^1S_0 \rightarrow ^3P_1$ | [206] |
| Bi$_4$Ge$_3$O$_{12}$ crystal | Bi$^{3+}$ | 250, 290 | $^1S_0 \rightarrow ^1P_1$, $^1S_0 \rightarrow ^3P_1$ | [212] |
| Sodium phosphate glass | Bi$^{3+}$ | 235 | $^1S_0 \rightarrow ^3P_1$ | [169] |
| Silica glass | Bi$^{5+}$† | 500, 700, 800 | - | [166] |
| Germanium sodium aluminate glass | Bi$^{5+}$† | 370, 500, 700, 800 | - | [215] |
| Lithium aluminosilicate glass | unassigned | 480, 700 | - | [174] |

The assignment of Bi$^{5+}$ for bismuth in various glasses has been discounted by Meng[168] and Peng[170, 173] because, according to the optical basicity theory proposed by Duffy,[217] the upper oxidation state of a dopant in glass is usually favourable to a higher basicity. The 5+ oxidation state of bismuth is known to exist in highly basic alkali oxides such as in NaBiO$_3$ or KBiO$_3$,[209, 218] since these are not present in GLS glass the presence of Bi$^{5+}$ is thought to be unfavourable. Therefore Bi$^+$ is proposed as the optically active ion in Bi:GLS.

The electron configuration of $Bi^+$ is $6s^2 6p^2$ which is split by spin-orbit coupling interaction into the ground state $^3P_0$ and excited states $^1S_0$, $^1D_2$, $^3P_2$ and $^3P_1$.[167] Energy matching the absorption peaks for Bi:GLS at 665 and 850 nm to the energy level scheme for $Bi^+$ proposed by Meng[167, 168] indicates that they are closest to the 700 and 800 nm absorptions of bismuth doped germanium aluminate glass and can be attributed to the $^3P_0 \rightarrow ^1D_2$ and $^3P_0 \rightarrow ^3P_2$ transitions of $Bi^+$. The $^3P_0 \rightarrow ^1S_0$ absorption transition may be obscured by the band-edge absorption of GLS and the $^3P_0 \rightarrow ^3P_2$ absorption transition may be very weak, as in bismuth doped germanium aluminate glass,[168] or not present as in bismuth doped phosphor aluminate glass.[168]

### 5.4.2 Photoluminescence of bismuth doped GLS

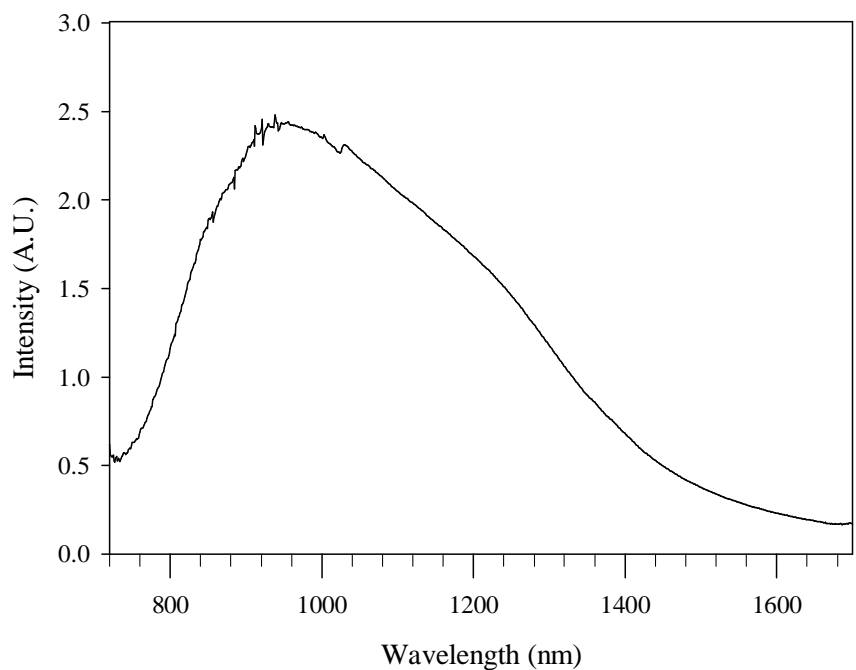

FIGURE 5.19 Photoluminescence spectrum of 1% bismuth doped GLS excited with a 5mW 633 nm laser source.

The photoluminescence spectrum of 1% bismuth doped GLS in figure 5.19 shows emission peaking at 950 nm with a shoulder at ~1240 nm and a FWHM of 540 nm (4650 $cm^{-1}$). The emission peaks of various oxidation states of bismuth in various glasses and crystals are given in table 5.4. Similarly to the absorption peaks, the emission peaks of $Bi^{2+}$ and $Bi^{3+}$ all occur at higher energy than Bi:GLS reinforcing the hypothesis that $Bi^{2+}$ and $Bi^{3+}$ do not contribute to observed optical transitions of Bi:GLS. The shoulder on the emission spectrum of Bi:GLS indicated that emission might be due to transitions from two energy levels, as in $Bi^{3+}$ doped $LiScO_2$.[211] Comparing the emission spectrum of Bi:GLS to that of $Bi^+$ in other glasses[167, 168] indicated that the shoulder at 1240 nm is a good energy match to the $^3P_1 \rightarrow ^3P_0$ transition of $Bi^+$, the main peak at 950 nm is therefore attributed to the $^3P_2 \rightarrow ^3P_0$ transition.

TABLE 5.4 Emission details for bismuth in a variety of glass and crystal hosts †Tentative assignment.

| Host | Ion | Emission peaks (nm) | Transition | Reference |
|---|---|---|---|---|
| Tantalum germanate glass | Bi clusters† | 1310 | - | [172] |
| Boron barium aluminate glass | Bi$^+$† | 1148 | $^3P_1 \rightarrow ^3P_0$ | [167] |
| Phosphor aluminate glass | Bi$^+$† | 1300 | $^3P_1 \rightarrow ^3P_0$ | [168] |
| Germanium aluminate glass | Bi$^+$† | 1300 | $^3P_1 \rightarrow ^3P_0$ | [168] |
| GLS glass | Bi$^+$† | 950, 1240 | $^3P_2 \rightarrow ^3P_0$, $^3P_1 \rightarrow ^3P_0$ | This work |
| SrB$_4$O$_7$ crystal | Bi$^{2+}$ | 585 | $^2P_{3/2}(1) \rightarrow ^2P_{1/2}$ | [216] |
| BaSO$_4$ crystal | Bi$^{2+}$ | 625 | $^2P_{3/2}(1) \rightarrow ^2P_{1/2}$ | [207] |
| BaSO$_4$ crystal | Bi$^{2+}$ | 640 | $^2P_{3/2}(1) \rightarrow ^2P_{1/2}$ | [207] |
| LiScO$_2$ crystal | Bi$^{3+}$ | 404 | $^3P_1 \rightarrow ^1S_0$, $^3P_0 \rightarrow ^1S_0$ | [211] |
| NaGdO$_2$ crystal | Bi$^{3+}$ | 384 | $^3P_1 \rightarrow ^1S_0$, $^3P_0 \rightarrow ^1S_0$ | [211] |
| YOCl crystal | Bi$^{3+}$ | 400 | $^3P_1 \rightarrow ^1S_0$ | [206] |
| LaOCl crystal | Bi$^{3+}$ | 345 | $^3P_1 \rightarrow ^1S_0$ | [206] |
| YOF crystal | Bi$^{3+}$ | 330 | $^3P_1 \rightarrow ^1S_0$ | [206] |
| Bi$_4$Ge$_3$O$_{12}$ crystal | Bi$^{3+}$ | 475 | $^3P_1 \rightarrow ^1S_0$ | [212] |
| Sodium phosphate glass | Bi$^{3+}$ | 400 | $^3P_1 \rightarrow ^1S_0$ | [169] |
| Silica glass | Bi$^{5+}$† | 750-1250 | - | [166] |
| Germanium sodium aluminate glass | Bi$^{5+}$† | 1220 | - | [215] |
| Lithium aluminosilicate glass | unassigned | 1100, 1350 | - | [174] |

In Bi$^{3+}$ doped LiScO$_2$ and NaGdO$_2$, excitation took place into the $^3P_1$ level but emission was observed from both the $^3P_1 \rightarrow 1S_0$ transition and the spin forbidden $^3P_0 \rightarrow ^1S_0$ transition. In bismuth doped silica glass infrared emission was only observed with aluminium codopant,[166] it also varied with excitation wavelength with two emission peaks at 750 and 1140 nm under 500 nm excitation, one at 1122 nm under 700 nm excitation and one at 1250 nm under 800 nm excitation. In bismuth doped lithium aluminosilicate glass, emission consisted of two Gaussian peaks at 1100 and 1350 nm, these peaks had lifetimes of 549 and 270 μs respectively. No infrared emission from Bi$^{3+}$ and no fluorescence lifetime from Bi$^{3+}$ longer than 5 μs has been reported.[173]

### 5.4.3 Photoluminescence excitation of bismuth doped GLS

Photoluminescence excitation spectra were taken using the setup described in section 3.3.3. The excitation spectrum of 1% Bi:GLS, in figure 5.20, shows a peak at 850 nm which can be related to the weak shoulder of the absorption spectrum in figure 5.18 and a peak at 665 nm which could not be resolved in the absorption measurement.

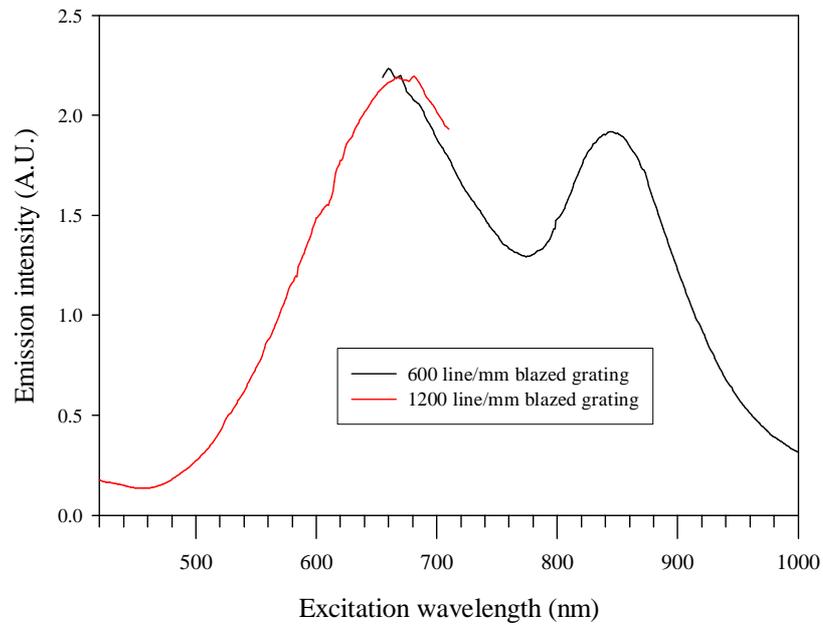

FIGURE 5.20 PLE spectra detecting emission at 1000-1700 nm of 1% bismuth doped GLS.

### 5.4.4 Fluorescence lifetime of bismuth doped GLS

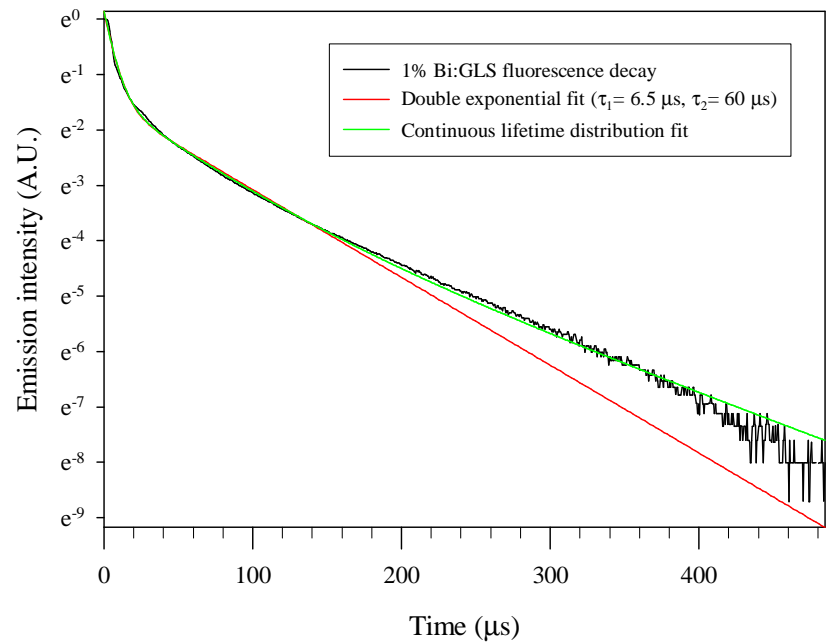

FIGURE 5.21 Fluorescence decay of 1% bismuth doped GLS exciting with a CW 10 mW, 658 nm laser source and fitted with a double exponential and the continuous lifetime distribution model.

Figure 5.21 shows the fluorescence decay of 1% bismuth doped GLS. Likewise to similar concentrations of vanadium and titanium doped GLS the decay did not follow stretched exponential behaviour and was more accurately described by a double exponential. The lifetimes of the double exponential fit were 6.5 and 60 μs. The decay was also fitted using the continuous lifetime distribution model (see section 4.10). The results are given in figure 5.22 and show two lifetime distributions centred at 7 and 47 μs, which is in reasonable agreement with the results of the double exponential fit. Two lifetime components have been reported for bismuth doped glasses[165, 171] which were attributed to the bismuth ion occupying different sites in the glasses. This may be the case for Bi:GLS since, as described in section 4.7, it is thought that transition metals occupy oxide and sulphide sites which give rise to long and short lifetime respectively. However, the shoulder in the emission spectrum indicates that two energy levels may be involved in the emission so the two lifetime components may be the lifetimes of the two energy levels as in $Bi^{3+}$ doped $LiScO_2$. [211]

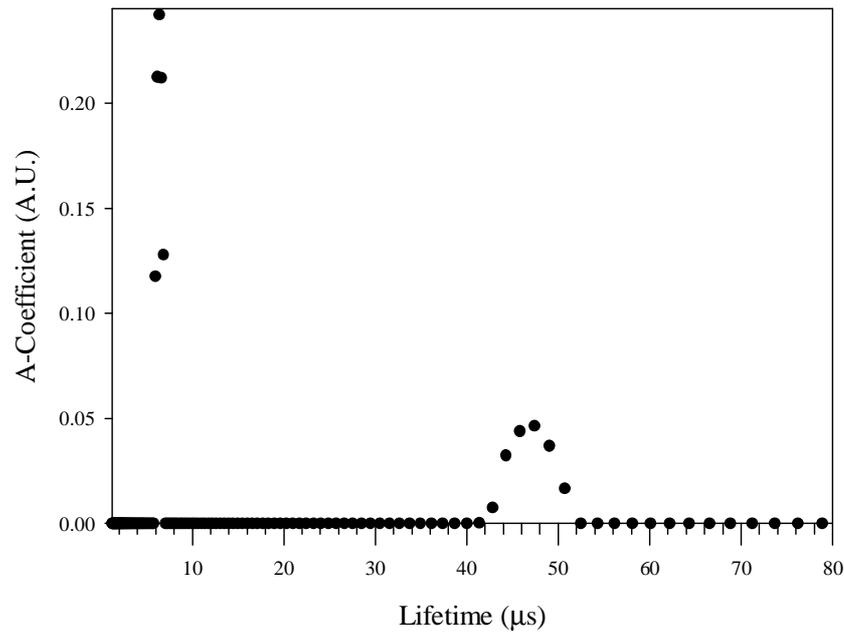

FIGURE 5.22 Lifetime distribution in the emission decay of 1% bismuth doped GLS.

The emission lifetimes of various oxidation states of bismuth in various glasses and crystals are given in table 5.5. The lifetime for Bi:GLS is longer than the reported lifetimes for $Bi^{3+}$, again reinforcing the hypothesis that $Bi^{3+}$ does not contribute to observed optical transitions of Bi:GLS. The lifetime for Bi:GLS is longer than the reported lifetimes for $Bi^{+}$ and bismuth doped glasses in which optical gain and lasing were demonstrated, however lowering the concentration and using GLSO host may significantly increase the lifetime.

The emission lifetime and emission cross section are important parameters for the characterisation of a laser material because the laser threshold is inversely proportional to the product of the emission lifetime and emission cross section.[130] However, calculation of the emission cross section requires determination of the quantum efficiency. This was not possible for Bi:GLS because of the loss of ORC facilities, see section 1.5, therefore quantum efficiency measurements of Bi:GLS is suggested as further work.

TABLE 5.5 Emission lifetime details for bismuth in a variety of glass and crystal hosts †Tentative assignment ‡Lasing demonstrated.

| Host | Ion | Lifetime (µs) | Transition | Reference |
|---|---|---|---|---|
| GLS glass | $Bi^+$† | 7, 47 | $^3P_2 \rightarrow ^3P_0$, $^3P_1 \rightarrow ^3P_0$ | This work |
| Tantalum germanate glass | Bi clusters† | 200 | - | [172] |
| Boron barium aluminate glass | $Bi^+$† | 350 | $^3P_1 \rightarrow ^3P_0$ | [167] |
| Aluminosilicate glass | unassigned‡ | 1000 | - | [84] |
| Phosphor aluminate glass | $Bi^+$† | 500 | $^3P_1 \rightarrow ^3P_0$ | [168] |
| $LiScO_2$ crystal | $Bi^{3+}$ | 0.045, 380 | $^3P_1 \rightarrow ^1S_0$, $^3P_0 \rightarrow ^1S_0$ | [211] |
| $NaGdO_2$ crystal | $Bi^{3+}$ | 0.1, 7 | $^3P_1 \rightarrow ^1S_0$, $^3P_0 \rightarrow ^1S_0$ | [211] |
| YOCl crystal | $Bi^{3+}$ | 1.4 | $^3P_1 \rightarrow ^1S_0$ | [206] |
| LaOCl crystal | $Bi^{3+}$ | 1.6 | $^3P_1 \rightarrow ^1S_0$ | [206] |
| $Bi_4Ge_3O_{12}$ crystal | $Bi^{3+}$ | 0.4 | $^3P_1 \rightarrow ^1S_0$ | [212] |
| Sodium phosphate glass | $Bi^{3+}$ | 3.9 | $^3P_1 \rightarrow ^1S_0$ | [169] |
| Germanium sodium aluminate glass | $Bi^{5+}$† | 434 | - | [215] |
| Silica glass | $Bi^{5+}$† | 630 | - | [166] |

## 5.5 Conclusions

Absorption measurements of Ti:GLS identified an absorption band at ~500-600 nm that could not be fully resolved because of its proximity to the band-edge of GLS. At concentrations of 0.5% and greater a shoulder at ~1000 nm is observed, there is also a weak and broad absorption centred at around 1800 nm. The second derivative absorption spectra identified an absorption peak at 980 nm in Ti:GLS but not in Ti:GLSO, absorption peaks at 615 and 585 nm are also identified for Ti:GLS and Ti:GLSO respectively. The excitation spectra of 0.1% titanium doped GLS and GLSO both show a single excitation peak at 580 nm The emission spectra of Ti:GLS and Ti:GLSO both peaked at 900 nm. It is proposed that the absorption at ~600 nm in Ti:GLS and Ti:GLSO is due to the $^2T_{2g} \rightarrow ^2E_g$ transition of octahedral $Ti^{3+}$ and the absorption at 980 nm in Ti:GLS is due to $Ti^{3+}$-$Ti^{4+}$ pairs. The 97 µs emission lifetime of Ti:GLSO compares very favourably to the lifetime of Ti:Sapphire of 3.1 µs. The optimum doping concentration for an active device based on Ti:GLSO may be lower than the lowest concentration of 0.05 % molar investigated in this chapter. Therefore the investigation of lower doping concentrations of Ti:GLSO is suggested as further work. The fabrication of a Ti:GLSO fibre at the optimum doping concentration for applications as a tuneable laser source is also suggested as further work.

The absorption of Ni:GLS is characterised by a red-shift of ~ 300 nm in the band-edge indicating a nickel absorption in the region 500-800 nm. There is also a very weak absorption peaking at ~1500 nm. The excitation spectra of Ni:GLS indicate a single absorption band centred at 690 nm. It is proposed that the 690 nm absorption of Ni:GLS is due to $Ni^+$ in octahedral coordination. The weak absorption at ~1500 nm is attributed to small amounts of $Ni^+$ in tetrahedral coordination. The photoluminescence spectrum peaks at 910 nm with a FWHM of 330 nm. The lifetimes of Ni:GLS and Ni:GLSO are 28 and 70 μs respectively. Fabrication of a range of concentrations of Ni:GLS and Ni:GLSO to further investigate its spectroscopic properties is suggested as further work.

A weak shoulder can be observed at ~850 nm in the Bi:GLS absorption spectrum. Further identification of bismuth absorptions cannot be made because dark patches in the sample were detrimental to its absorption. The excitation spectrum of Bi:GLS shows peaks at 665 and 850 nm Based on comparisons to other work the absorption peaks for Bi:GLS at 665 and 850 nm are attributed to the $^3P_0 \rightarrow ^1D_2$ and $^3P_0 \rightarrow ^3P_2$ transitions of $Bi^+$. The emission decay of Bi:GLS consisted of two lifetime distributions centred at 7 and 47 μs. The demonstration of lasing in bismuth doped aluminosilicate glass makes development of a Bi:GLS laser more favourable. Fabrication of a range of concentrations of Bi:GLS and Bi:GLSO to further investigate its spectroscopic properties is suggested as further work.

# Chapter 6

# Femtosecond laser written waveguides in chalcogenide glass

## 6.1 Introduction

In this chapter the fabrication and characterisation of buried waveguides written into GLS glass using 800 nm focused fs laser pulses is reported. The spectral broadening of 1550 nm fs laser pulses coupled into these waveguides is also reported.

### 6.1.1 Femtosecond laser material modification

Femtosecond (fs) lasers have several advantages over conventional laser systems for the micro-structuring of transparent dielectrics. These include the reduction of collateral damage[219] and sub-diffraction limited ablation.[220-222] In gold film 800 nm fs lasers have been used to ablate holes roughly 10% of the focus spot size.[223] This effect was attributed to the minimised thermal diffusion time of ultra-short pulses that have a peak laser fluence slightly above a well defined ablation threshold.[223] Femtosecond laser modification of transparent solids has many potential applications. The possibility of writing active devices has been demonstrated in Er-Yb doped silicate glass.[224] Plasma-induced bulk modification by the tightly focused fs lasers has been demonstrated as a tool for three-dimensional optical memory with data storage capacities of up to $10^{16}$ bits/m$^3$.[225-228] Tightly focused fs lasers have also been used for micro-structuring in transparent dielectrics.[229, 230] Infra-red femtosecond laser pulses have be used to permanently photo-reduce $Eu^{3+}$ to $Eu^{2+}$ in fluorozirconate glass,[231] photo-oxidation of $Mn^{2+}$ to $Mn^{3+}$ has also been observed in silicate glass[230]

### 6.1.2 Highly nonlinear glass

Highly nonlinear glass is an excellent candidate material for optical, ultrafast, nonlinear devices such as demultiplexers,[232] wavelength converters[233] and optical Kerr shutters.[234] This is because of its ability to cause nonlinear phase shifts over much shorter interaction lengths than conventional (silica based) devices. Various waveguiding structures such as fibres, proton beam written waveguides, continuous wave (CW) laser written waveguides and fs laser written waveguides could be used to realise such devices. Of these, fs laser writing is particularly attractive because, as well as having rapid processing times, waveguiding structures can be formed below the surface of the glass enabling 3-D structures to be fabricated. Optical components such as a Fresnel zone plate[235] and a fibre attenuator[230] have been fabricated using fs laser pulses. Several studies have described the fabrication and characterisation of waveguides using focused fs laser pulses in phosphate glass[236] chalcogenide glass[237] and heavy metal oxide glass.[238] Of these chalcogenide glasses are especially attractive because they have a high nonlinear refractive index and enhanced IR transmission, coupled with low maximum phonon energy. The nonlinear refractive index of chalcogenide glasses is in general higher than oxide glasses with the same linear refractive index. This is believed to be a consequence of the large polarizability of the chalcogen ions.[239] Of the chalcogenide glasses, gallium lanthanum sulphide (GLS) is probably the most notable, with respect to optical nonlinear devices, as it has a nonlinear figure of merit (FOM) of >7, which is believed to be the highest for any bulk glass reported to date,[240]

$$FOM = \frac{n_2}{2\beta_{TPA}\lambda} \qquad (6.1)$$

where $\lambda$ is the wavelength, $n_2$ is the real part of the nonlinear refractive index and $\beta_{TPA}$ is the two-photon absorption (TPA) coefficient.

### 6.1.3 Nonlinear optical devices

A potential use for waveguides written, in highly nonlinear glass such as GLS, is the realisation of nonlinear optical devices. A long term goal of the work presented in this chapter is the possibility of fabricating an optical chip using fs laser writing that could incorporate optical amplification and ultra fast switching integrated on a single device. What follows is a brief description of various optical devices that exploit the nonlinear properties of waveguiding structures.

### 6.1.3.1 Mach-Zehnder interferometer switch

The basic principle of a Mach-Zehnder interferometer switch is that guided light is split into two branches of a waveguiding structure and then recombined at an output in such a way that interference can occur. By inducing a $\pi$, or odd multiple of $\pi$, phase shift in one of the branches the recombined beams can interfere destructively. Devices have been demonstrated in which a $\pi$ phase shift is induced on a 1520 nm, 1 fJ signal pulse with a width of 150 fs, through the saturation of a semiconductor optical amplifier (SOA) by a 250 fJ control pulse with the same wavelength and pulse width as the signal pulse.[241] In other devices a $\pi$ phase shift was induced by applying a voltage to InGaAlAs/InAlAs waveguide structure.[242]

### 6.1.3.2 Optical Kerr shutters

Optical Kerr shutters exploit the nonlinear phase induced by the intensity dependent nonlinear birefringence to change the state of polarization (SOP) of a weak signal in a nonlinear medium.[243] These devices are often realised in a fibre geometry, in this case a linearly polarised signal is launched into a polarisation maintaining fibre polarized at 45° to both the two principal axes. The SOP of the signal changes periodically due to birefringence built into the fibre. The original SOP is restored by a quarter-wave plate at the output of the fibre and is then blocked by a crossed polarizer, this is the closed state of the optical Kerr shutter. To open the shutter, linearly polarized strong pump light is launched along one of the two principle axes of the fibre together with the signal. In this case the refractive indices for the parallel and perpendicular components of the signal become slightly different, with respect to the direction of the pump polarization, due to the pump induced birefringence. This nonlinear birefringence causes a nonlinear phase shift and also changes the SOP of the signal, thus the signal is transmitted through the polarizer.[243] A switching power of 3W has been demonstrated for an optical Kerr shutter based on 1.2 m of $As_2S_3$ fibre.[244] Optical Kerr shutters have been used for various applications such as intensity discrimination[245] and optical sampling.[234]

### 6.1.3.3 2R regenerator

An all-optical signal regeneration technique utilizing self-phase modulation (SPM) in fibre and subsequent filtering of the signal was first proposed by Mamyshev.[246] This is a 2R regenerator (where 2R stands for re-amplification and re-shaping) simply comprising of a nonlinear waveguide and an optical bandpass filter. This method is based on the effect of SPM of the data signal in a nonlinear medium with subsequent optical filtering at a frequency $\omega_f$, which is shifted with respect to the input data carrier frequency $\omega_0$.[246] Due to the effects of SPM the input pulse broadens to $\Delta\omega_{SPM}$, given by equation 6.2 (also given in section 6.4.4.2) [246]

$$\Delta\omega_{SPM} = \frac{\Delta\omega_0 \, 2\pi n_2 I_p L}{\lambda} \qquad (6.2)$$

Where $\Delta\omega_0$ is the bandwidth of the input pulse, $n_2$ is the nonlinear refractive index, $I_p$ is the pulse intensity, L is the waveguide length and $\lambda$ is the wavelength. After the nonlinear waveguide, the pulse passes through an optical filter with centre frequency $\omega_f$, shifted from the signal frequency $\omega_0$ given by: $\omega_f = \omega_0 + \Delta\omega_{shift}$. If $\Delta\omega_{SPM}/2 < \Delta\omega_{shift}$ the input pulse is rejected by the filter, this happens when the pulse intensity Ip is too small (noise in "zeros"). If the pulse intensity is high enough so that $\Delta\omega_{SPM}/2 \geq \Delta\omega_{shift}$, a part of the SPM broadened pulse passes through the filter.[246] Since the spectral density of the broadened spectrum at the filter pass band can be made to be relatively insensitive to the peak power of the input pulses, any amplitude fluctuations in "one" bits are reduced by the process.[247]

## 6.2 Waveguide fabrication and characterisation techniques

### 6.2.1 Waveguide fabrication

Waveguides were written in two different samples; these were a GLS and a GLSO sample, both having dimensions of ~ 12x12x5 mm. The GLS sample was prepared by mixing 65% gallium sulphide, 30% lanthanum sulphide and 5% lanthanum oxide (% molar) and the GLSO sample was prepared by mixing 77.5% gallium sulphide and 22.5% lanthanum oxide (% molar). The batching and melting details for these glasses are given in section 3.2.1. After the waveguides were written their enfaces were polished to a scratch-dig surface quality of 40-20 and a parallelism of < 0.03°. The first digits of the scratch-dig specification relate to the maximum width allowance of a scratch in μm, the next digits indicate the maximum diameter allowance for a dig in 1/100 mm.

A schematic of the waveguide writing process is shown in figure 6.1. Femto second laser radiation was generated using a Coherent system comprising of a Mira mode locked Ti:Sapphire oscillator which was pumped by a Verdi V10 frequency doubled ,10W, 532 nm diode pumped laser. The output of the Mira seeded a RegA 9000 regenerative amplifier which was tuned to 800 nm and emitted a train of pulses with a duration of 150 fs, a repetition rate of  250 KHz  and a pulse energy up to 6μJ. A mirror diverted the output of the RegA to the direct write setup where a computer controlled shutter regulated the irradiation time. Pulse energy was controlled using a variable neutral density filter and a half wave plate allowed rotation of the linear laser polarisation. The lenses L1 and L2 were used to reduce the beam radius so that it had a high coupling efficiency into the objective. The laser beam was focused via a 50x objective (NA=0.55) at 100-400 μm below the surface of the sample. The focus spot diameter was measured to be 1.5 μm in air and calculated to be around 2 μm inside the glass sample. The sample was mounted on a Aerotech computer controlled, linear motor, translation stage which could move in 3 axes with a resolution of a 20 nm in all three axes. A series of channels at various pulse energies and translation velocities was written in the sample by translating it perpendicularly to the propagation direction of the laser beam. A red glow attributed to plasma fluorescence, see section 6.3, was visible from the focal point as it passed through the sample. After processing, the end faces of the sample were polished for subsequent characterisation.

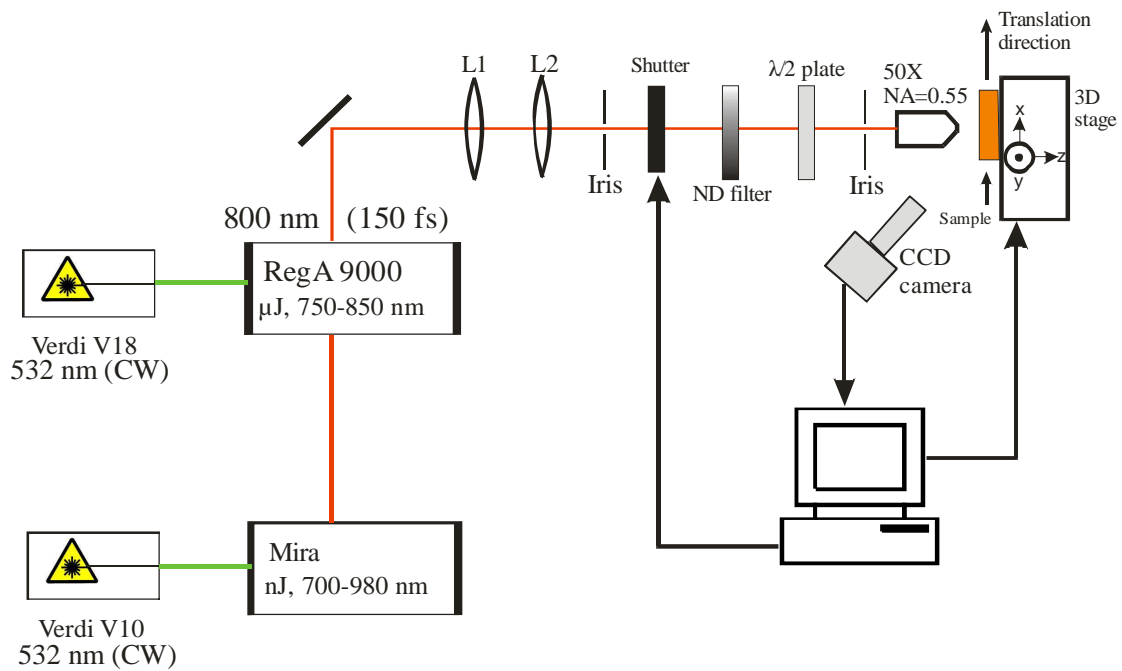

FIGURE 6.1 Schematic of waveguide writing process.

## 6.2.2 Guided mode profile and micrographs

To obtain guided mode profiles a vertically polarised, 633 nm, He-Ne laser was coupled into and out of the waveguides with 10x 0.25 NA objectives, the low magnification was needed because of the large mode size (~ 300 μm diameter) of some of the waveguides. The polarisation direction was changed with a half wave plate. The near field image was then captured by a charged coupled device (CCD) camera.

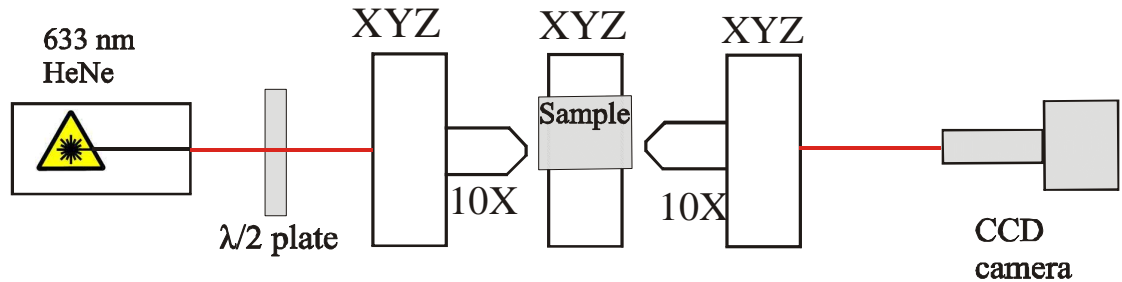

FIGURE 6.2 Guided mode profile setup.

Optical micrographs were taken on a Nikon Eclipse LV100 optical microscope in transmission and reflection mode.

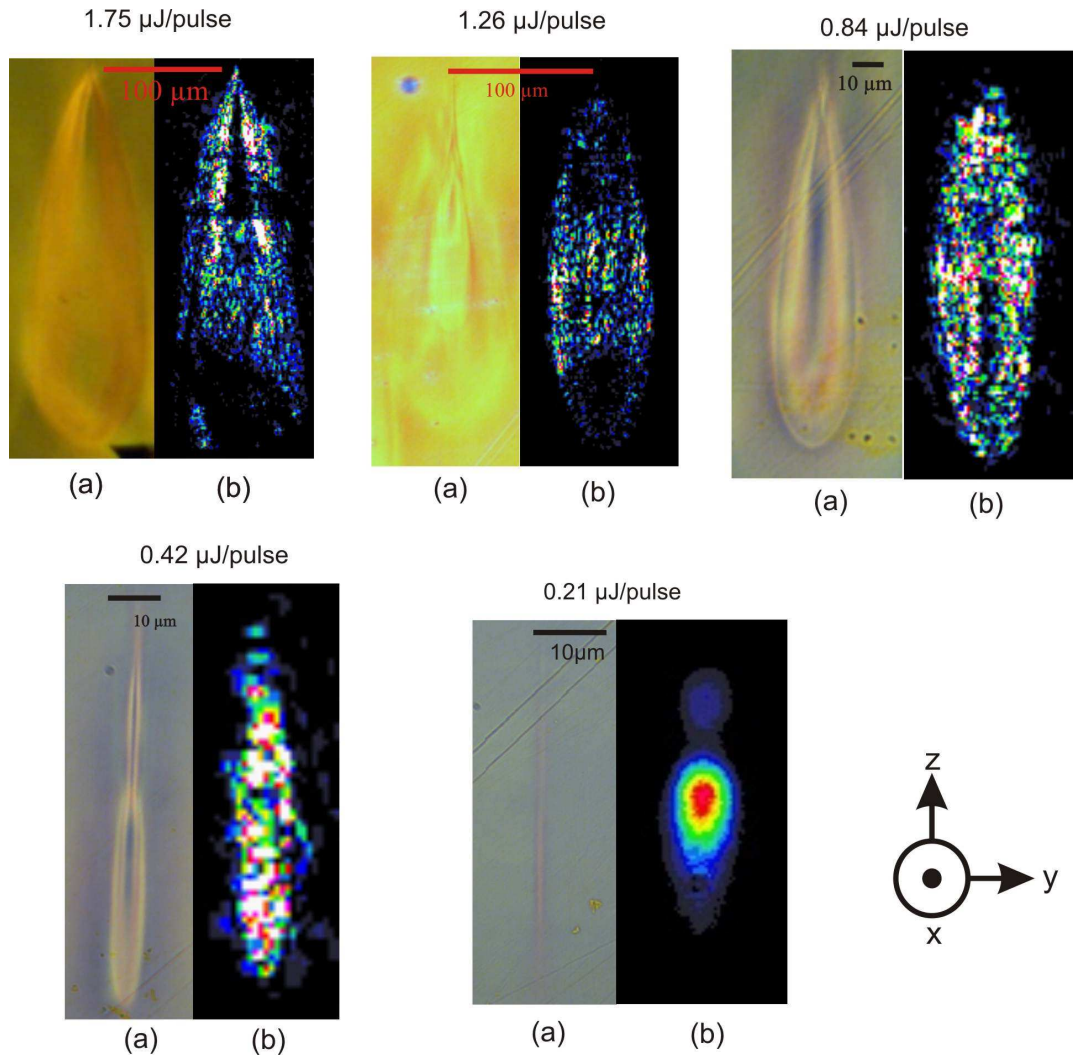

FIGURE 6.3 Transmission optical micrographs (a) and near-field guided mode at 633 nm (b) of waveguides written at a focus depth of ~400 μm into the GLS sample at various pulse energies and a translation speed of 200 μm/s. The arrow shows the propagation direction of the laser used to write the waveguides.

Figure 6.3 shows the transmission optical micrograph and near field mode profile of waveguides written in the GLS sample with pulse energies of 0.21-1.75 μJ and a scan speed of 200 μm/s, with the focus spot 400 μm below the surface of the sample. The figure shows that the form and cross-sectional size of the waveguides has a strong dependence on the writing pulse energy, which indicates that the waveguides are formed by a nonlinear process. The 800 nm wavelength of the laser used to write the waveguides is in the transmission window of GLS and there is no noticeable material modification occurring until the beam reaches its focus ~400 μm inside the glass. This also indicates that the waveguides are formed by a nonlinear process. It is noted that there was no apparent ablation or cracking of the glass even at the highest pulse energy used of 1.75 μJ, which is ~14 times the lowest pulse energy (0.12 μJ) at which material modification was observed (no guided mode could be found for this waveguide). For continuous wave (CW) ultra violet (UV) written waveguides in GLS damage occurred at ~6 times the lowest fluence used,[34] this suggests that a different formation mechanism is involved for CW UV written waveguides in GLS than for femtosecond writing. UV written waveguides in GLS are extremely delicate as they are written into the surface of the glass and hence the end faces are prone to cracking, the top surface is also easily damaged. Buried femtosecond laser written waveguides are inherently more robust than UV written waveguides as they are protected beneath the surface of the glass. This robustness was illustrated when the sample was accidentally dropped several times with no detrimental effect on the waveguides.

Figure 6.3 appears to show two distinct forms of waveguide. At pulse energies of 1.75, 1.26 and 0.84 μJ the waveguides have a distinctive "teardrop" shape with a dark central region and a brighter structure surrounding it; this type of waveguide is now referred to as A-type. The near field guided mode of the A-type waveguides appears to show that the dark central region does not guide light as well as the brighter surrounding structure. Rotating the polarisation of the guided 633 nm beam from vertical (E field along major axis of waveguide) to horizontal increased the transmitted power by ~10%. At a pulse energy of 0.21 μJ the waveguide structure is very different to the A-type waveguides and is characterised by a long and narrow filament like structure with no resolvable dark central region; this type of waveguide is now referred to as B-type. The guided mode of the B-type waveguide can be fitted to a Gaussian profile with a correlation coefficient 0.986 in the horizontal and 0.923 in the vertical, indicating the waveguide is single mode at 633 nm. However this waveguide could support higher order vertical modes by adjusting the input parameters. The waveguide written at a pulse energy of 0.42 μJ appears to show a transitional form with characteristics of both A-type and B-type waveguides.

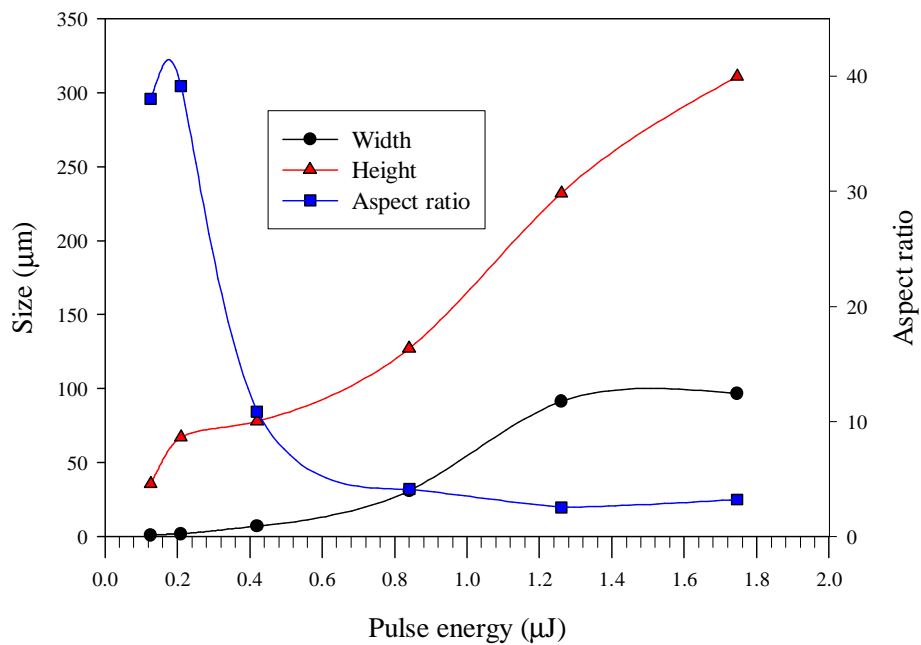

FIGURE 6.4 Height, width and aspect ratio of waveguides as a function of writing pulse energy for waveguides written at a depth of 400 μm in GLS. The lines are a guide for the eye.

The height (z dimension), width (y dimension) and aspect ratio (height/width) measured for these waveguides are shown in figure 6.4. Comparing the aspect ratio of the waveguides to their type indicates that A-type waveguides are characterised by an aspect ratio of 3-5 and B-type waveguides are characterised by an aspect ratio of ~40. The width of the waveguides reaches a maximum of ~100 μm at a pulse energy of ~1.2 μJ whereas the height of the waveguides continues to increase up to the maximum pulse energy used. The top of the waveguides were all at the same depth in the glass but as the pulse energy was increased the waveguides moved towards the source of the writing laser beam.

Figure 6.5 shows the transmission optical micrograph, near-field guided mode and reflection optical micrograph of waveguides written in the GLS sample at pulse energies between 0.28 and 0.4 μJ, translation speeds of 200, 100 and 50 μm/s at a focus depth of 100 μm below the surface of the sample. Similarly to figure 6.3, the transmission optical micrograph shows a "teardrop" shape with a dark central region (now referred to as region 1) and a brighter structure surrounding it (now referred to as region 2), however this differentiation of structure is clearer than for the waveguides in figure 6.3. It is also more apparent that region 1 does not actively guide light and it is only region 2 that guides light. This effect is probably related to the depth that the waveguides were written at and may arise from the lower resistance to expansion caused by exposure to the fs laser pulses at the lesser depth. It could also be caused by a greater aberration in the writing beam for the waveguides written at a depth of 400 μm than at 100 μm, caused by imperfections and inhomogeneities in the glass. The reflection optical micrograph shows that region 2 has a high reflectivity indicating that it has undergone a positive refractive index change; region 1 has a lower reflectivity indicating that it has undergone a lower refractive index change.

The waveguides in figure 6.5 are all A-type and comparing the size of the waveguides in figure 6.6 to those in figure 6.4 shows that for similar pulse energies the waveguides written at a depth of 100 μm are wider than those written at a depth of 400 μm, although they are not as high because the waveguides written at a depth of 400 μm are B-type or a transitional form. This indicates that there is some attenuation of the writing laser beam as it is transmitted through the glass before it reaches a focus and undergoes non-linear absorption.

Comparing the size and form of the waveguides in figure 6.5, written at translation speeds 200, 100 and 50 μm/s shows that the waveguides written at these speeds are virtually identical for the same pulse energy with a slight increase in size as the speed decreases. For the various pulse energies used there is a 10-20% increase in width and a 5-10% increase in height going from a translation speed of 200 μm/s to 50 μm/s. This small increase in size, with a quadrupling of the writing beam fluence on the sample, indicates that the absorption and thermalisation process caused by each pulse is almost complete when the next pulse arrives 4 μs later.

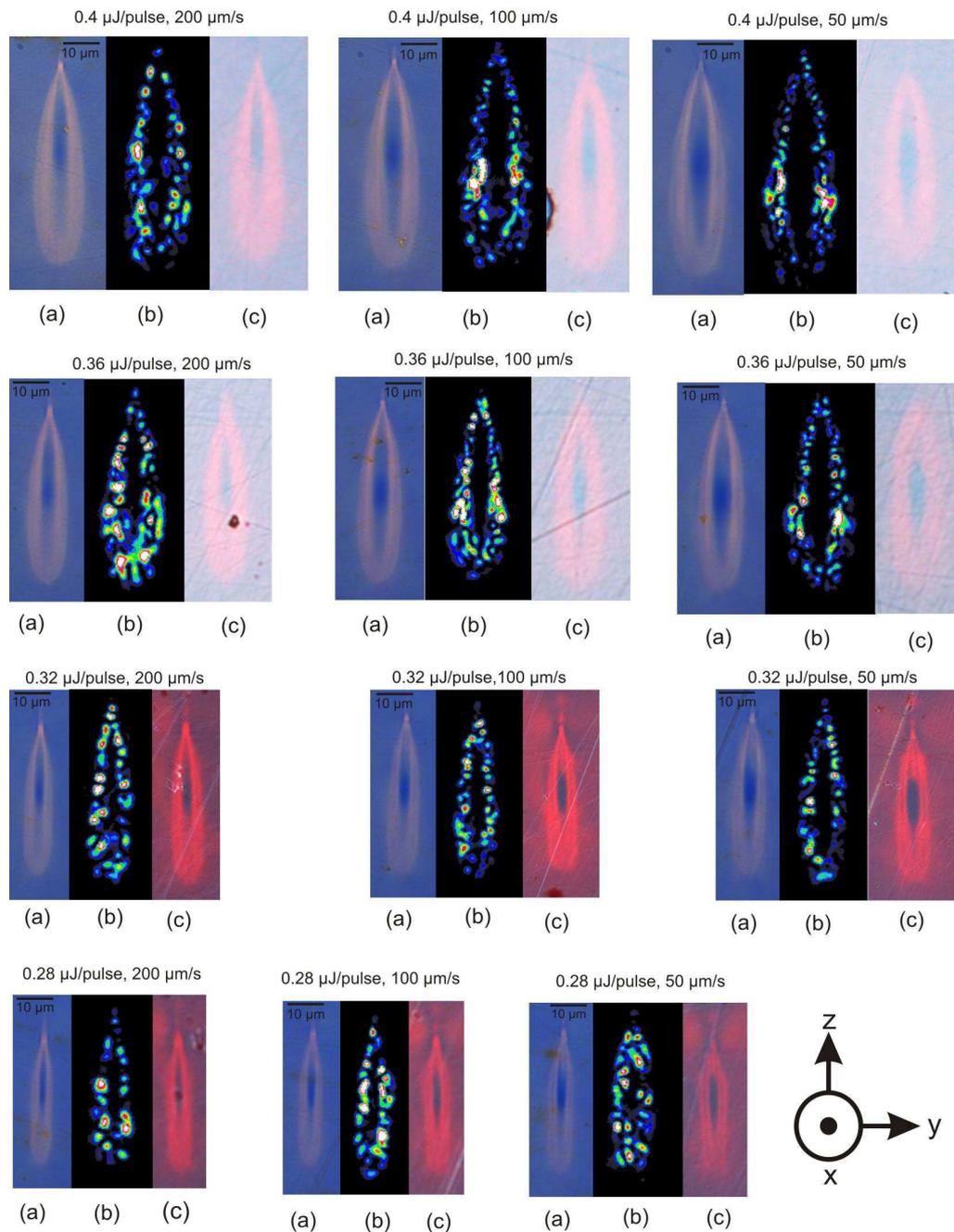

FIGURE 6.5 Transmission optical micrographs (a), guided mode at 633 nm (b) and reflection optical micrographs (c) of waveguides written at a focal depth of ~100 μm into GLS at various pulse energies and translation speeds.

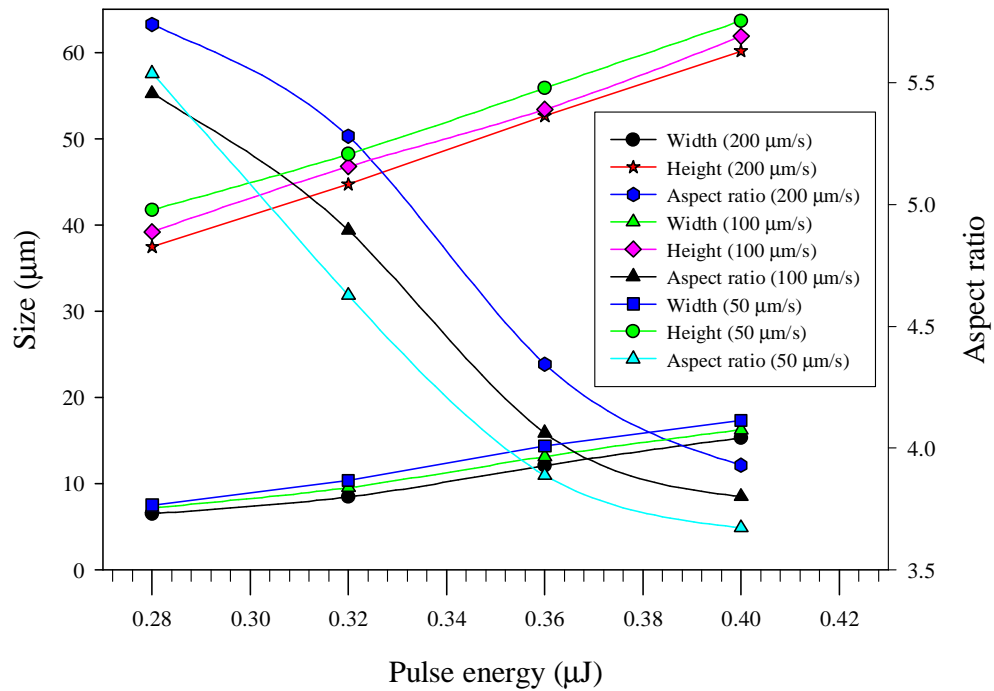

FIGURE 6.6 Height, width and aspect ratio of waveguides as a function of writing pulse energy for waveguides written at a depth of ~100 μm in GLS at various translation speeds.

Figure 6.7 shows the transmission optical micrograph, near field guided mode and reflection optical micrograph of waveguides written in the GLSO sample at pulse energies between 0.16 and 0.48 μJ, translation speeds of 100 and 50 μm/s and at a focal depth of 300 μm below the surface of the sample. The waveguides are similar in size to waveguides written at a similar depth and pulse energy in GLS. The transmission optical micrograph, near field guided mode and reflection optical micrograph show a similar pattern to that described previously for GLS waveguides. Waveguides written at pulse energies of 0.24 – 0.48 μJ are A-type and the waveguide written at 0.16 μJ/pulse is B-type. Figure 6.8 shows that, similarly to waveguides in GLS there is little dependence of the waveguide cross-section dimensions on the translation speed and A-type waveguides are characterised by an aspect ratio of 3-5 and B-type waveguides are characterised by an aspect ratio of ~30. The formation mechanism of these waveguides is discussed in section 6.3.

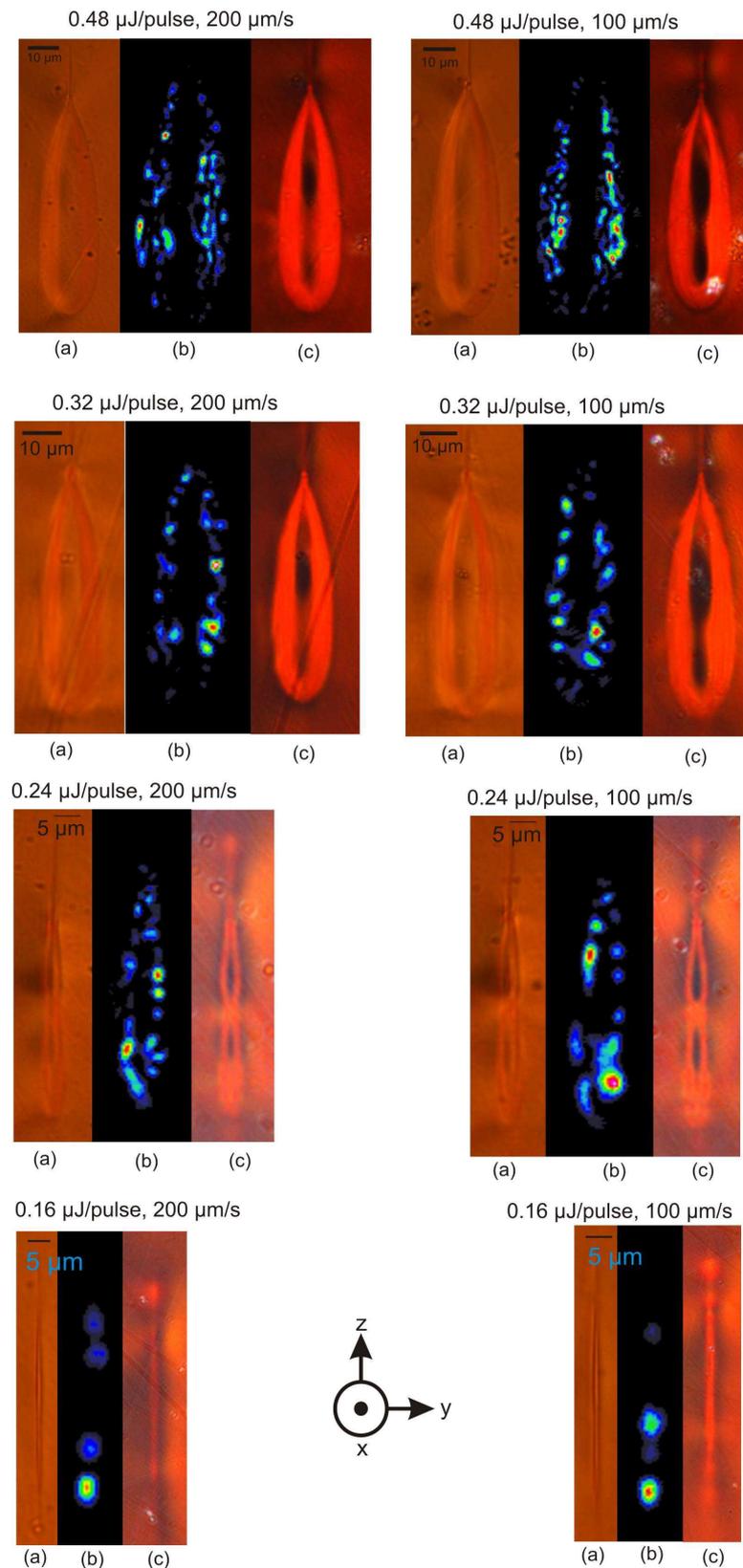

FIGURE 6.7 Transmission optical micrographs (a), guided mode at 633 nm (b) and reflection optical micrographs (c) of waveguides written at a focal depth of ~300 μm into GLSO at various pulse energies and translation speeds.

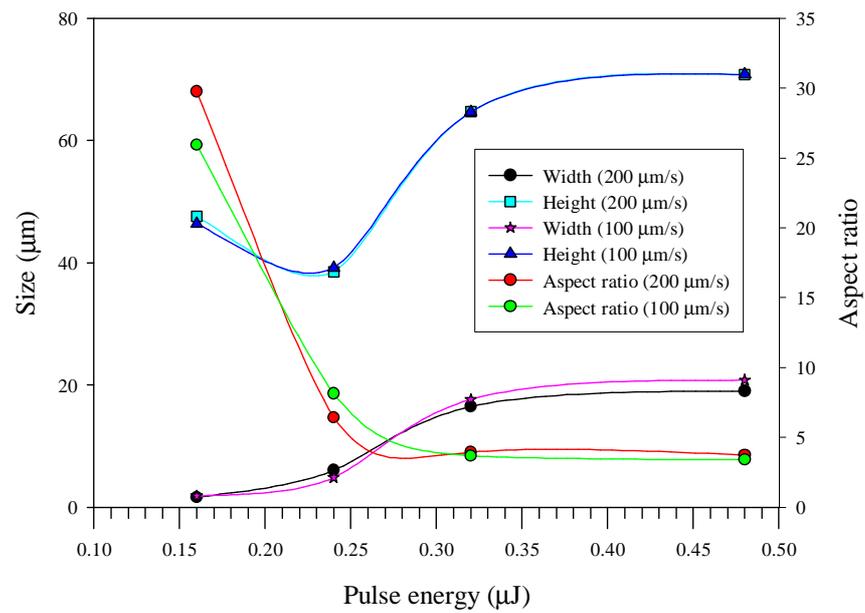

FIGURE 6.8 Height, width and aspect ratio of waveguides as a function of writing pulse energy for waveguides written at a depth of ~300 μm in GLSO at various translation speeds.

### 6.2.3 Refractive index change profile

The refractive index change ($\Delta$n) profile of the waveguides was deduced from a quantitative phase image taken in the axis that the waveguides were written, using quantitative phase microscopy (QPM). There are a number of methods available to recover the phase structure of an object. The most common of these is interferometry which can be implemented using the phase-stepping interferometric technique.[248, 249] However, interferometry requires radiation with a high degree of coherence and coherent imaging is not yet able to achieve the high resolution desired in optical microscopy. Images can have problems with speckle that prevent the formation of high quality images.[250] Phase structure can be obtained using a technique called differential interference contrast (DIC), which works by separating a polarized light source into two beams which take slightly different paths through the sample.[251-253] The two beams then recombine and interfere. However, this method is not quantitative and requires a relatively complex lighting scheme.

In order to explain the principle of quantitative phase microscopy, consider a beam of light propagating nominally in the +z direction. It can be shown[254] that the amplitude ($U_z(\mathbf{r})$) of the light beam satisfies approximately the parabolic equation 6.3.

$$\left( i\frac{\partial}{\partial z} + \frac{\nabla^2}{2k} + k \right) u_z(\mathbf{r}) = 0 \qquad (6.3)$$

Where $\nabla^2 = \dfrac{\partial^2}{\partial x^2} + \dfrac{\partial^2}{\partial y^2}$ , $k = \dfrac{2\pi}{\lambda}$ and $\mathbf{r}$ = (x,y) is a two dimensional vector in the transverse direction. By assuming that when the light passes through an object it undergoes both phase retardation and absorption and that it is imaged with a perfect, aberration free optical microscope and is coherently illuminated; then the field leaving the object ($O(\mathbf{r})$) can be described by equation 6.4

$$O(\mathbf{r}) = A(\mathbf{r})e^{i\phi(\mathbf{r})} \qquad (6.4)$$

Where $A(\mathbf{r})$ and $\phi(\mathbf{r})$ are the objects' absorption and phase profile respectively.[255] The intensity distribution in the image plane ($I_{image}(\mathbf{r})$) is given by 6.5

$$I_{image}(\mathbf{r}) = \left| A\left( \frac{\mathbf{r}}{M} \right) \right|^2 I_{Illum}(\mathbf{r}) \qquad (6.5)$$

Where $I_{illum}(\mathbf{r})$ is the intensity distribution in the absence of the object and M is the magnification of the microscope. The introduction of a small amount of defocus is mathematically equivalent to a differential propagation of the field and may be described by the transport of intensity equation.[255]

$$\frac{2\pi}{\lambda}\frac{\partial I_{\mathrm{Image}}(\mathbf{r})}{\partial z} = -\nabla \cdot \left( I_{\mathrm{Image}}(\mathbf{r})\nabla \phi\left( \frac{\mathbf{r}}{M} \right) \right) \qquad (6.6)$$

Given the availability of positively and negatively defocused images, equation 6.6 can be solved for the phase using Fourier transform methods.[256] This analysis is based on coherent illumination, however, the optical microscopes used in QPM systems use a partially coherent illumination source. It can be shown[255] that a partially coherent source gives an identical result to the coherent case, in equation 6.6, provided that the irradiance distribution of the illumination source shows inversion symmetry.[255] After the phase has been found, the refractive index change ($\Delta$n) of the waveguides can be calculated using equation 6.7.[257]

$$\Delta n = \frac{\phi\lambda}{2\pi d} \qquad (6.7)$$

Where $\lambda$ is the wavelength of 550 nm used by IATIA software to calculate the quantitative phase image and d is the height (z dimension) of the waveguides.

The QPM system used here is illustrated in figure 6.9 and incorporated an Olympus BX51 optical microscope equipped with a Physik Instrument P721K039, nano-focusing, Z drive with a resolution of < 1 nm to take in-focus and very slightly positively and negatively defocused images. IATIA software was then used to calculate the quantitative phase image.

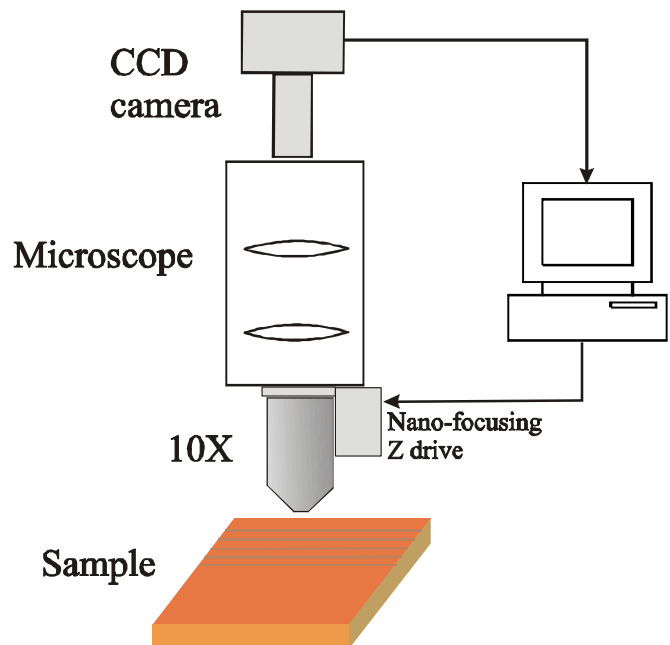

FIGURE 6.9 Quantitative phase microscopy setup.

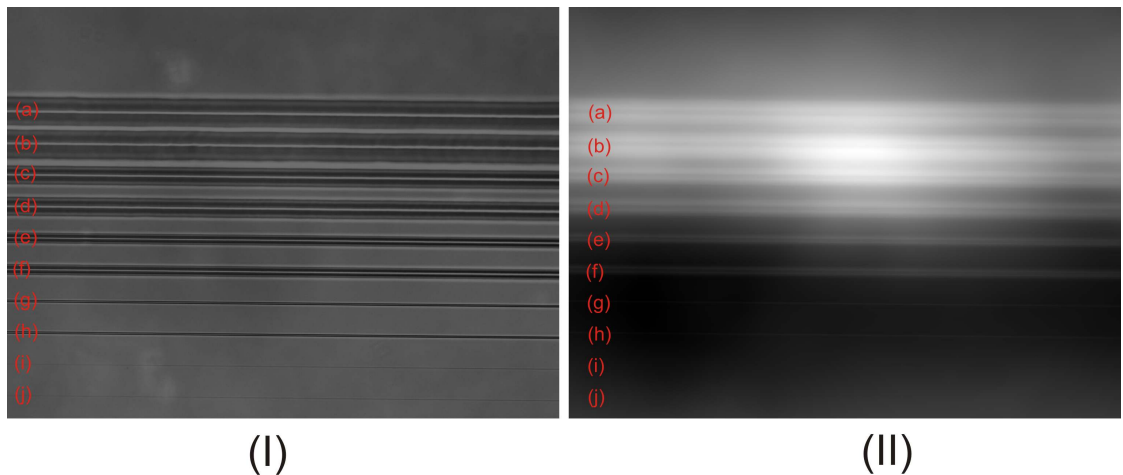

FIGURE 6.10 Optical micrograph (I) and quantitative phase image (II) of waveguides written at a translation speed of 200 μm/s and pulse energies of 1.74 μJ (a and b), 1.26 μJ (c and d), 0.84 μJ (e and f), 0.42 μJ (g and h), 0.21 μJ (i and j). The images were taken in the axis the waveguides were written.

Figure 6.10 (I) shows the transmission optical micrograph of waveguides written at pulse energies of 1.74 – 0.21 μJ, a translation speed of 200 μm/s and a depth of 400 μm. Its corresponding quantitative phase image, shown in figure 6.11 (II), was calculated from the image in figure 6.11 (I) positively and negatively defocused by 30 μm. The index change profiles of the waveguides were calculated from vertical cross-sections of figure 6.10 (II) in areas of good contrast against the background phase information. A transmission optical micrograph and quantitative phase image is also shown in figure 6.12 (I) and (II) respectively but for a set of waveguides written at a depth of 100 μm.

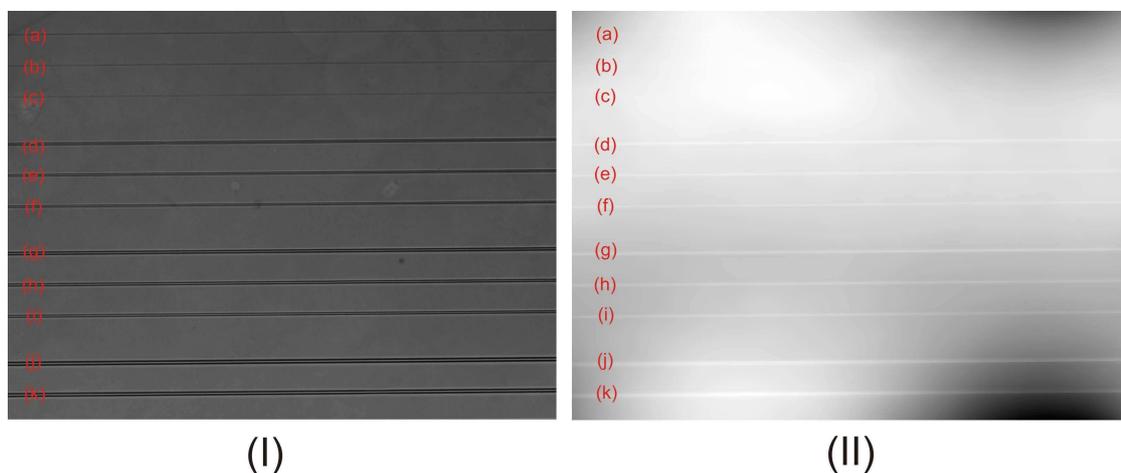

FIGURE 6.11 Optical micrograph (I) and quantitative phase image (II) of waveguides written at a depth of 100 μm with pulse energies and speeds of 0.28 μJ and 50 μm/s (a), 0.28 μJ and 100 μm/s (b), 0.28 μJ and 200 μm/s (c), 0.32 μJ and 50 μm/s (d), 0.32 μJ and 100 μm/s (e), 0.32 μJ and 200 μm/s (f), 0.36 μJ and 50 μm/s (g), 0.36 μJ and 100 μm/s (h), 0.36 μJ and 200 μm/s (i), 0.4 μJ/ and 50 μm/s (j), 0.4 μJ and 100 μm/s (k) respectively.

The quantitative phase map in figure 6.11 (II) was used to emulate a differential interference contrast (DIC) image, shown in figure 6.12, which illustrates the excellent image quality available through DIC. Although the DIC image does not contain quantitative phase information it has enhanced contrast and resolution and a reduced number of artefacts.

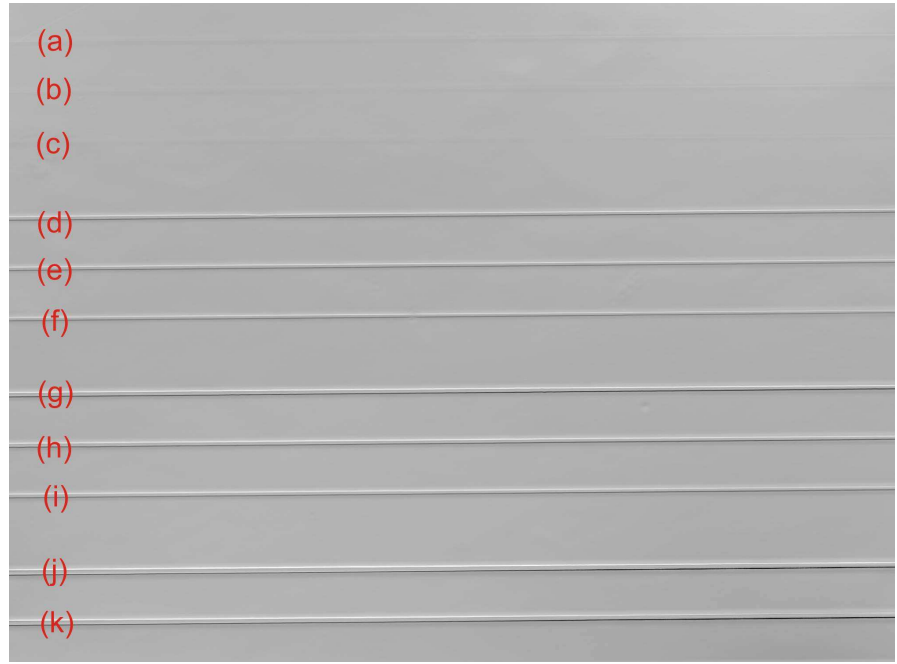

FIGURE 6.12 Emulated differential interference contrast image from quantitative phase image in figure 6.12 II.

An example of the phase profiles extracted from figure 6.10 (II) and 6.11 (II) is shown in figure 6.13. The background phase profile was assumed to vary linearly over the width of the waveguide. This linear approximation was then subtracted from the phase change profile to give the phase profile for the waveguide alone.

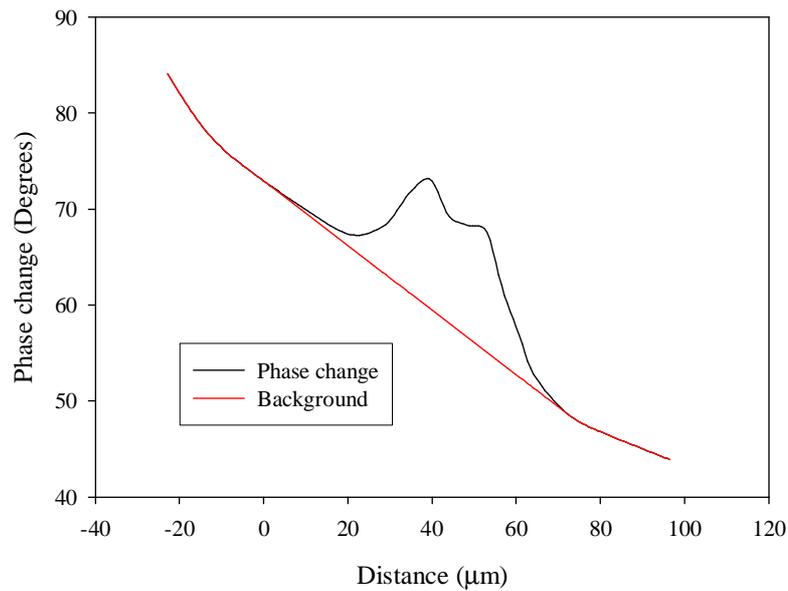

FIGURE 6.13 Phase change profile of a waveguide written with 0.84 μJ/pulse, 200 μm/s translation speed and a depth of 400 μm together with the (assumed to be linear) background phase change that was subtracted from the phase change data.

The index change profile of each waveguide was then calculated from its phase change profile using equation 6.7.

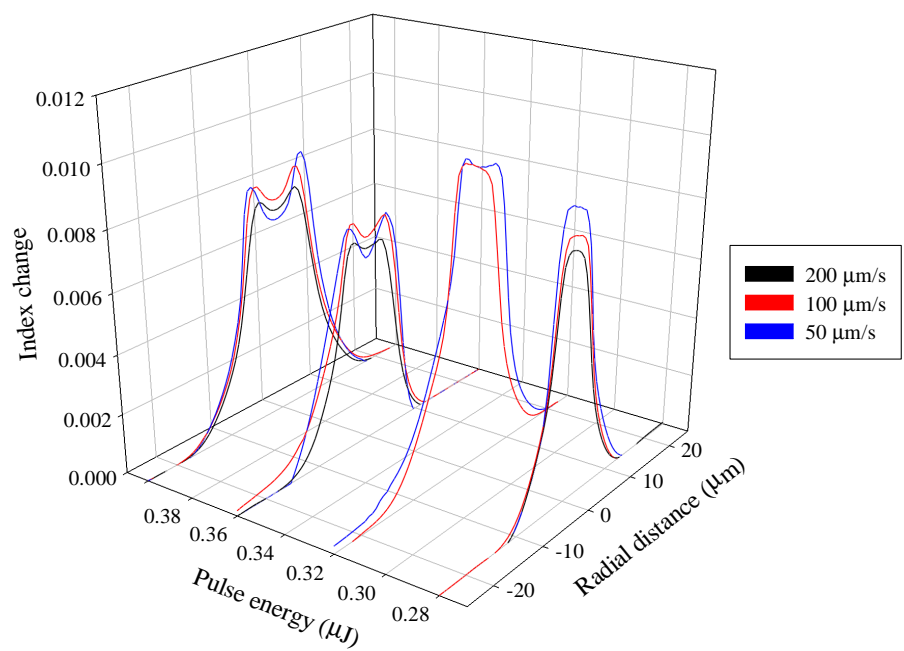

FIGURE 6.14 Refractive index change profile as a function of writing pulse energy for waveguides written into the GLS sample at a depth of ~100 μm at translation speeds of 50, 100 and 200 μm/s.

Figure 6.14 shows the refractive index change ($\Delta$n) profile as a function of writing pulse energy for waveguides written into the GLS sample at a depth of ~100 μm at translation speeds of 50, 100 and 200 μm/s; the figure shows a double peak structure with a trough at the centre for waveguides written at pulse energies of 0.36 and 0.4 μJ. Comparing the $\Delta$n profile of the waveguides written at pulse energies of 0.36 and 0.4 μJ to their optical micrographs, in figure 6.5, demonstrates that region 1 has undergone a lower refractive index change than region 2 but whether the index change of region 1 is negative is still unclear. Figure 6.15 shows the $\Delta$n profile of a waveguide written with a pulse energy of 0.4 μJ and a translation speed of 50 μm/s superimposed onto its transmission optical micrograph. This clarifies that region 1 has undergone a lower refractive index change than region 2, it also indicates that the index variation extends beyond the visible boundary of the waveguide. Examining the micrographs of the waveguides (figures 6.3, 6.5 and 6.7) shows that the waveguides tend to get narrower moving away from the centre of the waveguide (zero radial distance), the index change profiles will therefore be an underestimated distance away from the centre of the waveguide. Because the visible boundary of the waveguide may not represent the boundary of the index modulation, taking into account the change in waveguide thickness with radial distance may be unproductive.

The $\Delta$n profile in the middle of the waveguide includes the index change of region 1 and 2, decoupling the index change for both these regions to determine if region 1 has undergone a negative index change required the boundaries of these regions to be accurately defined. As previously discussed, defining the boundaries of the different waveguide regions is difficult. However, using the waveguide shown in figure 6.15 and approximating the waveguide regions by eye, the waveguide thickness at the peak index change of $8.6 \times 10^{-3}$ is ~ 69% of the maximum waveguide thickness, giving a peak index change of $1.24 \times 10^{-2}$, all of this index change is due to region 2. The thickness of region 2 in the centre of the waveguide is ~ 38% of the maximum waveguide thickness, therefore the index change due to region 2 in the centre of the waveguide should be $38/69 \times 1.24 \times 10^{-2} = 7.0 \times 10^{-3}$. This value is slightly higher than the observed index change of $6.7 \times 10^{-3}$ and indicates that region 1 has undergone a negative index change of $3 \times 10^{-4}$. However, region 1 could show a positive index change, depending on how the boundaries of region 1 and 2 are defined, therefore a more systematic way of defining the boundaries is required. Ellipsometry measurements of the waveguide end faces should be able to directly measure the index change of regions 1 and 2, these measurements were attempted but they failed to reveal any index variation, the reason for this is unclear.

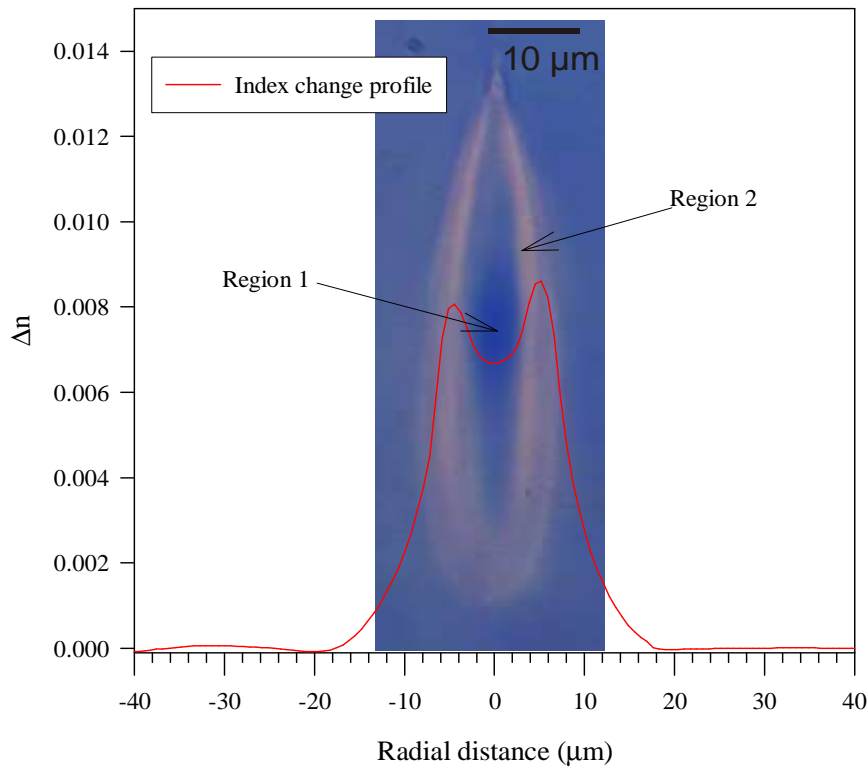

FIGURE 6.15 Index change profile of a waveguide written with a pulse energy of 0.4 μJ and a translation speed of 50 μm/s scaled and superimposed on to its transmission optical micrograph.

Figure 6.14 shows that, at a pulse energy of 0.32 μJ and a translation speed of 100 μm/s, the double peak structure of the index change profile is no longer apparent. Examining the micrograph of this waveguide, in figure 6.5, indicates that this is due to a reduction in the relative size of region 1, which has a lower index change than region 2. The increase in maximum index change, as the pulse energy decreases from 0.36 to 0.32 μJ (also show in figure 6.17), is also attributed to this effect. The refractive index change may also have saturated, as happens in fs laser written waveguides in fused silica where the index change for a speed of 10 μm/s saturated at ~ 0.75 μJ, for borosilicate glass the corresponding value was 0.5 μJ.[50] Figure 6.14 also shows that reducing the translation velocity increases the contrast of the double-peak structure, for example for waveguides written at 0.4 μJ/pulse the ratio of the double peaks to the central trough is 1.07, 1.11 and 1.29 for translation speeds of 200, 100 and 50 μm respectively, which indicates that region 1 is larger and/or has a lower index in waveguides written with lower translation speeds.

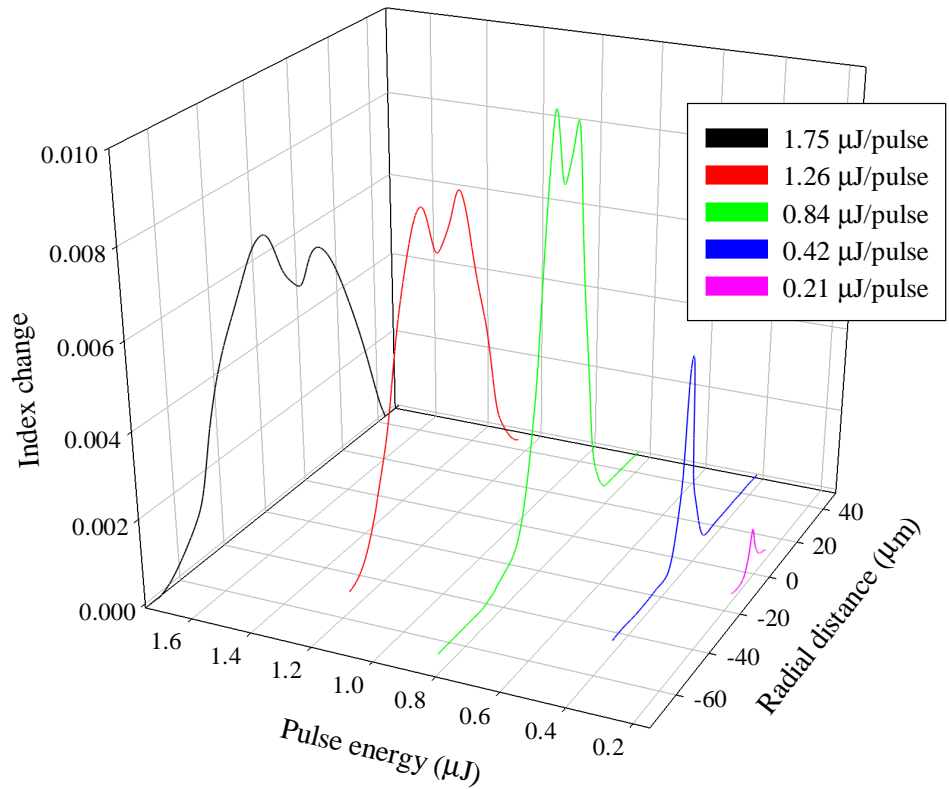

FIGURE 6.16 Refractive index change profile as a function of writing pulse energy for waveguides written into the GLS sample at a depth of ~400 µm.

Figure 6.16 shows the refractive index change (Δn) profile as a function of writing pulse energy for waveguides written into the GLS sample at a depth of ~400 µm. Similarly to figure 6.14 there is a deviation from a double peak structure at 0.4 µJ/pulse, however, examining the micrograph of this waveguide in figure 6.3 indicates that this originates from a transition from A-type to B-type waveguides.

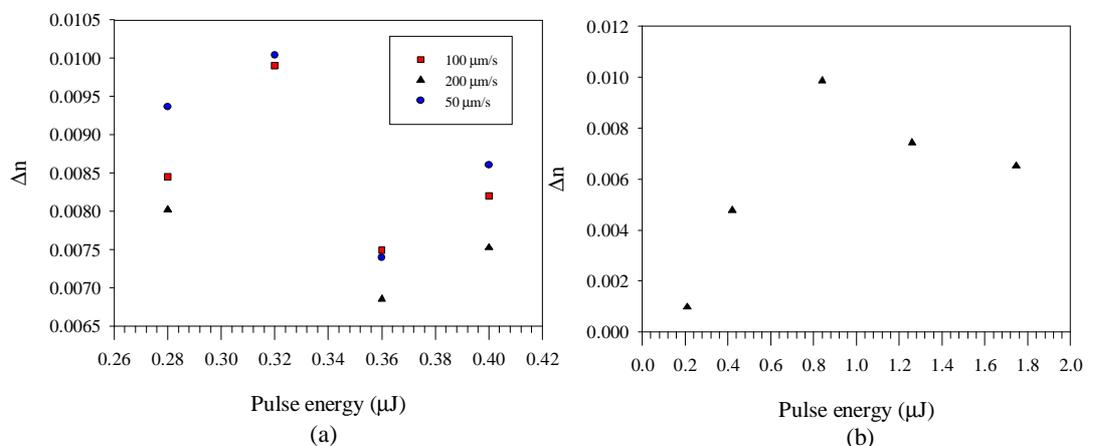

FIGURE 6.17 Peak index change as a function of writing pulse energy for waveguides written at a depth of 100 µm (a) and 400 µm (b).

Figure 6.17 shows the peak index changes taken from figure 6.14 and 6.16. The index change appears to reaches a maximum of ~ 0.01 at 0.32 μJ/pulse and 0.84 μJ/pulse for waveguides written at a depth of 100 and 400 μm respectively. This is attributed to a saturation of the index change and an increase of the relative size of region 1 with increasing pulse energy. The index change for waveguides written with ~0.4 μJ/pulse and 200 μm/s at a depth of 100 and 400 μm is $7.5 \times 10^{-3}$ and $4.8 \times 10^{-3}$ respectively. This decrease, with increasing focal depth, is attributed to linear absorption of the pulse before it undergoes nonlinear absorption at its focus, as discussed for the relative decrease in waveguide size at a depth of 400 μm in section 6.2.2. The mechanisms by which refractive index modification occurs due to fs laser exposure are discussed in section 6.3.3.

## 6.2.4 Micro Raman spectra

Raman spectroscopy has been used by other authors to determine the structural modification caused by fs laser exposure in $As_2S_3$ glass [237, 258], borosilicate glass[50] and aluminosilicate glass[259]. Raman spectra of GLS fs written waveguides were taken with a Renishaw Ramascope with a 10mW, 633 nm, HeNe laser and a 50× objective lens, described in detail in section 3.3.6.

Micro-Raman measurements were made of the endfaces of waveguides written at the highest energy used in this study (1.75 μJ) so that any differences from the unexposed glass would be as clear as possible. The Raman measurements of region 1 and 2 of waveguides written at pulse energies of 1.75 μJ and 1.26 μJ and two regions of unexposed glass are shown in figure 6.18. The image at the top of the figure shows the position at which each Raman spectrum was taken, the insets show the two regions of interest expanded. Figure 6.18 shows that the variations in the Raman spectra between the various waveguide structures and the bulk glass is no greater than the variations between two different areas of unexposed glass. The Raman spectra of GLS consists of two broad bands located at 150 and 340 cm$^{-1}$ [16] making any structural variations difficult to distinguish. We therefore propose that any structural modifications that occur are subtle, such as a bond angle change. There were also no significant variations in the Raman spectra of fs laser exposed and unexposed regions in heavy metal oxide glasses.[238]

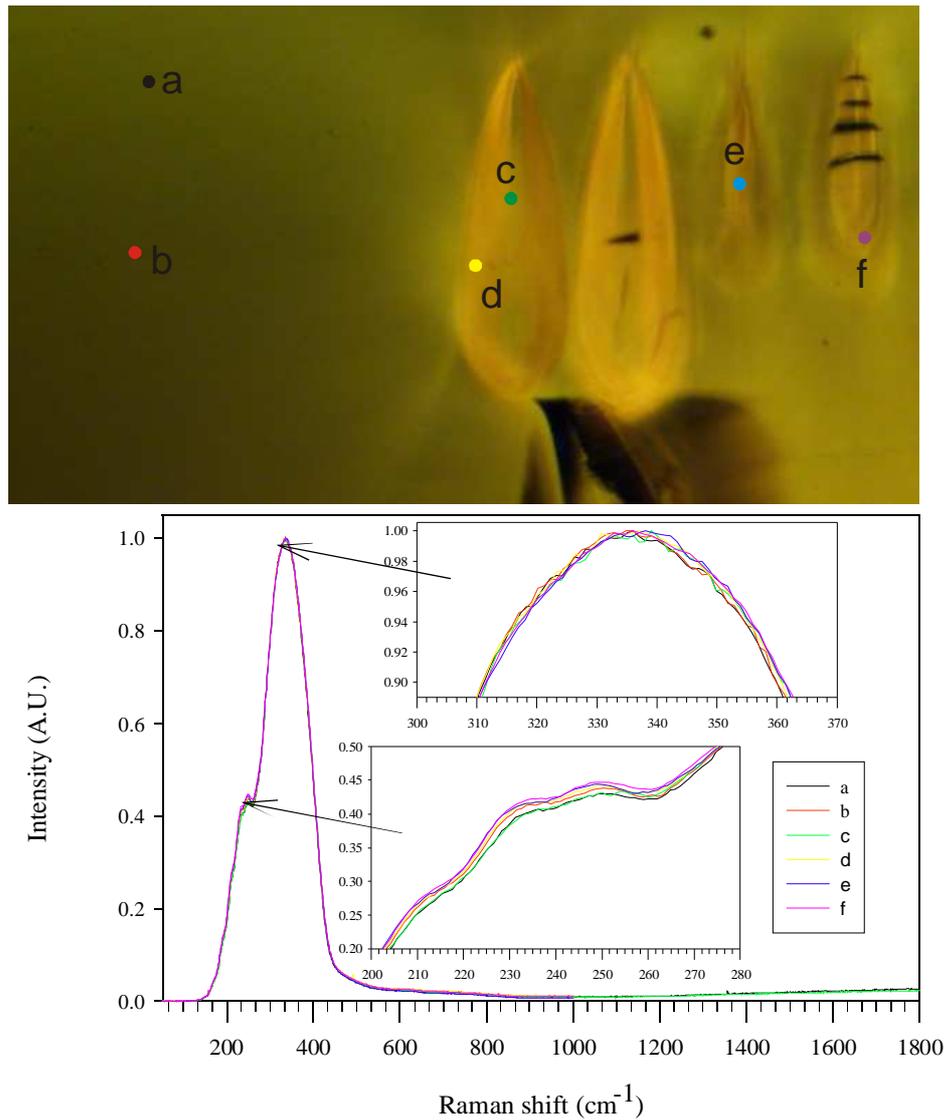

FIGURE 6.18 Micro-Raman spectra of waveguides written at pulse energies of 1.75 μJ and 1.26 μJ and two regions of unexposed glass. The top image shows the position at which each Raman spectrum was taken.

Energy dispersive X-Ray (EDX) measurements of the region 1, region 2 and the unexposed glass found no compositional variation between these regions greater than the system detection limit of around 1%.

## 6.2.5 Waveguide transmission

Waveguide transmission measurements are used here to examine the change in the absorption band-edge of the waveguides compared to bulk glass. This can be used to examine effects related to photo-darkening and colour centre formation.

Waveguide transmission measurements were made using the setup shown in figure 6.5. The output from a tungsten halogen, white light source was attenuated with an iris and a neutral density filter before being collimated with two lenses and then modulated with a mechanical chopper. The collimated white light beam was then coupled into and out of the waveguides with a 10x microscope objectives. The guided modes were then imaged onto an iris and the selected guided mode was focused into a monochromator; the signal was then detected with a silicon detector using standard phase sensitive detection. In order to correct for the spectral dependence of the various optical components in the system a transmission measurement of the bulk glass ($I_{syst}(\lambda)$) was made. Then using the transmission measurement ($I_{ref}(\lambda)$) of the same sample, measured using the Varian Cary 500 spectrophotometer, detailed in section 3.3.1, a correction spectrum ($C(\lambda)$) was calculated using equation 6.8.

$$C(\lambda) = \frac{I_{ref}(\lambda)}{I_{syst}(\lambda)} \qquad (6.8)$$

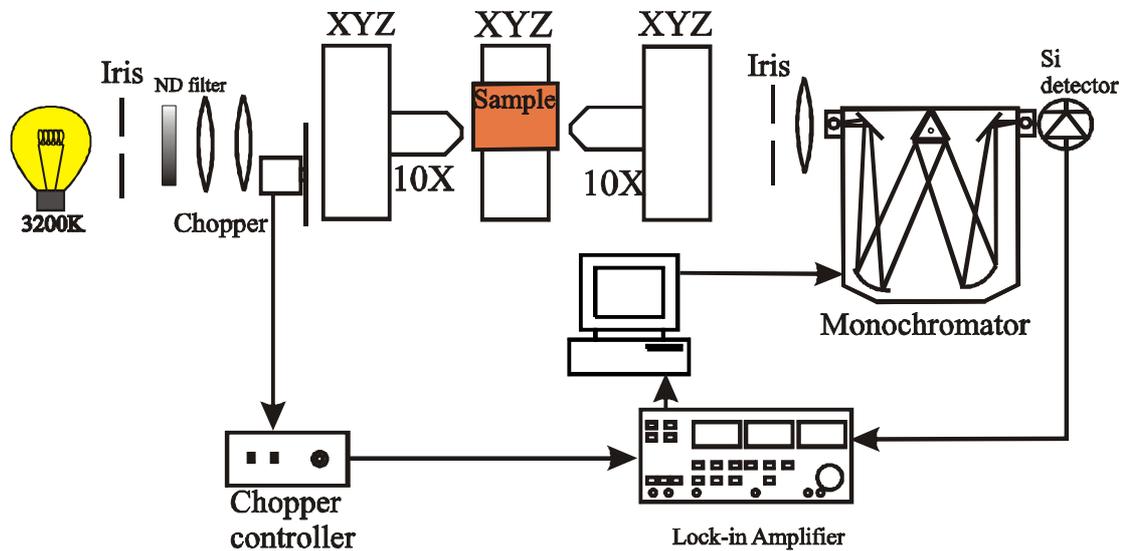

FIGURE 6.19 Waveguide transmittance measurement setup.

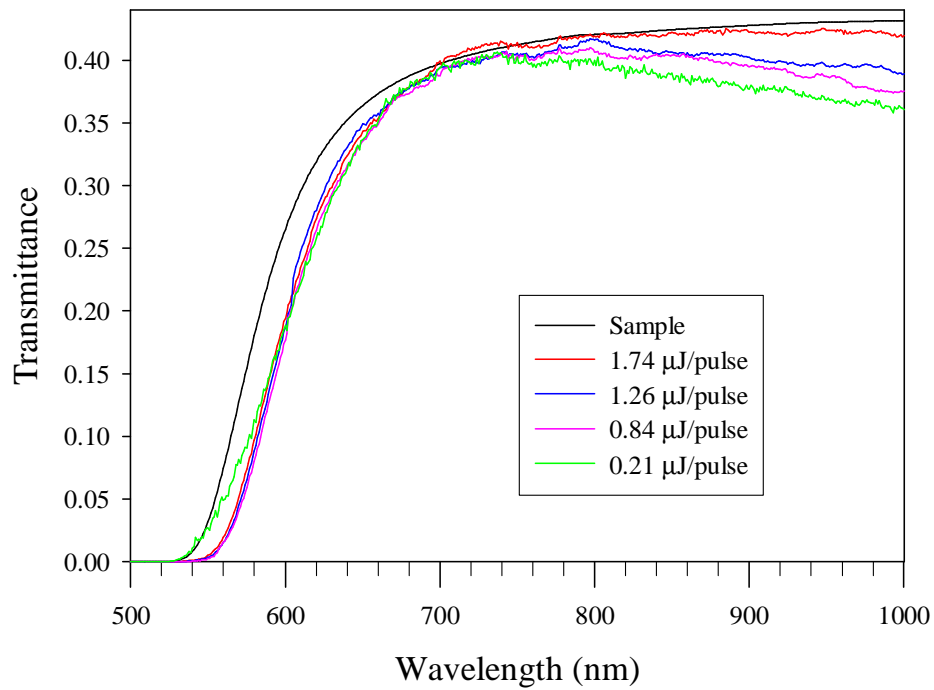

FIGURE 6.20 Transmission spectra of waveguides written in GLS at a depth of 400 μm and a translation speed of 200 μm/s.

Figure 6.20 shows the transmission spectra of waveguides written at various pulse energies. The measurements show a red shift in the band-gap of around 15 nm for all the waveguides. Femtosecond laser written waveguides in $As_{40}S_{60}$ glass also showed a 15 nm red shift of the band edge, [237] this was related, by the Kramer-Kronig relation, to a refractive index increase of $5x10^{-4}$. The band-edge and refractive index of $As_{40}S_{60}$ glass are approximately the same as GLS.[240] Therefore the index variation in the GLS waveguides in figure 6.16 that could be attributed to the Kramer-Kronig processes is expected to be approximately $5x10^{-4}$. The transmission spectra of the waveguide, written at 0.21 μJ/pulse, deviates from that of the sample by peaking at around 720 nm, as the pulse energy is increased the transmission spectra of the waveguides matches that of the sample more closely. This may be due to a decrease in coupling efficiency at longer wavelengths.

## 6.2.6 Waveguide optical loss

The attenuation of a waveguide is an important characteristic to be measured as the requirement for most applications is for low loss. There are several methods for measuring waveguide loss; these include the scattering technique,[260] cutback method,[230] prism coupling,[261] comparison of input and output power[34] and the Fabry-Perot resonance method.[262] The scattering technique requires the measurement of the intensity of scattered light coupled into the waveguide as a function of propagation distance. The relatively high reflectivity of GLS and the relatively short length of the waveguides meant that light reflected from the waveguide end faces made this method impractical. The cut-back method is destructive and therefore unsuitable. The prism coupling technique could not be used because the waveguides are buried below the surface of the glass. The comparison of input and output power method requires the estimation of coupling efficiency, which could be prone to error because of the asymmetry of the waveguides.

Waveguide loss measurements were therefore taken using the Fabry-Perot resonance method,[262] with the experimental setup shown in figure 6.21. By taking into account reflections from its end faces the waveguide structure may be regarded as a resonant cavity. By varying the wavelength of the input light source the output will reach periodic maxima ($I_{max}$) and minima ($I_{min}$), defining $\zeta = I_{max}/ I_{min}$ it can be shown[262] that the loss coefficient ($\alpha$) of a waveguide can be calculated using equation 6.9.

$$\alpha = -\frac{1}{L}\ln\left(\frac{1}{R}\frac{\sqrt{\xi}-1}{\sqrt{\xi}+1}\right)$$                (6.9)

Where R is the reflectivity of the end faces and L is the length of the waveguides. To carry out the experiment a fibre coupled Photonetics Tunics tuneable external cavity laser was coupled into and out of the waveguides with 10x 0.25 NA objectives with the output detected by a power meter. The guided mode of each waveguide was found with a CCD camera and any unguided radiation was blocked with an iris. The external cavity laser was scanned from 1550 to 1550.5 nm in steps of 0.001 nm and the waveguide output power plotted as a function of input wavelength. Error bounds were calculated from the variance of $\zeta$ and the accuracy to which R was known.

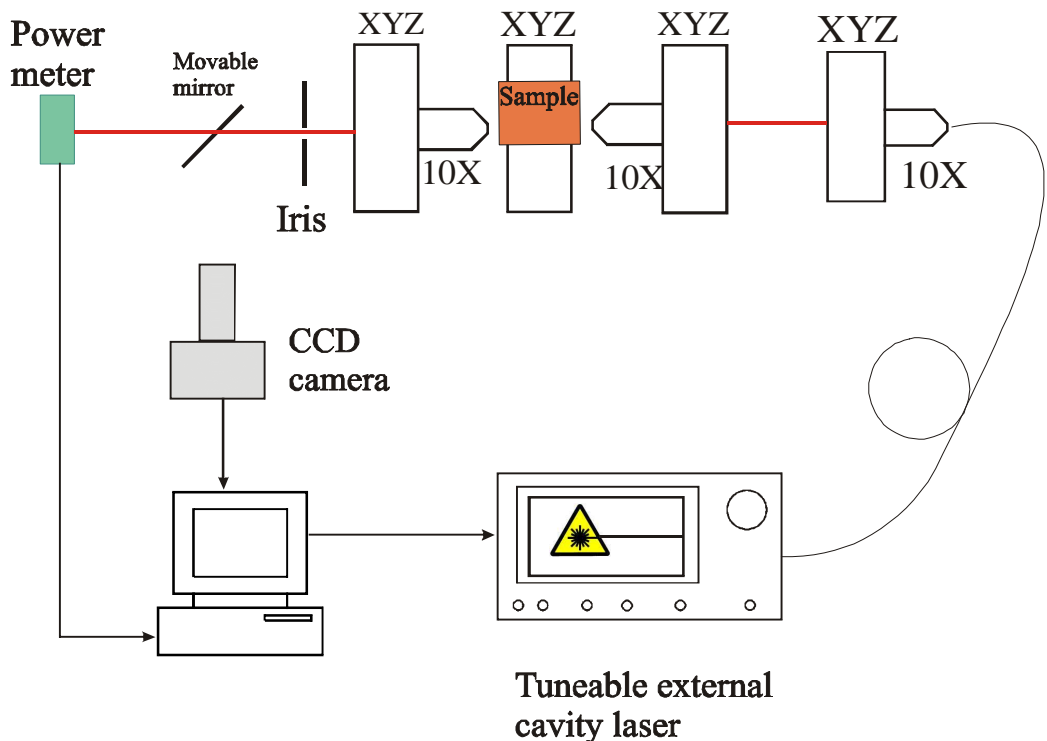

FIGURE 6.21 Waveguide loss measurements using the Fabry-Perot resonance method.

Figure 6.22 shows an example of a Fabry-Perot scan which shows a relatively clean and undistorted scan with regular maxima and minima. This indicates that a fundamental mode has been excited, although it does not confirm that the waveguide is single mode at 1550 nm. The slight periodic variation of the maxima and minima may indicate the presence of a higher order mode although it is not thought to be significant enough to interfere with the measurement. The Fabry-Perot effect was not observed in waveguides written at pulse energies > 0.5 μJ, this is attributed to their large cross-section area which makes them highly multimode. In order to estimate the loss of these waveguides the comparison of input and output power method was the best remaining technique. The loss of waveguides written at pulse energies of 1.75 and 1.26 μJ was required for spectral broadening measurements in section 6.4. To calculate the loss of these waveguides a lens matched to the NA of the waveguides was used to couple a 1550 nm, 200 fs pulse width laser source (see section 6.4.2) into the waveguides. The numerical aperture of the fabricated waveguides can be estimated from equation 6.10[263]

$$NA \approx \sqrt{2n\Delta n} \tag{6.10}$$

From the $\Delta$n calculated in section 6.2.3 the NA of waveguides written at 1.75 and 1.26 μJ/pulse was 0.18 and 0.19 respectively. Therefore the 0.1 NA lens used to couple into the waveguides was within the acceptance cone of the waveguides. By measuring the power coupled into and out of the waveguides, then taking into account reflections from the end faces, losses of 1.7 and 1.6 dB/cm were calculated for the waveguides written at 1.75 and 1.26 μJ/pulse respectively. This estimate is a maximum since it assumes 100% coupling efficiency, however the relatively low loss calculated indicates that the actual coupling efficiency is close to 100%.

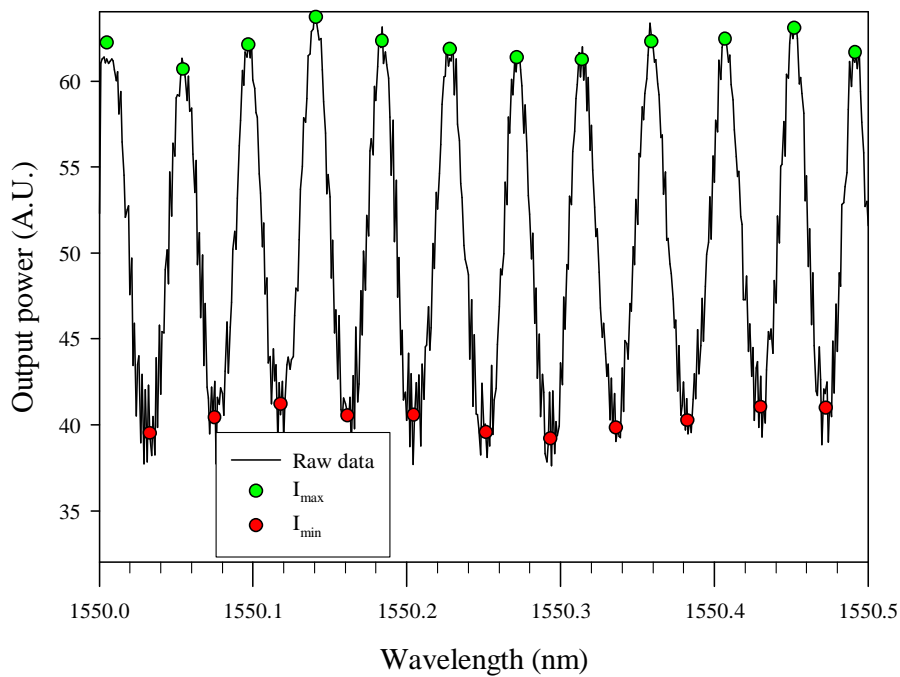

FIGURE 6.22 Fabry-Perot scan of a waveguide written with at 0.36 μJ/pulse and 50 μm/s translation speed.

Figure 6.23 shows the loss measurements, taken with the setup in figure 6.21, of the waveguides in the GLS sample as a function of pulse energy at various translation speeds. For waveguides written at 200 and 50 μm/s, an apparent optimum pulse energy was reached at 0.32 and 0.36 μJ/pulse giving a loss of 2.08 and 1.47 dB/cm respectively. However for waveguides written at 100 μm/s there is no apparent optimum writing pulse energy, this may have been because one of the waveguides was damaged or written through a flaw or crystal in the glass. We propose that the optimum writing parameters occur because at higher pulse energies damage occurs to the glass which increases loss. At the lowest pulse energy a B-type waveguide is formed, this indicates that B-type waveguides have a higher intrinsic loss than A-type waveguides. The higher loss of B-type waveguides may be caused by the lower index change in these waveguides, see figure 6.16 and 6.17. This would mean that the guided mode is not as tightly confined to the waveguide structure and may be attenuated by scattering centres in the bulk glass. A similar dependence of waveguide loss on writing pulse energy was observed in fs laser written waveguides in erbium-doped oxyfluoride silicate glass,[264] this was attributed to a similar effect.

As discussed earlier as the writing pulse energy is increased further to 1.75 and 1.26 μJ the loss appears to fall again. This is attributed to the very large cross-section of these waveguides which could mean that guided light is not confined to damaged regions of glass. The minimum reported loss in this work of 1.47 dB/cm compares to minimum losses in fs laser written waveguides of 1.0 dB/cm in aluminosilicate glass,[259] 0.8 dB/cm in fused silica,[265] 0.25 dB/cm in phosphate glass,[266] and 0.8 dB/cm in heavy metal oxides glass[238]. It is believed that the loss reported in this work could be improved by further optimisation of pulse energy and scan speed, using double pulse fs lasers[265], astigmatically shaping the writing beam to reduce the asymmetry of the waveguide cross-section[266] and annealing to reduce any internal stresses in the glass.

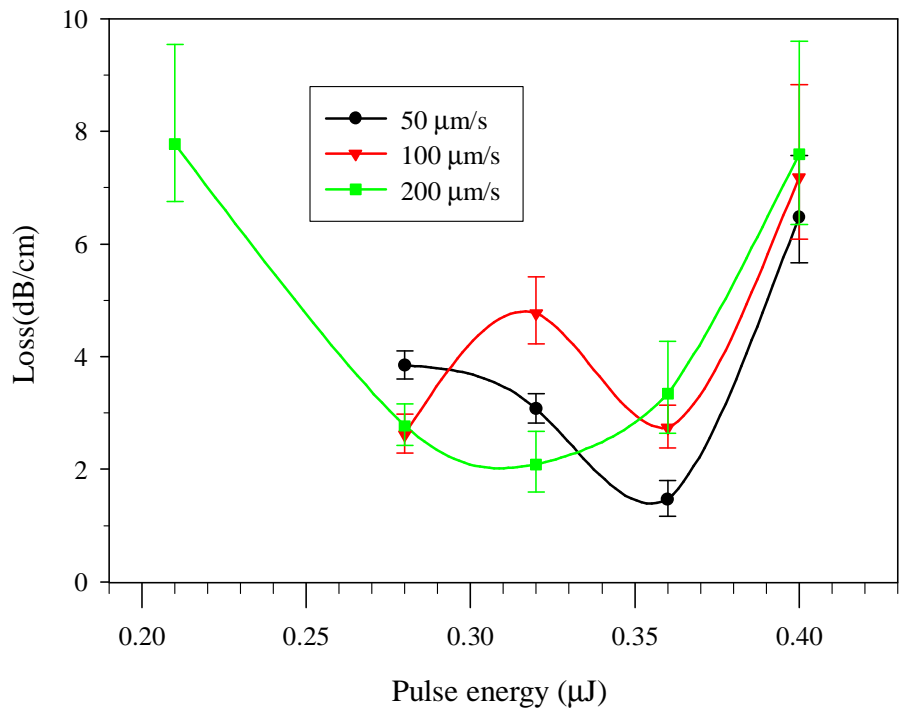

FIGURE 6.23 Waveguide losses in the GLS sample as a function of writing pulse energy at various translation speeds. Lines are a guide for the eye.

Figure 6.24 shows the loss measurements of the waveguides in the GLSO sample as a function of pulse energy at various translation speeds. Similarly to the GLS sample an apparent optimum is reached at 0.24 μJ/pulse and 100 μm/s translation speed giving a loss of 7.8 dB/cm. This optimum loss is far higher than the optimum loss in the GLS sample and for the same writing parameters the losses are far higher. This difference is unexpected as the absorption coefficients of GLS and GLSO at 1550 nm are very similar. The difference is attributed to the reduction in tunnelling rate in GLSO predicted by equation 6.18. Calculation of the tunnelling rate requires further investigation.

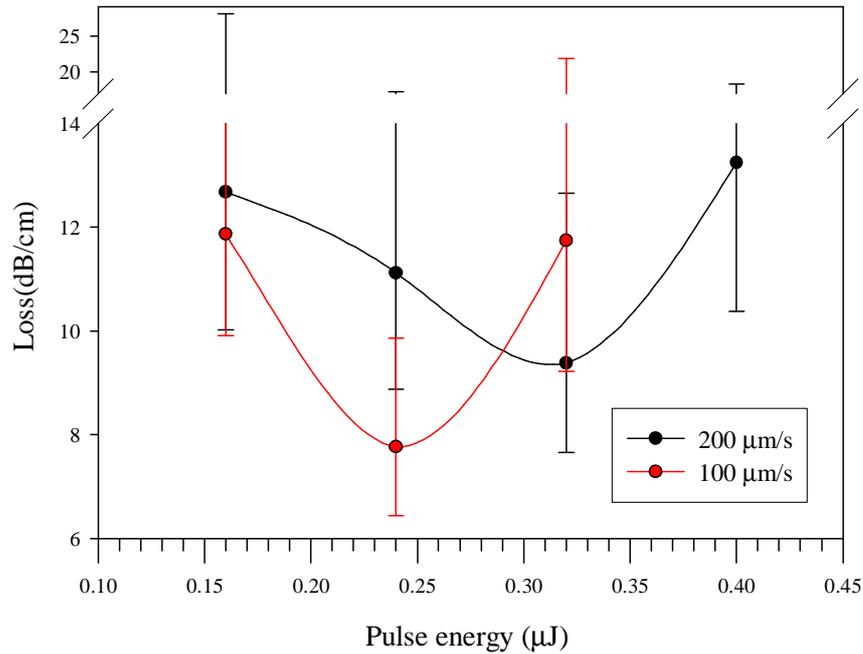

FIGURE 6.24 Waveguide losses in GLSO sample as a function of writing pulse energy at various translation speeds. Lines are a guide for the eye.

## 6.3 Discussion of waveguide formation mechanism

In this section the observed waveguide structure and refractive index change profile presented in section 6.2.2 and 6.2.3 are discussed and analysed in terms of formation and non-linear absorption mechanisms.

### 6.3.1 Waveguide asymmetry

The inherent asymmetry of the A-type waveguides could be explained as follows: perpendicular to the propagation direction (y axis) of the writing laser the waveguide dimension ($L_y$) is given approximately by the beam focal diameter $2\omega_0$,

$$L_y \approx 2\omega_0 \qquad (6.11)$$

while along the propagation direction (z axis) the waveguide dimension ($L_z$) is given by the confocal parameter b,

$$L_z \approx b = \frac{2\pi\omega_0^2}{\lambda} \qquad (6.12)$$

where $\omega_0$ is the focus waist. This results in a large difference in the waveguide sizes in the two directions.[266] Given that the focus waist of the 800 nm beam used to write the waveguides was around 1 μm, this corresponds to $L_y \approx 2$ μm and $L_z \approx 8$ μm. These dimensions do not correspond to either the A or B-type waveguides, however the aspect ratio is similar to A-type waveguides and the dimensions are much closer (although still roughly half) to those of the central region of the A-type waveguides. Therefore further explanation of the waveguide structure is required.

## 6.3.2 Self focusing and plasma defocusing

The phenomenon of self focusing, otherwise known as Kerr lensing, is a result of the intensity dependent non-linear refractive index given in equation 6.13[267]

$$n = n_0 + n_2 I \tag{6.13}$$

Where $n_0$ is the linear refractive index, $n_2$ is the non-linear refractive index and I is the laser irradiance. A Gaussian beam that is more intense at its centre than at its edges will experience a larger index change at its centre, so the beam effectively passes through a graded index lens which focuses the pulse. As the power of the laser pulse is increased further the self focusing becomes stronger until at some point it overcomes diffraction and the pulse undergoes a catastrophic collapse,[268] see figure 6.25(a). Given no other stabilisation mechanisms the beam is expected to form a singularity.[17, 269] However, in reality, as the pulse self focuses and the intensity rises it eventually becomes sufficient to non linearly ionize electrons producing an electron gas or plasma. This plasma contributes a negative refractive index change that prevents further self focusing.[18] The complex refractive index variation in the created plasma is quantitatively described by means of the laser induced plasma index modulation ($\Delta n_{pl}$) in equation 6.14[270]

$$\Delta n_{pl} = \frac{2}{n_0} \frac{e^2 \tau_e}{m_e \omega} \left( \frac{(1 - \omega^3 \tau_e^{\ 3}) + i(\omega^2 \tau_e^{\ 2} + \omega \tau_e)}{\omega^4 \tau_e^{\ 4} + 1} \right) N_0 \exp\left( \int_0^t \eta(E) dt \right) \tag{6.14}$$

Where $\tau_e$ is the free electron collision time, $\omega$ is the laser angular frequency, $N_0$ is the initial plasma density, $m_e$ is the electron mass, e is the electronic charge and $\eta(E)$ is the probability per unit time for an electron to undergo an ionising collision. The realisation of the dynamic equilibrium between self focusing and plasma defocusing is a phenomenon known as filamentation. Filaments of light are robust light guides that are self confined without requiring a waveguide.

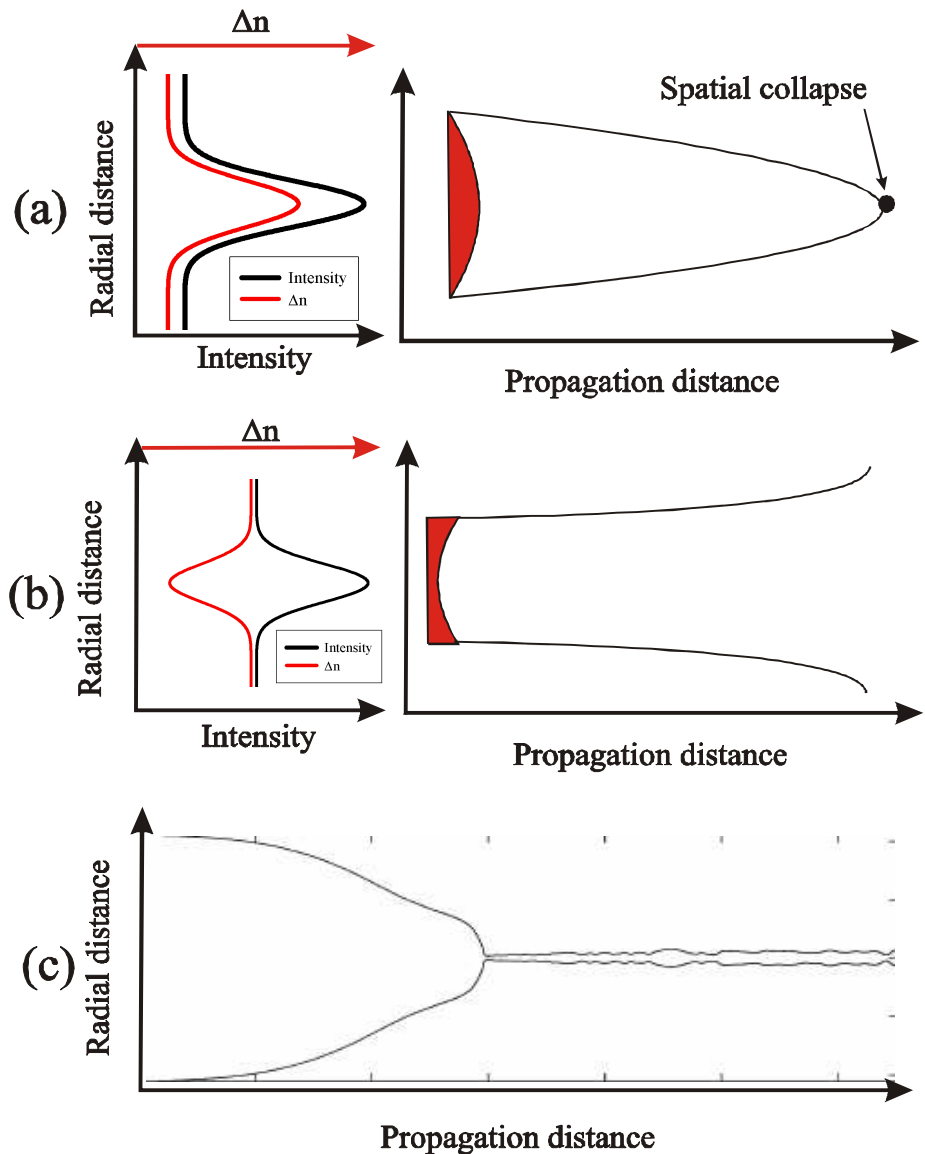

FIGURE 6.25 Illustration of self-focusing (a), plasma defocusing (b) and a numerical simulation of filamentation (c).[17]

It can be shown that, as a result of the intensity dependent index, self-focusing exhibits a power threshold rather than an intensity threshold by considering a collimated beam with sufficient power to self focus inside a transparent material. If the diameter of the laser beam is doubled, the laser intensity drops by a factor of four resulting in a four-fold reduction of refractive index change. However, the area of the self focusing lens has also increased by a factor of four. This increase in area compensates for the decrease in refractive index change, giving the same refractive power.[18, 271] The power for critical self-focusing ($P_{cr}$) can be calculated using the expression given by Marburger[18, 272]

$$P_{cr} = \frac{3.77\lambda^2}{8\pi n_0 n_2} \qquad (6.15)$$

GLS has $n_0$ and $n_2$ of 2.41 and $2.16 \times 10^{-14}$ cm$^2$/W respectively[240] which gives $P_{cr}$ = 1.84 kW. This compares to $P_{cr}$ = 1.5 MW calculated for borosilicate glass.[271] The irradiance required for the optical breakdown ($I_{ob}$) of most dielectrics is around $1 \times 10^{13}$ W/cm$^2$.[238, 268] Assuming that $I_{ob}$ for GLS is comparable to that of other dielectrics, the relation $I_{ob} = P_{cr}/\pi\omega_0^2$ allows an estimation of the maximum focus waist acceptable to induce optical breakdown without critical self-focusing occurring, which gives a focus waist of 242 nm. This is much smaller than the actual focus waist inside the glass and indicates that the self focusing threshold is lower than the optical breakdown threshold. The inability to obtain material modification without self-focusing presents a problem for the formation of symmetric waveguides. The new beam waist produced by self-focusing is expected to be at a distance $z_f$ from the original focus point.[220]

$$z_f = \frac{2n_0\omega_0^2}{0.61\lambda\sqrt{P/P_{cr}}} \qquad (6.16)$$

Where $\omega_0$ is the focus beam waist and P is the laser power. The lowest pulse energy used, 0.21 μJ, corresponds to a peak power of ~ 1.4 MW which in turn corresponds to $z_f$ of ~ 1 μm. This indicates that self focusing should cause little change in the position of the focus waist even at the lowest pulse energy. This is confirmed by the observation of no change in the focus waist position with varying pulse energy.

In order to avoid self-focusing a longer wavelength could be used to increase the power for critical self-focusing, see equation 6.15. Because the threshold for optical breakdown depends on laser intensity and the threshold for self-focusing depends on laser power, tighter focusing of the beam may allow the intensity for optical breakdown to be reached before the critical power for self-focusing. However it is unclear whether non-linear absorption will occur at longer wavelengths and the high refractive index of GLS limits the focus waist that can be achieved. A possible way of avoiding the inherent asymmetry of waveguides, that happens when self focusing occurs would be to use parallel writing geometry. In parallel writing geometry the sample is translated in a direction parallel to the propagation direction of the writing laser beam. This method has been used to fabricate symmetric waveguides in aluminosilicate glass,[259] phosphate glass,[236] silicate and borosilicate glass.[50] This method is therefore proposed as further work. However, the length of the waveguides may be limited because of the linear absorption of the fs pulse that is believed to occur before it reached a focus, as discussed in section 6.2.1. Another method to fabricate symmetric waveguides could be to use two writing beams and form the waveguide from a densified region in-between the writing beams.

### 6.3.3 Refractive index change

The physical origin of refractive index change produced in glass by fs laser exposure is not well understood. [50, 273] It is assumed that in the focal spot local rapid heating occurs. Heating the glass and freezing it while its temperature is still high could produce a lowering of the refractive index.[50] Is has been proposed that the plasma induced by a fs pulse in glass expands rapidly, creating a shock wave or micro-explosion.[274] This shockwave could compress the glass and increase the refractive index.

A rapid quenching model has been proposed to explain index modification in glasses after fs laser exposure.[275] This model assumes that, after a high temperature plasma is formed by the laser pulse it cools rapidly, freezing the glass in the same structure that it had at high temperature. The temperature at which a glass structure becomes frozen is called the fictive temperature ($T_f$). The concept of fictive temperature was proposed by Tool[276] who suggested that, depending on the cooling rate, glass has a frozen in structure that corresponds to some temperature (the fictive temperature) of an equilibrium liquid.[277]

In silica glass fs laser exposure induces a positive index change [50, 263] which is explained by the rapid quenching model as rapidly quenched silica is known to have a positive index change.[41, 277] In phosphate glass, fs laser exposure induces a negative index change[236] which is explained by the rapid quenching model because rapidly quenched phosphate glass is known to have a negative index change.[236] Negative index change after fs laser exposure has also been observed in borosilicate, heavy metal oxide and lanthanum-borate glasses.[273] The dependence of refractive index on fictive temperature has not been measured in GLS. However, since measurements in section 6.2.3 indicate that fs exposure of GLS can result in a negative index change it is proposed that the refractive index of GLS has a negative dependence on fictive temperature; in other words the higher the temperature GLS is rapidly quenched from, the lower the resulting refractive index will be. Also according to the Lorenz-Lorentz relationship, compaction increases and expansion decreases the refractive index of glasses.[278, 279]

Other authors have suggested a colour centre model to explain index modification in glasses after fs laser exposure.[50, 221, 280] This model stipulates that fs laser pulses introduce defects called colour centres in sufficient numbers and strength to alter the index through a Kramers-Kronig mechanism. Waveguide transmission measurements, see section 6.2.5, indicate that the maximum index change due to a Kramers-Kronig mechanism in fs laser written GLS waveguides is $\sim 5 \times 10^{-4}$ which is a factor of 20 smaller than the index change measured by QPM in section 6.2.3. This indicates that colour centre formation does not play a significant roll in the index modification of fs laser written GLS waveguides.

Positive index change resulting from densification has been attributed to many fs laser written waveguides in glasses[238, 273, 280, 281]. Laser induced compaction of GLS glass has been shown to result in a positive index change,[282] which has been related to the rather open three-dimensional structure of GLS that can be rearranged to a more compact structure by laser exposure.[278]

Based on the above discussion it is proposed that the negative index change of region 1 in A-type waveguides is the result of it having a high fictive temperature; the positive index change of region 2 in A-type waveguides is the result of densification. Because of the lower intensity used in their formation, the positive index change in B-type waveguides may simply be analogous to the positive index change observed in CW, UV written waveguides in GLS.[282]

The maximum observed index change of ~ 0.01 in fs laser written waveguides in GLS (this work) compares to 0.08 in $As_2S_3$ glass[258], $3.5 \times 10^{-3}$ in silica glass [50] and $4 \times 10^{-3}$ in borosilicate glass[50]

### 6.3.4 Waveguide formation mechanism

The form and structure of B-type waveguides indicates that they are formed by filamentation. However, as the pulse energy is increased and type-A waveguides are formed filamentation is no longer observed. This indicates that there is a threshold at the transition between the formation of B-type and A-type waveguides ($E_{BA}$) where a large increase in plasma density breaks the equilibrium between self focusing and plasma defocusing; $E_{BA}$ is estimated to be around 0.2 µJ/pulse.

This indicates, along with the guided mode and reflection optical micrograph findings in section 6.2.2 and the discussion of index modification in section 6.2.3, that the A-type waveguide structure is formed by a similar mechanism that occurs in phosphate glass[236] and sodium calcium silicate glass[275]. In these glasses the fs laser beam induces a modified region in its focal volume that has a lower density and refractive index than the initial glass; this exposed region did not guide light and it was only regions peripheral to the exposed region that guided light. It is therefore proposed that region 1 was formed by exposure to the focused fs laser beam where a high density plasma was formed. The waveguide structure that actively guides light (region 2) was formed by movement of glass from the fs laser exposed region in a shock wave that resulted in a region of higher density and higher refractive index. The B-type waveguide structure was formed through filamentation and the plasma density was not high enough to induce a negative refractive index change. These mechanisms for the formation of A-type and B-type waveguides are illustrated in figure 6.26.

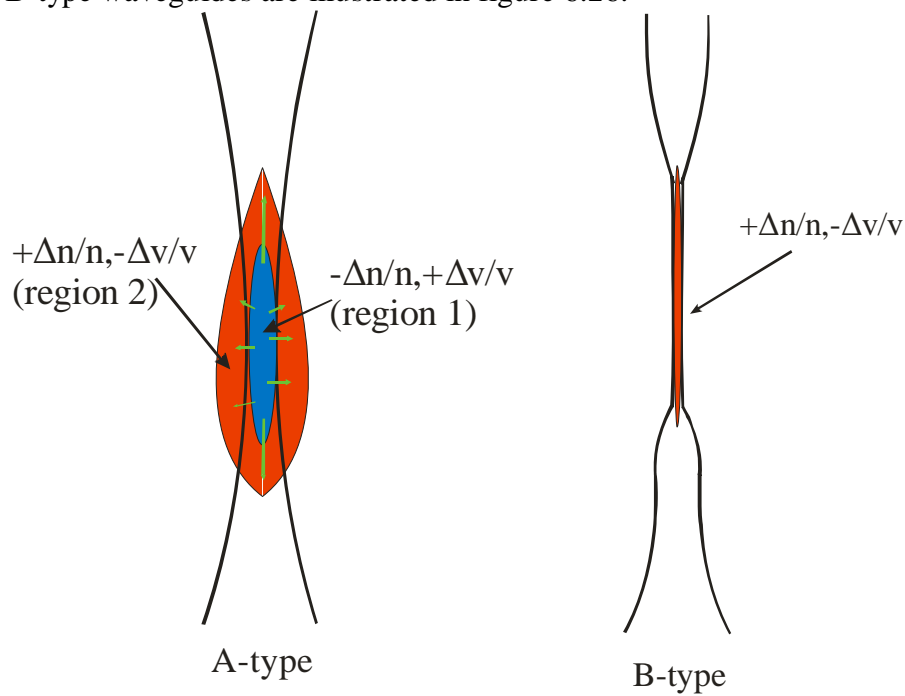

FIGURE 6.26 Waveguide formation mechanism in GLS at pulse energies > ~0.2 µJ/pulse (A-type) and GLS at pulse energies < ~0.2 µJ/pulse (B-type).

### 6.3.5 Non-linear absorption

Several mechanisms for nonlinear absorption of fs pulses have been considered in the literature; the principal ones are multiphoton ionisation (MPI), tunnelling and avalanche ionisation.

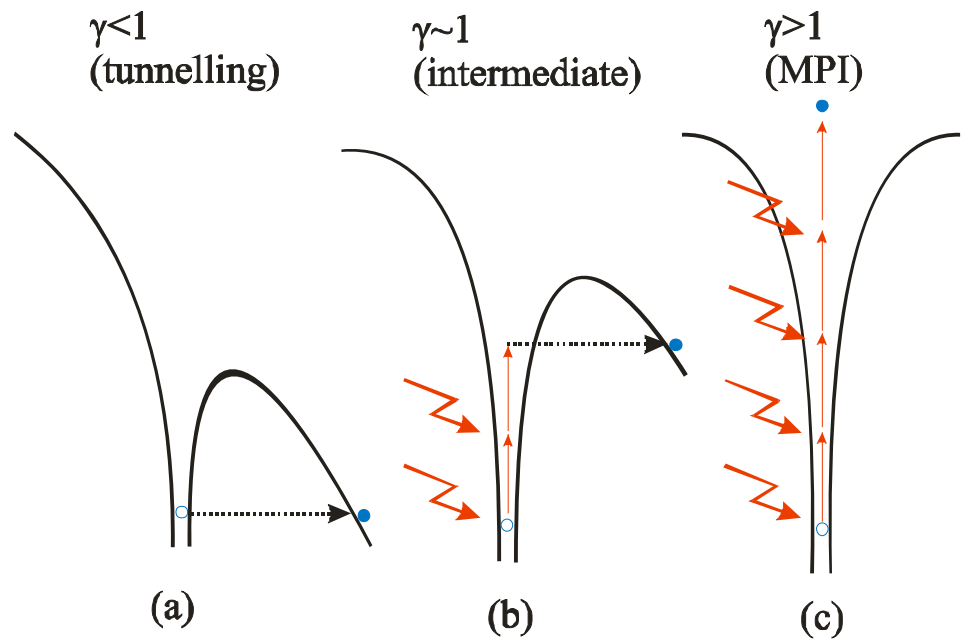

FIGURE 6.27 Schematic diagram of the photoionisation of an electron in a Coulomb well for different values of the Keldysh parameter leading to tunnelling (a), an intermediate scheme (b) and multi-photon ionisation (c); after [18].

In tunnelling ionisation, the high electric field of the laser pulse deforms the Coulomb well that binds a valence electron to its parent atom. If the electric field is strong enough the Coulomb well can be deformed enough that the bound electron tunnels through the short barrier and becomes free, as shown in figure 6.27(a). In a solid the electron is promoted from the valence to the conduction band, rather than ionised.[18] In multi-photon ionisation (MPI), shown in figure 6.27(c) a valence electron is promoted to the conduction band via the simultaneous absorption of several photons having satisfied the condition that the number of photons absorbed (m) times the photon energy is equal to or greater than the band-gap of the material.[18] In the intermediate case tunnelling and MPI occur simultaneously, as shown in figure 6.27(b). The ionisation rate in the multi-photon ionisation regime ($W_{MPI}$) is given by equation 6.17[50]

$$W_{MPI} = \sigma_m I^m \qquad (6.17)$$

Where $\sigma_m$ is the multi-photon absorption coefficient for absorption of m photons and I is the laser intensity. The ionisation rate in the tunnelling regime is given by equation 6.18[283]

$$W_{tun} = \frac{2}{9\pi^2} \frac{\Delta}{\hbar} \left( \frac{m_{red}\Delta}{\hbar^2} \right)^{3/2} \left( \frac{e\hbar E}{m_{red}^{1/2}\Delta^{3/9}} \right)^{5/2} \times \exp \left\{ -\frac{\pi}{2} \frac{m_{red}^{1/2}\Delta^{3/2}}{e\hbar E} \left( 1 - \frac{1}{8} \frac{m_{red}\Omega^2\Delta}{e^2 E^2} \right) \right\} \; (6.18)$$

Where $\Omega$ is the laser frequency, $m_{red}$ is the reduced effective mass, $\Delta$ is the bandgap energy which is 2.28 eV for GLS and 2.48 eV for GLSO,[240] e is the electron charge and E is the peak electric field intensity. The reduced effective mass is related to the effective mass in the conduction band ($m_c$) and valence ($m_v$) band, which are both assumed to be equal to the free electron mass, by: $1/m_{red} = 1/m_c + 1/m_v$.

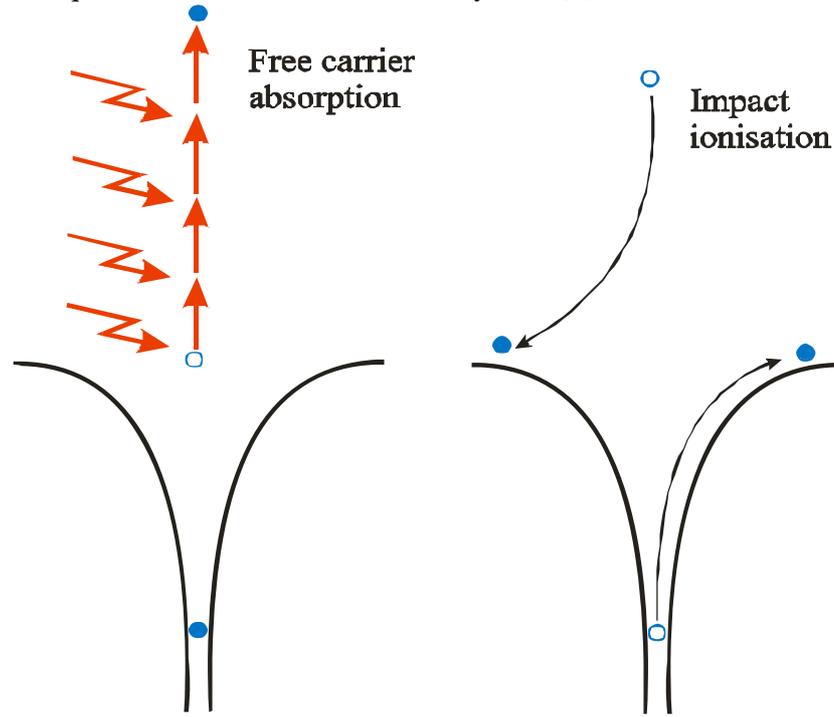

FIGURE 6.28 Schematic diagram of avalanche ionisation, after[18].

Avalanche ionisation involves free carrier absorption followed by impact ionisation. In free carrier absorption, illustrated in the left panel of figure 6.28, an electron already in the conduction band moves to higher energy states in the conduction band by linearly absorbing several laser photons sequentially.[18] Once the electrons' energy exceeds the conduction band minimum by more than the band-gap energy, the electron can collisionally ionise another electron from the valence band such than both electrons end up in the conduction band;[18] this is illustrated in the right panel of figure 6.28. As long as the laser field is present, the electron density (N) due to avalanche ionisation in the conduction band grows according to[284]

$$\frac{dN}{dt} = \alpha_{ai} I(t) N \qquad (6.19)$$

where $\alpha_{ai}$ is the avalanche ionisation coefficient and I is the laser intensity. Thus the total ionisation rate is given by[18, 248, 285]

$$\frac{dN}{dt} = W_{PI}(I, \Omega, \Delta) + \alpha I(t)N - W_{Loss} \qquad (6.20)$$

Where $W_{PI}$ is the photoionisation rate, generalised for either multi photon absorption or tunnelling, and $W_{Loss}$ is the loss rate due to electron diffusion and recombination. It has been proposed by several authors that avalanche ionisation is of little importance in the absorption of sub-picosecond pulses in transparent materials.[50, 285, 286] However, other authors have proposed that it does[18, 284]. Therefore, in the absence of another explanation, it is proposed that the expected rapid increase in plasma density that breaks down the dynamic equilibrium between self focusing and plasma defocusing (discussed in section 6.3.2) and causes a transition from A-type to B-type waveguides is instigated by avalanche ionisation.

In order to determine whether MPI or tunnelling dominate the nonlinear absorption the Keldysh parameter (γ) can be calculated using equation 6.21.[50, 283, 286, 287]

$$\gamma = \frac{\Omega\sqrt{m_{red}\Delta}}{eE} \qquad (6.21)$$

For γ>1 MPI is dominant, for γ < 1 tunnelling is dominant. The waveguides investigated in this study were written with pulse energies of 0.21 to 1.74 µJ which corresponds to a Keldysh parameter (γ) of 0.11 to 0.04; this indicates that tunnelling was the dominant nonlinear absorption process in the formation of all the waveguides investigated in this study.

## 6.4 Spectral broadening

### 6.4.1 Introduction

Optical communication channel rates exceeding 40 Gbit/s are difficult to achieve with high speed electronics. Higher channel data rates (> 100 Gbit/s) require optical time division multiplexing (OTDM) solutions. Such OTDM systems rely on all optical switching (AOS). Ultra-fast optical switches require a highly nonlinear material with an extremely fast response time.[35] The nonlinear material in an ultra-fast optical switch is typically required to produce a nonlinear phase shift near $\pi$ rad, with a switching energy $E_s$ of less than 1 pJ, while maintaining a low loss.[35, 288] This section details the spectral broadening of 1550 nm 200 fs pulses in GLS waveguides and demonstrates that these waveguides may have applications for AOS applications.

### 6.4.2 Experimental setup

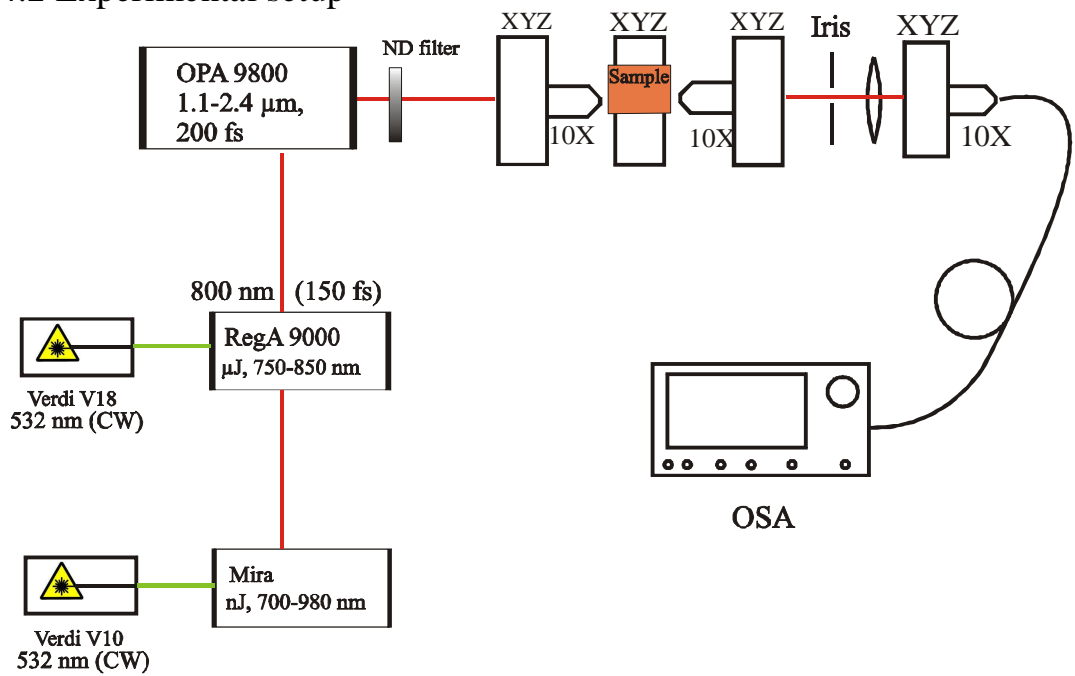

FIGURE 6.29 Schematic of the experimental setup used to measure ultra short pulse broadening in GLS waveguides.

Figure 6.29 shows the experimental setup used to take spectral broadening measurements. The operation of the Coherent RegA 9000 is described in more detail in section 6.2.1. The output of the Coherent RegA 9000 was coupled into a Coherent optical parametric amplifier (OPA) 9800 which has a tuning range of 1.1 to 2.4 µm, producing pulses with a duration of around 200 fs. In the OPA 9800 ~µJ seed pulses from the RegA 9000 are split into two beams. One beam is used to produce white light continuum seed pulses from a sapphire crystal, these are then amplified by a phase matched β barium borate (BBO) OPA crystal pumped by the other beam from the RegA 9000. The pulse energy of the output from the OPA was varied with a variable neutral density filter then the average power was measured with a Coherent Powermax PM10 thermal power meter head. The beam was coupled into the waveguides with 0.1 or 0.25 NA objective lenses, and then coupled out of the waveguides with a 0.25 NA objective lens. The guided modes were isolated with an iris then coupled into a single mode silica fibre with a 0.1 NA lens and a 0.25 NA objective lens. The beam was then detected with an Ando AQ6315A optical spectrum analyser (OSA) which had a detection range of 350 to 1750 nm and was set to a resolution of 5 nm.

Measurement of the beams' average power before it was coupled into the waveguides, i.e. the power incident on the waveguide ($P_{inc}$), was found not to accurately represent the power coupled into the waveguides. This is because the NA, cross section area and asymmetry of the various waveguides examined varied, this in turn varied the coupling efficiency of the waveguides considerably. Therefore it was decided to measure the output power from the waveguides ($P_{out}$). The power coupled into the waveguides ($P_{in}$) was estimated from $P_{out}$ by taking into account reflections from the waveguide's output end face and the loss of the waveguide. The waveguide losses are calculated in section 6.2.6. Since the beam shifted in wavelength and broadened after passing through the waveguides it was decided to calculate the output power from the waveguide output spectra $I_{out}(\lambda)$ from the OSA. This is because the available power meters required calibration to one particular wavelength and they had a lower sensitivity than the OSA. $P_{out}$ was calculated from the ratio of the number of photons in $I_{out}(\lambda)$ to the number of photons in a spectrum of the laser beam that had not passed through the waveguides ($I_{ref}(\lambda)$) which had a power measured with a thermal power meter ($P_{ref}$). The number of photons (n) detected at a particular wavelength is proportional to the measured irradiance $I(\lambda)$ multiplied by the wavelength ($\lambda$), therefore $P_{out}$ was calculated with equation 6.22.

$$P_{out} = P_{ref} \frac{\int \lambda I_{out}(\lambda) d\lambda}{\int \lambda I_{ref}(\lambda) d\lambda} \qquad (6.22)$$

As illustrated in figure 6.30 $P_{out}$ was found to be a very nearly a linear function of $P_{inc}$, indicating that no two photon absorption was occurring. Taking into account the reflectivity (R) of GLS, which was 0.1632, and the propagation loss of the waveguides ($\Gamma$), $P_{in}$ was calculated using equation 6.23.

$$P_{in} = \frac{\Gamma P_{out}}{1 - R} \qquad (6.23)$$

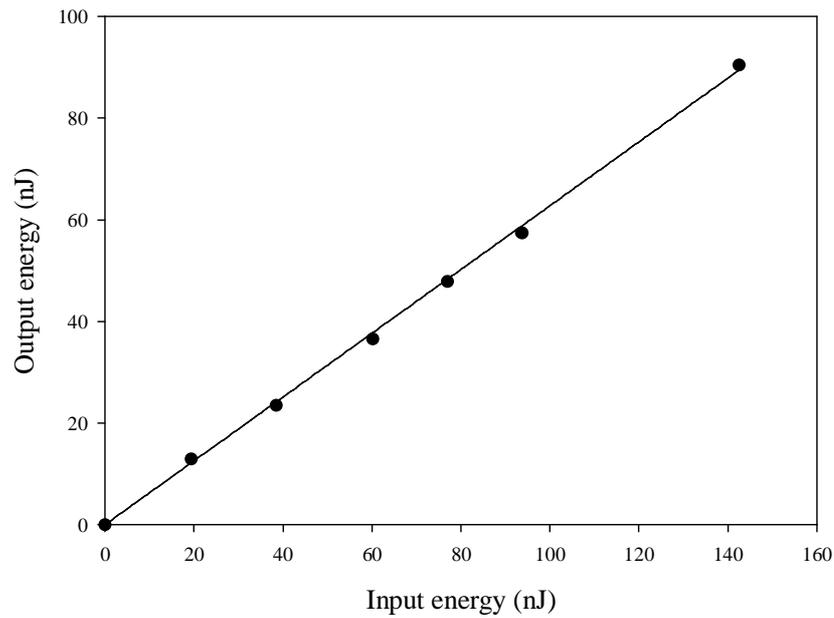

FIGURE 6.30 Output energy as a function of input energy for a waveguide written with a pulse energy of 1.26 μJ. The high linearity indicates negligible two photon absorption.

## 6.4.3 Broadened spectra

Figure 6.31 shows the spectra of 1540 nm laser pulses, with a 200 fs duration, after passing through 12 mm of GLS fs written waveguide, as a function of the pulse energy coupled into the waveguide. The waveguide was written with a pulse energy of 1.75 μJ and a translation speed of 200 μm/s and had a cross section size of ~ 100x300 μm, 0.25 NA objectives were used to couple the beam into and out of the waveguide. The spectrum of the laser beam before it had passed through the waveguide is also shown. At pulse energies up to 4.7 nJ the output spectrum is very similar to the input spectrum except for a small blue shift in peak position. At 18.8 nJ/pulse the peak position is red shifted and at 31.3 nJ/pulse the spectrum changes dramatically from a peak still resembling the input spectrum to a very broad almost flat spectrum. As the pulse energy is increased further broad double peak spectra are formed with one peak at roughly the same position as the input spectrum and the other at shorter wavelengths. The long wavelength and short wavelength peaks are referred to as peak 1 and 2 respectively. The apparent decrease in signal to noise ratio with increasing pulse energy is believed to be related to the broadening mechanism.

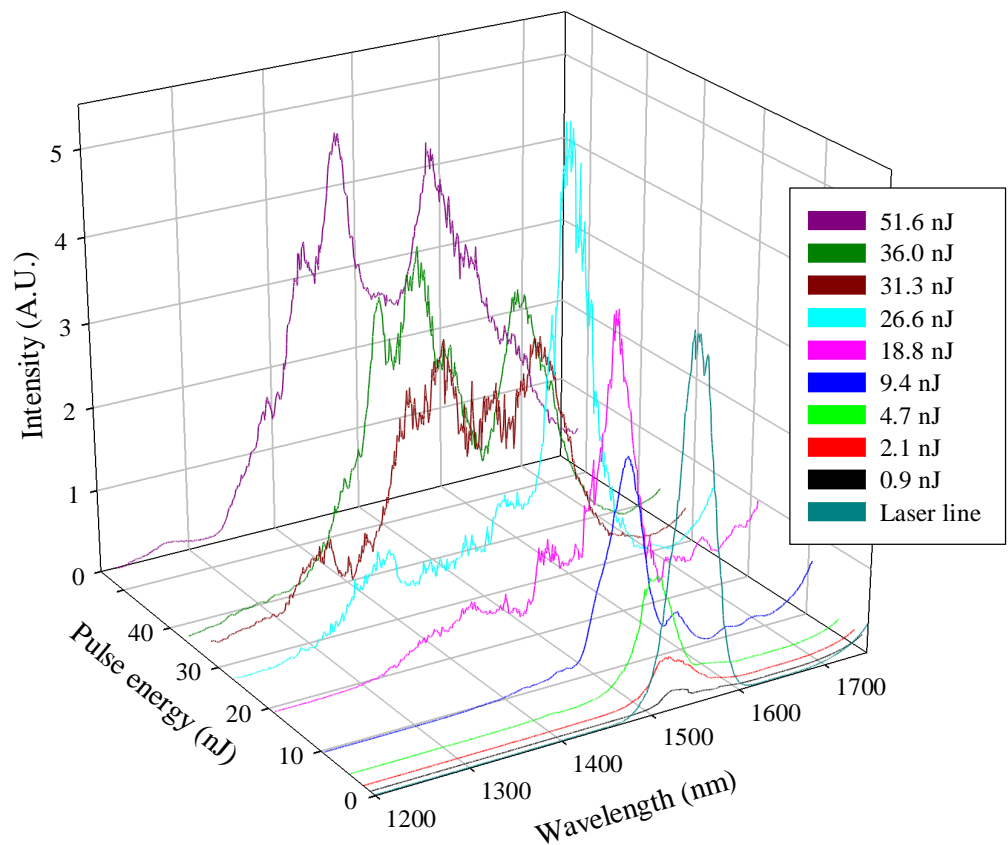

FIGURE 6.31 Spectra of 1540 nm 200 fs laser beam coupled into an 12 mm GLS waveguide, that was written with a pulse energy of 1.75 µJ, as a function of input beam pulse energy.

Figure 6.32 shows the peak positions and full width at half maxima (FWHM) of the spectra shown in figure 6.31, error bars for wavelength and pulse energy were estimated from the resolution of the OSA and the power meter respectively. The figure indicates that at ~ 20 nJ/pulse there is a red shift in peak position of around 40 nm. The FWHM remains at a relatively constant 50 nm up to a pulse energy of ~ 30 nJ/pulse where there is a relatively sudden increases in FWHM to ~200 nm. As the pulse energy is increased further, FWHM falls. This is attributed to the emergence of a slightly asymmetric double peak structure. At pulse energies > ~ 30 nJ/pulse, a double peak structure is evident with peaks at ~ 1570 and 1460 nm.

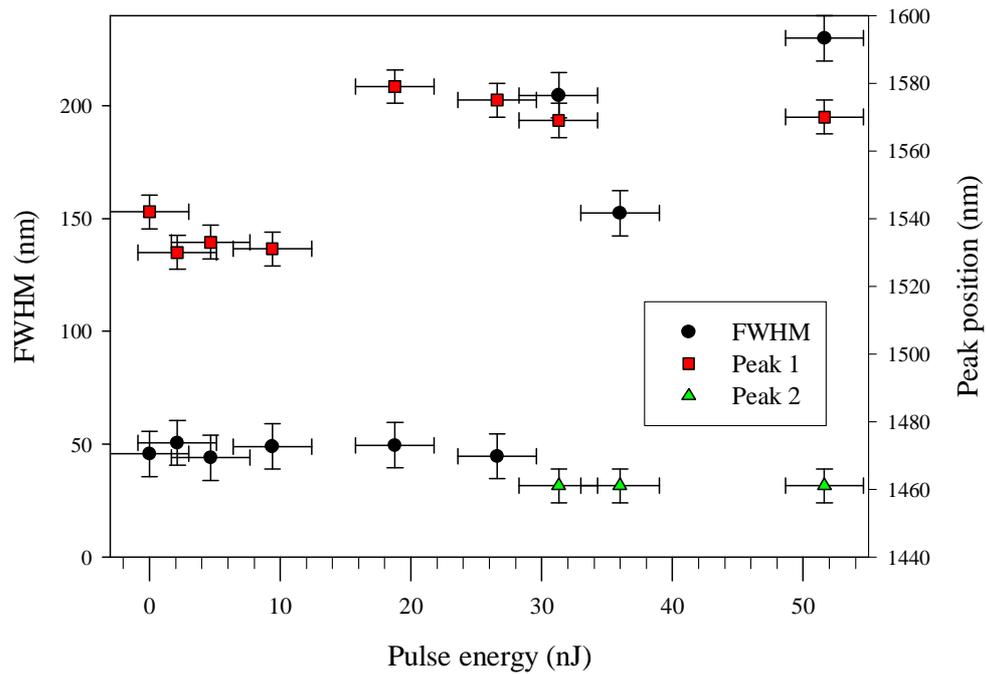

FIGURE 6.32 FWHM and peak position of spectra given in figure 6.31. The FWHM is attributed to the left vertical axis and the peak positions are attributed to the right vertical axis.

Figure 6.33 shows the spectra of 1540 nm laser pulses with a 200 fs duration, after passing through 12 mm of GLS fs written waveguide, as a function of the pulse energy coupled into the waveguide. The waveguide was written with a pulse energy of 1.26 μJ and a translation speed of 200 μm/s and had a cross-sectional size of ~ 100x220 μm. Comparisons with figure 6.31 firstly show that higher pulse energies were coupled into this waveguide, this is attributed to the higher symmetry and maximum index change of this waveguide. Unlike the spectra in figure 6.31 a double peak structure is evident at the lowest pulse energy of 1.1 nJ. At 36.8 nJ/pulse the double peak structure disappears then reappears at 56.9 nJ/pulse. At 87 nJ/pulse an asymmetric triple peak structure is evident. The triple peak is indicative of a 2.5π phase shift[267, 289, 290] and its weighting towards the long wavelength side of the pump is indicative of stimulated Raman scattering (SRS).[35, 291]

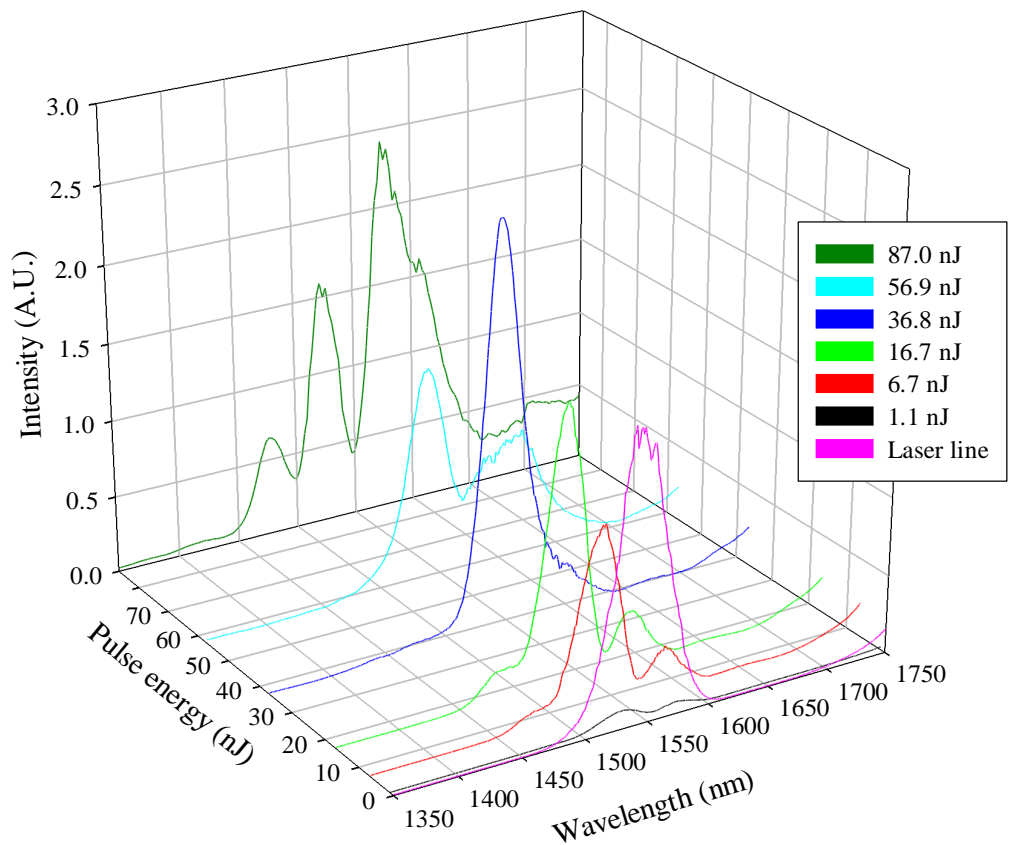

FIGURE 6.33 Spectra of 1540 nm 200 fs laser beam coupled into an 12 mm GLS waveguide written with a pulse energy of 1.26 µJ as a function of input beam pulse energy.

Figure 6.34 shows the peak positions and full width at half maxima of the spectra shown in figure 6.33. Compared to figure 6.32 there is no significant peak shift of the input spectrum. The onset of broadening occurs at ~ 60 nJ/pulse compared to ~30 nJ/pulse in figure 6.32. The onset of broadening also appears to be more gradual and reaches a maximum of ~ 100 nm compared to ~220 nm. The triple peak spectrum at 87 nJ/pulse has peaks at ~ 1530 nm (peak 1), 1580 nm (peak 2) and 1480 nm (peak 3). The stronger broadening effect in the waveguide in figure 6.32 is attributed to its larger cross section which means that it could support more higher order modes, that may take longer path lengths through the waveguide.

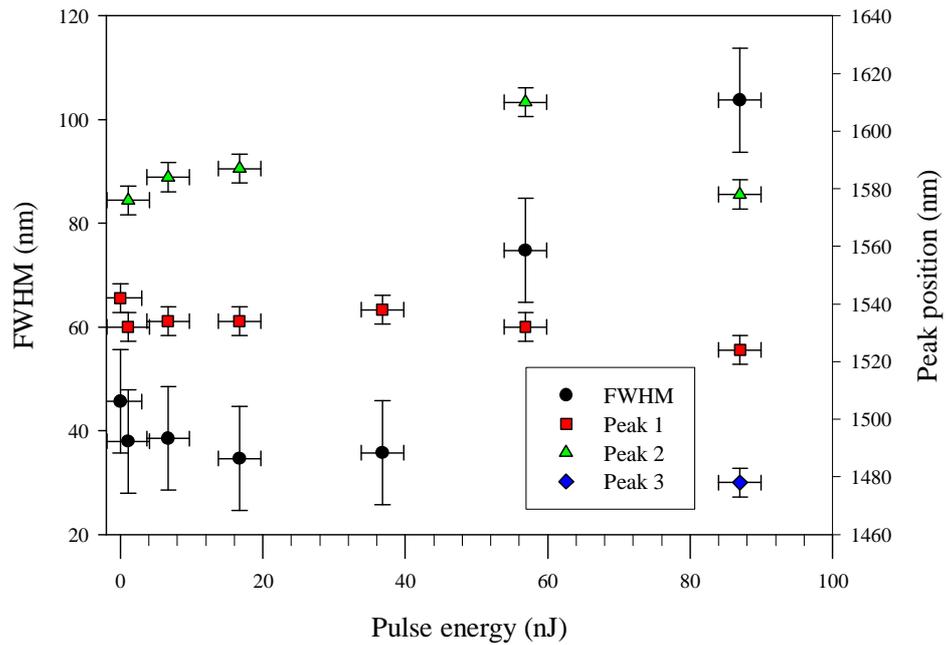

FIGURE 6.34 FWHM and peak position of spectra given in figure 6.33 The FWHM is attributed to the left vertical axis and the peak positions are attributed to the right vertical axis.

The spectra in figure 6.35 are from the same waveguide as in figure 6.33, except the input wavelength is 1600 nm instead of 1540 nm. A 0.1 NA, instead of a 0.25 NA, objective was used to couple in the beam, this allowed a higher pulse energy to be coupled into the waveguide. Similarly to figure 6.33, a triple peak spectrum is observed at 88 nJ/pulse indicating a $2.5\pi$ phase shift, however it is more symmetric indicating that SRS is lower at this wavelength. Compared to figure 6.33 the evolution of the triple peak spectra is clearer.

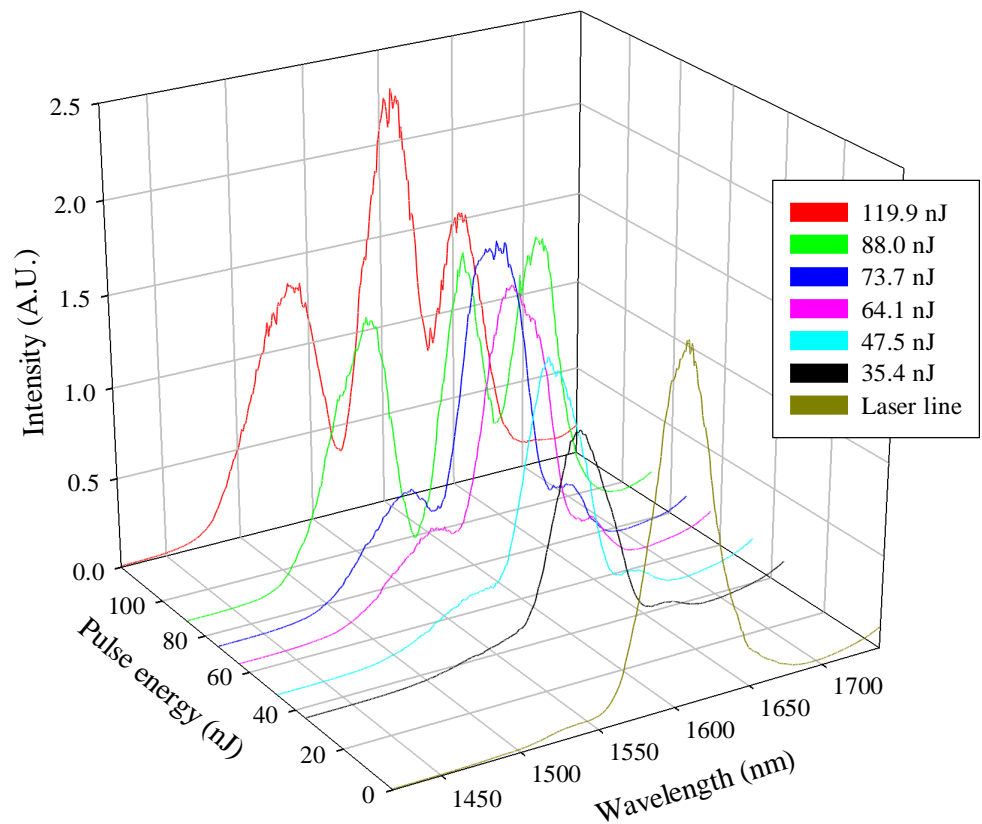

FIGURE 6.35 Spectra of 1600 nm, 300 fs laser beam coupled into an 11 mm GLS waveguide, that was written with a pulse energy of 1.26 μJ, as a function of input beam pulse energy.

Figure 6.36 shows the peak positions and full width at half maxima of the spectra shown in figure 6.35. The onset of broadening occurs at ~90 nJ/pulse compared to ~60 nJ/pulse for the 1540 nm input beam. Broadening measurements in 50 μm core diameter GLS fibre and single mode, UV written, GLS waveguides were attempted but sufficient pulse energies could not be coupled into these waveguides to make a comparison with the broadening observed in fs written waveguides.

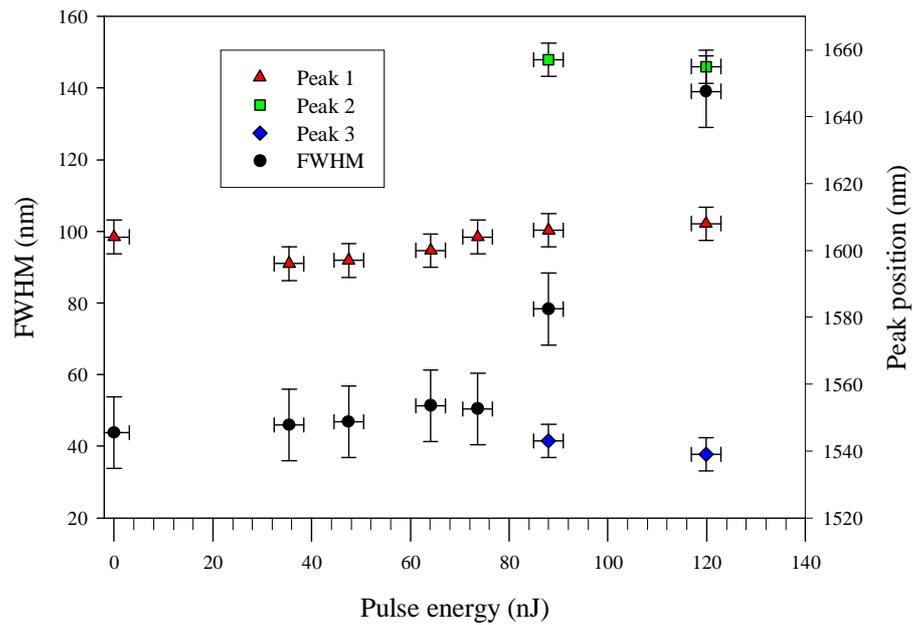

FIGURE 6.36 FWHM and peak position of spectra given in figure 6.35.

## 6.4.4 Discussion of spectral broadening

### 6.4.4.1 Switching energy

Significant broadening was only observed in waveguides written at 1.75 and 1.26 μJ, which had very large cross-sections. This is because the maximum coupling efficiency that could be achieved with waveguides written at lower pulse energies was up to 10 times lower than that achieved for the larger waveguides. This low coupling efficiency is attributed to the lower NA, smaller cross-sectional area and higher asymmetry of waveguides written at lower pulse energies. The lowest input pulse energies for the onset of peak shift and broadening are ~20 nJ which is much higher than the sub pJ pulse energy required for all optical switching applications. However the switching energy $E_s$ can be calculated using equation 6.24[288]

$$E_s = \frac{C\lambda\tau A_{eff}}{n_2 L} \qquad (6.24)$$

Where $A_{eff}$ is the effective cross-sectional area of the waveguide, $\lambda$ is the wavelength of light, $\tau$ is the switching pulse width (FWHM), C is a constant that depends on pulse shape (0.56 for sech-shaped pulses) and L is the waveguide length. The high $n_2$ of GLS indicates that it is a good candidate material for all optical switching applications. The effective cross-sectional area of the waveguides examined in this work could be significantly reduced and their length could be significantly increased, this could bring $E_s$ closer to that required for AOS applications.

### 6.4.4.2 Self phase modulation

Self phase modulation (SPM) is a manifestation of the intensity dependence of the refractive index in nonlinear media and leads to the spectral broadening of optical pulses.[267] SPM is the temporal analogue of self focusing (see section 6.3.2). SPM causes a phase shift between the peak of an optical pulse and its low intensity leading and trailing edges, resulting in the spectral broadening of the pulse. The nonlinear phase shift induced by SPM is given by[243]

$$\phi_{SPM} = \frac{2\pi L_{eff}}{\lambda} \frac{n_2 P_0}{A_{\text{mod}}} \qquad (6.25)$$

Where $P_0$ is the peak pulse power, $n_2$ is the nonlinear refractive index, $A_{\text{mod}}$ is the effective mode area and $L_{eff}$ is the effective length, defined as

$$L_{eff} = \frac{1 - e^{-\alpha L}}{\alpha} \qquad (6.26)$$

Where $\alpha$ is the loss coefficient and L is the waveguide length. For the maximum pulse energy of ~ 100nJ, used in figure 6.35, and approximating $A_{\text{mod}}$ to the cross-sectional area of the waveguide, the phase shift calculated using equation 6.25 is ~$\pi$. However, the value of $A_{\text{mod}}$ may be an overestimate because the guided mode is confined to certain regions of the waveguide therefore the phase shift calculated using equation 6.25 may be an underestimate.

The number of peaks and the extent of the broadening in an SPM broadened spectrum are dependent on the magnitude of the nonlinear phase shift and increase with it linearly.[243] The oscillatory structure of the broadened pulses in figures 6.31, 6.33 and 6.35 are indicative of SPM. The maximum phase shift $\Phi_{\text{max}}$ of an SPM broadened pulse is given approximately by [267, 289]

$$\phi_{\text{max}} \approx \left( M - \frac{1}{2} \right) \pi \qquad (6.27)$$

Where M is the number of peaks in an SPM broadened spectrum. Equation 6.27 and a comparison with other SPM broadened spectra[267, 290] indicate that the double peak structure in figure 6.31 demonstrates a 1.5$\pi$ phase shift, at 36 nJ/pulse, and the triple peak structure in figure 6.33 and 6.35 demonstrates a 2.5$\pi$ phase shift, at 87 and 88 nJ/pulse respectively. These nonlinear phase shifts compare to a maximum phase shift of 1.6$\pi$ and 3.5$\pi$ in 2.03 cm[288] and 2.8 cm[35] long photodarkened planar $Ge_{0.25}Se_{0.75}$ glass waveguides respectively. The spectral bandwidth of pulses broadened by SPM ($\Delta\omega_{SPM}$) is given by[246]

$$\Delta\omega_{SPM} = \frac{\Delta\omega_0 2\pi n_2 P_0 L}{\lambda A_{eff}} \qquad (6.28)$$

Where $\Delta\omega_0$ is the bandwidth of the input pulse. For the maximum pulse energy of 50 nJ, used in figure 6.31, the bandwidth for pulses broadened by SPM is ~100 nm, however the observed bandwidth of the broadened pulse was ~ 200 nm. This extra broadening is attributed to stimulated Raman scattering.

For pulse widths < 100 ps the combined effect of group velocity dispersion (GVD) and SPM can become significant.[267] GVD causes a short pulse of light to spread in time as a result of different frequency components of the pulse traveling at different velocities.[267] An effect of the interplay between GDV and SPM on the pulse spectrum is to reduce the depth of the minima that would be expected in a pulse spectrum broadened by SPM alone.[267] This is evident in the spectra in figure 6.31, but not in figure 6.33 and 6.35. A measurement of the pulse width after passing through the waveguides would provide more information on the effect of GVD, however this measurement requires an autocorrelator which was not available at the time. Therefore pulse-width measurement is suggested as further work.

### 6.4.4.3 Stimulated Raman scattering

Stimulated Raman scattering (SRS) is an important nonlinear process that can be exploited to construct broadband fibre based Raman amplifiers and tuneable Raman lasers.[243] At low intensities, spontaneous Raman scattering can transfer a small fraction (typically ~$10^{-6}$) of power from one optical field to another, whose frequency is downshifted by an amount determined by the vibration modes of the medium.[267] The downshifted radiation is called the Stokes wave. The spontaneous Raman effect is also discussed in section 3.3.6. At high pump intensities the nonlinear phenomenon of SRS can occur in which the Stokes wave grows rapidly inside the medium such that most of the pump energy is transferred to it.[267] The SRS generated Stokes wave intensity $I_s$ for a CW, or quasi CW, condition is given by[243]

$$I_s(L) = I_s(0)\exp(g_r I_p(0)L_{eff} - \alpha_s L), L_{eff} = \frac{1 - e^{-\alpha_p L}}{\alpha_p} \qquad (6.29)$$

Where L is the waveguide length, $I_s(0)$ is the Stokes intensity at L=0, $\alpha_s$ and $\alpha_p$ are the waveguides' loss coefficients at the Stokes and pump wavelengths respectively, $L_{eff}$ is the effective length, defined by $\alpha_p$, and $g_r$ is the Raman gain coefficient. In general $g_r$ depends on composition and dopants in the waveguide material and is inversely proportional to pump wavelength.[243] For pulses with widths below 100 ps, the length of the waveguide can exceed the walk-off length, $L_w$, defined as[267]

$$L_w = \frac{T_0}{\left| v_{gp}^{-1} - v_{gs}^{-1} \right|} \qquad (6.30)$$

Where $v_{gp}$ and $v_{gs}$ are the group velocities of the pump and Stokes pulses respectively and $T_0$ is the duration of the pump pulse.

The Raman threshold is defined as the pump power at which the Stokes power becomes equal to the pump power at the waveguide output. Assuming a Lorentzian shape for the Raman gain spectrum and a CW or quasi CW condition, the critical power $P_{cr}$ required to reach the Raman threshold, to a good approximation, is given by.[292]

$$P_{cr} \approx \frac{16 A_{mod}}{g_r L_{eff}} \qquad (6.31)$$

For the short pulse condition, equation 6.31 holds provided the effective length is taken to be: $L_{eff} = \sqrt{\pi} L_w$.[267]

In silica fibre the SPM broadened pump spectra became broadened with an asymmetry weighted towards the long wavelength end of the pump spectrum once the pump power passes the Raman threshold. This is similar to the effect observed at the highest pulse energy in figure 6.33. It is therefore proposed that the asymmetry of the triple peak structure in figure 6.33 when the input beam was 1540 nm is cause by the peak power exceeding the Raman threshold. The higher symmetry of the SPM broadened triple peak structure, in figure 6.35, when the input beam was 1600 nm indicates that the Raman threshold was not reached in this case. This can be explained by examining equation 6.31 and considering that $g_r$ is inversely proportional to input beam wavelength, therefore at longer input beam wavelengths the Raman threshold will be higher.

## 6.4.4.4 Device applications

Two of the nonlinear optical devices, mentioned in section 6.1.3, are considered here as possible applications of the pulse broadening effect discussed in this section. Because the birefringent properties of the waveguides have not been studied, optical Kerr shutters are not discussed here. The different broadening behaviours observed for waveguides written at different pulse energies indicate that waveguide writing parameters could be used to customise the broadening properties of a waveguide for a specific application. The fabrication of passive optical components such as Fresnel zone plates and fibre attenuators, as mentioned in section 6.1.2, has been demonstrated in silica but not in a highly nonlinear glass such as GLS, this is therefore suggested as further work. The methods of waveguide fabrication presented in this chapter, could be used to construct a ring resonator[293] which could have applications as optical filters[294] and all optical switching devices.[295] This is therefore suggested as further work.

## 6.4.4.5 Mach-Zehnder interferometer switch

The operation of these devices is described in section 6.1.3.1; as discussed here a $\pi$ phase shift is required for the operation of these devices. Calculations, in section 6.4.4.2, indicate that a 2.5 $\pi$ phase shift was demonstrated with a 87 nJ pulse. Therefore the phase shift demonstrated in these waveguides is more than adequate for the operation of a Mach-Zehnder interferometer switch. Many of the Mach-Zehnder interferometer (MZI) switches demonstrated so far use a semiconductor optical amplifier (SOA) to obtain a $\pi$ phase-shift. The slow relaxation time (several hundred picoseconds) of the SOAs has been cancelled out by using a symmetric MZI switching arrangement. In this switch, two SOAs, one in each arm of the interferometer, are excited by short control pulses with an appropriate time delay.[241] Switching times of 1.5 and 8 ps have been demonstrated for symmetric MZIs using InGaAsP [241] and GaAs [296] SOAs respectively. The nonlinearity of the GLS waveguides is expected to be characterised by a nearly instantaneous (fs) response time.[247] This inherent high switching speed means that MZI switches based on GLS waveguides would not be limited to a symmetric arrangement, as SOA based MZI switches are for high switching speeds. The nonlinear FOM required for a MZI is ~5, [297] this indicates that GLS is one of the few glasses suitable for a MZI, based on this constraint.

### 6.4.4.6 2R regenerator

The operation of these devices is described in section 6.1.3.3; this device relies on SPM broadening of a signal pulse. Equation 6.28 indicates that SPM bandwidth should increase linearly with pulse energy, however, figure 6.32 indicates a relatively sharp threshold for the onset of broadening. The reason for this is unclear, but it indicates that the transfer function for a device using this waveguide would have a closer to ideal transfer function than a device using waveguides that had a linear increase in SPM broadening with pulse energy; an ideal transfer function is a step increase in output intensity with input intensity. 2R regenerators have been demonstrated using 3.3 m of silica holey fibre with a peak power x length product of 0.13 W km[298] and in 2m of bismuth oxide fibre.[28] The large broadening observed at 30 nJ/pulse in figure 6.32 indicates that a 2R regenerator based on this particular GLS waveguide could operate with a peak power x length product of 1.5 W km, however this is expected to be greatly improved by reducing the cross-sectional area of the waveguide.

### 6.5 Conclusions

A formation mechanism is presented for fs laser written waveguides in GLS glass, based on optical characterisation and comparisons to previous work. Two different forms of waveguide have been identified and are referred to as A-type and B-type. B-type waveguides have a characteristic long narrow structure and are formed through filamentation. A-type waveguides have a characteristic "teardrop" structure, with a central region (region 1) that has undergone a negative reflective index change through exposure to the focused fs laser beam, and an outer region (region 2) that has undergone a positive index change. A-type waveguides are formed at pulse energies > ~0.2 μJ and B-type waveguides are formed at pulse energies < ~0.2 μJ. The negative index change, in region 1, results from rapid quenching of a high temperature plasma formed by the fs laser pulse which resulted in this region having a high fictive temperature. The positive index change, in region 2, resulted from movement of glass from the region 1 in a shock wave, that resulted in a region with a higher density and refractive index than unexposed glass. Only region 2 was found to actively guide light.

Tunnelling has been identified as the nonlinear absorption mechanism in the formation of the waveguides, by calculation of the Keldysh parameter. Single mode operation at 633 nm has been demonstrated. The writing parameters for the minimum achieved loss of 1.47 dB/cm are 0.36 μJ pulse energy and 50 μm/s scanning speed. A maximum index change of 0.01 has been observed.

Spectral broadening, from an initial FWHM of 50 nm to 200 nm, has been demonstrated with a 1540 nm, 200 fs pulse, at a pulse energy of 30 nJ, in a waveguide written at a pulse energy of 1.75 μJ. A change in peak position to 1580 nm was observed at 20 nJ/pulse. A maximum phase shift of 2.5π has been demonstrated at a pulse energy of 88 nJ/pulse. The broadening has been attributed to self phase modulation, with an asymmetry in some of the broadened spectra attributed to stimulated Raman scattering.

The high nonlinearity of GLS makes it a promising material for nonlinear optical devices. An interesting effect of this nonlinearity is the spectral broadening presented in this chapter. This broadening indicates that these waveguides may have applications in nonlinear optical devices, such as a Mach-Zehnder interferometer switch or a 2R regenerator. However, the high nonlinearity of GLS frustrates the fabrication of symmetric waveguides using fs pulses because the threshold for critical self focusing appears to be higher than the threshold for material modification.

## 6.6 Further work

Several methods for overcoming the inherent asymmetry of fs laser written waveguides in GLS have been proposed. These include using a higher NA objective, augmenting the beam profile using a slit or cylindrical lenses, using two writing beams and using a parallel writing geometry. Several methods for improving the minimum loss have been suggested. These include investigation of a greater range of translation speeds and pulse energies, using a double pulse fs laser to write the waveguides, varying the wavelength of the writing beam and annealing the sample after waveguide writing. Impressive spectral broadening was restricted to highly multimode waveguides, written at high pulse energies. Coupling sufficient pulse energy into smaller waveguides in order to observe spectral broadening is suggested as further work. A method for achieving this could be to use a tapered fibre to couple into the waveguide. Fabrication of other structures such as Fresnel zone plates and ring resonators is also suggested as further work.

# Chapter 7

# Summary and further work

## 7.1 Chalcogenide glasses

Chalcogenide glasses transmit to longer wavelengths in the IR than silica and fluoride glasses, they also often exhibit a low phonon energy, this allows the observation of certain transition in rare earth dopant that are not observed in silica. The low phonon energy of chalcogenides can be thought of as resulting from the relatively large mass of there constituent atoms and the relatively weak bonds between them. Chalcogenide glasses have a nonlinear refractive index around two orders of magnitude higher than silica. This makes them suitable for ultra-fast switching in telecommunication systems. The glass forming ability of gallium sulphide and lanthanum sulphide (GLS) was discovered in 1976 by Loireau-Lozac'h *et al.* GLS glasses have a wide region of glass formation centred about the $70Ga_2S_3 : 30La_2S_3$ composition and can readily accept other modifiers into their structure. This means that GLS can be compositionally adjusted to give a wide variety of optical and physical responses. GLS has a high refractive index of ~2.4, a transmission window of ~0.5-10 µm and a low maximum phonon energy of ~425 cm$^{-1}$. They also have a high dn/dT and low thermal conductivity, causing strong thermal lensing, thus they are not suitable for bulk lasers. However, the high glass transition temperature of GLS makes it resistant to thermal damage, it has good chemical durability and its glass components are non-toxic. Because of its high lanthanum content GLS has excellent rare-earth solubility. This property motivated much of the original interest in GLS in the quest for a rare-earth host for solid state lasers.

## 7.2 Vanadium doped chalcogenide glass

Vanadium doped GLS (V:GLS) was optically characterised to investigate its suitability as an active material for an optical device. Absorption measurements of V:GLS unambiguously identified one absorption band at 1100 nm, with evidence of a spin forbidden transition around 1000 nm, and two further higher energy absorption bands that could not be resolved. Derivative analysis of the absorption measurements clarified the identification of the spin forbidden transition and was able to resolve the second highest, but not the highest, energy absorption band at 750 nm. PLE measurements were able to resolve all three absorption bands, peaking at 1160, 760 and 580 nm. However there was a preferential detection of ions in low crystal field strength sites. XPS measurements indicated the presence of vanadium in a broad range of oxidation states from $V^+$ to $V^{5+}$. Excitation into each of the three absorption bands produced the same characteristic emission spectrum, peaking at 1500 nm with a FWHM of ~500nm. The decay lifetime and decay profile were also the same. This was a strong indication that only one of the vanadium oxidation states was responsible for the observed absorption bands. The quantum efficiency of 0.0023 % V:GLS was 4.2 %. Out of the possible vanadium oxidation states, only $V^{2+}$ and $V^{3+}$ is expected to exhibit three spin allowed transitions. Tanabe-Sugano analysis indicates that out of the possible configurations of coordination and oxidation state only tetrahedral $V^{3+}$ and octahedral $V^{2+}$ had a crystal field strength in the expected low field region. Out of these configurations only octahedral $V^{2+}$ had a C/B value in the expected range of 4-5. The configuration of the optically active vanadium ion in V:GLS is therefore proposed to be octahedral $V^{2+}$. The crystal field strengths (Dq/B) calculated from absorption measurements of V:GLS and V:GLSO are 1.84 and 2.04 respectively.

Lifetime measurements of V:GLS found that the decay was non exponential and at low concentrations could be modelled with the stretched exponential function. Analysis of the coefficient of determination of stretched and double exponential functions and results from a continuous lifetime distribution analysis of the emission decay, at various vanadium concentrations, indicated that at concentrations < 0.1% there was one lifetime component centred ~30 µs. At concentrations > 0.1 two lifetime components centred ~ 30 µs and 5 µs are present. This was argued to be caused by a preferentially filled, high efficiency, oxide site that gives rise to characteristic long lifetimes and a low efficiency sulphide site that gives rise to characteristic short lifetimes.

Comparisons of the $\sigma_{em}\tau$ product of V:GLS to that in other laser materials indicates the best possibility for demonstration laser action in V:GLS is in a fibre geometry. Modelling of laser action an a V:GLS fibre is not presented because the number of assumptions to be made about such a device is too great.

Unlike rare earth ions the optically active orbitals of transition metals are not shielded from the surrounding glass ligands. Because of this the optical properties of transition metal ions in glass is strongly affected by the local bonding environment experienced by the ion, including the ligands nature, distance from the ion, coordination and symmetry. This fundamental difference with rare earths makes transition metals less suitable for active optical devices in certain respects. However this means that the optical characterisation on transition metals can be used to deduce more information about the local bonding environment in the glass; which, negating optical device applications, justifies characterisation of transition metal doped glass.

## 7.3 Titanium nickel and bismuth doped chalcogenide glass

Absorption measurements of Ti:GLS identified an absorption band at ~500-600 nm that could not be fully resolved because of its proximity to the band-edge of GLS. At concentrations of 0.5% and greater a shoulder at ~1000 nm is observed, there is also a weak and broad absorption centred at around 1800 nm. The second derivative absorption spectra identified an absorption peak at 980 nm in Ti:GLS but not in Ti:GLSO, absorption peaks at 615 and 585 nm are also identified for Ti:GLS and Ti:GLSO respectively. The excitation spectra of 0.1 % titanium doped GLS and GLSO both show a single excitation peak at 580 nm The emission spectra of Ti:GLS and Ti:GLSO both peaked at 900 nm. It is proposed that the absorption at ~600 nm in Ti:GLS and Ti:GLSO is due to the $^2T_{2g} \rightarrow ^2E_g$ transition of octahedral $Ti^{3+}$ and the absorption at 980 nm in Ti:GLS is due to $Ti^{3+}$-$Ti^{4+}$ pairs. The 97 µs emission lifetime of Ti:GLSO compares very favourably to the lifetime of Ti:Sapphire of 3.1 µs. The optimum doping concentration for an active device based on Ti:GLSO may be lower than the lowest concentration of 0.05 % molar investigated in this chapter.

The absorption of Ni:GLS is characterised by a red-shift of ~ 300 nm in the band-edge indicating a nickel absorption in the region 500-800 nm. There is also a very weak absorption peaking at ~1500 nm. The excitation spectra of Ni:GLS indicate a single absorption band centred at 690 nm. It is proposed that the 690 nm absorption of Ni:GLS is due to $Ni^+$ in octahedral coordination. The weak absorption at ~1500 nm is attributed to small amounts of $Ni^+$ in tetrahedral coordination. The photoluminescence spectrum peaks at 910 nm with a FWHM of 330 nm. The lifetimes of Ni:GLS and Ni:GLSO are 28 and 70 μs respectively.

A weak shoulder can be observed at ~850 nm in the Bi:GLS absorption spectrum. Further identification of bismuth absorptions cannot be made because dark patches in the sample were detrimental to its absorption. The excitation spectrum of Bi:GLS shows peaks at 665 and 850 nm  Based on comparisons to other work the absorption peaks for Bi:GLS at 665 and 850 nm are attributed to the $^3P_0 \rightarrow ^1D_2$ and $^3P_0 \rightarrow ^3P_2$ transitions of $Bi^+$. The emission decay of Bi:GLS consisted of two lifetime distributions centred at 7 and 47 μs. The demonstration of lasing in bismuth doped aluminosilicate glass makes development of a Bi:GLS laser more favourable.

## 7.4 Femtosecond laser written waveguides in chalcogenide glass

A formation mechanism is presented for fs laser written waveguides in GLS glass, based on optical characterisation and comparisons to previous work. Two different forms of waveguide have been identified and are referred to as A-type and B-type. B-type waveguides have a characteristic long narrow structure and are formed through filamentation. A-type waveguides have a characteristic "teardrop" structure, with a central region (region 1) that has undergone a negative reflective index change through exposure to the focused fs laser beam, and an outer region (region 2) that has undergone a positive index change. A-type waveguides are formed at pulse energies > ~0.2 μJ and B-type waveguides are formed at pulse energies < ~0.2 μJ. The negative index change, in region 1, results from rapid quenching of a high temperature plasma formed by the fs laser pulse which resulted in this region having a high fictive temperature. The positive index change, in region 2, resulted from movement of glass from the region 1 in a shock wave, that resulted in a region with a higher density and refractive index than unexposed glass. Only region 2 was found to actively guide light.

Tunnelling has been identified as the nonlinear absorption mechanism in the formation of the waveguides, by calculation of the Keldysh parameter. Single mode operation at 633 nm has been demonstrated. The writing parameters for the minimum achieved loss of 1.47 dB/cm are 0.36 μJ pulse energy and 50 μm/s scanning speed. A maximum index change of 0.01 has been observed.

Spectral broadening, from an initial FWHM of 50 nm to 200 nm, has been demonstrated with a 1540 nm, 200 fs pulse, at a pulse energy of 30 nJ, in a waveguide written at a pulse energy of 1.75 μJ. A change in peak position to 1580 nm was observed at 20 nJ/pulse. A maximum phase shift of 2.5π has been demonstrated at a pulse energy of 88 nJ/pulse. The broadening has been attributed to self phase modulation, with an asymmetry in some of the broadened spectra attributed to stimulated Raman scattering.

The high nonlinearity of GLS makes it a promising material for nonlinear optical devices. An interesting effect of this nonlinearity is the spectral broadening presented in this chapter. This broadening indicates that these waveguides may have applications in nonlinear optical devices, such as a Mach-Zehnder interferometer switch or a 2R regenerator. However, the high nonlinearity of GLS frustrates the fabrication of symmetric waveguides using fs pulses because the threshold for critical self focusing appears to be higher than the threshold for material modification.

## 7.5 Further work

The immediate further work suggested for transition metal doped chalcogenide glasses centres on the fabrication and characterisation of glass with doping concentrations that could not be investigated in this work because of the loss of ORC glass melting facilities. Some important measurements such as the quantum efficiency of Ti, Ni and Bi doped GLS could not be completed, this is therefore suggested as further work. In particular bismuth doped GLS warrants further investigation because lasing has been demonstrated in other glass hosts. Having determined the optimum doping concentration by measuring the lifetime of a range of doping concentration the fabrication of optical fibres based on V, Ti, Ni, and Bi doped GLS for possible demonstration of laser action is suggested as further work. Comparison of the spectroscopic properties of theses dopants in other chalcogenide glasses such as germanium sulphide could provide more understanding on how the phonon energy and crystal field strength of the host glass affects the quantum efficiency. This is therefore suggested as further work.

Several methods for overcoming the inherent asymmetry of fs laser written waveguides in GLS have been proposed. These include using a higher NA objective, augmenting the beam profile using a slit or cylindrical lenses, using two writing beams and using a parallel writing geometry. Several methods for improving the minimum loss have been suggested. These include investigation of a greater range of translation speeds and pulse energies, using a double pulse fs laser to write the waveguides, varying the wavelength of the writing beam and annealing the sample after waveguide writing. Impressive spectral broadening was restricted to highly multimode waveguides, written at high pulse energies. Coupling sufficient pulse energy into smaller waveguides in order to observe spectral broadening is suggested as further work. A method for achieving this could be to use a tapered fibre to couple into the waveguide. Fabrication of other structures such as Fresnel zone plates and ring resonators is also suggested as further work.

# Appendix A MATLAB code used for continuous lifetime distribution analysis

This code is used in chapters 4 and 5. The x and y vectors are the time and intensity of the fluorescence decay data respectively. The logspace parameters are user specified. The distance function is the Marquardt-Levenberg least squares fitting and is called from a separate M-file. The 'fmincon' function requires the optimisation toolbox to be installed.

```matlab
%%
format long g
% Vector of time decay constants
n = 120;
tau = logspace(0.3,1.97,n);
% Build the exp(-t/tau) matrix.
[TAU,T] = meshgrid(tau,x);
M = exp(-T./TAU);
% Compute the coefficients a.
A = fmincon(@(a) distance(a,M,y),ones(n,1),-eye(n),zeros(n,1));
% Plot the fitted data
yFit = M*A;
figure(1);
clf;
semilogy(x,y,'b',x,yFit,'r');
figure(2);
clf;
plot(tau,A,'o-');

function d = distance(a,M,y)
d = sum((M*a-y).^2);
```

# Appendix B Area under spectra and calculated quantum efficiencies

This data is used in section 4.12

TABLE B1 Area under spectra and quantum efficiency of LD1175-C 0.0023% V:GLS

| | Number of photons (arbitrary number) | | | | | |
|---|---|---|---|---|---|---|
| | Run1 | Run2 | Run3 | Run4 | Run5 | Run6 |
| Emission | 2.05E-05 | 1.91E-05 | 2.42E-05 | 2.49E-05 | 4.05E-05 | 4.17E-05 |
| Laser line a | 1.60E-03 | 1.60E-03 | 1.50E-03 | 1.50E-03 | 4.92E-03 | 4.92E-03 |
| Laser line b | 1.61E-03 | 1.61E-03 | 1.49E-03 | 1.49E-03 | 4.95E-03 | 4.95E-03 |
| Laser line c | 1.61E-03 | 1.61E-03 | 1.49E-03 | 1.49E-03 | 4.97E-03 | 4.97E-03 |
| Laser line average | 1.61E-03 | 1.61E-03 | 1.49E-03 | 1.49E-03 | 4.95E-03 | 4.95E-03 |
| Laser line no sample a | 2.07E-03 | 2.07E-03 | 2.07E-03 | 2.07E-03 | 6.05E-03 | 6.05E-03 |
| Laser line no sample b | 2.04E-03 | 2.04E-03 | 2.04E-03 | 2.04E-03 | 6.03E-03 | 6.03E-03 |
| Laser line no sample c | 2.06E-03 | 2.06E-03 | 2.06E-03 | 2.06E-03 | 5.96E-03 | 5.96E-03 |
| Laser line  no sample average | 2.06E-03 | 2.06E-03 | 2.06E-03 | 2.06E-03 | 6.01E-03 | 6.01E-03 |
| Efficiency (photons emitted/photons absorbed) | | | | | | |
| Quantum efficiency | 4.56E-02 | 4.23E-02 | 4.28E-02 | 4.39E-02 | 3.80E-02 | 3.91E-02 |
| Quantum efficiency Average | 4.19E-02 | | | | | |
| Quantum efficiency Standard Deviation | 2.88E-03 | | | | | |

TABLE B2 Area under spectra and quantum efficiency of LD1257-1 0.0944% V:GLS

| | Number of photons (arbitrary number) | | |
|---|---|---|---|
| | Run1 | Run2 | Run3 |
| Emission | 7.7E-05 | 4.62E-05 | 4.09E-05 |
| Laser line a | 0.001219 | 0.004048 | 0.003825 |
| Laser line b | 0.001213 | 0.004017 | 0.003738 |
| Laser line c | 0.001209 | 0.004056 | |
| Laser line average | 0.001214 | 0.00404 | 0.003781 |
| Laser line no sample a | 0.006124 | 0.006599 | 0.005934 |
| Laser line no sample b | 0.006119 | 0.006567 | 0.005897 |
| Laser line no sample c | 0.006246 | | |
| Laser line no sample average | 0.006163 | 0.006583 | 0.005916 |
| Efficiency (photons emitted/photons absorbed) | | | |
| Quantum efficiency | 0.015564 | 0.01816 | 0.019175 |
| Quantum efficiency Average | 0.017633 | | |
| Quantum efficiency Standard Deviation | 0.001862 | | |

TABLE B3 Area under spectra and quantum efficiency of LD1285-2 0.0616% V:GLS *outlier rejected

| | Number of photons (arbitrary number) | | | |
|---|---|---|---|---|
| | Run1 | *Run2** | Run3 | Run4 |
| Emission | 4.8362e-5 | *1.6533e-5* | 3.2780e-5 | 3.0972e-5 |
| Laser line a | 2.6642e-3 | *5.4154e-3* | 4.5759e-3 | 4.5759e-3 |
| Laser line b | 2.6822e-3 | *5.5583e-3* | 4.5222e-3 | 4.5222e-3 |
| Laser line c | 2.6479e-3 | *5.4663e-3* | 4.4862e-3 | 4.4862e-3 |
| Laser line average | 2.6648e-3 | *5.4800e-3* | 4.5281e-3 | 4.5281e-3 |
| Laser line no sample a | 5.8567e-3 | *5.8567e-3* | 5.8567e-3 | 5.8567e-3 |
| Laser line no sample b | 5.8567e-3 | *5.8567e-3* | 5.8567e-3 | 5.8567e-3 |
| Laser line no sample c | 5.9034e-3 | *5.9034e-3* | 5.9034e-3 | 5.9034e-3 |
| Laser line no sample average | 5.8723e-3 | *5.8723e-3* | 5.8723e-3 | 5.8723e-3 |
| Efficiency (photons emitted/photons absorbed) | | | | |
| Quantum efficiency | 0.0151 | *0.0421* | 0.0244 | 0.0230 |
| Quantum efficiency Average | 0.0208 | | | |
| Quantum efficiency Standard Deviation | 5.0309e-3 | | | |

TABLE B4 Area under spectra and quantum efficiency of LD1284-5 0.0242% V:GLSO

| | Number of photons (arbitrary number) | | |
|---|---|---|---|
| | Run1 | Run2 | Run3 |
| Emission | 3.9330e-5 | 1.4210e-5 | 1.9086e-5 |
| Laser line a | 2.4306e-3 | 3.4956e-3 | 3.3815e-3 |
| Laser line b | 2.3863e-3 | 3.4328e-3 | 3.3510e-3 |
| Laser line c | 2.3980e-3 | 3.4786e-3 | 3.3661e-3 |
| Laser line average | 2.4050e-3 | 3.4690e-3 | 3.3662e-3 |
| Laser line no sample a | 3.9734e-3 | 3.9734e-3 | 3.9734e-3 |
| Laser line no sample b | 3.9245e-3 | 3.9245e-3 | 3.9245e-3 |
| Laser line no sample c | 3.9795e-3 | 3.9795e-3 | 3.9795e-3 |
| Laser line no sample d | 3.9296e-3 | 3.9296e-3 | 3.9296e-3 |
| Laser line no sample e | 3.9203e-3 | 3.9203e-3 | 3.9203e-3 |
| Laser line no sample f | 3.9582e-3 | 3.9582e-3 | 3.9582e-3 |
| Laser line  no sample average | 3.9476e-3 | 3.9476e-3 | 3.9476e-3 |
| Efficiency (photons emitted/photons absorbed) | | | |
| Quantum efficiency | 0.0255 | 0.0297 | 0.0328 |
| Quantum efficiency Average | 0.0293 | | |
| Quantum efficiency Standard Deviation | 3.6783e-3 | | |

TABLE B5 Area under spectra and quantum efficiency of LD1284-4 0.0087% V:GLSO

| | Number of photons (arbitrary number) | | | |
|---|---|---|---|---|
| | Run1 | Run2 | Run3 | Run4 |
| Emission | 3.2383e-5 | 9.3056e-6 | 6.8218e-6 | 2.2048e-5 |
| Laser line a | 3.0896e-3 | 3.6146e-3 | 3.7073e-3 | 3.3512e-3 |
| Laser line b | 3.1169e-3 | 3.6285e-3 | 3.7961e-3 | 3.3423e-3 |
| Laser line c | 3.1258e-3 | 3.6050e-3 | 3.7081e-3 | 3.4037e-3 |
| Laser line average | 3.1107e-3 | 3.6160e-3 | 3.7372e-3 | 3.3657e-3 |
| Laser line no sample a | 3.9734e-3 | 3.9734e-3 | 3.9734e-3 | 3.9734e-3 |
| Laser line no sample b | 3.9245e-3 | 3.9245e-3 | 3.9245e-3 | 3.9245e-3 |
| Laser line no sample c | 3.9795e-3 | 3.9795e-3 | 3.9795e-3 | 3.9795e-3 |
| Laser line no sample c | 3.9296e-3 | 3.9296e-3 | 3.9296e-3 | 3.9296e-3 |
| Laser line no sample e | 3.9203e-3 | 3.9203e-3 | 3.9203e-3 | 3.9203e-3 |
| Laser line no sample f | 3.9582e-3 | 3.9582e-3 | 3.9582e-3 | 3.9582e-3 |
| Laser line no sample average | 3.9476e-3 | 3.9476e-3 | 3.9476e-3 | 3.9476e-3 |
| Efficiency (photons emitted/photons absorbed) | | | | |
| Quantum efficiency | 0.0387 | 0.0281 | 0.0324 | 0.0379 |
| Quantum efficiency Average | 0.0343 | | | |
| Quantum efficiency Standard Deviation | 4.9872e-3 | | | |

TABLE B6 Area under spectra and quantum efficiency of LD1284-3 0.0489% V:GLSO

| | Number of photons (arbitrary number) | | |
|---|---|---|---|
| | Run1 | Run2 | Run3 |
| Emission | 3.0275e-5 | 9.8348e-6 | 2.9948e-5 |
| Laser line a | 2.2483e-3 | 3.4551e-3 | 2.2240e-3 |
| Laser line b | 2.2075e-3 | 3.4786e-3 | 2.2196e-3 |
| Laser line c | 2.1957e-3 | 3.5058e-3 | 2.2023e-3 |
| Laser line average | 2.2172e-3 | 3.4798e-3 | 2.2153e-3 |
| Laser line no sample a | 3.9734e-3 | 3.9734e-3 | 3.9734e-3 |
| Laser line no sample b | 3.9245e-3 | 3.9245e-3 | 3.9245e-3 |
| Laser line no sample c | 3.9795e-3 | 3.9795e-3 | 3.9795e-3 |
| Laser line no sample d | 3.9296e-3 | 3.9296e-3 | 3.9296e-3 |
| Laser line no sample e | 3.9203e-3 | 3.9203e-3 | 3.9203e-3 |
| Laser line no sample f | 3.9582e-3 | 3.9582e-3 | 3.9582e-3 |
| Laser line  no sample average | 3.9476e-3 | 3.9476e-3 | 3.9476e-3 |
| Efficiency (photons emitted/photons absorbed) | | | |
| Quantum efficiency | 0.0175 | 0.0210 | 0.0173 |
| Quantum efficiency Average | 0.0186 | | |
| Quantum efficiency Standard Deviation | 2.0998e-3 | | |

TABLE B7 Area under spectra and quantum efficiency of LD1284-1 0.0608% V:GLSO

| | Number of photons (arbitrary number) | | |
|---|---|---|---|
| | Run1 | | |
| Emission | 1.5826e-5 | | |
| Laser line a | 2.8121e-3 | | |
| Laser line b | 2.8376e-3 | | |
| Laser line c | 2.8381e-3 | | |
| Laser line average | 2.8293e-3 | | |
| Laser line no sample a | 3.9734e-3 | | |
| Laser line no sample b | 3.9245e-3 | | |
| Laser line no sample c | 3.9795e-3 | | |
| Laser line no sample d | 3.9296e-3 | | |
| Laser line no sample e | 3.9203e-3 | | |
| Laser line no sample f | 3.9582e-3 | | |
| Laser line  no sample average | 3.9476e-3 | | |
| Efficiency (photons emitted/photons absorbed) | | | |
| Quantum efficiency | 0.0142 | | |
| Quantum efficiency Average | 0.0142 | | |
| Quantum efficiency Standard Deviation | | | |

# Appendix C Energy matrices and energy terms for $d^2$ and $d^3$ ions

These are used in section 4.16, the matrix elements are taken from reference [59].

## Matrix elements for the $d^2$ tetrahedral configuration

| $^1A_1(^1G,^1S)$ | |
|---|---|
| $t_2^2$ | $e^2$ |
| 10B | $\sqrt{6}(2B+C)$ |
| $\sqrt{6}(2B+C)$ | -20Dq+8B+4C |

| $^1E(^1D,^1G)$ | |
|---|---|
| $t_2^2$ | $e^2$ |
| B+2C | $-2\sqrt{3}B$ |
| $-2\sqrt{3}B$ | -20Dq+2C |

| $^1T_2(^1D,^1G)$ | |
|---|---|
| $t_2^2$ | $t_2e$ |
| B+2C | $2\sqrt{3}B$ |
| $2\sqrt{3}B$ | -10Dq+2C |

| $^3T_1(^3F,^3P)$ | |
|---|---|
| $t_2^2$ | $t_2e$ |
| -5B | 6B |
| 6B | -10Dq+4B |

$t_2e\ ^1T_1(^1G)$ -10Dq+4B+2C

$t_2e\ ^3T_2(^3G)$ -10Dq-8B

$e^2\ ^3A_2(^3F)$ -20Dq-8B

## Energy terms for the $d^2$ tetrahedral configuration

$$E\left(^1A_1(^1G)\right)=\frac{1}{2}\left(18B+9C-20Dq-\sqrt{100B^2+100BC+25C^2+80BDq+40CDq+400Dq^2}\right)$$

$$E\left(^1A_1(^1S)\right)=\frac{1}{2}\left(18B+9C-20Dq+\sqrt{100B^2+100BC+25C^2+80BDq+40CDq+400Dq^2}\right)$$

$$E\left(^1E(^1D)\right)=\frac{1}{2}\left(B+4C-20Dq-\sqrt{49B^2+40DqB+400Dq^2}\right)$$

$$E\left(^1E(^1G)\right)=\frac{1}{2}\left(B+4C-20Dq+\sqrt{49B^2+40DqB+400Dq^2}\right)$$

$$E\left(^1T_2(^1D)\right)=\frac{1}{2}\left(B+4C-10Dq-\sqrt{49B^2+20DqB+100Dq^2}\right)$$

$$E\left(^1T_2(^1G)\right)=\frac{1}{2}\left(B+4C-10Dq+\sqrt{49B^2+20DqB+100Dq^2}\right)$$

$$E\left(^3T_1(^3F)\right)=\frac{1}{2}\left(-B+5Dq-\sqrt{225B^2-180DqB+100Dq^2}\right)$$

$$E\left(^3T_1(^3P)\right)=\frac{1}{2}\left(-B+5Dq+\sqrt{225B^2-180DqB+100Dq^2}\right)$$

$$E\left(^1T_1(^1G)\right)=-10Dq+4B+2C$$
$$E\left(^3T_2(^3G)\right)=-10Dq-8B$$

$E\left(^{3}A_{2}(^{3}F)\right) = -20Dq - 8B$

**Matrix elements for the d² octahedral configuration**

| $^{1}A_{1}(^{1}G, ^{1}S)$ | |
| --- | --- |
| $t_{2}^{2}$ | $e^{2}$ |
| 10B | $\sqrt{6}(2B+C)$ |
| $\sqrt{6}(2B+C)$ | 20Dq+8B+4C |

| $^{1}E(^{1}D, ^{1}G)$ | |
| --- | --- |
| $t_{2}^{2}$ | $e^{2}$ |
| B+2C | $-2\sqrt{3}B$ |
| $-2\sqrt{3}B$ | 20Dq+2C |

| $^{1}T_{2}(^{1}D, ^{1}G)$ | |
| --- | --- |
| $t_{2}^{2}$ | $t_{2}e$ |
| B+2C | $2\sqrt{3}B$ |
| $2\sqrt{3}B$ | 10Dq+2C |

| $^{3}T_{1}(^{3}F, ^{3}P)$ | |
| --- | --- |
| $t_{2}^{2}$ | $t_{2}e$ |
| -5B | 6B |
| 6B | 10Dq+4B |

$t_{2}e$  $^{1}T_{1}(^{1}G)$  10Dq+4B+2C

$t_{2}e$  $^{3}T_{2}(^{3}G)$  10Dq-8B

$e^{2}$  $^{3}A_{2}(^{3}F)$  20Dq-8B

**Energy terms for the d² octahedral configuration**

$$E\left(^{1}A_{1}(^{1}G)\right) = \frac{1}{2}\left(18B + 9C + 20Dq - \sqrt{100B^{2} + 100BC + 25C^{2} - 80BDq - 40CDq + 400Dq^{2}}\right)$$

$$E\left(^{1}A_{1}(^{1}S)\right) = \frac{1}{2}\left(18B + 9C + 20Dq + \sqrt{100B^{2} + 100BC + 25C^{2} - 80BDq - 40CDq + 400Dq^{2}}\right)$$

$$E\left(^{1}E(^{1}D)\right) = \frac{1}{2}\left(B + 4C + 20Dq - \sqrt{49B^{2} - 40DqB + 400Dq^{2}}\right)$$

$$E\left(^{1}E(^{1}G)\right) = \frac{1}{2}\left(B + 4C + 20Dq + \sqrt{49B^{2} - 40DqB + 400Dq^{2}}\right)$$

$$E\left(^{1}T_{2}(^{1}D)\right) = \frac{1}{2}\left(B + 4C + 10Dq - \sqrt{49B^{2} - 20DqB + 100Dq^{2}}\right)$$

$$E\left(^{1}T_{2}(^{1}G)\right) = \frac{1}{2}\left(B + 4C + 10Dq + \sqrt{49B^{2} - 20DqB + 100Dq^{2}}\right)$$

$$E\left(^{3}T_{1}(^{3}F)\right) = \frac{1}{2}\left(-B + 10Dq - \sqrt{225B^{2} + 180DqB + 100Dq^{2}}\right)$$

$$E\left(^{3}T_{1}(^{3}P)\right) = \frac{1}{2}\left(-B + 10Dq + \sqrt{225B^{2} + 180DqB + 100Dq^{2}}\right)$$

$$E\left(^{1}T_{1}(^{1}G)\right) = 10Dq + 4B + 2C$$

$$E\left(^{3}T_{2}(^{3}G)\right) = 10Dq - 8B$$

$$E\left(^{3}A_{2}(^{3}F)\right) = 20Dq - 8B$$

## Matrix elements for the $d^3$ tetrahedral configuration

| $^2T_2(A^2D, B^2D, ^2F, ^2G, ^2H)$ | | | | |
|---|---|---|---|---|
| $t_2^3$ | $t_2^2(^3T_1)e$ | $t_2^2(^1T_2)e$ | $t_2e^2(^1A_1)$ | $t_2e^2(^1E)$ |
| 12Dq+5C | -3√3B | -5√3B | 4B+2C | 2B |
| -3√3B | 2Dq-6B+3C | 3B | -3√3B | -3√3B |
| -5√3B | 3B | 2Dq+4B+3C | -3√3B | √3B |
| 4B+2C | -3√3B | -3√3B | 6B+5C-8Dq | 10B |
| 2B | -3√3B | √3B | 10B | -2B+3C-8Dq |

| $^2T_1(^2P, ^2F, ^2G, ^2H)$ | | | | |
|---|---|---|---|---|
| $t_2^3$ | $t_2^2(^3T_1)e$ | $t_2^2(^1T_2)e$ | $t_2e^2(^1A_1)$ | $t_2e^2(^1E)$ |
| 12Dq-6B+3C | -3B | 3B | 0 | -2√3B |
| -3B | 2Dq+3C | -3B | 3B | 3√3B |
| 3B | -3B | 2Dq-6B+3C | -3B | -√3B |
| 0 | 3B | -3B | -8Dq-6B+3C | 2√3B |
| -2√3B | 3√3B | -√3B | 2√3B | -8Dq-2B+3C |

| $^2E(A^2D,B^2D, ^2G, ^2H)$ | | | |
|---|---|---|---|
| $t_2^3$ | $t_2^2(^1A_1)e$ | $t_2^2(^1E)e$ | $e^3$ |
| -6B+3C+12Dq | -6√2B | -3√2B | 0 |
| -6√2B | 2Dq+8B+6C | 10B | √3(2B+C) |
| -3√2B | 10B | 2Dq-B+3C | 2√3B |
| 0 | √3(2B+C) | 2√3B | -18Dq-8B+4C |

| $^4T_1(^4P, ^4F)$ | |
|---|---|
| $t_2^2(^3T_1)e$ | $t_2e^2(^3A_2)$ |
| 2Dq-3B | 6B |
| 6B | -8Dq-12B |

$t_2^3\ ^4A_2(^4F)$          12Dq-15B
$t_2^2(^3T_1)e\ ^4T_2\ (^4F)$     2Dq-15B
$t_2^2(^1E)e\ ^2A_1\ (^2G)$     2Dq-11B+3C
$t_2^2(^1E)e\ ^2A_2\ (^2F)$     2Dq+9B+3C

## Energy terms for the d³ tetrahedral configuration

$$E\left({}^{2}E({}^{2}G)\right)=1^{st}\ root\ \left(\det\left(\begin{bmatrix} -6B+3C+12Dq & -6\sqrt{2}B & -3\sqrt{2}B & 0 \\ -6\sqrt{2}B & 8B+6C+2Dq & 10B & \sqrt{3}(2B+C) \\ -3\sqrt{2}B & 10B & -1B+3C+2Dq & 2\sqrt{3}B \\ 0 & \sqrt{3}(2B+C) & 2\sqrt{3}B & -8B+4C-18Dq \end{bmatrix}-\lambda I\right)\right)=0$$

$$E\left({}^{2}E(a^{2}D)\right)=2^{nd}\ root\ \left(\det\left(\begin{bmatrix} -6B+3C+12Dq & -6\sqrt{2}B & -3\sqrt{2}B & 0 \\ -6\sqrt{2}B & 8B+6C+2Dq & 10B & \sqrt{3}(2B+C) \\ -3\sqrt{2}B & 10B & -1B+3C+2Dq & 2\sqrt{3}B \\ 0 & \sqrt{3}(2B+C) & 2\sqrt{3}B & -8B+4C-18Dq \end{bmatrix}-\lambda I\right)\right)=0$$

$$E\left({}^{2}E({}^{2}H)\right)=3^{rd}\ root\ \left(\det\left(\begin{bmatrix} -6B+3C+12Dq & -6\sqrt{2}B & -3\sqrt{2}B & 0 \\ -6\sqrt{2}B & 8B+6C+2Dq & 10B & \sqrt{3}(2B+C) \\ -3\sqrt{2}B & 10B & -1B+3C+2Dq & 2\sqrt{3}B \\ 0 & \sqrt{3}(2B+C) & 2\sqrt{3}B & -8B+4C-18Dq \end{bmatrix}-\lambda I\right)\right)=0$$

$$E\left({}^{2}E(b^{2}D)\right)=4^{th}\ root\ \left(\det\left(\begin{bmatrix} -6B+3C+12Dq & -6\sqrt{2}B & -3\sqrt{2}B & 0 \\ -6\sqrt{2}B & 8B+6C+2Dq & 10B & \sqrt{3}(2B+C) \\ -3\sqrt{2}B & 10B & -1B+3C+2Dq & 2\sqrt{3}B \\ 0 & \sqrt{3}(2B+C) & 2\sqrt{3}B & -8B+4C-18Dq \end{bmatrix}-\lambda I\right)\right)=0$$

$$E\left({}^{2}T_{1}({}^{2}G)\right)=1^{st}\ root\ \left(\det\left(\begin{bmatrix} -6B-3C+12Dq & -3B & 3B & 0 & -2\sqrt{3}B \\ -3B & 3C+2Dq & -3B & 3B & 3\sqrt{3}B \\ 3B & -3B & -6B+3C+2Dq & -3B & -\sqrt{3}B \\ 0 & 3B & -3B & -6B+3C-8Dq & 2\sqrt{3}B \\ -2\sqrt{3}B & 3\sqrt{3}B & -\sqrt{3}B & 2\sqrt{3}B & -2B+3C-8Dq \end{bmatrix}-\lambda I\right)\right)=0$$

$$E\left({}^{2}T_{1}({}^{2}H)\right)=2^{nd}\ root\ \left(\det\left(\begin{bmatrix} -6B-3C+12Dq & -3B & 3B & 0 & -2\sqrt{3}B \\ -3B & 3C+2Dq & -3B & 3B & 3\sqrt{3}B \\ 3B & -3B & -6B+3C+2Dq & -3B & -\sqrt{3}B \\ 0 & 3B & -3B & -6B+3C-8Dq & 2\sqrt{3}B \\ -2\sqrt{3}B & 3\sqrt{3}B & -\sqrt{3}B & 2\sqrt{3}B & -2B+3C-8Dq \end{bmatrix}-\lambda I\right)\right)=0$$

$$E\left(^2T_1(^2F)\right)=3^{rd}\,root\,\det\left(\begin{bmatrix}-6B-3C+12Dq & -3B & 3B & 0 & -2\sqrt{3}B \\ -3B & 3C+2Dq & -3B & 3B & 3\sqrt{3}B \\ 3B & -3B & -6B+3C+2Dq & -3B & -\sqrt{3}B \\ 0 & 3B & -3B & -6B+3C-8Dq & 2\sqrt{3}B \\ -2\sqrt{3}B & 3\sqrt{3}B & -\sqrt{3}B & 2\sqrt{3}B & -2B+3C-8Dq\end{bmatrix}-\lambda I\right)=0$$

$$E\left(^2T_1(^2P)\right)=4^{th}\,root\,\det\left(\begin{bmatrix}-6B-3C+12Dq & -3B & 3B & 0 & -2\sqrt{3}B \\ -3B & 3C+2Dq & -3B & 3B & 3\sqrt{3}B \\ 3B & -3B & -6B+3C+2Dq & -3B & -\sqrt{3}B \\ 0 & 3B & -3B & -6B+3C-8Dq & 2\sqrt{3}B \\ -2\sqrt{3}B & 3\sqrt{3}B & -\sqrt{3}B & 2\sqrt{3}B & -2B+3C-8Dq\end{bmatrix}-\lambda I\right)=0$$

$$E\left(^2T_2(^2G)\right)=1^{st}\,root\,\det\left(\begin{bmatrix}5C+12Dq & -3\sqrt{3}B & -5\sqrt{3}B & 4B+2C & 2B \\ -3\sqrt{3}B & -6B+3C+2Dq & 3B & -3\sqrt{3}B & -3\sqrt{3}B \\ -5\sqrt{3}B & 3B & 4B+3C+2Dq & -3\sqrt{3}B & \sqrt{3}B \\ 4B+2C & -3\sqrt{3}B & -3\sqrt{3}B & 6B+5C-8Dq & 10B \\ 2B & -3\sqrt{3}B & \sqrt{3}B & 10B & -2B+3C-8Dq\end{bmatrix}-\lambda I\right)=0$$

$$E\left(^2T_2(a^2D)\right)=2^{nd}\,root\,\det\left(\begin{bmatrix}5C+12Dq & -3\sqrt{3}B & -5\sqrt{3}B & 4B+2C & 2B \\ -3\sqrt{3}B & -6B+3C+2Dq & 3B & -3\sqrt{3}B & -3\sqrt{3}B \\ -5\sqrt{3}B & 3B & 4B+3C+2Dq & -3\sqrt{3}B & \sqrt{3}B \\ 4B+2C & -3\sqrt{3}B & -3\sqrt{3}B & 6B+5C-8Dq & 10B \\ 2B & -3\sqrt{3}B & \sqrt{3}B & 10B & -2B+3C-8Dq\end{bmatrix}-\lambda I\right)=0$$

$$E\left(^2T_2(^2H)\right)=3^{rd}\,root\,\det\left(\begin{bmatrix}5C+12Dq & -3\sqrt{3}B & -5\sqrt{3}B & 4B+2C & 2B \\ -3\sqrt{3}B & -6B+3C+2Dq & 3B & -3\sqrt{3}B & -3\sqrt{3}B \\ -5\sqrt{3}B & 3B & 4B+3C+2Dq & -3\sqrt{3}B & \sqrt{3}B \\ 4B+2C & -3\sqrt{3}B & -3\sqrt{3}B & 6B+5C-8Dq & 10B \\ 2B & -3\sqrt{3}B & \sqrt{3}B & 10B & -2B+3C-8Dq\end{bmatrix}-\lambda I\right)=0$$

$$E\left(^2T_2(^2F)\right)=4^{th}\,root\,\det\left(\begin{bmatrix}5C+12Dq & -3\sqrt{3}B & -5\sqrt{3}B & 4B+2C & 2B \\ -3\sqrt{3}B & -6B+3C+2Dq & 3B & -3\sqrt{3}B & -3\sqrt{3}B \\ -5\sqrt{3}B & 3B & 4B+3C+2Dq & -3\sqrt{3}B & \sqrt{3}B \\ 4B+2C & -3\sqrt{3}B & -3\sqrt{3}B & 6B+5C-8Dq & 10B \\ 2B & -3\sqrt{3}B & \sqrt{3}B & 10B & -2B+3C-8Dq\end{bmatrix}-\lambda I\right)=0$$

$$E\left(^2T_2(a^2D)\right)=5^{th}\,root\,\det\left(\begin{bmatrix}5C+12Dq & -3\sqrt{3}B & -5\sqrt{3}B & 4B+2C & 2B \\ -3\sqrt{3}B & -6B+3C+2Dq & 3B & -3\sqrt{3}B & -3\sqrt{3}B \\ -5\sqrt{3}B & 3B & 4B+3C+2Dq & -3\sqrt{3}B & \sqrt{3}B \\ 4B+2C & -3\sqrt{3}B & -3\sqrt{3}B & 6B+5C-8Dq & 10B \\ 2B & -3\sqrt{3}B & \sqrt{3}B & 10B & -2B+3C-8Dq\end{bmatrix}-\lambda I\right)=0$$

$$E\left(^4T_1(^3F)\right)=\frac{1}{2}\left(-15B-6Dq-\sqrt{225B^2+180DqB+100Dq^2}\right)$$

$$E\left(^4T_1(^3P)\right)=\frac{1}{2}\left(-15B-6Dq+\sqrt{225B^2+180DqB+100Dq^2}\right)$$

$$E\left(^4A_2(^2F)\right)=12Dq-15B$$

$$E\left(^4T_2(^4F)\right)=2Dq-15B$$

$$E\left(^2A_1(^2G)\right) = 2Dq - 11B + 3C$$
$$E\left(^2A_2(^2F)\right) = 2Dq + 9B + 3C$$

## Matrix elements for the d³ octahedral configuration

| $^2T_2(A^2D, B^2D, ^2F, ^2G, ^2H)$ | | | | |
|---|---|---|---|---|
| $t_2^3$ | $t_2^2(^3T_1)e$ | $t_2^2(^1T_2)e$ | $t_2e^2(^1A_1)$ | $t_2e^2(^1E)$ |
| -12Dq+5C | -3√3B | -5√3B | 4B+2C | 2B |
| -3√3B | -2Dq-6B+3C | 3B | -3√3B | -3√3B |
| -5√3B | 3B | -2Dq+4B+3C | -3√3B | √3B |
| 4B+2C | -3√3B | -3√3B | 6B+5C+8Dq | 10B |
| 2B | -3√3B | √3B | 10B | -2B+3C+8Dq |

| $^2T_1(^2P, ^2F, ^2G, ^2H)$ | | | | |
|---|---|---|---|---|
| $t_2^3$ | $t_2^2(^3T_1)e$ | $t_2^2(^1T_2)e$ | $t_2e^2(^1A_1)$ | $t_2e^2(^1E)$ |
| -12Dq-6B+3C | -3B | 3B | 0 | -2√3B |
| -3B | -2Dq+3C | -3B | 3B | 3√3B |
| 3B | -3B | -2Dq-6B+3C | -3B | -√3B |
| 0 | 3B | -3B | 8Dq-6B+3C | 2√3B |
| -2√3B | 3√3B | -√3B | 2√3B | 8Dq-2B+3C |

| $^2E(A^2D, B\ ^2D, ^2G, ^2H)$ | | | |
|---|---|---|---|
| $t_2^3$ | $t_2^2(^1A_1)e$ | $t_2^2(^1E)e$ | $e^3$ |
| -6B+3C-12Dq | -6√2B | -3√2B | 0 |
| -6√2B | -2Dq+8B+6C | 10B | √3(2B+C) |
| -3√2B | 10B | -2Dq-B+3C | 2√3B |
| 0 | √3(2B+C) | 2√3B | 18Dq-8B+4C |

| $^4T_1(^4P, ^4F)$ | |
|---|---|
| $t_2^2(^3T_1)e$ | $t_2e^2(^3A_2)$ |
| -2Dq-3B | 6B |
| 6B | 8Dq-12B |
| $t_2^3\ ^4A_2(^4F)$ | -12Dq-15B |
| $t_2^2(^3T_1)e\ ^4T_2\ (^4F)$ | -2Dq-15B |
| $t_2^2(^1E)e\ ^2A_1\ (^2G)$ | -2Dq-11B+3C |
| $t_2^2(^1E)e\ ^2A_2\ (^2F)$ | -2Dq+9B+3C |

## Energy terms for the d³ octahedral configuration

$$E\left(^2E(^2G)\right) = 1^{st} \; root \; \det\left(\begin{bmatrix} -6B+3C-12Dq & -6\sqrt{2}B & -3\sqrt{2}B & 0 \\ -6\sqrt{2}B & 8B+6C-2Dq & 10B & \sqrt{3}(2B+C) \\ -3\sqrt{2}B & 10B & -1B+3C-2Dq & 2\sqrt{3}B \\ 0 & \sqrt{3}(2B+C) & 2\sqrt{3}B & -8B+4C+18Dq \end{bmatrix} - \lambda I\right) = 0$$

$$E\left(^2E(a^2D)\right) = 2^{nd} \; root \; \det\left(\begin{bmatrix} -6B+3C-12Dq & -6\sqrt{2}B & -3\sqrt{2}B & 0 \\ -6\sqrt{2}B & 8B+6C-2Dq & 10B & \sqrt{3}(2B+C) \\ -3\sqrt{2}B & 10B & -1B+3C-2Dq & 2\sqrt{3}B \\ 0 & \sqrt{3}(2B+C) & 2\sqrt{3}B & -8B+4C+18Dq \end{bmatrix} - \lambda I\right) = 0$$

$$E\left(^2E(^2H)\right) = 3^{rd} \; root \; \det\left(\begin{bmatrix} -6B+3C-12Dq & -6\sqrt{2}B & -3\sqrt{2}B & 0 \\ -6\sqrt{2}B & 8B+6C-2Dq & 10B & \sqrt{3}(2B+C) \\ -3\sqrt{2}B & 10B & -1B+3C-2Dq & 2\sqrt{3}B \\ 0 & \sqrt{3}(2B+C) & 2\sqrt{3}B & -8B+4C+18Dq \end{bmatrix} - \lambda I\right) = 0$$

$$E\left(^2E(b^2D)\right) = 4^{th} \; root \; \det\left(\begin{bmatrix} -6B+3C-12Dq & -6\sqrt{2}B & -3\sqrt{2}B & 0 \\ -6\sqrt{2}B & 8B+6C-2Dq & 10B & \sqrt{3}(2B+C) \\ -3\sqrt{2}B & 10B & -1B+3C-2Dq & 2\sqrt{3}B \\ 0 & \sqrt{3}(2B+C) & 2\sqrt{3}B & -8B+4C+18Dq \end{bmatrix} - \lambda I\right) = 0$$

$$E\left(^2T_1(^2G)\right) = 1^{st} \; root \; \det\left(\begin{bmatrix} -6B-3C-12Dq & -3B & 3B & 0 & -2\sqrt{3}B \\ -3B & 3C-2Dq & -3B & 3B & 3\sqrt{3}B \\ 3B & -3B & -6B+3C-2Dq & -3B & -\sqrt{3}B \\ 0 & 3B & -3B & -6B+3C+8Dq & 2\sqrt{3}B \\ -2\sqrt{3}B & 3\sqrt{3}B & -\sqrt{3}B & 2\sqrt{3}B & -2B+3C+8Dq \end{bmatrix} - \lambda I\right) = 0$$

$$E\left(^2T_1(^2H)\right) = 2^{nd} \; root \; \det\left(\begin{bmatrix} -6B-3C-12Dq & -3B & 3B & 0 & -2\sqrt{3}B \\ -3B & 3C-2Dq & -3B & 3B & 3\sqrt{3}B \\ 3B & -3B & -6B+3C-2Dq & -3B & -\sqrt{3}B \\ 0 & 3B & -3B & -6B+3C+8Dq & 2\sqrt{3}B \\ -2\sqrt{3}B & 3\sqrt{3}B & -\sqrt{3}B & 2\sqrt{3}B & -2B+3C+8Dq \end{bmatrix} - \lambda I\right) = 0$$

$$E\left(^2T_1(^2F)\right) = 3^{rd} \; root \; \det\left(\begin{bmatrix} -6B-3C-12Dq & -3B & 3B & 0 & -2\sqrt{3}B \\ -3B & 3C-2Dq & -3B & 3B & 3\sqrt{3}B \\ 3B & -3B & -6B+3C-2Dq & -3B & -\sqrt{3}B \\ 0 & 3B & -3B & -6B+3C+8Dq & 2\sqrt{3}B \\ -2\sqrt{3}B & 3\sqrt{3}B & -\sqrt{3}B & 2\sqrt{3}B & -2B+3C+8Dq \end{bmatrix} - \lambda I\right) = 0$$

$$E(^2T_1(^2P)) = 4^{th}\,root\ \det\left(\begin{bmatrix} -6B-3C-12Dq & -3B & 3B & 0 & -2\sqrt{3}B \\ -3B & 3C-2Dq & -3B & 3B & 3\sqrt{3}B \\ 3B & -3B & -6B+3C-2Dq & -3B & -\sqrt{3}B \\ 0 & 3B & -3B & -6B+3C+8Dq & 2\sqrt{3}B \\ -2\sqrt{3}B & 3\sqrt{3}B & -\sqrt{3}B & 2\sqrt{3}B & -2B+3C+8Dq \end{bmatrix} -\lambda I\right) = 0$$

$$E(^2T_2(^2G)) = 1^{st}\,root\ \det\left(\begin{bmatrix} 5C-12Dq & -3\sqrt{3}B & -5\sqrt{3}B & 4B+2C & 2B \\ -3\sqrt{3}B & -6B+3C-2Dq & 3B & -3\sqrt{3}B & -3\sqrt{3}B \\ -5\sqrt{3}B & 3B & 4B+3C-2Dq & -3\sqrt{3}B & \sqrt{3}B \\ 4B+2C & -3\sqrt{3}B & -3\sqrt{3}B & 6B+5C+8Dq & 10B \\ 2B & -3\sqrt{3}B & \sqrt{3}B & 10B & -2B+3C+8Dq \end{bmatrix} -\lambda I\right) = 0$$

$$E(^2T_2(a^2D)) = 2^{nd}\,root\ \det\left(\begin{bmatrix} 5C-12Dq & -3\sqrt{3}B & -5\sqrt{3}B & 4B+2C & 2B \\ -3\sqrt{3}B & -6B+3C-2Dq & 3B & -3\sqrt{3}B & -3\sqrt{3}B \\ -5\sqrt{3}B & 3B & 4B+3C-2Dq & -3\sqrt{3}B & \sqrt{3}B \\ 4B+2C & -3\sqrt{3}B & -3\sqrt{3}B & 6B+5C+8Dq & 10B \\ 2B & -3\sqrt{3}B & \sqrt{3}B & 10B & -2B+3C+8Dq \end{bmatrix} -\lambda I\right) = 0$$

$$E(^2T_2(^2H)) = 3^{rd}\,root\ \det\left(\begin{bmatrix} 5C-12Dq & -3\sqrt{3}B & -5\sqrt{3}B & 4B+2C & 2B \\ -3\sqrt{3}B & -6B+3C-2Dq & 3B & -3\sqrt{3}B & -3\sqrt{3}B \\ -5\sqrt{3}B & 3B & 4B+3C-2Dq & -3\sqrt{3}B & \sqrt{3}B \\ 4B+2C & -3\sqrt{3}B & -3\sqrt{3}B & 6B+5C+8Dq & 10B \\ 2B & -3\sqrt{3}B & \sqrt{3}B & 10B & -2B+3C+8Dq \end{bmatrix} -\lambda I\right) = 0$$

$$E(^2T_2(^2F)) = 4^{th}\,root\ \det\left(\begin{bmatrix} 5C-12Dq & -3\sqrt{3}B & -5\sqrt{3}B & 4B+2C & 2B \\ -3\sqrt{3}B & -6B+3C-2Dq & 3B & -3\sqrt{3}B & -3\sqrt{3}B \\ -5\sqrt{3}B & 3B & 4B+3C-2Dq & -3\sqrt{3}B & \sqrt{3}B \\ 4B+2C & -3\sqrt{3}B & -3\sqrt{3}B & 6B+5C+8Dq & 10B \\ 2B & -3\sqrt{3}B & \sqrt{3}B & 10B & -2B+3C+8Dq \end{bmatrix} -\lambda I\right) = 0$$

$$E(^2T_2(a^2D)) = 5^{th}\,root\ \det\left(\begin{bmatrix} 5C-12Dq & -3\sqrt{3}B & -5\sqrt{3}B & 4B+2C & 2B \\ -3\sqrt{3}B & -6B+3C-2Dq & 3B & -3\sqrt{3}B & -3\sqrt{3}B \\ -5\sqrt{3}B & 3B & 4B+3C-2Dq & -3\sqrt{3}B & \sqrt{3}B \\ 4B+2C & -3\sqrt{3}B & -3\sqrt{3}B & 6B+5C+8Dq & 10B \\ 2B & -3\sqrt{3}B & \sqrt{3}B & 10B & -2B+3C+8Dq \end{bmatrix} -\lambda I\right) = 0$$

$$E(^4T_1(^3F)) = \frac{1}{2}\left(-15B + 6Dq - \sqrt{225B^2 - 180DqB + 100Dq^2}\right)$$

$$E(^4T_1(^3P)) = \frac{1}{2}\left(-15B + 6Dq + \sqrt{225B^2 - 180DqB + 100Dq^2}\right)$$

$$E(^4A_2(^2F)) = -12Dq - 15B$$
$$E(^4T_2(^4F)) = -2Dq - 15B$$
$$E(^2A_1(^2G)) = -2Dq - 11B + 3C$$
$$E(^2A_2(^2F)) = -2Dq + 9B + 3C$$

# Appendix D

# Publications

## Refereed publications

- <u>M. Hughes</u>, H. Rutt, D. Hewak, and R. Curry, *Spectroscopy of vanadium (III) doped gallium lanthanum sulphide glass.* Applied Physics Letters, 2007. **90**(3): p. 031108.
- <u>M. Hughes</u>, W. Yang, and D. Hewak, *Fabrication and characterization of femtosecond laser written waveguides in chalcogenide glass.* Applied Physics Letters, 2007. **90**(13): p. 131113.

## Conference presentations, proceedings and other publications

- <u>M. Hughes</u>, D.W. Hewak, and R.J. Curry. *Concentration dependence of the fluorescence decay profile in transition metal doped chalcogenide glass.* in *Photonics West.* 2007. San Jose, USA: SPIE.
- <u>M. Hughes</u>, R.J. Curry, A. Mairaj, J.E. Aronson, W.S. Brocklesby, and D.W. Hewak. *Transition metal doped chalcogenide glasses for broadband near-infrared sources.* in *SPIE Symposium on Optics and Photonics in Security and Defence.* 2004. London.
- R.J. Curry, <u>M. Hughes</u>, J. Aronson, W.S. Brocklesby, and D. Hewak. *Vanadium doped chalcogenide glasses for broadband near-infrared sources.* in *ACerS, Glass & Optical Materials Division Fall Meeting.* 2004. Florida.
- A.K. Mairaj, R.J. Curry, <u>M. Hughes</u>, R. Simpson, K. Knight, and D.W. Hewak. *Towards a compact optical waveguide device for active infrared applications.* in *SPIE Symposium on Optics and Photonics in Security and Defence.* 2004. London.